%% file: global_energy_astroph.tex
\documentclass[preprint]{aastex} 

\newif\ifimage
\imagetrue

\usepackage{graphicx}
\usepackage{natbib}

\defcitealias{Ianj2012}{I2012}

\newcommand{\um}{$\mu$m}
\newcommand{\hi}{\mbox{\rm H\,{\sc i}}}

\newcommand{\m}[1]{\ensuremath{M_\mathrm{#1}}}
\newcommand{\msun}{\ensuremath{\mathrm{M}_{\odot}}}
\newcommand{\mbaryon}{\ensuremath{M_\mathrm{baryon}}}
\newcommand{\mstar}{\ensuremath{M_{\star}}}
\newcommand{ \kms}{km~s$^{-1}$}

\newcommand{\htwo}{\mbox{\rm H$_2$}}
\newcommand{\halpha}{\mbox{\rm H$\alpha$}}
\newcommand{\degrees}{$^{\circ}$}
\newcommand{\gh}{Gauss-Hermite}
\newcommand{\mhi}{\ensuremath{M_{\mathrm{HI}}}}
\newcommand{\sfrsd}{\ensuremath{\Sigma_{\mathrm{SFR}}}}
\newcommand{\mgas}{\ensuremath{M_{\mathrm{gas}}}}

\newcommand{\hisd}{\ensuremath{\Sigma_\mathrm{HI}}}

\newcommand{\sd}[1]{\ensuremath{\Sigma_{[1]}}}
\newcommand{\ave}[1]{\ensuremath{\langle #1 \rangle}}

\newcommand{\per}[2][1]{#2$^{-#1}$}
\newcommand{\sumlim}[2][]{\displaystyle\sum\limits_{#2}^{#1}}

\newcommand{\citeeg}[1]{\citep[e.g.,][]{#1}}

\newcommand{\vdisp}{\ensuremath{ \sigma_{\mathrm{HI}}}}
\newcommand{\vp}{\ensuremath{v_{\mathrm{peak}}}}

\newcommand{\snarrow}{\ensuremath{\sigma_n}}
\newcommand{\sbroad}{\ensuremath{\sigma_b}}
\newcommand{\anab}{\ensuremath{\mathrm{A}_n / \mathrm{A}_b}}

\newcommand{\scentral }{\ensuremath{\sigma_{\mathrm{central}}}}

\newcommand{\swing}{\ensuremath{ \sigma_{\mathrm{wings}}}}
\newcommand{\swingsq}{\ensuremath{ \sigma^2_{\mathrm{wings}}}}
\newcommand{\fw}{\ensuremath{f_{\mathrm{wings}}}}

\newcommand{\aw}{\ensuremath{a}}

\shorttitle{Global \hi{} Kinematics in Dwarf Galaxies}

\begin{document}

\title{Global \hi{} Kinematics in Dwarf Galaxies}

\author{Adrienne M. Stilp\altaffilmark{1},
Julianne J. Dalcanton\altaffilmark{1},
Steven R. Warren\altaffilmark{2,3},
Evan Skillman\altaffilmark{2},
J\"{u}rgen Ott\altaffilmark{4},
B\"{a}rbel Koribalski\altaffilmark{5}}

\altaffiltext{1}{Department of Astronomy, University of Washington, Box 351580, Seattle, WA 98195,
USA}

\altaffiltext{2}{Minnesota Institute for Astrophysics, University of Minnesota, 116 Church St. SE,
Minneapolis, MN 55455, USA}

\altaffiltext{3}{Department of Astronomy, University of Maryland, CSS Bldg., Rm. 1024, Stadium Dr., College Park, MD 20742-2421}

\altaffiltext{4}{National Radio Astronomy Observatory, P.O. Box O, 1003
  Lopezville Road, Socorro, NM 87801, USA}

\altaffiltext{5}{Australia Telescope National Facility, CSIRO Astronomy and Space Science, PO Box 76,
Epping NSW 1710, Australia}

\begin{abstract}

\hi{} line widths are typically interpreted as a measure of ISM turbulence, which is potentially driven by star formation.
In an effort to better understand the possible connections between line widths and star formation, we have characterized \hi{} kinematics in a sample of nearby dwarf galaxies by co-adding line-of-sight spectra after removing the rotational velocity to produce an average, global \hi{} line profile. 
These ``superprofiles'' are composed of a central narrow peak ($\sim6-10$ \kms{}) with higher-velocity wings to either side that contain $\sim 10-15$\% of the total flux.
The superprofiles are all very similar, indicating a universal global \hi{} profile for dwarf galaxies. 
We compare characteristics of the superprofile to various galaxy properties, such as mass and measures of star formation (SF), with the assumption that the superprofile represents a turbulent peak with energetic wings to either side. 
We use these quantities to derive average scale heights for the sample galaxies.
When comparing to physical properties, we find that the velocity dispersion of the central peak is correlated with \ave{\hisd{}}.
The fraction of mass and characteristic velocity of the high velocity wings are correlated with measures of SF, consistent with the picture that SF drives surrounding \hi{} to higher velocities.
While gravitational instabilities provide too little energy, the SF in the sample galaxies does provide enough energy through supernova, with realistic estimates of the coupling efficiency, to produce the observed superprofiles.
\end{abstract}


%%%%%%
%%%%%% INTRODUCTION
%%%%%%
\section{Introduction}
\label{intro}

The neutral hydrogen (\hi{}) component of the interstellar medium (ISM) is an ideal tracer of kinematics in disk galaxies.
\hi{} is observable in some galaxy disks far beyond optical emission, making it a superb tool for probing large-scale kinematics speeds. 
On smaller scales, the \hi{} velocity dispersions offer a way to measure the random turbulence velocities on scales of $10 - 200$ pc. 
By connecting the \hi{} velocity dispersion with possible drivers of turbulence, we can study the sources of energy in the ISM.

\hi{} velocity dispersions typically vary between $5 - 15$ \kms{} across a wide range of disk galaxy types, and generally decrease in the outskirts of galaxies to $\sim6 - 10$ \kms{} \citeeg{Tamburro2009}. 
Generally, these line widths are thought to be due to turbulent velocities rather than thermal broadening. 
For example, \citet{Wolfire1995} found two stable temperatures for \hi{} gas: $\sim 150$ K for the cold neutral medium (CNM) and $\sim 7000$ K for the warm neutral medium (WNM). 
These temperatures correspond to velocity dispersions of $\sim 1$ \kms{} and $\sim 7$ \kms{} at typical ISM pressures, which is often smaller than the observed line widths in nearby galaxies. 
This mismatch suggests that the line widths are set primarily by turbulence. 
However, the time scale for dissipating turbulent energy is $\sim 10^7$ yr \citep{MacLow1999}. 
Energy must therefore be continually injected in order to maintain the \hi{} line widths we see in galaxies.

The sources of energy that drive turbulence are still debated.
A number of studies have suggested that star formation can provide the necessary energy to generate \hi{} turbulence in the inner regions of galaxies \citeeg{Kim1998, Tamburro2009, Joung2009}. 
However, \hi{} velocity dispersions are still substantial at large radii, whereas the majority of star formation in galaxies is contained within $r_{25}$, defined as the radius where the B-band surface brightness drops below 25 mag \per[2]{arcsec} \citeeg{Kennicutt1989, Bigiel2010}. 
Beyond $r_{25}$, the star formation rate (SFR) falls off much more quickly than the \hi{} velocity dispersion, implying that it cannot be the only contribution to the \hi{} velocity dispersion in disk galaxies.

Other proposed drivers of turbulence are the magneto-rotational instability \citep[MRI;][]{Sellwood1999}, shear from rotation curves \citeeg{Schaye2004}, or gravitational instabilities \citeeg{Wada2002}.  
One can potentially distinguish among these various mechanisms by comparing the observed energy in turbulence to the energy available from the possible drivers. 
Many of these processes should be effective in spiral galaxies, which have spiral arms and exhibit differential rotation, but should be less strong in their lower mass dwarf counterparts, which lack spiral structure and show solid-body rotation. 
However, the observed \hi{} velocity dispersions of spirals are surprisingly similar to those of dwarfs.

In this paper we take a different approach to study the global behavior of the \hi{} velocity dispersion and its relationship to possible drivers of turbulence. 
By working on global scales, we can not only increase the signal-to-noise of individual spatially-resolved line-of-sight spectra but also remove the assumption that input energy must necessarily couple to \hi{} in the same spatial region. 
Our work extends the many previous studies of turbulence in the ISM but uses better data over a wider baseline in galaxy mass.
We also improve the characterization of the average velocity dispersion.
In contrast, most earlier papers typically use the intensity-weighted second velocity moment as a proxy for intrinsic \hi{} velocity dispersion \citeeg{Tamburro2009}, fit single Gaussians to line-of-sight spectra \citeeg{Dickey1990, Petric2007}, or both \citeeg{vanZee1999}.
However, the second moment can be artificially increased by gas with anomalous velocities, such as bulk inward or outward flows or expanding \hi{}, while single Gaussian fits are unable to represent asymmetric line-of-sight spectra.
In addition, most literature studies of \hi{} turbulence have not used a uniform sample of observations, as few such samples have been available until recently.
Instead, studies focused on a single galaxies \citeeg{Petric2007} or worked at the instrumental resolution for each galaxy \citeeg{Tamburro2009}.
The combination of such studies means that \hi{} turbulence is sampled on different physical scales in each galaxy. Since turbulence is larger on larger physical scales \citeeg{Zhang2012}, this mismatch in spatial resolution makes galaxy-to-galaxy comparisons dubious.

A better, uniform measurement of the typical underlying \hi{} turbulence is necessary to accurately constrain the detailed kinematics of the ISM.
By co-adding \hi{} line-of-sight spectra after removal of the rotational velocity, we can obtain an average measurement of \hi{} turbulent velocities.
A small number of previous studies have followed a similar approach as we undertake here. 
\citet{Dickey1990} found relatively constant Gaussian line widths in the face-on spiral NGC~1058, with median profiles at some radii exhibiting wings larger than expected from a simple Gaussian profile. These high-velocity wings were shown to exist in the average line profiles regardless of the average FWHM of the contributing line-of-sight profiles \citep{Petric2007}. Similar results were found by \citet{Boulanger1992} and \citet{Kamphuis1993} in NGC 6946. 
\citet{Braun1997} also found \hi{} gas at higher velocities compared to the average central \hi{} line width in a number of other spirals by studying average \hi{} line profile shapes.  However, the existence of high-velocity wings superimposed on a Gaussian center may not be ubiquitous; no evidence of such wings is seen in NGC~5457 \citep{Rownd1994} or the outer regions of NGC~1232 \citep{vanZee1999}. Unfortunately, the majority of the studies of average \hi{} line profiles had poor velocity resolution \citep[$\geq 5.2$ \kms{}; ][]{Braun1997}, coarse spatial resolution \citep[$\gtrsim 1$~kpc; ][]{Dickey1990, Petric2007, vanZee1999}, or both \citep{Boulanger1992, Rownd1994}. 
A recent study by \citet[hereafter I2012]{Ianj2012}, generated average \hi{} profiles (``superprofiles'') using a similar approach as this paper, but for a number of more massive spirals within $D \sim 10$ Mpc. 
They found the same basic line profile structure as we see here, and proposed that they may be comprised of emission from the CNM and WNM.

Recently, a number of \hi{} synthesis observation surveys of nearby galaxies have greatly improved the available data.
Compared with the numerous published \hi{} studies of single galaxies, these surveys can provide better spatial and spectral resolution as well as a uniform observing setup.
The \hi{} Nearby Galaxy Survey \cite[THINGS;][]{Walter2008} pioneered this new era of high-resolution \hi{} surveys by observing 34 nearby spiral galaxies with high spatial ($6 - 10$\arcsec) and spectral ($1.3 - 5.2$ \kms{}) resolution.
The Very Large Array ACS Nearby Galaxy Survey Treasury Project \cite[VLA-ANGST;][]{Ott2012} followed in its footsteps, extending THINGS to smaller galaxy masses at a similar sensitivity and better spectral resolution ($0.6 - 2.6$ \kms{}).
Other surveys, such as FIGGS \citep[``Faint Irregular Galaxy GMRT Survey'';][]{Begum2008} and LITTLE THINGS \citep[``Local Irregulars That Trace Luminosity Extremes-THINGS'';][]{Hunter2012} have also sought to provide a uniform sample of \hi{} observations of dwarfs with similar observing setups.

In this paper we focus on the global measurements of \hi{} kinematics in a wide range of dwarf galaxies chosen from VLA-ANGST and THINGS.
We present a method to measure the intrinsic \hi{} kinematics in low-mass disk galaxies by co-adding flux-weighted \hi{} line profiles after removal of the rotational velocity.
The combined sample  covers a wide range of galaxy properties, allowing us to examine the overall \hi{} kinematics in a broader range of environments than previously studied. 
In \S~\ref{sec:data} we describe the data we use to determine galaxy global properties.
We next explain our method of characterizing the global \hi{} gas kinematics in \S~\ref{sec:superprofiles}.
We then discuss our parameterization and the physical interpretation of these superprofiles in \S~\ref{sec:analysis}.
In \S~\ref{sec:correlations}, we investigate significant correlations between superprofile parameters and galaxy physical properties. 
In \S~\ref{sec:disc} we discuss the potential physical cause behind correlations with each parameter;
give energy estimates for driving kinematics in the different components;
assess limits on \hi{} scale heights; and
examine the possibility of a universal \hi{} velocity profile for dwarfs. 
Finally, we summarize our results in \S~\ref{sec:conclusions} .

%%%%%%
%%%%%% DATA
%%%%%%
\section{Sample and Data}
\label{sec:data}

In the following sections, we briefly describe the \hi{} data available from VLA-ANGST and THINGS as well as our sample selection criteria. 
We then discuss our conversion from the data to the physical properties of our sample galaxies. 
If available, we use the ANGST TRGB distance from \citet{Dalcanton2009}. 
Otherwise, we use the distances compiled in \citet{Karachentsev2004}. 
When necessary, we correct all published quantities to our adopted distances, and we include published or estimated distance uncertainties in the uncertainties for all our calculated quantities. 
In \S~\ref{sec:sample--selection} we discuss the criteria that we use to select the galaxies for our sample.

The 9 galaxies from THINGS and 14 galaxies from VLA-ANGST in our sample are listed in Table~\ref{tab:sample} in order of decreasing baryonic mass along with their basic physical properties. We list
(1) the galaxy name;
(2) alternate names;
(3) \hi{} survey (VLA-ANGST or THINGS);
(4-5) right ascension and declination;
(6) distance;
(7) inclination $i$, (see \S~\ref{sec:data--inclination});
(8) total \hi{} mass taken from \citet{Walter2008} or \citet{Ott2012};
(9) $r_{25}$, from \citet{Karachentsev2004};
(10) inclination-corrected width at 20\% of the total line profile, $w_{20}$, from \citet{Walter2008} or \citet{Ott2012} and corrected using $i$ as listed in this table;
and
(11) de Vaucouleurs T-type taken from \citet{Walter2008} or \citet{Ott2012}. If necessary, previously-published quantities such as mass are corrected for the adopted distance.

\subsection{HI Data}
\label{sec:sample--hi-data}

We use a combination of NRAO Very Large Array (VLA) \hi{} data from both THINGS and VLA-ANGST for our analysis. 
The two surveys provide complementary information, as THINGS targets are primarily large spiral galaxies while VLA-ANGST probes gas-rich galaxies at lower mass scales. 
Our final sample is composed primarily of dwarfs that span a wide range of galaxy properties, including absolute magnitude; stellar, gas, and baryonic masses; SFRs; and rotation speeds.

We use the robust-weighted data cubes published in \cite{Ott2012} and \cite{Walter2008} from VLA-ANGST and THINGS, respectively.
This weighting scheme offers $\sim$30\% better spatial resolution and a well-behaved synthesized beam with only a moderate decrease in sensitivity.
Both VLA-ANGST and THINGS provide two sets of robust-weighted data cubes, the standard and flux-rescaled cubes.
Standard data cubes have uniform noise properties but incorrect fluxes, while the flux-rescaled cube has been scaled to have correct fluxes in exchange for more complicated noise properties.
Because we use both cubes in our analysis, we specify which cube we are using in each step.
To ensure that we are sampling the same physical scales in the galaxies' ISM, we work at a common physical resolution of 200 pc as discussed in \S \ref{sec:sample--hi-data-prep}.

The parameters of the \hi{} observations are listed in Table~\ref{tab:hi-obs}. We list
(1) the galaxy name;
(2) velocity resolution, $\Delta v$;
(3) FWHM beam corresponding to 200~pc physical resolution, $\theta_\mathrm{200pc}$; 
and
(4) \emph{rms} noise in a single channel of the 200 pc convolved standard cube, $\sigma_\mathrm{chan}$.

\subsubsection{VLA-ANGST}
\label{sec:sample--hi-data--vla-angst}

The ACS Nearby Galaxy Survey Treasury (ANGST) Program obtained multi-color HST photometry of a volume limited sample of galaxies within $4$ Mpc, excluding the Local Group, and provides spatially-resolved star formation histories for its sample \citep{Dalcanton2009}.
As a followup, the VLA-ANGST survey targeted all galaxies in ANGST that were visible with the VLA ($\delta \gtrsim -30$\degrees), showed signs of having observable \hi{} reservoirs, and lacked adequate previous \hi{} observations. 
Many of the galaxies in the sample of 35 are therefore low-mass, low-luminosity dwarfs.

The velocity resolution of the survey is $0.6 - 2.6$ \kms{}, which is necessary to study the detailed \hi{} line profiles in galaxies with the low peak rotation speeds characteristic of the sample \citeeg{Warren2012}. 
The typical instrumental spatial resolutions at the median galaxy distance is $\sim 7$\arcsec{}, corresponding to $\sim100$ pc at the median distance of 2.8 Mpc.

\subsubsection{THINGS}
\label{sec:sample--hi-data--things}

THINGS provides a complementary sample of 34 large, gas-rich spirals chosen mainly from the \emph{Spitzer} Infrared Nearby Galaxy Survey (SINGS). 
Since THINGS galaxies have characteristically higher masses and rotation speeds, coarser velocity resolutions were often required to fully cover the \hi{} emission. 
Therefore, the velocity resolution is often coarser than VLA-ANGST, and the majority of THINGS observations have velocity resolutions of either 2.6 or 5.2 \kms{}. 
THINGS resolves spatial scales of $\sim7$\arcsec{}, corresponding to $\sim200$~pc at the median distance of $\sim6$ Mpc.

\subsubsection{Sample Selection}
\label{sec:sample--selection}

To increase the robustness of our results, we select a high-quality subset of the 63 detected galaxies in VLA-ANGST and THINGS surveys for our analysis. 
We consider only disk-dominated galaxies (de Vaucouleurs T-type $> 3$) to avoid confusion with the bulge; all selected galaxies are $7 \leq \mathrm{T} \leq 10$.

We further exclude galaxies that suffer from one or more of the following problems, with the number of galaxies eliminated due to each criterion given in parentheses.

\begin{enumerate}

\item Instrumental physical resolution larger than our working resolution of 200~pc (18 galaxies). %n3521, n3351, n3198, n3621, n4449, n3627, n5055, n7331, n2841, n3184, n5194, n628, n4826, n925, n5236, n5457, n2903, mcg9

\item Velocity resolution $\Delta v \geq 5.2$ \kms{}, which complicates determination of the peak velocity and approaches the width of turbulent regions in these galaxies (16 galaxies). % n3521, n4736, n3351, n3198, n3621, n4449, n2976, n3627, n5055, n7331, n2841, n2403, n5194, n4826, n5457, n2903

\item Inclinations $> 70$\degrees{}, which could lead to artificially broadened line profiles due to beam smearing (7 galaxies). % n3521, n3198, n247, n7731, n2841, n3109, ddo183

\item Noticeable contamination from the Milky Way or from a companion, which would hinder separation of the galaxy \hi{} emission from its companion (8 galaxies).%n1569, n6946, n2976, n3077, bk3n, a0925, n3031, n404

\item Fewer than 10 independent beams above the signal-to-noise threshold where we can accurately measure \vp{} ($\mathrm{S/N} > 5$; see \S~\ref{sec:superprofiles--signal-to-noise}) at our working resolution. Galaxies with fewer independent beams show very noisy co-added profiles and have more than $50$\% of \hi{} flux in pixels below our S/N threshold (8 galaxies). % antlia, kk230, kkh86, ddo6, ddo82, kkh98, kdg73, m81dwa

\item A lack of ancillary far-ultraviolet (FUV) imaging, needed to determine the current average star formation rate (10 galaxies). % n1569, n2841, n3077, n3184, n3198, n3621, n6946, n7731, n925, u8508

% \item Kinematically disturbed galaxies, such as those with counter-rotating inner disks.

\end{enumerate}

These cuts eliminate 40 potential galaxies from our sample; the majority of these were cut because they failed item (1) or (2) of the above criteria, and a number of galaxies failed more than one of the criteria. 
The final sample for analysis has 23 galaxies, with 14 from VLA-ANGST and 9 from THINGS.

\subsection{\hi{} Data Preparation}
\label{sec:sample--hi-data-prep}

To provide the best galaxy-to-galaxy comparison of \hi{} kinematics, we must take the spatial resolution into account. 
Since the velocity dispersion is typically larger on larger spatial scales \citeeg{Zhang2012}, we must ensure that we are sampling the \hi{} kinematics on the same spatial scale in our sample galaxies. 
Since the instrumental angular resolution is roughly the same but the galaxies are at different distances, we must apply spatial smoothing to some of the cubes to ensure that the same spatial resolution is sampled from galaxy to galaxy. 
This is essentially equivalent to placing all galaxies at the same distance.

We choose a spatial resolution of 200~pc to match that used by \citet{Warren2012}, allowing for future comparison of our results. 
This resolution is also a good compromise between potential sample size and physical resolution at the distance of each galaxy, which is limited primarily by galaxy distance. 
It also matches results from \citet{Joung2009}, who find that most turbulence is contained on spatial scales of 200 pc or less.

To apply the spatial smoothing to the data cubes, we first calculate the beam size that yields 200~pc resolution at each galaxy's distance. 
We produce spatially-smoothed versions of both the standard and flux-rescaled data cubes at 200~pc resolution using the AIPS task \textsc{convl}, which accounts for the original beam major axis, minor axis, and position angle.

To generate a mask for these 200~pc data cubes, we first convolve the original standard cube to 45\arcsec{} resolution (\textsc{convl}).
Next, we measure $\sigma_\mathrm{chan,45\arcsec}$, the \emph{rms} noise in the 45\arcsec{} cube, and mask all emission below $3\sigma_\mathrm{chan, 45\arcsec}$ using the AIPS task \textsc{blank}. 
Finally, we remove any remaining non-emission regions by hand. 
To regenerate the moment maps for the 200~pc resolution data, we blank the convolved, flux-rescaled data cube in regions outside of the mask. 
We then use the AIPS task \textsc{xmom} to produce zeroth, first, and second moment maps. 
For the remainder of the paper, all mention of data cubes or moment maps refer to the 200~pc data sets described in this section, unless otherwise specified.

We note that convolution to a circular beam means that inclined galaxies have slightly larger physical resolution along their minor axis than along their major axis. 
However, the uncertainty of inclinations and position angles for the majority of galaxies in our sample makes it difficult to correct. 
Therefore, we choose to use the simplest option of a circular beam.

Finally, we include in our final analysis only line-of-sight spectra with a signal-to-noise (S/N) $ > 5$, where S/N is defined as the ratio between our fits to the peak divided by the \emph{rms} noise in the line-free channels of the data cube. 
The reasons for this choice are discussed further in \S~\ref{sec:superprofiles--signal-to-noise}.

\subsection{Converting Data to Physical Quantities}
\label{sec:sample--other-data}

In this section we describe the methods we use to measure the physical quantities discussed in the paper so that we are able to compare them to the \hi{} superprofile properties of the sample.
For many of the quantities, such as the star formation rate (SFR) or the \hi{} surface density $\Sigma_\mathrm{HI}$, we calculate the global properties using only the pixels whose \hi{} line-of-sight spectra contribute to the superprofile (S/N$ > 5$).
This choice provides a matched aperture measurement that allows us to consider only regions that are able to directly affect the \hi{} measured by the superprofiles.
A notable exception is the total baryonic mass of the galaxy, \m{baryon,tot}, which we use as a proxy for halo mass (\S~\ref{sec:data--other-data--mhalo}).
In this case, including only pixels above the S/N threshold would artificially underestimate the total baryonic mass, and therefore the halo mass, of galaxies in the low signal-to-noise regime.
The halo mass is not expected to directly influence \hi{} velocity dispersions, but it is useful to first characterize how the \hi{} superprofile properties behave as a function of total halo mass before exploring their connection with other physical properties.

We list our derived quantities in Table~\ref{tab:derived-properties}. We give these quantities, followed by the relevant section:
(1) the galaxy name;
(2) \m{baryon,tot}, \S~\ref{sec:data--other-data--mhalo};
(3) \hi{} mass, \mhi{}, \S~\ref{sec:data--other-data--mhi};
(4) stellar mass, \mstar{}, \S~\ref{sec:data--other-data--mstar}; 
(5) SFR, \S~\ref{sec:data--other-data--sfr};
(6) SFR / \mhi{}, \S~\ref{sec:data--other-data--sfr-per-mhi};
(7) average star formation rate surface density, \ave{\sfrsd{}}, \S~\ref{sec:data--other-data--sfr}; 
and
(8) \ave{\hisd{}}, \S~\ref{sec:data--other-data--mhi}.
We measure the quantities listed in columns $3 - 7$ using only pixels above our S/N threshold.

\subsubsection{Galaxy Inclination}
\label{sec:data--inclination}

%The inclinations of dwarf galaxies are very uncertain, due to their irregular shapes and velocity fields. 
%For disk galaxies, the best inclination indicator is usually rotation curves inferred from tilted ring models fit to the \hi{} velocity field. 
For disk galaxies, the best inclination is usually the one inferred from tilted ring model fits to the \hi{} velocity field.
Inclinations derived this way are available for 12 galaxies in our sample from a variety of sources in the literature \citeeg{Skillman1988, Begum2005, deBlok2008, Swaters2009, Oh2011}; we use these inclinations if available. 
We note that for strictly solid body rotation, as is common in dwarfs, the velocity field does not contain any information about the inclination angle of the disk.

%However, 11 galaxies in our sample do not have inclinations derived from rotation curve analysis. 
However, 11 galaxies in our sample do not have previously-derived rotation curves and are therefore lacking these inclination estimates.
In the absence of velocity field analysis, the traditional method is to measure ellipticity from B-band observations and then to derive an inclination after assuming an intrinsic disk thickness. 
This method often fails dramatically in dwarf galaxies, since it relies on the assumption that the intrinsic disk structure is well-traced by the B-band surface brightness. 
Unfortunately, SFRs in dwarf galaxies are much lower than in spiral galaxies, such that the massive stars that dominate the B-band surface brightness are formed stochastically across the disk. 
Since their light does not smoothly trace the galactic disk, we must turn to another indicator to measure the projected galactic disk.

As opposed to B-band observations, near-infrared observations are dominated by flux from the older stars and should therefore provide a better measurement of the projected shape of a galaxy's disk. 
All sample galaxies without inclinations derived from rotation curves are part of the Local Volume Legacy survey \citep[LVL;][]{Dale2009}, which provides photometric infrared observations of galaxies within $11$ Mpc. 
For these galaxies, we fit ellipses to isophotes in the the LVL 3.6\um{} Spitzer images. We then calculate inclination by assuming:
\begin{equation}
\sin^2 i = \frac{ 1 - (b / a)^2  } {  1 - q_0^2   },
\end{equation}
where $(b/a)$ is the measured axial ratio and $q_0$ is the intrinsic disk thickness. 
We use the values for $q_0$ provided in \citet{Karachentsev2004} for different galaxy types, which range between $0.12 - 0.2$. 
The derived inclinations only change by $\lesssim 5$\degrees{} when we increase $q_0$ to a fixed value of 0.3. 
We estimate an uncertainty $\Delta b/a \sim 0.05$ from repeated measurements of ``best-fit by eye'' ellipses to the 3.6\um{} surface brightness distribution, which leads to uncertainties in the inclination of $\sim 5$\degrees{}. 
When we compare inclinations derived using this method to those of galaxies with tilted ring inclinations, we find that the inclinations typically differ by less than 10\degrees{}. 
The inclinations derived using \hi{} morphology \citep{Begum2008} for the five galaxies that overlap both samples are within 5\degrees{}, with the exception of DDO~187.
We therefore estimate our total uncertainty on the inclination as $\sigma_i \sim 10$\degrees{}.

We denote the galaxies whose inclinations have been derived using this method with $^\ast$ in column 7 of Table~\ref{tab:sample}.

\subsubsection{Halo Mass}
\label{sec:data--other-data--mhalo}

In large spiral galaxies, the inclination-corrected widths of the \hi{} integrated line profile at 20\% or 50\% of the peak ($w_{20}$ and $w_{50}$) are good tracers of the total halo mass, as the rotation curve flattens to approximately the circular velocity \citeeg{Verheijen2001}. 
In low-mass dwarf galaxies, however, rotation curves often continue rising past the extent of the observable \hi{}, so any measured velocity provides only a lower limit on the circular velocity of the halo. 
Second, the global profiles are generally Gaussian and, due to the small rotational velocities, are more affected by turbulent motions \citep{Begum2006}. 
The ability to derive the intrinsic velocity width also requires an accurate knowledge of the galaxy inclination, which is uncertain for many of the low-mass dwarfs in our sample.

However, detailed studies of the baryonic Tully-Fisher relation in dwarf galaxies \citeeg{Geha2006, Stark2009} indicate a strong correlation between baryonic mass, as measured with gas and stars, and halo mass, as measured with either $w_{20}$ or $w_{50}$ when inclinations are well-known.
This correlation is expected if all halos in our sample have approximately the same baryon fraction. % \citeeg{Crain2007}. 
In the absence of reliable inclination-corrected $w_{20}$ measurements, however, it is preferable to use the baryonic mass as a proxy for halo mass in dwarf galaxies. We therefore use \m{baryon,tot} to indirectly measure halo mass in our sample.

Unlike the other measurements described in this section, we use the total baryonic mass instead of only the mass contained in the same pixels as we use to derive the superprofiles.
Because this measurement is simply a proxy for halo mass and does not have a direct causal connection to the \hi{} properties, we include the entire baryonic mass for each galaxy.
Including only pixels above our S/N threshold would underestimate the baryonic mass, and therefore the halo mass, of galaxies in the low signal-to-noise regime.

We calculate $\m{baryon, tot} = 1.36 \m{HI, tot} + \m{\star, tot}$, using listed values for \m{HI, tot} from \citet{Walter2008} and \citet{Ott2012}; and $L_\mathrm{3.6}$ values from \citet{Dale2009} and then apply the $L_\mathrm{3.6} \rightarrow \m{\star}$ conversion in \S \ref{sec:data--other-data--mstar}.
The factor of 1.36 accounts for helium; we neglect metals as the low metallicities of our sample imply that $< 1$\% of the gas is composed of metals (see \S~\ref{sec:data--other-data--mgas}).
In most cases, \m{baryon, tot} is higher than the aperture-matched measurement of \m{baryon}, primarily due to \hi{} in low S/N pixels that do not contribute to the superprofiles.

\subsubsection{\hi{} Mass}
\label{sec:data--other-data--mhi}

We calculate the average \hi{} surface density, \ave{\hisd{}}, and \hi{} mass, \m{HI}, from the convolved \hi{} total intensity maps. We first convert this map to a de-projected surface density in M$_{\odot}$ \per[2]{pc}:
\begin{equation}
\Sigma_\mathrm{HI} \left( \mathrm{M}_{\odot} \; \mathrm{pc}^{-2} \right) = 12.14 \, \frac{ S \Delta v}{\mathrm{FWHM}_\mathrm{maj} \; \mathrm{FWHM}_{\mathrm{min}} } \, \cos{i}
\end{equation}
where $S \Delta v$ is the \hi{} surface brightness in Jy \per{beam} \kms{}, FWHM$_\mathrm{maj}$ and FWHM$_\mathrm{min}$ are the beam major and minor axes in arcsec, and $i$ is the inclination.

To calculate the average $\langle \Sigma_{\mathrm{HI}} \rangle$ for each galaxy, we average the $\Sigma_{\mathrm{HI}}$ map using all the pixels above our S/N threshold (see \S~\ref{sec:superprofiles--signal-to-noise}). 
We then calculate \mhi{} by summing $\Sigma_\mathrm{HI}$ times the de-projected physical area of each pixel. 
We note that the mass we calculate in this step is less than the total \hi{} mass of each galaxy given in Table~\ref{tab:sample} due to our S/N threshold.

\subsubsection{Gas Mass}
\label{sec:data--other-data--mgas}

We calculate the gas mass by assuming $\m{gas} = 1.36 \, \mhi$. 
We include the factor of 1.36 to account for the presence of helium. 
As most of our galaxies are dwarfs with currently-undetectable \htwo{}, we neglect possible contributions from molecular gas. 
Only two galaxies in our sample have detected \htwo{}, and in both cases the molecular gas contributes $< 10$\% to the gas mass. 
We also do not correct for metals, since the metallicities of our sample are likely to be low; the 4.5\um{} luminosity-metallicity relations given in \citet{Berg2012} imply metallicities of $12 + \log{ (\mathrm{O / H})} < 8$ for all but two galaxies (NGC 4214 and NGC 7793). 
Corrections for the masses of heavy elements in the ISM are therefore less than 1\% for all of our galaxies, and thus much less than the uncertainties in the gas masses.

\subsubsection{FUV + 24\um{} Star Formation Rate}
\label{sec:data--other-data--sfr}

We measure the SFR by combining \emph{GALEX} FUV and \emph{Spitzer} 24\um{} luminosities following the prescription from \citet{Leroy2008}. 
FUV emission primarily traces unobscured star formation that has occurred within the past $\sim10-100$ Myr, while the 24\um{} emission traces warm gas that has been heated by embedded star formation on timescales of $3-10$ Myr \citep{Calzetti2007}. 
Therefore, the combination of FUV with 24\um{} provides a measurement of both the embedded and the unobscured recent star formation.

We generate pixel-by-pixel maps of SFR following the formalism described in \citet{Leroy2008}. 
The empirically-calibrated relationship between FUV emission, 24\um{} emission, and SFR is given by:
\begin{equation}
\sfrsd{} = \left( 8.1 \times 10^{-2} I_\mathrm{FUV} + 3.2 \times 10^{-3} I_{24} \right) \cos i,
\label{eqn:sfr}
\end{equation}
where \sfrsd{} is in \msun{} \per[2]{kpc} \per{yr} and the FUV and 24\um{} intensities are in MJy \per{ster}. 
The conversion assumes a \cite{Kroupa2001} IMF with a maximum mass of 120 \msun{} as implemented in STARBURST99 \citep{Leitherer1999}. 
The measured SFRs are smaller by a factor of 1.59 compared to a \citet{Salpeter1955} IMF with a mass range of $0.1 - 100$ \msun{} when normalized for the same number of ionizing photons. 
At $\sfrsd{} < 10^{-4}$ \msun{} \per{yr} \per[2]{kpc}, the 24\um{} emission is an upper limit to the SFR due to the diffuse dust component of the disk. 
However, only two of our galaxies have average $\langle \sfrsd{} \rangle < 10^{-4}$ \msun{} \per{yr} \per[2]{kpc}, and in these cases, over 90\% of the SFR is determined by the FUV component alone. 
The 24\um{} tracer contributes less than 20\% of the total SFR in all but four galaxies (DDO 53, NGC 2366, NGC 4214, and NGC 7793).

We start with publicly available FUV and 24\um{} images from LVL \citep{Dale2009}. 
These maps have resolutions of $\sim 5$\arcsec{} and $\sim1.6$\arcsec{}, respectively. 
For the FUV images, we subtract a small sky background and correct for Galactic extinction using the dust maps from \citet{Schlegel1998} and $A_\mathrm{FUV} / \mathrm{E(B-V)} = 8.376$. 
We mask foreground stars, identified by pixels that have NUV / FUV flux ratios $> 15$. 
The 24\um{} maps have already been background-subtracted, so no additional correction is applied. 
We then convolve both the 24\um{} and FUV data to our working resolution, place both maps on the astrometric grid defined by the \hi{} data, and apply Equation~\ref{eqn:sfr}.

To calculate the global SFR for each galaxy, we sum the \sfrsd{} contribution from only the pixels above our \hi{} S/N threshold multiplied by the de-projected physical area of each pixel. 
To estimate \ave{\sfrsd{}}, we divide the SFR by the de-projected physical area covered by pixels above our S/N threshold.

We have attempted to calculate \halpha{}-based star formation rates using LVL \halpha{} maps, as they trace star formation on timescales $\sim 10$~ Myr, shorter than the FUV+24\um{} measurement.
However, the \sfrsd{} values implied by \halpha{} observations are often below the $10^{-3}$ \msun{} \per{yr} \per[2]{kpc} level where \halpha{} maps are no longer reliable \citeeg{Leroy2012}.
%Additionally, \halpha{} luminosities underpredict the star formation rate compared to FUV measurements by factors of 2 - 10 at $\mathrm{SFR} \lesssim 0.003$ \msun{} \per{yr}, a regime inhabited by many of our galaxies \citeeg{Lee2011}.
Additionally, at the low SFR typical of our sample, FUV tracers of SFR appear to be more robust than \halpha{} tracers \citeeg{Lee2011, Leroy2012}.
We therefore use the FUV+24\um{} as our only SFR tracer.

\subsubsection{Star Formation Rate per \mhi{}}
\label{sec:data--other-data--sfr-per-mhi}

The star formation rate per unit \mgas{} is often taken to be the star formation efficiency (SFE). 
However, such an interpretation can be problematic in dwarf galaxies. 
The \htwo{} component of the total gas mass is notoriously difficult to measure in dwarfs but must exist if our current understanding of star formation is correct \cite[although see][]{Krumholz2012}.
Using dust as a proxy for \htwo{}, \citet{Bolatto2011} find molecular gas fractions of $\Sigma_\mathrm{H_2} / \Sigma_\mathrm{HI} \sim 0.1$ on 200 pc scales in the SMC. 
However, the typical Kennicutt-Schmidt relation for standard spirals tends to overpredict the star formation rate for a given total gas mass even on a pixel-by-pixel basis \citeeg{Leroy2008, Bolatto2011}.
Such observations could be explained if dwarf galaxies have a fundamentally different SFE, or if the \hi{} surface density in dwarfs is simply less directly connected to star formation than in larger spirals. 
Since the \sfrsd{}-$\Sigma_\mathrm{H_2}$ relation is comparable to that found in larger spirals, the latter is likely the case. 
Therefore, our SFR / \mhi{} measurement more likely traces the ability of recent star formation to affect the \hi{} gas than any true SFE effects.

To calculate the global average $\langle \mathrm{SFR} \, / \, \mhi{} \rangle$, we simply calculate:
\begin{equation}
\langle \mathrm{SFR} / \mhi{} \rangle = \frac{\mathrm{SFR}}{\mhi{}},
\end{equation}
where SFR and \mhi{} are the star formation rate and \hi{} mass from contributing pixels as derived in \S~\ref{sec:data--other-data--sfr} and \S~\ref{sec:data--other-data--mhi}.

\subsubsection{Stellar Mass}
\label{sec:data--other-data--mstar}

We use 3.6\um{} Spitzer data from LVL to estimate the stellar mass, 
using the method in Appendix C of \cite{Leroy2008}. 
This band primarily traces the light from older stellar populations. In more massive galaxies, the 3.6\um{} band can also contain emission from hot dust and polycyclic aromatic hydrocarbons (PAHs). 
However, dwarf galaxies show reduced PAH emission compared to larger galaxies of the same color \citeeg{Hogg2005, Engelbracht2005, Madden2006, Rosenberg2006, Jackson2006}. 
Since our sample is primarily composed of dwarfs, we do not account for PAH emission when converting from 3.6\um{} intensity to stellar mass.

We use the empirically-derived conversion from \citet{Leroy2008}:
\begin{equation}
\Sigma_\star = \Upsilon_{\star,\mathrm{K}} \langle \frac{ I_K }{ I_{3.6} } \rangle I_{3.6} \; \cos{i}  = 280 \; I_{3.6} \, \cos{i} ,
\label{eqn:sigma-star}
\end{equation}
where $\Upsilon_{\star, \mathrm{K}} \sim 0.5$ is the mass-to-light ratio in the K-band, $\Sigma_\star$ is in \msun{} \per[2]{pc}, 
$I_{3.6}$ is the 3.6\um{} intensity in MJy \per{ster}
 and $\langle I_\mathrm{K} / I_{3.6} \rangle \sim 1.81$ is the 3.6\um{}-to-K-band conversion as derived by \citet{Leroy2008}. 
 The conversion assumes a \citet{Kroupa2001} IMF. The mass-to-light ratio has a scatter of $\sim 0.1$ dex. 
 Further discussion of the conversion is given in \citet{Leroy2008}.

We start with the LVL point-subtracted 3.6\um{} maps from \citet{Dale2009}. 
In a few cases we have extended the point-subtracted mask to the outskirts of the galaxy, as the \hi{} covers a larger area than the LVL aperture in which point-subtraction was initially performed. 
We place the maps on the same astrometric grid as the \hi{} data and convolve to our 200 pc resolution.

To calculate the total stellar mass, \m{star}, we sum the contribution from only the $\Sigma_\star$ pixels above our \hi{} S/N threshold and multiply by the de-projected physical area of a single pixel. 
To calculate the global average $\langle \mathrm{\Sigma}_\star \rangle$, we divide \m{\star} by the de-projected area covered by pixels above our S/N threshold.

\subsubsection{Baryonic Mass and Surface Density}
\label{sec:data--other-data--mbaryon}

We combine our $\Sigma_\mathrm{gas}$ maps with the $\Sigma_{\star}$ maps to find the total baryonic surface density, $\Sigma_\mathrm{baryon} = 1.36 \, \Sigma_\mathrm{HI} + \Sigma_{\star}$. 
We calculate the aperture-matched baryonic mass, \m{baryon}, by summing only the pixels above our \hi{} S/N threshold times the de-projected area of a single pixel. 
We note that this mass is typically smaller than the \m{baryon,tot} used as a proxy for halo mass in \S~\ref{sec:data--other-data--mhalo} due to eliminating low S/N pixels in this measurement.
The global average \ave{\Sigma_\mathrm{baryon}} is calculated by dividing \m{baryon} by the de-projected area covered by pixels above our S/N threshold.

\subsubsection{Global \hi{} Second Velocity Moment}
\label{sec:data--other-data--global-m2}

To facilitate comparisons with velocity dispersions in the literature, which often use the second moment map as a proxy, we use the intensity-weighted \hi{} velocity dispersion (second moment) maps to calculate a global second moment for the entire galaxy:
\begin{equation}
  \langle \sigma_\mathrm{m_2} \rangle = \frac{ \sumlim{i,j} \Sigma_{\mathrm{HI},i,j} \; \sigma_{m_2, i,j} }
	  {\sumlim{i,j} \Sigma_{\mathrm{HI}, i,j}}.
\label{eqn:m2-global}
\end{equation}
for every pixel $(i, j)$ above our \hi{} S/N threshold. 
We weight each pixel's second moment value, $\sigma_{m_2, i,j}$, by the \hi{} surface density $\Sigma_{\mathrm{HI},i,j}$. 
We include flux weighting in this calculation because the co-added profiles discussed in \S~\ref{sec:superprofiles} are also flux-weighted. 
It therefore allows for a more meaningful comparison between the velocity dispersions derived from the superprofiles presented in this paper and those derived from the second moment maps as in the literature.

\subsubsection{Correlations between properties}
\label{sec:data--other-data--correlations}

A number of the above physical properties are correlated with each other. 
As discussed further in \S~\ref{sec:correlations}, we use the Spearman rank correlation coefficient, $r_s$, to determine whether two properties are correlated.
This statistic also yields $p_s$, the probability of a random sample having an $r_s$ value that is equal or more extreme than the measured $r_s$ value.
We choose $p_s \leq 0.01$ as a conservative threshold for correlation.

\begin{figure}
  \centering
  \ifimage
    \includegraphics[height=7in]{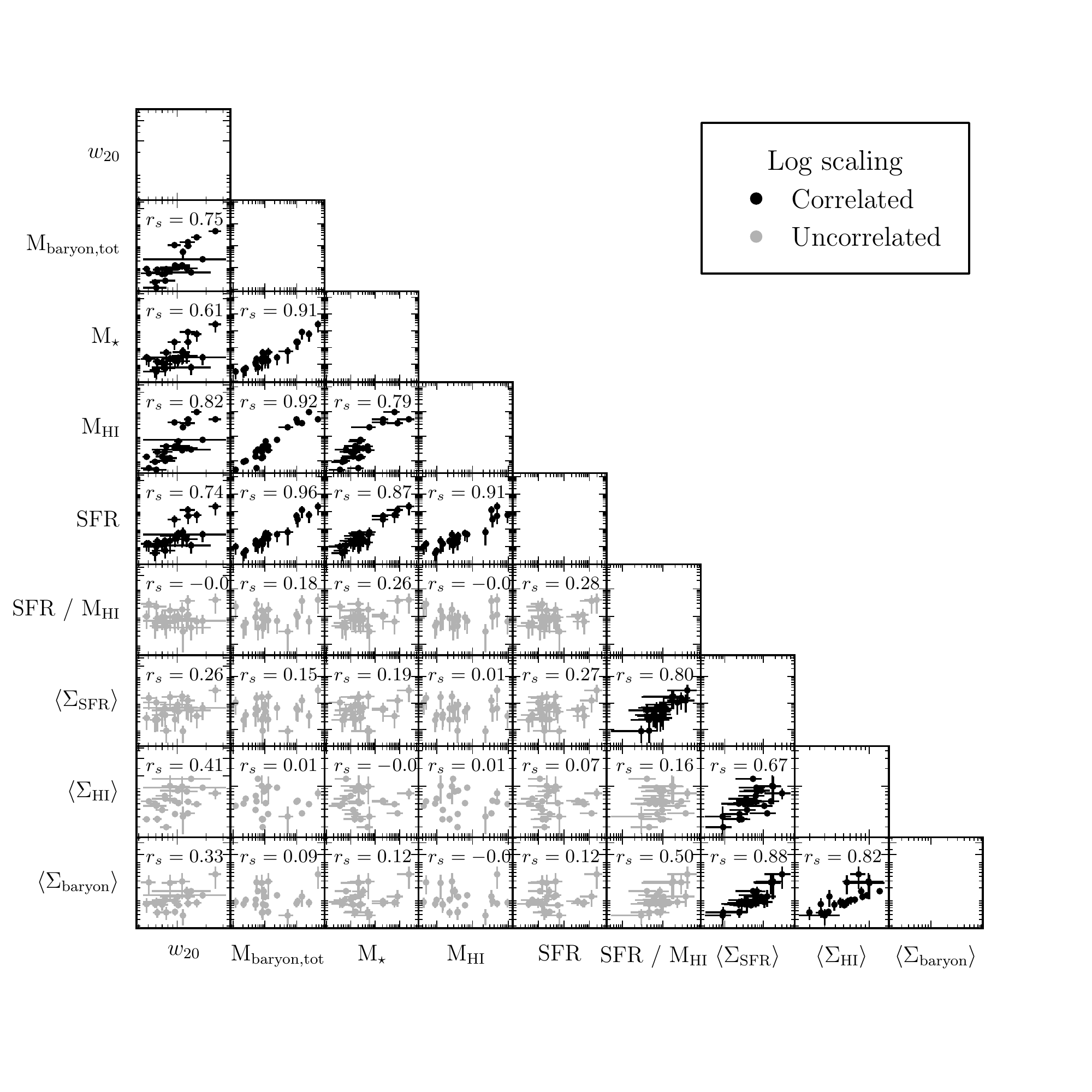}%{{figures/phys_prop.correlations}.pdf}
  \fi
  \caption{Correlations between the globally-averaged physical properties for our sample. Each panel shows the correlation between two different properties. Within each panel, each point represents the globally-averaged value for a single galaxy. Points are colored black if the two properties are significantly correlated ($p_s \leq 0.01$, or $r_S \gtrsim 0.53$) or grey if they are uncorrelated. %Correlation coefficients are listed in Table~\ref{tab:phys-correlations}.
\label{fig:property-correlations} }
\end{figure}

In Figure~\ref{fig:property-correlations}, we show the correlations between many of the above physical properties. 
Within one panel, each point represents the globally-averaged properties for a single galaxy. 
These points are then colored either black if the two properties are significantly correlated or grey if they are uncorrelated. 
The correlation coefficient, $r_s$, is shown in each panel.
It is immediately clear that many of the mass tracers (\mhi{}, \mstar{}, \mbaryon{}, $w_{20}$) are correlated, as expected. 
The total SFR is also strongly correlated with mass for the reason that more massive disk galaxies simply tend to have more material for star formation.
We also see correlations among many of the more local surface density quantities.
The correlation between \ave{\sfrsd{}} and \ave{\hisd{}} is expected due to the Kennicutt-Schmidt relation, although this relation begins to break down in dwarf galaxies for gas masses measured only using \hi{} \citeeg{Bolatto2011}.
The baryonic surface density \ave{\Sigma_\mathrm{baryon}} also correlates with these quantities; given that the galaxies are gas rich with a median gas fraction of $f_\mathrm{gas} = 0.74$, \ave{\hisd{}} makes a large contribution to \ave{\Sigma_\mathrm{baryon}}.

%%%%%%
%%%%%% ANALYSIS
%%%%%%
\section{Global \hi{} Superprofiles}
\label{sec:superprofiles}

The global properties of the \hi{} velocity dispersion are not necessarily well-characterized by \hi{} second moment maps, as the second moment can be artificially increased by bulk motions of small amounts of gas at anomalous velocities. 
Instead, we co-add individual line-of-sight profiles after removal of the rotational velocity. 
This method produces an average, high S/N \hi{} line spectrum, which allows us to characterize the average velocity structure of the ISM.

The basic outline of the procedure is as follows. 
We first measure the rotational line-of-sight velocity from each profile using the standard, 200 pc resolution data cube (\S~\ref{sec:superprofiles--vpeak}). 
After applying a S/N cut (\S~\ref{sec:superprofiles--signal-to-noise}), we recenter each line-of-sight profile in the flux-rescaled data cube so that the peak is at zero and then sum all recentered line-of-sight profiles into a single, global superprofile (\S~\ref{sec:superprofiles--generation}). %Finally, we give an overview of the superprofiles  (\textbf{say this differently}) (\S~\ref{sec:superprofiles--overview}).

\subsection{Determining the Peak Velocity}
\label{sec:superprofiles--vpeak}

To calculate the \hi{} superprofiles, we must first find the velocity by which to shift each line-of-sight spectrum in the data cube.
For undisturbed, idealized \hi{} line-of-sight profiles, this velocity is simply the velocity where the spectrum reaches its maximum.
However, non-circular motions and instrumental effects can influence the location of the peak.
Initially, \citet{Braun1997} determined this velocity by simply finding the velocity at which the line profile reached its maximum.
In the comparatively higher S/N regime of Braun's data, the peak is unlikely to be strongly affected by noise. The peak position is also affected by the velocity resolution, such that observations with coarse velocity resolution cannot be used to identify the peak to better than the velocity resolution.
 In lower S/N spectra with high velocity resolution, however, the peak of observed profiles can be artificially shifted either to neighboring channels or even to a completely arbitrary value by noise spikes.
 Other median line profile studies used the velocity field derived from the first moment map \citeeg{Dickey1990, Boulanger1992} or from single Gaussian fitting \citep{Kamphuis1993, Rownd1994} to determine the velocity shift.
These velocity fields, however, are often affected by asymmetric \hi{} line-of-sight profiles.
Because any offset in velocity can introduce artificial broadening into the superprofile, we must find a more robust method for determining \vp{} by using information from the entire line-of-sight spectrum.

Toward this goal, \citet{deBlok2008} tested a variety of methods to determine \vp{} for rotation curve calculation: the intensity-weighted mean velocity; the velocity of the peak flux; a single or a multiple Gaussian fit; and a \gh{} polynomial fit \citeeg{VanderMarel1993}.
They concluded that the most robust function in the low S/N regime is a \gh{} polynomial that includes an $h_3$ term.
This method has already been used to generate velocity fields used in rotation curve analysis for a number of galaxies \citeeg{Noordermeer2007, deBlok2008}. 
Double Gaussian functions also tend to accurate fit line-of-sight spectra, but must be subjected to stringent parameter and S/N constraints to avoid fitting noise spikes \citeeg{Warren2012}.

\begin{figure}
  \centering
  \ifimage
  \includegraphics[width=8cm]{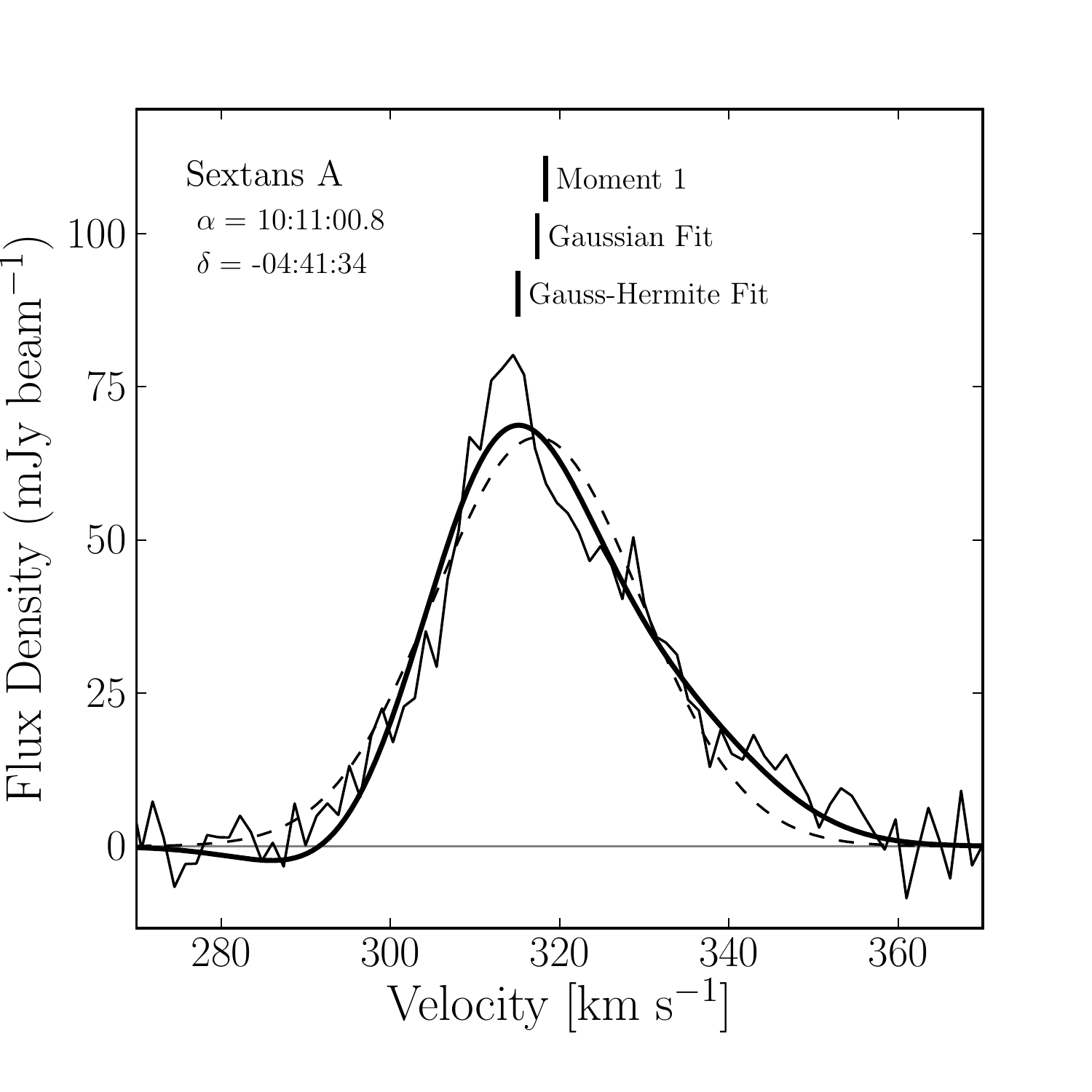}%{{figures/sexa.ro.phys200.peak-fit-methods}.pdf}
  \includegraphics[width=8cm]{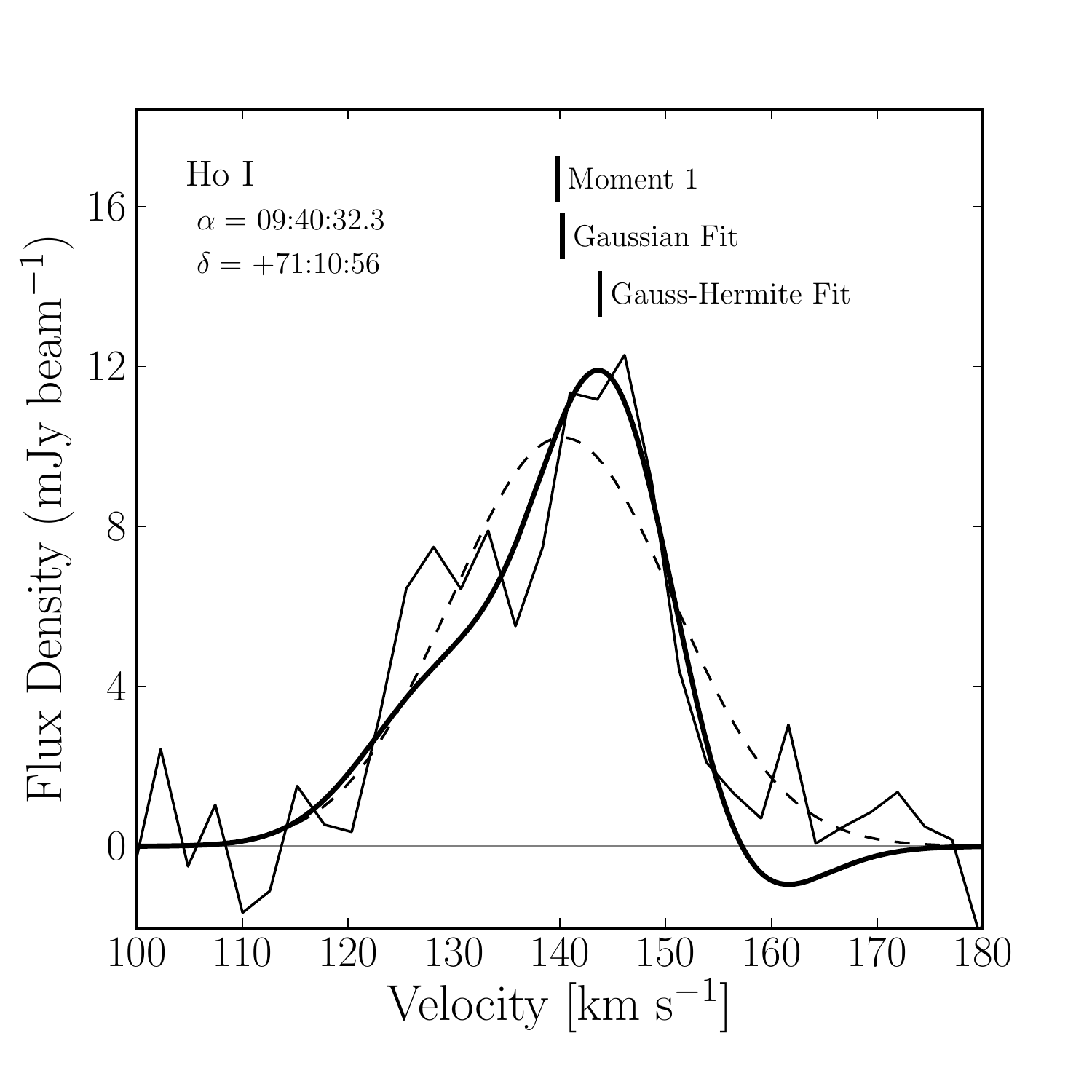}%{{figures/hoi.ro.phys200.peak-fit-methods}.pdf}
  \fi
  \caption{Two observed line-of-sight spectra from standard cubes with \vp{} methods: the first moment map, a Gaussian fit, and a \gh{} fit. The upper panels show the spectra and various fits, while the lower panels are the residuals. The thin black line in the spectrum, the dashed line is the Gaussian fit, and the thick black line is the \gh{} fit. The plotted \hi{} profiles are for line-of-sight spectra with higher than average S/N ratios, to better show the adopted functional form. The \gh{} polynomials are primarily to find \vp{} and are not  meant to characterize the detailed line profile structure. Left: Sextans A, for a line-of-sight spectrum with $\mathrm{S/N} = 16.9$. Right: Holmberg I, for a line-of-sight spectrum with $\mathrm{S/N} = 9.7$. \label{fig:peak-fit-methods}}
\end{figure}

In individual line-of-sight spectra, we find that the first moment, single Gaussian fits, and the velocity of the peak can be strongly influenced by asymmetries or noise, as result similar to that found by \citet{deBlok2008}.
We show a comparison of these various peak-determining methods for two individual line-of-sight spectra in Figure~\ref{fig:peak-fit-methods}. 
Both line-of-sight spectra show clear asymmetries, and the \gh{} polynomial best approximates the peak of the line-of-sight spectra.
In comparison, the single Gaussian fit and the first moment value are both shifted due to the asymmetry. 
Because we are interested only in determining \vp{} and not the detailed underlying structure of each individual line-of-sight spectrum, we adopt \gh{} polynomials when deriving \vp{}.

We use the unmasked, standard data cube to generate \vp{} maps for each galaxy.
This cube provides the correct, uniform noise properties necessary for \gh{} fitting, even though the fluxes are not accurate. %due to difficulties in approximating the synthesized beam.
Nonetheless, we expect \vp{} to be the same in both the standard cube and the flux-rescaled cube.
As explained in \citet{Ott2012} and \citet{Walter2008}, and references therein, the correction applied to the flux-rescaled cube rescales the intensity of the residuals of deconvolution to the same beam area as the intensity measured from the clean components.
However, the \hi{} emission in channels near the peak of the profile is primarily in the clean components and is not rescaled.
The low level of flux in the residuals should not affect the location of the highest intensity emission that determines \vp{}.

To generate the \vp{} maps for each galaxy, we fit a \gh{} polynomial with an $h_3$ term to each pixel in the unmasked, standard data cube as given by:
\begin{equation}
\phi (v) = A e ^{-y^2 / 2} \left[ 1 + \frac { h_3 } { \sqrt{6} } \left( 2 \sqrt{2} y^3 - 3 \sqrt{2} y \right) \right]
\label{eqn:gh}
\end{equation}
where $y \equiv (v - \mu) / \sigma_\mathrm{GH}$ and the $h_3$ component measures an asymmetric deviation from a Gaussian with amplitude $A$, offset $\mu$, and standard deviation $\sigma_\mathrm{GH}$ \citep{VanderMarel1993}. 
We note that the value of $\sigma_\mathrm{GH}$ is not necessarily the same as the standard deviation of a best-fit Gaussian without the $h_3$ term. 
The \gh{} polynomial fits are not intended to be a representation of the underlying \hi{} distribution, and we attach no physical significance to the parameters other than the determination of \vp{}.

For the fitting process itself, we use a Python implementation of the Levenberg-Marquardt fitting algorithm\footnote{mpfit.py; available at http://code.google.com/p/astrolibpy/} with uniform weight given by $1 / \sigma_\mathrm{chan}$ on each channel. 
We require that the peak velocity be within $\pm 20$ \kms{} of the first moment to ensure that the peak falls in the range of true \hi{} emission. 
The width of the profile is forced to be greater than the velocity resolution to prevent the algorithm from fitting individual noise spikes. 
While individual line-of-sight profiles can show double peaks indicative of expanding structures, recent studies of spectra in these galaxies have shown that they comprise only a small fraction of all spectra \citep{Warren2012}; thus, they do not contribute large amounts of flux to the superprofiles.
We assess the accuracy of our \vp{} measurements in \S \ref{sec:superprofiles--signal-to-noise}. 
We further discuss the the effects of \vp{} uncertainties on our results in Appendix \ref{appendix:uncertainties--vp}.

\begin{figure}
  \centering
  \ifimage
  \includegraphics[height=7in]{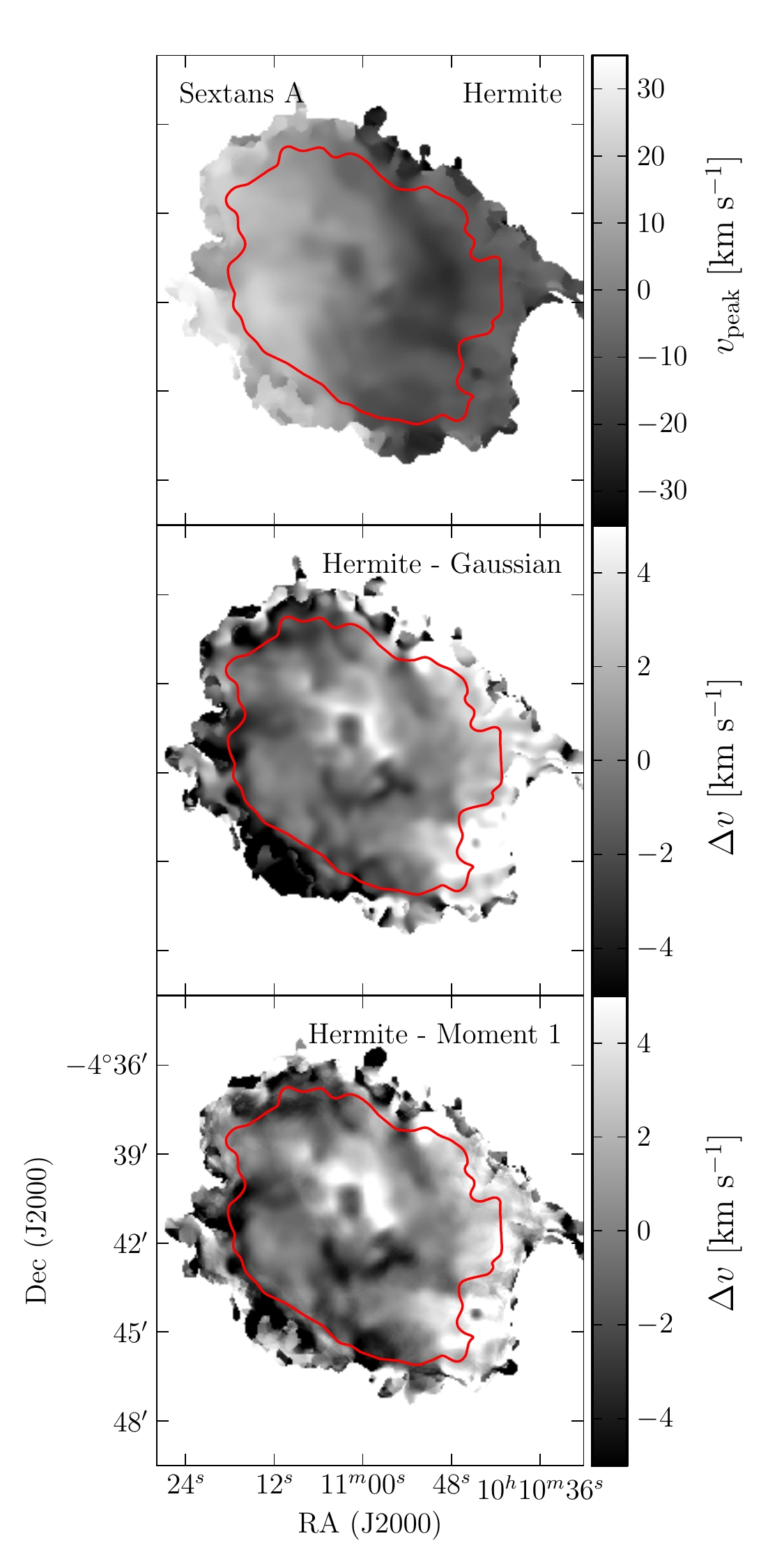}%{{figures/sexa.ro.phys200.vpeaks}.pdf}
  \fi
  \caption{The \vp{} map generated from the \gh{} fits with the systemic velocity removed (top) and differences from both a single Gaussian fit (middle) and the first moment map (bottom). The red contour line shows regions with S/N$ > 5$.  \label{fig:vpeak-map}}
\end{figure}

As an example, the final \vp{} map for Sextans~A is shown in the upper panel of Figure~\ref{fig:vpeak-map}. 
The middle and lower panels illustrate the differences between the \gh{} determination of \vp{} and what would have been determined by fitting a single Gaussian (middle panel) or using the first moment map (lower panel). 
In each map, we have overlaid a line to indicate where S/N $> 5$. 
The absolute differences between the \gh{} velocity field and the first moment and Gaussian velocity fields are small in a global sense (i.e. $\sim 4$ \kms). 
However, the differences are often only a factor of two smaller than the second moment values themselves, which could lead to spurious broadening of the global superprofiles if we use either the first moment map or a single Gaussian fit to determine \vp{}

\subsection{Signal-to-Noise Threshold Selection}
\label{sec:superprofiles--signal-to-noise}

\begin{figure}
  \centering
  \ifimage
  \includegraphics{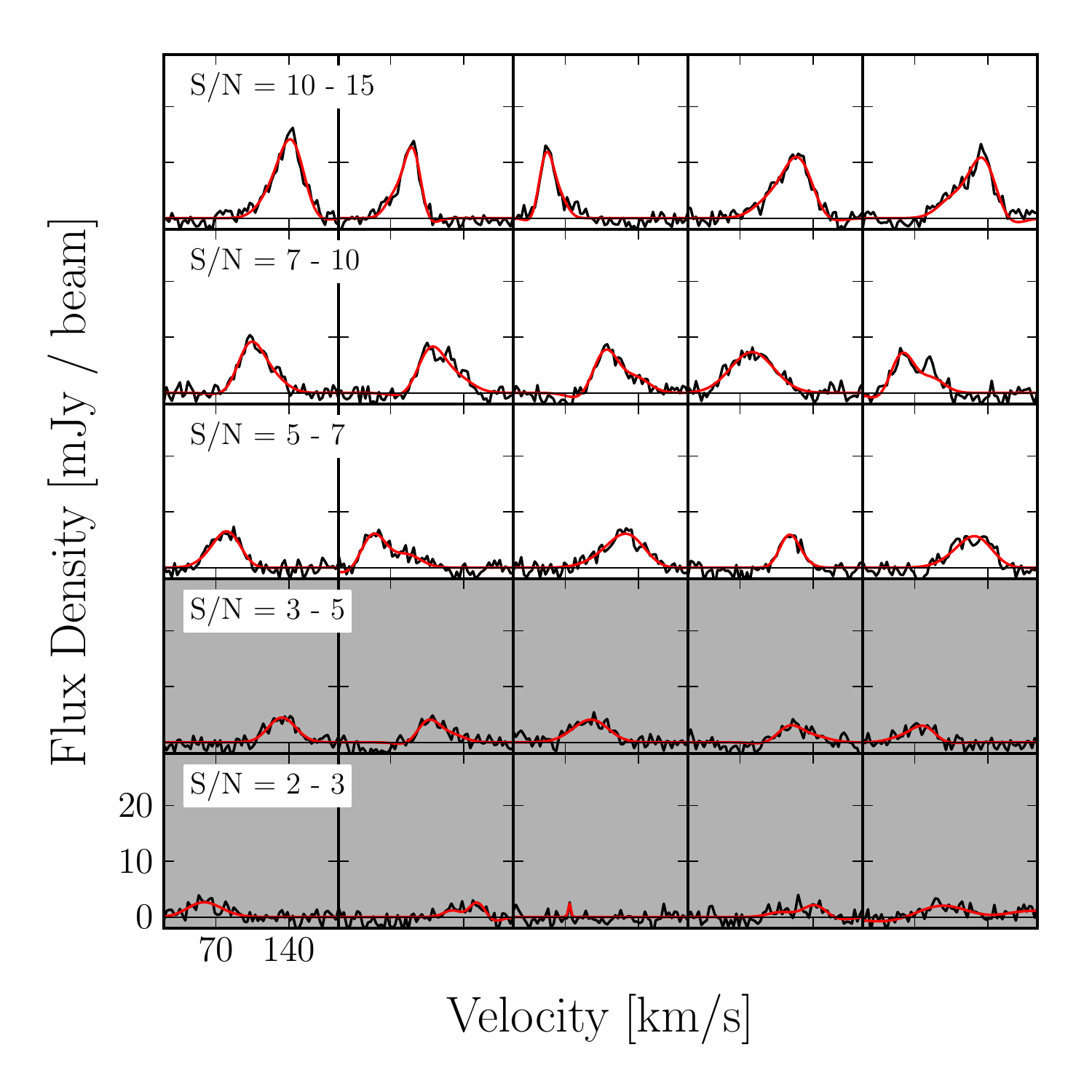}%{{figures/n2366.ro.phys200.sn_fits}.pdf}
  \fi
  \caption{Example \gh{} polynomial fits to various \hi{} line-of-sight spectra for NGC~2366, a galaxy with a velocity resolution of $2.6$ \kms{}. The line-of-sight profiles are sorted into rows based on their S/N; spectra with S/N$ < 5$ have grey backgrounds. The general asymmetry of the line-of-sight spectra is readily apparent. For spectra with S/N$ > 5$, the \gh{} polynomials generally do a good job at finding the peak. At lower S/N, the peak is more difficult to determine and may even be due solely to noise spikes. In our analysis, we only use pixels with S/N$ > 5$.  \label{fig:sn-fits-n2366}}
\end{figure}

\begin{figure}
  \centering
  \ifimage
  \includegraphics{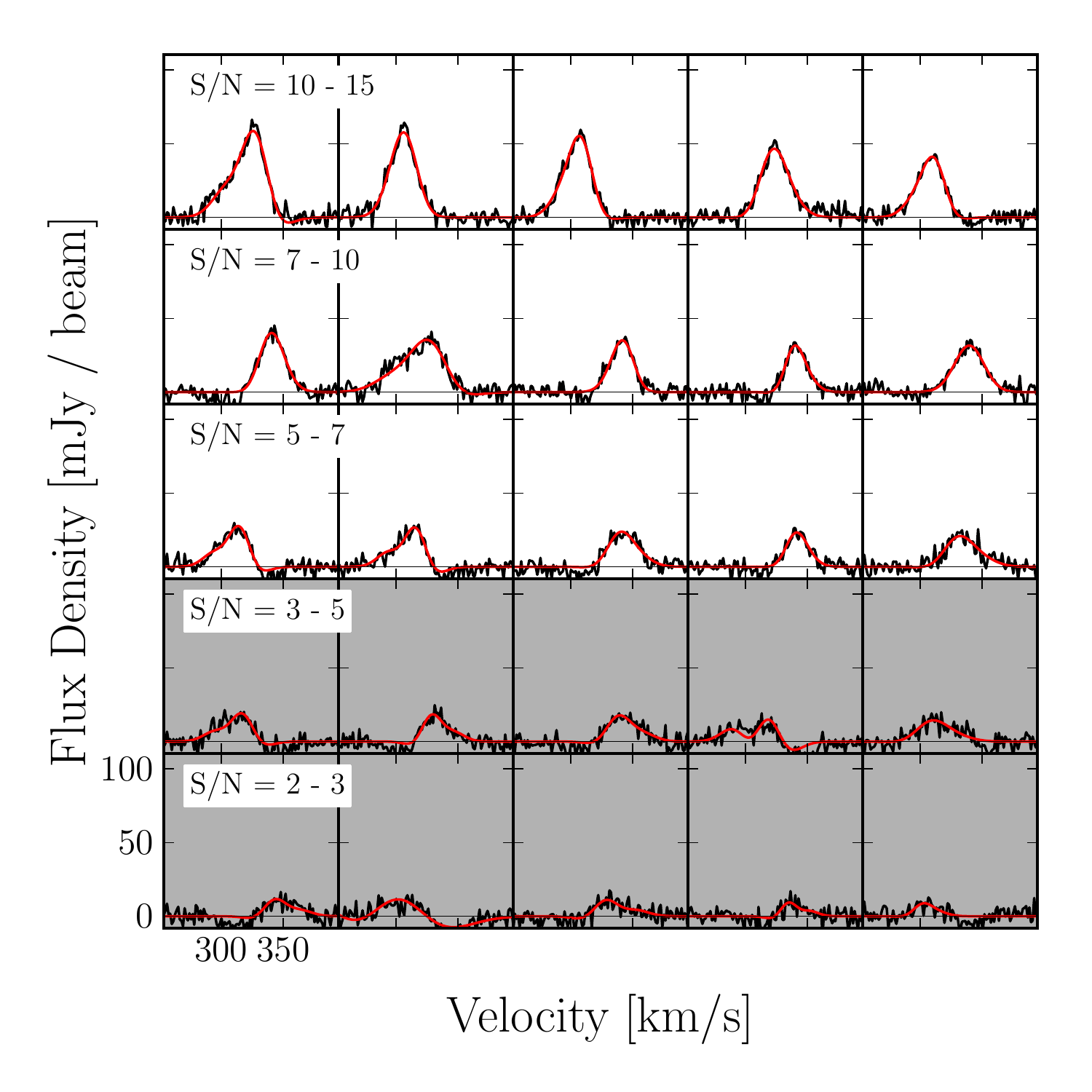}%{{figures/sexa.ro.phys200.sn_fits}.pdf}
  \fi
  \caption{Same as Figure~\ref{fig:sn-fits-n2366}, for Sextans A, a galaxy with a velocity resolution of 1.3 \kms{}.  \label{fig:sn-fits-sexa}
}
\end{figure}

To ensure that we have accurately measured \vp{} for each line-of-sight spectrum we include in the superprofile, we select only pixels for which the peak of the profile fit in Equation~\ref{eqn:gh} is above a specific S/N threshold. 
For each pixel, we define S/N as the ratio between the maximum of the \gh{} polynomial fit and $\sigma_\mathrm{chan}$. 
Figures~\ref{fig:sn-fits-n2366} and \ref{fig:sn-fits-sexa} show a number of individual line profiles with various S/N from NGC~2366 and Sextans~A, with the \gh{} fits overlaid in red. 
It is qualitatively clear that the \gh{} polynomials provide better fits as the S/N increases. For spectra with $\mathrm{S/N} > 5$, the fitting routine does a good job at identifying the peak. 
For spectra with $3 < \mathrm{S/N} < 5$, the fitting routine does a reasonable job, and spectra with $\mathrm{S/N} < 3$ have decidedly questionable fits.

\begin{figure}
  \centering
  \ifimage
  \includegraphics[width=3in]{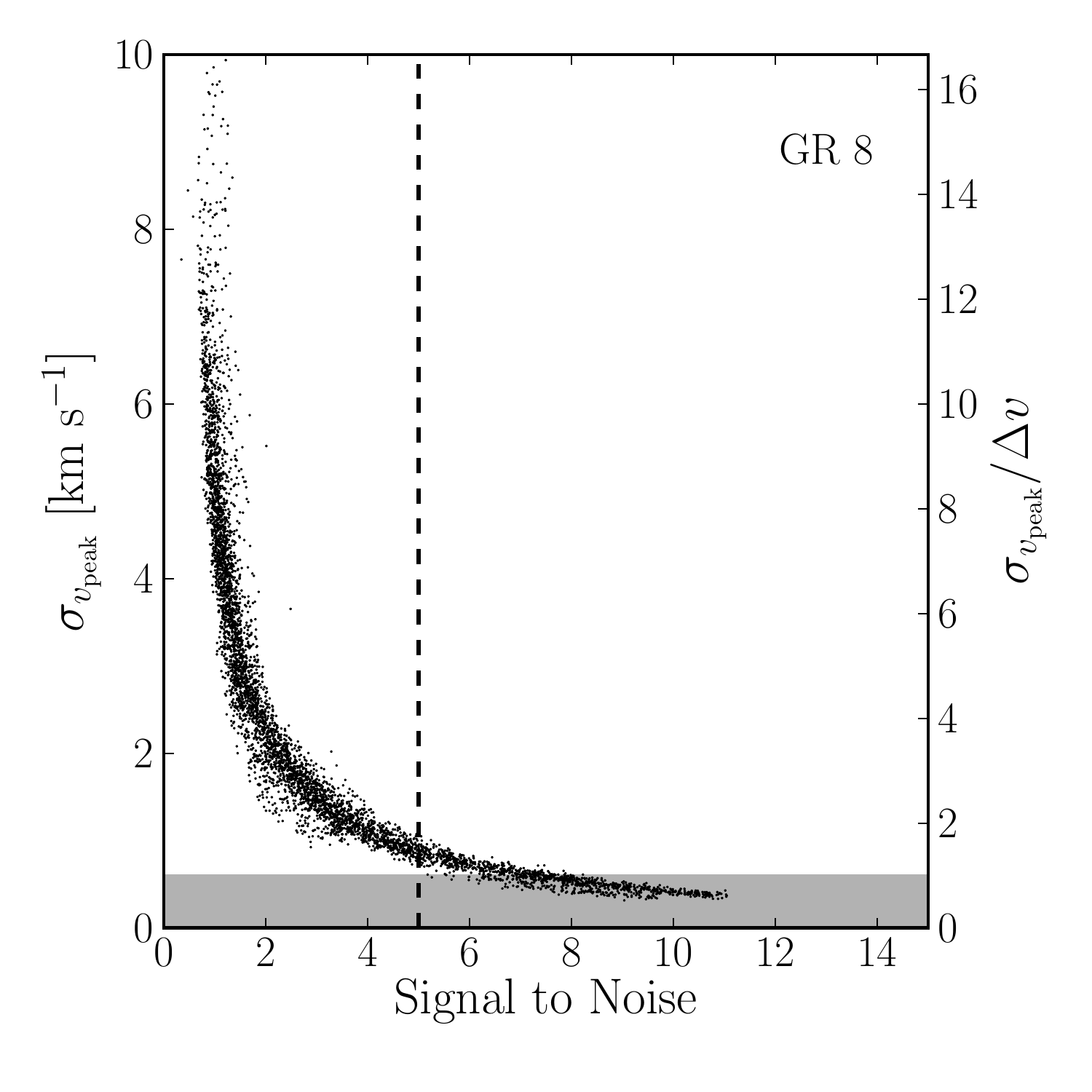}%{{figures/gr8.ro.phys200.sn.vmean}.pdf}
  \includegraphics[width=3in]{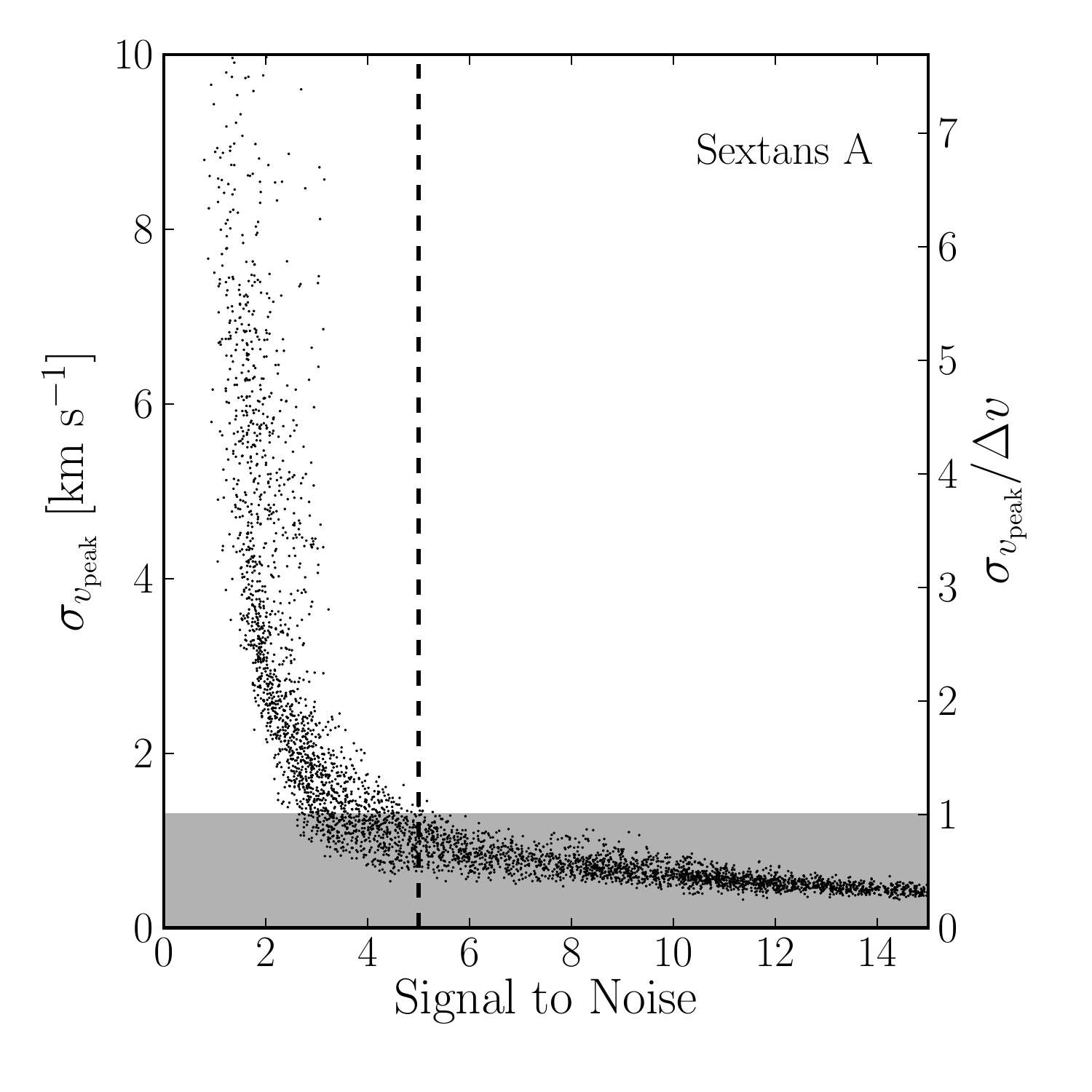}%{{figures/sexa.ro.phys200.sn.vmean}.pdf}
  \\
  \includegraphics[width=3in]{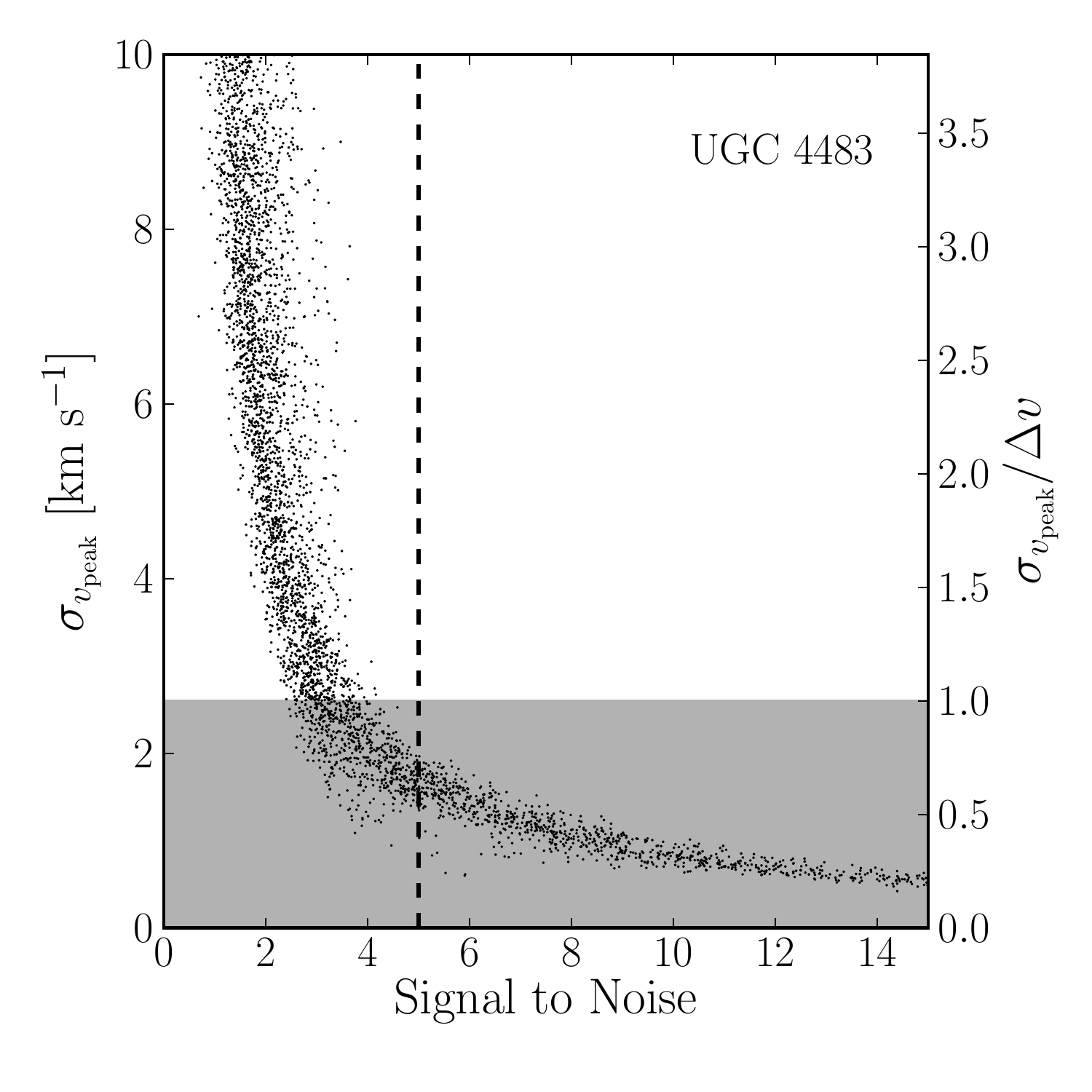}%{{figures/u4483.ro.phys200.sn.vmean}.pdf}
  \includegraphics[width=3in]{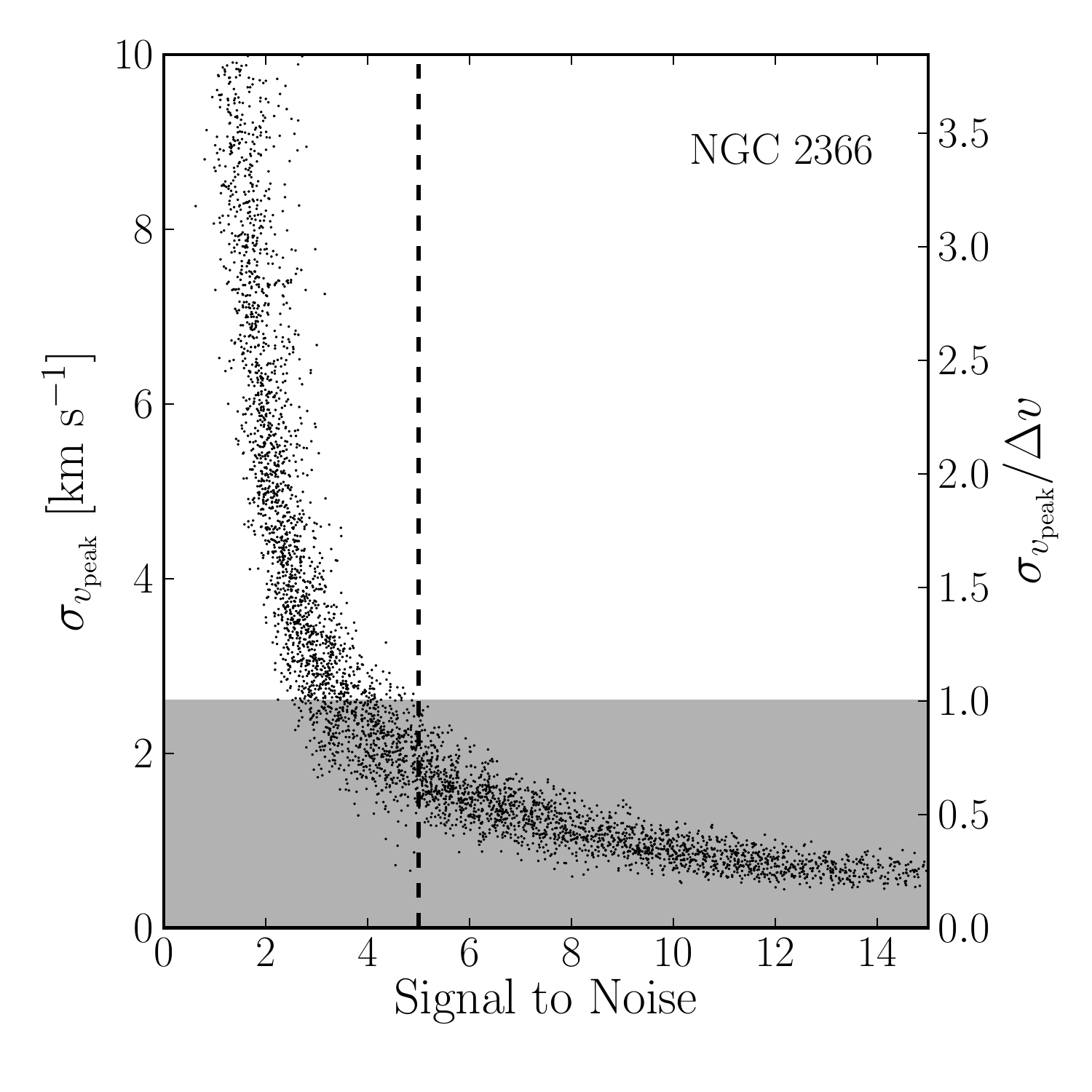}%{{figures/n2366.ro.phys200.sn.vmean}.pdf}
  \fi
  \caption{The average error on finding \vp{} ($\sigma_{v_\mathrm{peak}}$) as a function of S/N for three different simulated galaxy data cubes with three different velocity resolutions. The grey box indicates regions where $\sigma_{v_\mathrm{peak}}$ is below the observation's velocity resolution. The dashed black line is our adopted S/N = 5 threshold.
  At our S/N threshold, we find an average uncertainty of $\lesssim 1$ \kms{} for $\Delta v = 0.6$ \kms{} (GR 8), $\lesssim 1$ for $\Delta v = 1.3$ \kms{} (Sextans A), and $\lesssim 1.5$ for $\Delta v = 2.6$ \kms{} (UGC 4483).  \label{fig:mc-vp-offsets} }
\end{figure}

To quantify this behavior, we run Monte Carlo (MC) simulations for four representative galaxies that span the range in velocity resolution and S/N (GR~8, Sextans~A, UGC~4483, and NGC~2366).
We create a simulated data cube from the \gh{} polynomial fits to the data, add noise at the appropriate level, and run our \vp{}-finding algorithm. 
We then repeat this process 100 times. 
While the \gh{} fits are not necessarily representative of the underlying \hi{} distribution, they provide a known input for our tests. 
The output \vp{} maps from all the MC realizations allow us to calculate the average uncertainty in measuring \vp{} as a function of S/N. 
In Figure~\ref{fig:mc-vp-offsets}, we show the standard deviation of MC \vp{} offsets versus S/N for the four test galaxies. 
Each point represents the standard deviation around the input \vp{} of the 100 repeated \vp{} measurements for a single pixel. 
For clarity, only 5000 random points are shown for each galaxy. Regions below the cube's velocity resolution $\Delta v$ are shown in grey. %reword.

At smaller velocity resolutions, we are better able to determine \vp{}, for a fixed S/N. 
We overlay a vertical dashed line at $\mathrm{S/N} = 5$, where the standard deviations of the coarsest velocity resolution data (2.6 \kms{}) start to flatten. 
In all cases, we reproduce the input \vp{} with an error of $< 2$ \kms{}. Based on these tests, we adopt a S/N $> 5$ as our threshold.

\subsection{Co-addition of Line-of-Sight Spectra}
\label{sec:superprofiles--generation}

We use the flux-rescaled cubes to generate the final global superprofiles. 
These cubes provide correct flux properties and have been corrected for primary beam attenuation. %Further discussion of the flux rescaling procedure can be found in \citet{Walter2008}, \citet{Ott2012}, and references therein.
The advantages of using the flux-rescaled cubes is that they allow us to calculate accurate estimates of \hi{} mass and energy, at the cost of having more complicated noise properties than superprofiles generated from standard cubes.

We apply a mask to the data cubes so that only channels with real \hi{} emission contribute. 
We use the 45\arcsec{} resolution masks described in \S~\ref{sec:sample--hi-data-prep} for each galaxy but extend them by 15 \kms{} on either side in velocity. 
This extension includes any low-level \hi{} emission from gas in the surrounding channels that is below our masking threshold, but mostly eliminates spurious signals from instrumental effects such as sidelobes and clean bowls, which often occur further from the true \hi{} emission in velocity space. 
The final superprofiles are not strongly changed when using unmasked spectra in data cubes that do not show detectable instrumental artifacts such as sidelobes or negative bowls due to missing short spacings.

To create the final global line profile from the masked, flux-rescaled data cubes, we first recenter the selected (S/N $> 5$) individual line-of-sight spectra such that \vp{} is at zero. 
Because \vp{} is often located in the middle of a channel, simply shifting all profiles by an integer number of channels and then summing can artificially broaden the final global line profile. 
Instead, we linearly interpolate across each line-of-sight profile by a factor of 10 before shifting in velocity space. 
Finally, we co-add the shifted line-of-sight profiles with equal weight to obtain the intensity-weighted \hi{} superprofile.

The final global line profiles are shown in Figure~\ref{fig:superprofiles}, ordered by decreasing \m{baryon,tot}. They will be discussed in detail in \S~\ref{sec:analysis--overview}.

\subsection{Uncertainties}
\label{sec:superprofiles--uncertainties}

We define the noise on each point of the superprofile, $\sigma_\mathrm{SP}$, as:
\begin{equation}
\sigma_\mathrm{SP} = \sigma_\mathrm{chan} 
\times \sqrt{ N_\mathrm{pix} / N_\mathrm{pix/beam}}
\times \frac{F_\mathrm{rescaled}}{F_\mathrm{standard}}.
\label{eqn:noise-jvm}
\end{equation}
Here, $\sigma_\mathrm{chan}$ is the \emph{rms} noise in a single channel. 
$N_\mathrm{pix}$ is the number of channels contributing to a superprofile point, and $N_\mathrm{pix/beam}$ is the number of pixels per resolution element; 
this term represents the approximate number of independent profiles contributing to each superprofile point.
We count each interpolated point contributing to a single superprofile point as one pixel.
The $F_\mathrm{rescaled} / F_\mathrm{standard}$ term is the flux ratio between the total measured flux in the superprofile generated from the rescaled cube to that from the standard cube.
It is included to approximate the rescaling process, and maintains the same fractional noise between the standard cube and the rescaled cube.
Typical values for $F_\mathrm{rescaled} / F_\mathrm{standard}$ are $0.4 \pm 0.13$. We discuss the details of this estimate further in Appendix~\ref{appendix:noise}.

 The noise for each superprofile is shown in Figure~\ref{fig:superprofiles} as the grey shaded region around each measured superprofile. 
 In some cases, the S/N of the final superprofile is high enough that the uncertainties are smaller than the black line showing the superprofile itself.

\section{Characterizing the Superprofiles}
\label{sec:analysis}

In this section we discuss our analysis of the superprofiles. 
We first give an overview of their general properties (\S~\ref{sec:analysis--overview}). 
We then discuss the parameterization chosen to characterize the superprofiles and to quantify the observed asymmetry (\S~\ref{sec:analysis--parameterization}). 
We then discuss the physical interpretation of the parameterization (\S~\ref{sec:analysis--interpretation}) and provide a comparison with other studies of \hi{} kinematics (\S~\ref{sec:analysis--previous-studies}).

\subsection{Overview of the Final Global Line Profiles}
\label{sec:analysis--overview}

In Figure~\ref{fig:superprofiles} we show the superprofile for each galaxy, overlaid with a Gaussian scaled to the amplitude and to the half-width half-maximum (HWHM) of the superprofiles. 
We also shade in transparent red the regions in the wings where there is more \hi{} than expected compared to the Gaussian. 
The superprofiles show similar structures from galaxy to galaxy --- namely, a central narrow peak with additional wings to either side.

To compare the overall shape of the superprofiles from galaxy to galaxy, we plot all the superprofiles together in Figure~\ref{fig:superprofiles-scaled}. 
We normalize each superprofile so that its maximum flux is 1, and we scale the velocity axis by the HWHM of the superprofile. 
Regions where the scaled superprofiles overlap are darker. 
We also overplot a Gaussian with the same scaling, shown as the dashed line. 
Residuals from the scaled Gaussian are shown in the bottom panel. 
Because we have plotted each individual line with some transparency, regions where the superprofiles overlap are darker.

The superprofiles in Figure~\ref{fig:superprofiles-scaled} exhibit remarkably similar shapes, especially in the central regions. 
Typically, the profiles are peakier than a Gaussian in the central regions, and show wider wings whose residual amplitude peaks at approximately 2 $\times$ HWHM.
The amplitude of the non-Gaussian wings varies from galaxy to galaxy, but the general shape does not change.

\subsection{Superprofile Decomposition}
\label{sec:analysis--parameterization}

We first parameterize the superprofiles with a single Gaussian. 
Although a Gaussian is not the optimal match to the detailed shape of the profile, it is a widely used parameterization and provides an estimate of the average profile width. 
Because the Gaussian shape is a poor match to the overall profile, especially in the wing regions, we do not perform a traditional $\chi^2$ minimization of the fit, but instead scale the width of the Gaussian to match the HWHM of the superprofile and the amplitude to match the peak of the superprofile.  
If we had instead fit the superprofile with uniform or noise-based weighting, the width of the Gaussian would increase to compensate for the wings. 
In contrast, the HWHM scaling provides a simple estimate of the average \hi{} kinematics without relying on fitting details.

\begin{figure*}
  \centering
  \ifimage
  \includegraphics[width=5cm]{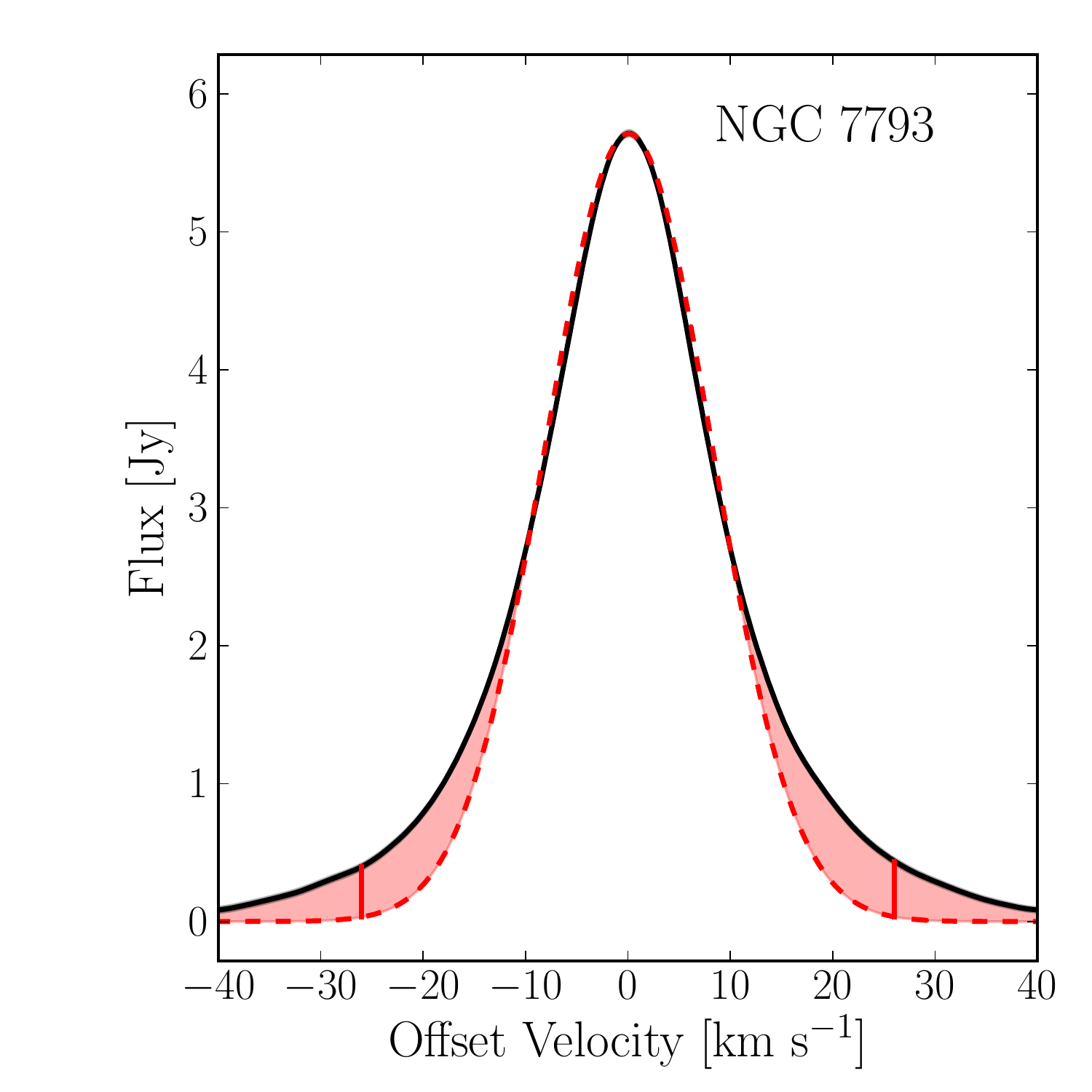}%{{figures/n7793.ro.phys200.superprofile.hwhm}.pdf}
  \includegraphics[width=5cm]{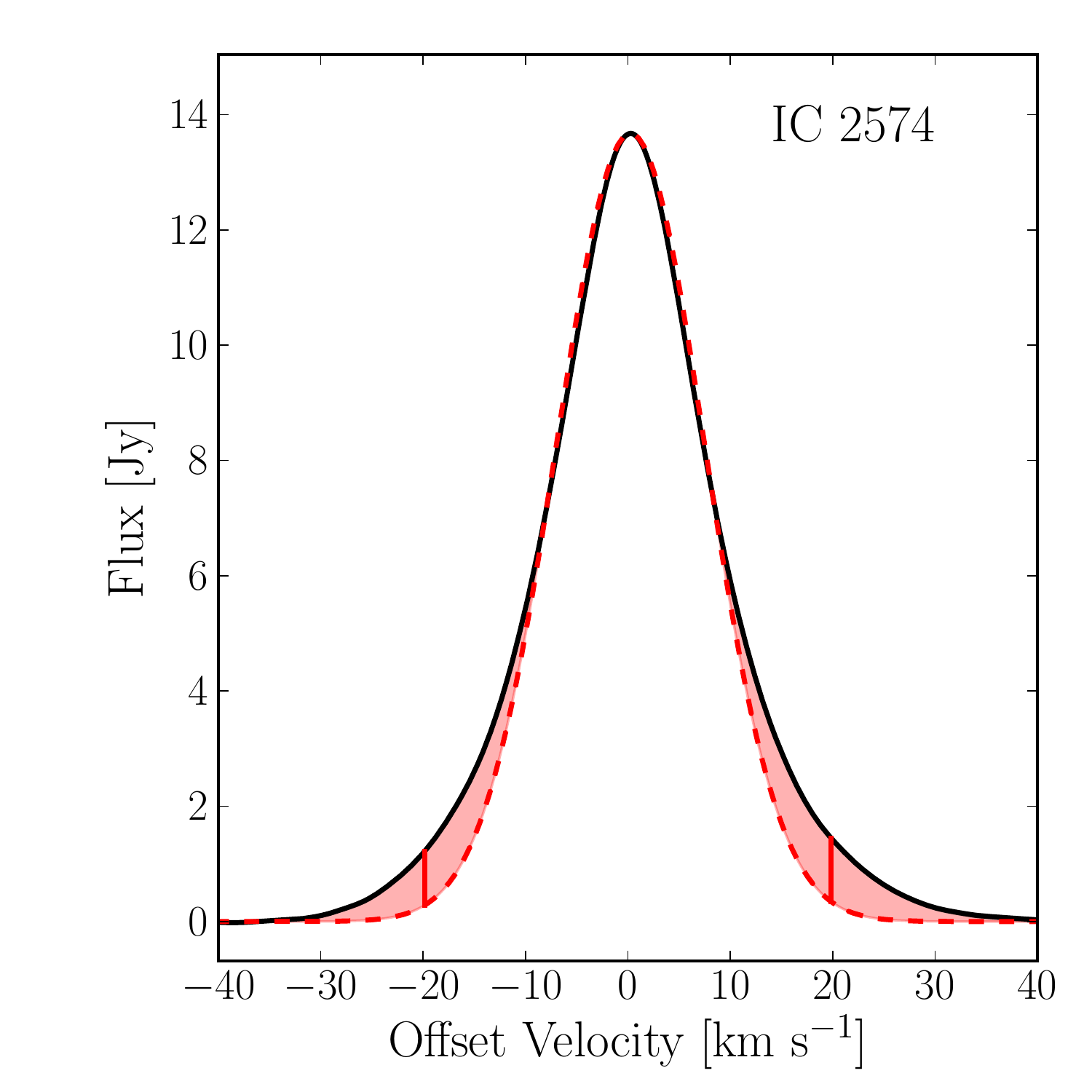}%{{figures/ic2574.ro.phys200.superprofile.hwhm}.pdf}
  \includegraphics[width=5cm]{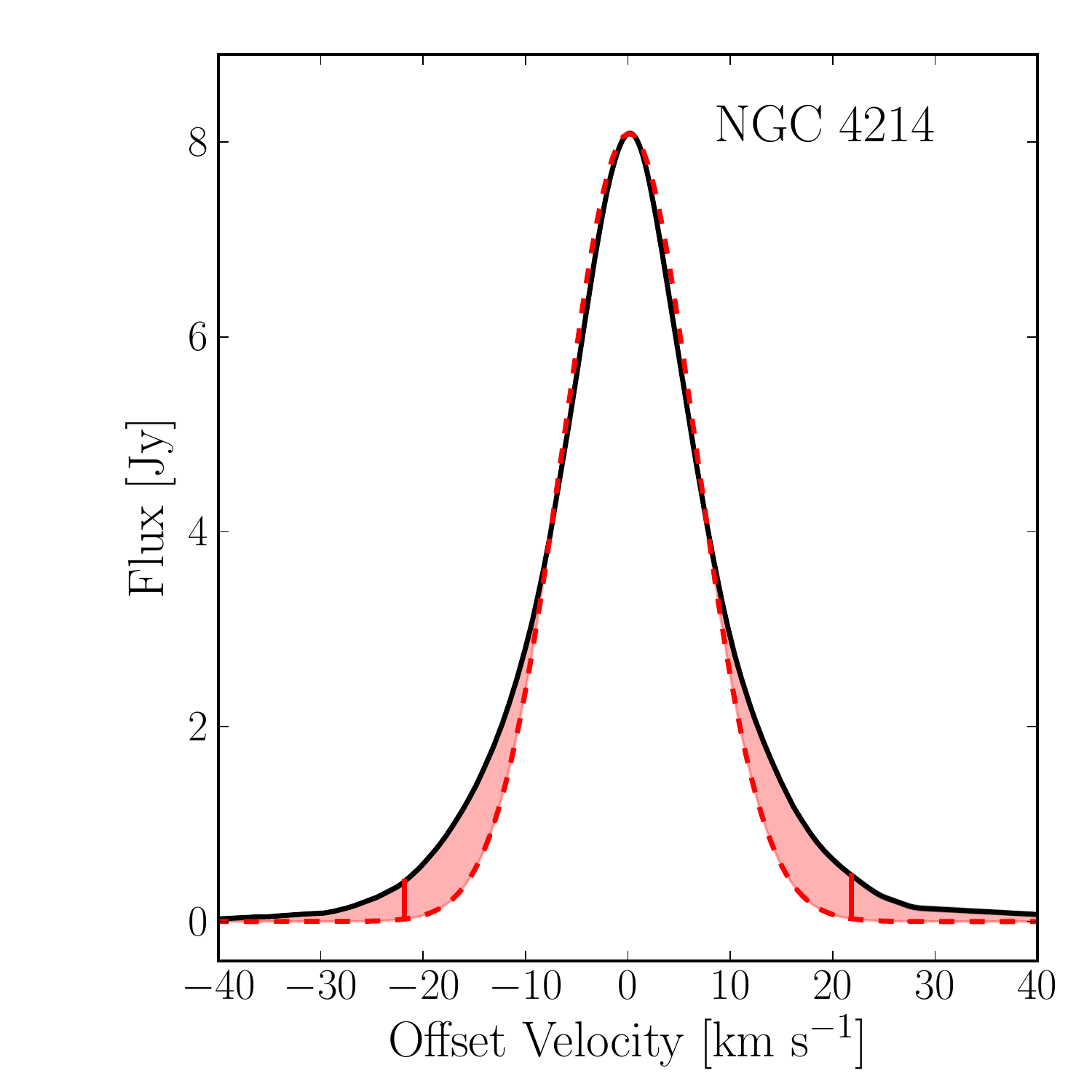}%{{figures/n4214.ro.phys200.superprofile.hwhm}.pdf}
  \\
  \includegraphics[width=5cm]{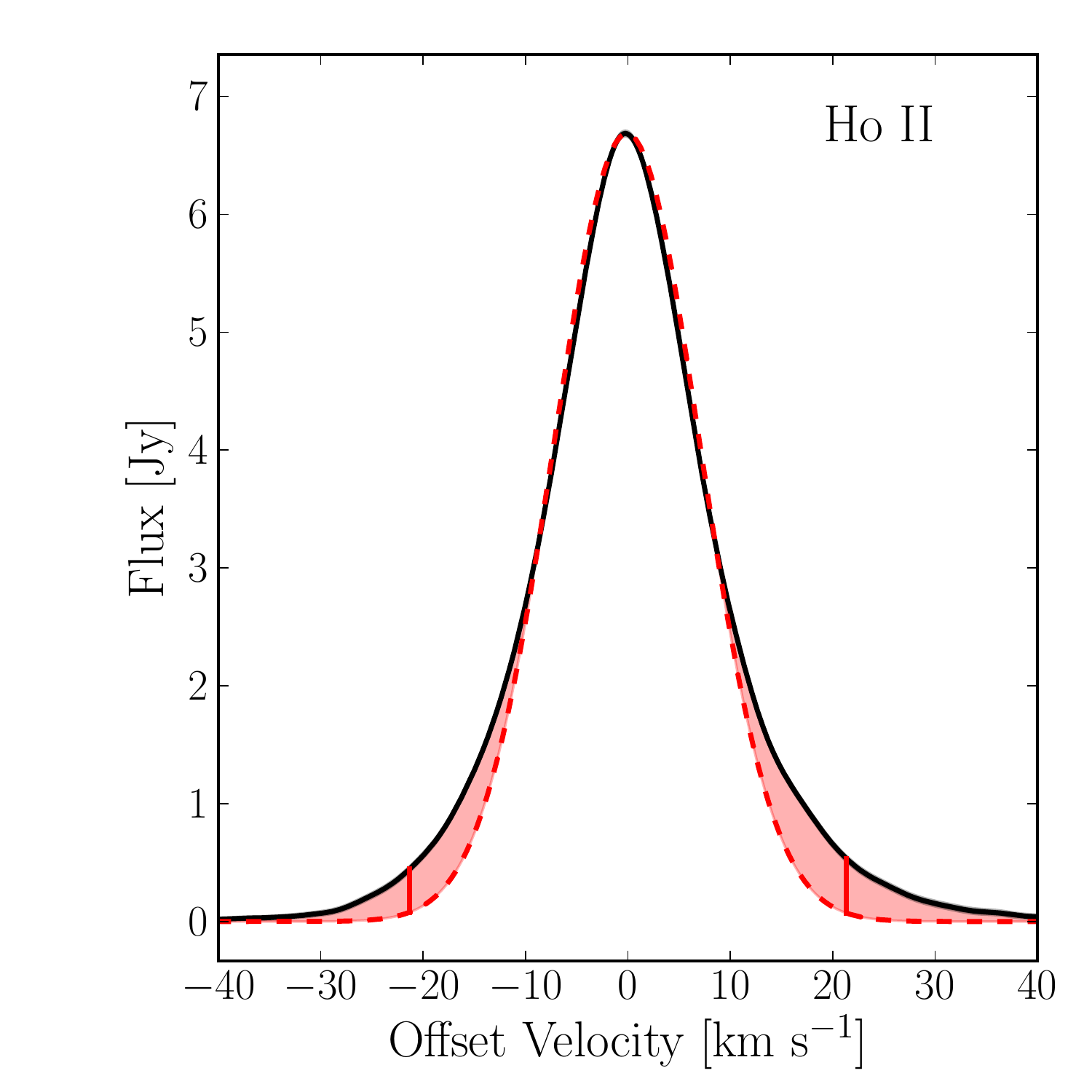}%{{figures/hoii.ro.phys200.superprofile.hwhm}.pdf}
  \includegraphics[width=5cm]{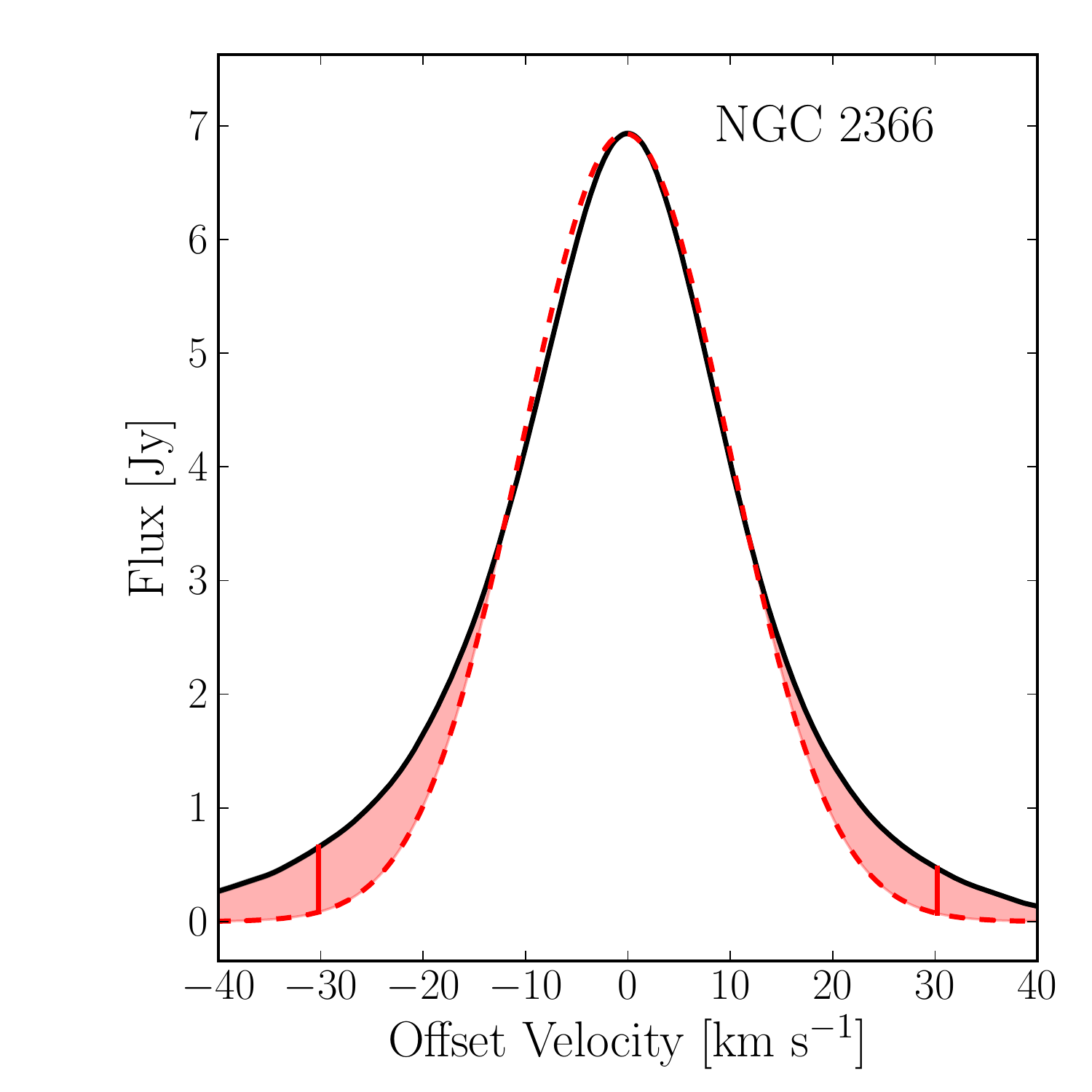}%{{figures/n2366.ro.phys200.superprofile.hwhm}.pdf}
  \includegraphics[width=5cm]{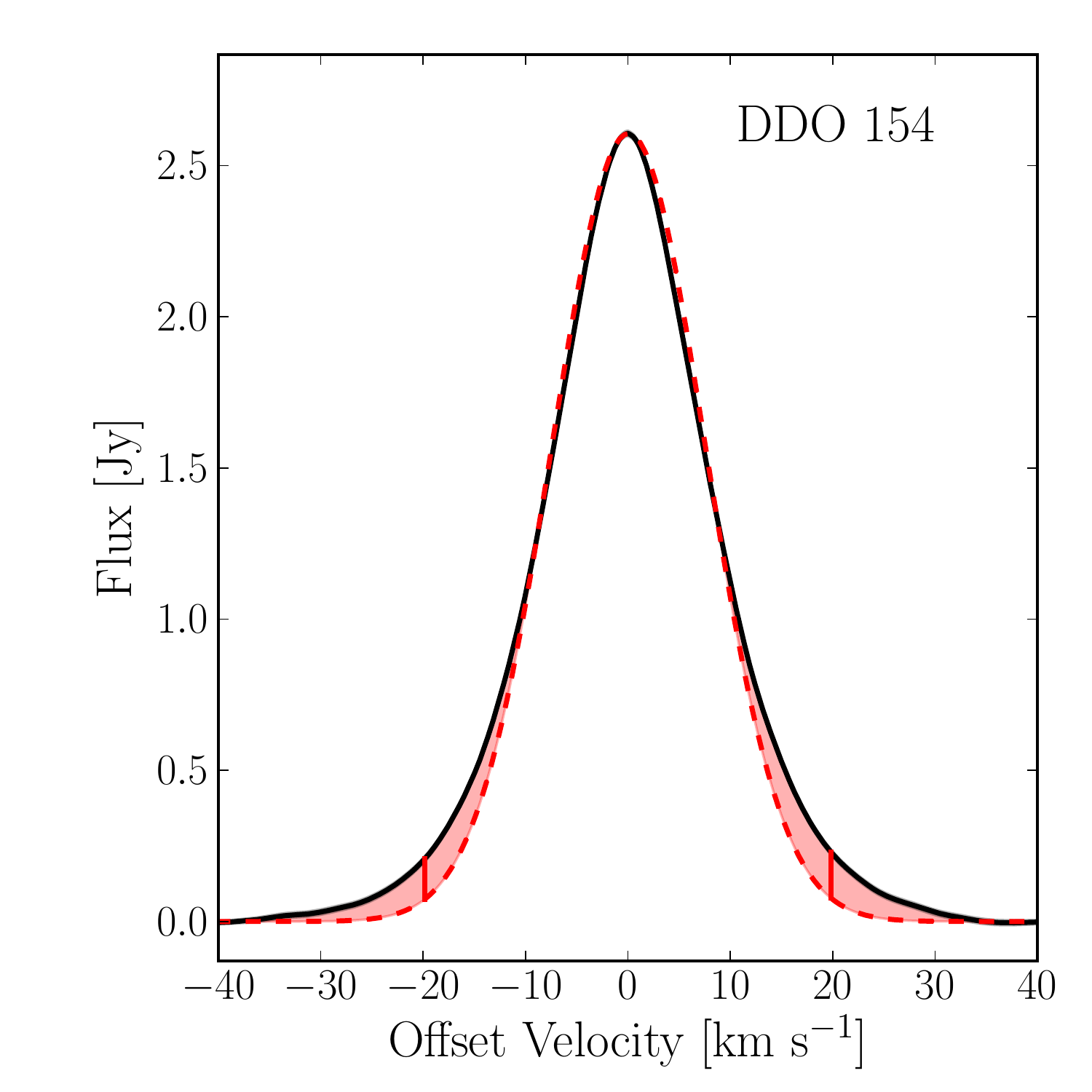}%{{figures/ddo154.ro.phys200.superprofile.hwhm}.pdf}
  \\
  \includegraphics[width=5cm]{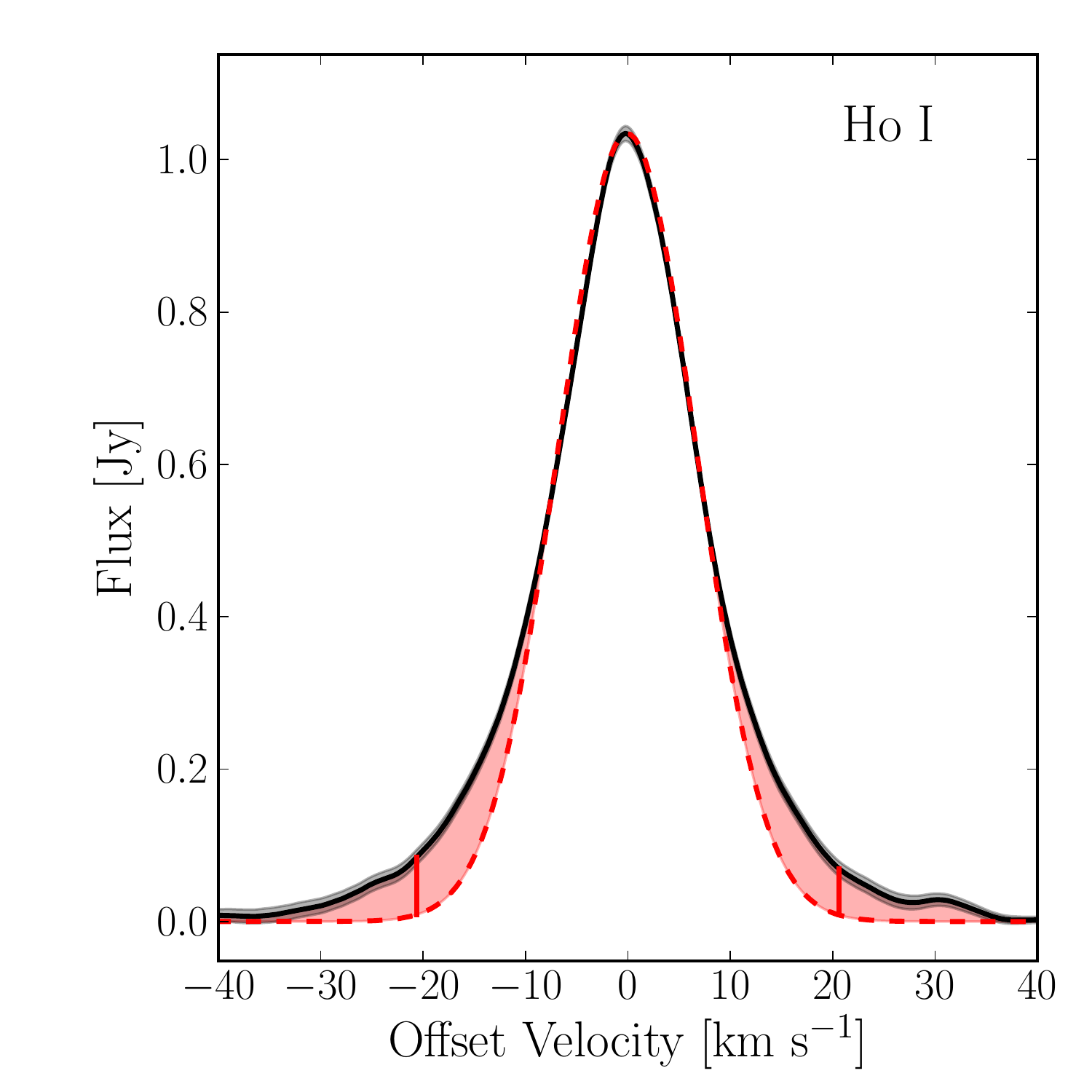}%{{figures/hoi.ro.phys200.superprofile.hwhm}.pdf}
  \includegraphics[width=5cm]{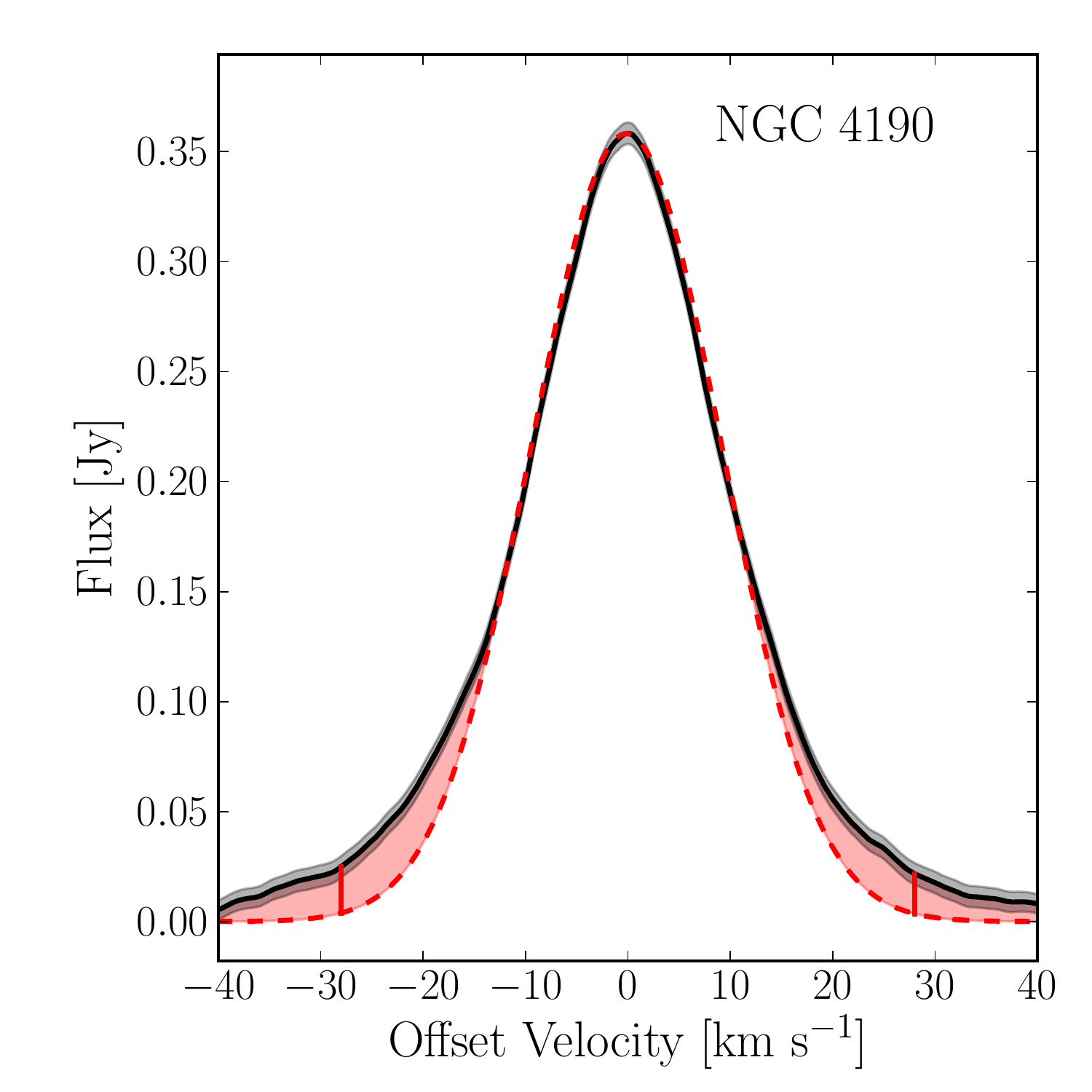}%{{figures/n4190.ro.phys200.superprofile.hwhm}.pdf}
  \includegraphics[width=5cm]{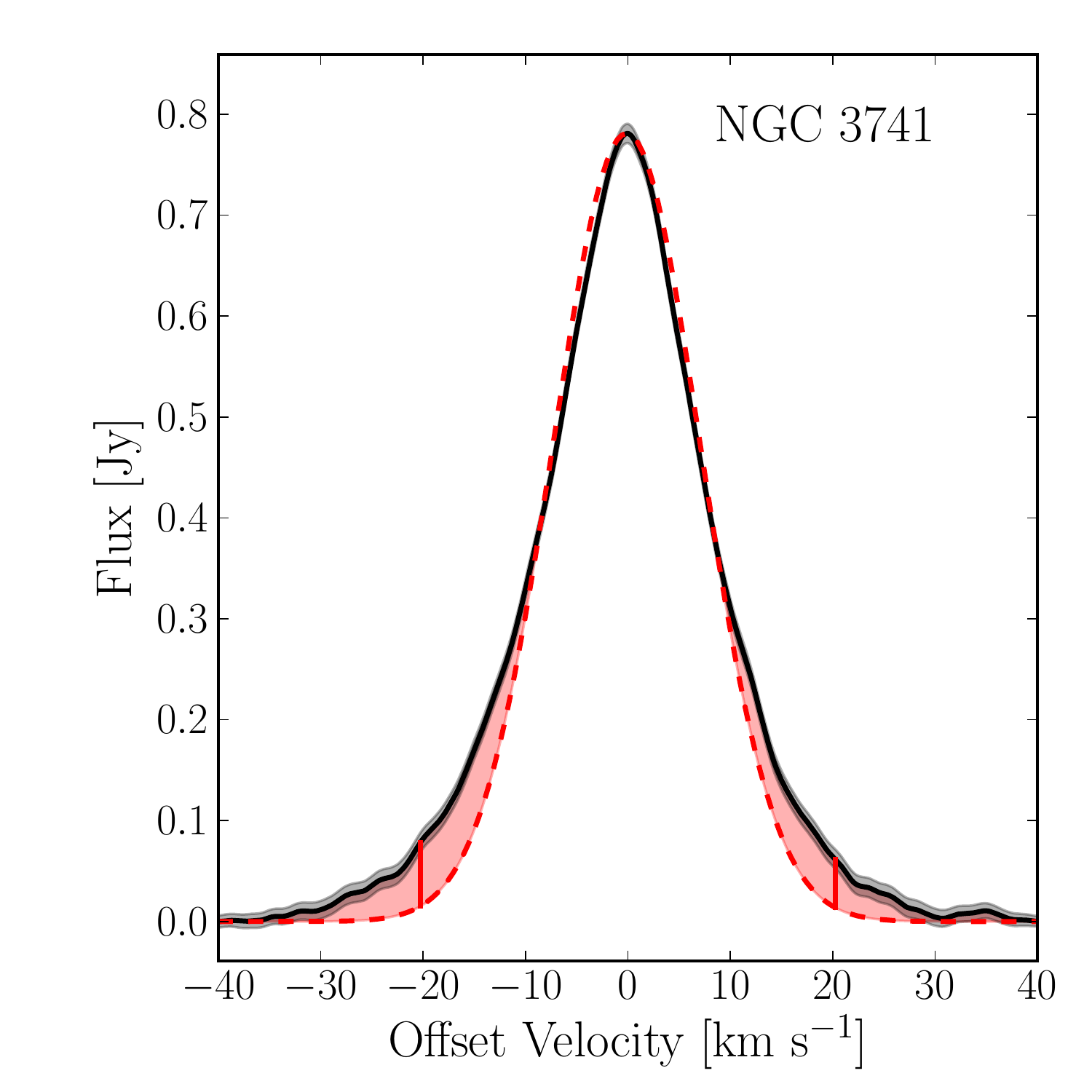}%{{figures/n3741.ro.phys200.superprofile.hwhm}.pdf}

  \fi
  \caption{\hi{} superprofiles. In each panel,
  the black line represents the measured superprofile for each galaxy. Grey regions around the superprofile are the 1-$\sigma$ uncertainties on the flux. The dashed red line shows a Gaussian scaled to the amplitude and the half-width at half-maximum of the superprofile. Shaded red regions between this line and the superprofile are the ``wing'' regions and represent \fw{}. The vertical red line is the characteristic velocity of the wings, \swing{}.
  As the superprofiles are the analogue of integrated \hi{} spectra, but with the rotational velocity removed, we plot flux in Jy versus offset velocity. However, the Jy value is not indicative of our signal in a single channel. Galaxies are ordered by decreasing \m{baryon,tot}. \label{fig:superprofiles} }
\end{figure*}

The HWHM-scaled Gaussian is shown in Figure~\ref{fig:superprofiles} as a dashed red line overlaid on the superprofile for each galaxy. We measure four parameters using this HWHM-scaled fit.

First, we measure \scentral{}, the width of the central peak scaled to match the HWHM of the superprofile. 
This parameter characterizes the average \hi{} line width in each galaxy. 
The use of \scentral{} instead of the HWHM value is chosen to facilitate comparison with other studies, which often describe line widths in terms of a Gaussian $\sigma$. 
We find that the median $\scentral{} = 7.7$ \kms{} with interquartile range $7.2 - 8.5$ \kms{}. 
These widths typically are $\gtrsim 3$ times larger than the coarsest velocity resolution of the observations.

Next, we measure the fraction of \hi{} in the wings:
\begin{equation}
\fw{} = \frac{ \sumlim{|v| > \mathrm{HWHM}} \left[ S(v) - G(v) \right]} { \sumlim{|v| > 0} S (v) },
\label{eqn:fw}
\end{equation}
where $v$ is the offset velocity, $S(v)$ is the superprofile, and $G(v)$ is the single Gaussian scaled to the superprofile HWHM. 
The \fw{} parameter measures the fraction of gas moving at velocities faster than expected compared to the bulk of \hi{}. 
We find median values of $\fw{} = 0.11$ with an interquartile range $0.1 - 0.13$. 
The measured \fw{} values are quite small, which indicates that the majority of the \hi{} is contained in the central peak. 
In Figure~\ref{fig:superprofiles}, the regions of the superprofiles that contribute to \fw{} are shown as transparent red regions.

Third, we measure the root mean squared velocity of excess flux in the wings, weighted by the fraction of gas moving faster expected based on the HWHM-scaled Gaussian:
\begin{equation}
\swingsq{} = \frac{ \sumlim{ | v | > \mathrm{HWHM}}  \left[ S (v) - G(v) \right] v^2 }  { \sumlim{ | v | > \mathrm{HWHM}} S (v) - G(v) } .
\label{eqn:swing}
\end{equation}
The \swing{} parameter measures the excess energy in the wings of the profile per unit \hi{} mass. 
It is also equal to the characteristic velocity of excess gas. 
We find median values of $\swing{} = 21.8$ \kms{} with an interquartile range $20.2 - 25.0$ \kms{}. 
These values are typically $\sim 8$ times the coarsest velocity resolution of our data, and a factor of $\sim 2$ smaller than the median characteristic inclination-corrected rotational velocity $w_{20} / 2$. 
In Figure~\ref{fig:superprofiles}, $\swing{}$ is shown as a solid vertical red line on other side of the superprofile.

Finally, we quantify the asymmetry of the residuals in the wing regions:
\begin{equation}
\aw{} = \frac{ \sumlim{ | v | > \mathrm{HWHM} } \sqrt{ \left( S(v) - S(-v) \right) ^ 2 } }  { \sumlim{ | v | > \mathrm{HWHM} } S(v) - G(v)  },
\label{eqn:aw}
\end{equation}
where $S(-v)$ is simply the mirror image of $S(v)$ around the peak.
This parameter ranges between 0 in the case of complete symmetry to 1 if all the excess flux was concentrated on one side of the superprofile. We find a median value of $\aw{} = 0.22$ with an interquartile range $0.17 - 0.30$.

We have also evaluated the asymmetry of the superprofiles around the peak:
\begin{equation}
a_\mathrm{global} = \frac{ \sumlim{v}{ \sqrt{ \left( S(v) - S ( - v) \right)^2 } } }  { \sumlim{v}{ S(v)} }.
\label{eqn:aglobal}
\end{equation}
To first order, the superprofiles are very symmetric. 
We find a median global asymmetry of only 0.05 with an interquartile range between $0.03-0.08$ for the superprofiles. 
In all cases, approximately 80\% of the asymmetry is from the wings of the profile, while regions with velocities less than the HWHM are considerably more symmetric. We therefore do not include $a_\mathrm{global}$ in our parameters as it is less sensitive to asymmetries compared to \aw{}.

In Table~\ref{tab:measurements} we report the measured quantities from our single Gaussian fits. We list
(1) the galaxy name;
(2) \scentral{} and associated uncertainty of the HWHM Gaussian;
(3) \swing{} and associated uncertainty;
(4) \fw{} and associated uncertainty;
(5) the asymmetry parameter \aw{} and associated uncertainty; 
(6) the global asymmetry $a_\mathrm{global}$;
and
(7) the number of independent resolution elements comprising each superprofile, $N_\mathrm{beams} = N_\mathrm{pix, \, S/N > 5} \; / \, N_\mathrm{pix / beam}$. 
We show the distribution of \scentral{}, \swing{}, \fw{}, and \aw{} in Figure~\ref{fig:parameter-histograms}.

We discuss the determination of uncertainties on these parameters in Appendix~\ref{appendix:uncertainties}.

\subsection{Physical Interpretation of the Superprofiles}
\label{sec:analysis--interpretation}

The parameterization described in \S~\ref{sec:analysis--parameterization} implicitly assumes that the majority of the \hi{} has a velocity dispersion that is reasonably well-described by \scentral{}.
The central peak can then be taken to represent the average kinematics of widespread well-thermalized \hi{} gas across the galaxy.
Thermal temperatures implied by the measured \scentral{} values are $\sim 4,000 - 12,000$ K, a range that brackets the predicted stable ISM temperature of $T \sim 7000$ K \citep{Wolfire1995}, but with a larger range.
Gas in this temperature range can radiate its energy away in $\sim 10^3$ years, a timescale too short to replenish the lost energy from external sources.
Therefore, the value of \scentral{} may be better interpreted as random turbulent velocities, which decay more slowly but still require an energy source to maintain over the galaxy lifetime. 
The deviations from a Gaussian profile can then potentially be explained by the fact that the central peak is likely to be the sum of Gaussians with a range of velocity widths due to the decline of velocity dispersion of warm \hi{} with radius \citep{Tamburro2009,Warren2012}.
Additionally, cold \hi{} with velocity dispersions $< 6$ \kms{} has been identified in some of the sample galaxies along individual lines of sight, but it makes up a small fraction of the \hi{}, with typical fractions of only $\lesssim 20$\% \citeeg{Young2003, Warren2012}.
Because cold \hi{} is present along some lines of sight that contribute to the superprofiles, its small velocity dispersion could explain why the observed superprofiles are peakier than a Gaussian profile.

Gas in the wings of the superprofiles can then be interpreted as localized regions where \hi{} is moving faster than expected compared to the average velocity dispersion. 
These anomalous motions likely require additional energy input to drive gas from its undisturbed state into the wings of the profile.
The superposition of this energetic gas atop the turbulent component of the central peak produces superprofile wings with amplitudes higher than expected from the turbulent component alone.

An alternative physically-motivated decomposition of the superprofiles is to consider them to be the sum of the CNM and WNM, characterized by narrow and broad velocity components.
This approach has been recently pursued in a similar study of THINGS galaxies \citepalias{Ianj2012}, motivated by the fact that \hi{} can exist at two stable temperatures and that some individual line-of-sight spectra show evidence of these two phases.
In this scenario, the ratio of \hi{} flux in the narrow component to that in the broad component provides a measurement of the relative amounts of cold and warm \hi{}.
However, this interpretation also implies that the two Gaussian components are well-thermalized but independent, and that only two distinct warm and cold gas populations exist in a single galaxy.
It also presumes that the line widths are directly connected to the thermal temperatures of the gas, which is difficult to reconcile with both the short thermal timescales and the mismatch between the inferred kinetic temperatures and the predicted thermal temperatures of the CNM and WNM.
We have found that double Gaussian fits to the superprofiles are indeed a good representation of the overall shape, but we believe that interpreting the Gaussian components as representative of the CNM and WNM is not necessarily convincing for the global profiles, although it can be valid along individual lines of sight.
We discuss the reasons we have chosen not to use this method in more detail in Appendix~\ref{appendix:double-gaussian}.

To facilitate comparison with \citetalias{Ianj2012}, however, we have also fit the superprofiles with a double Gaussian using the \citetalias{Ianj2012} methodology. 
As in \citetalias{Ianj2012}, the double Gaussian fits are weighted by the inverse of the approximate uncertainty due to noise on each point.
The results of these fits are given in Table~\ref{tab:parameters--double-gaussian}.
We list
(1) the galaxy  name;
(2) the width of the narrow component, $\sigma_n$;
(3) the width of the broad component, $\sigma_b$;
(4) the area of the narrow component relative to that of the broad component, $A_n / A_b$;
and
(5) the width of the narrow component relative to that of the broad component, $\sigma_n / \sigma_b$.

%We have also attempted to fit the superprofiles with a Voigt profile, which provides a reasonable prediction for the line profile of gas that is subject to both Doppler and Lorenzian broadening mechanisms.
%The Voigt profile is defined for gas a single temperature and density. However, there is no reason to expect \hi{} in a single galaxy to have the same density and temperature everywhere, thus rendering the physical interpretation of Voigt fits unclear.

For the reasons given in \S~\ref{sec:analysis--parameterization}, \ref{sec:analysis--interpretation}, and Appendix~\ref{appendix:double-gaussian}, we have chosen to parameterize the superprofiles based on the simple HWHM scaling discussed above.
We interpret the superprofiles physically as exhibiting a central turbulent peak with more energetic gas in the high-velocity wings to either side.

\subsection{Comparison with Other Studies}
\label{sec:analysis--previous-studies}

The \hi{} velocity dispersion and associated energy have traditionally been estimated either by fitting single Gaussian profiles to line-of-sight spectra \citeeg{Petric2007} or by using the second moment map \citeeg{Tamburro2009}.
Since our method is an uncommonly-used estimate of \hi{} turbulent velocity, we provide comparisons between our work and previous studies from \citet{Petric2007} and \citet{Tamburro2009}.

\cite{Petric2007} determined \vdisp{} by fitting single Gaussians to each line-of-sight spectrum in the face-on spiral NGC~1058 at a spatial resolution of $\sim1.3$ kpc, much larger than the 200~pc scales studied in this paper. 
The typical range of velocity dispersion in the disk was 4 - 14 \kms{}, with a majority at $\sim 7$ \kms{}. 
They also found that the median profile shape, after normalization for line-of-sight profile width, was similar to the shape observed in our superprofiles, characterized by a central narrow peak with wider wings. 
Many other studies have found similar shapes for the average \hi{} line profile \citeeg{Dickey1990, Boulanger1992, Kamphuis1993}, but quantitative comparisons with these studies are hampered by large differences in velocity resolution or physical resolution.

\citet{Tamburro2009} used the second moment as a a proxy for \hi{} velocity dispersion. They found that galaxies have turbulent components with amplitudes of $\sim 10$ \kms{}. 
When we compare the four galaxies that overlap both samples (Ho~II, NGC~4214, IC~2574, NGC~7793), we measure consistently lower \scentral{} values than the global second moment by $\sim 2$ \kms{}.
We do not believe that the difference results from the fact that the \citet{Tamburro2009} sample focused mainly on large spirals compared to the low-mass dwarfs that dominate our sample, as all galaxies in our sample have \scentral{} values that are smaller than their global second moments.
The difference is more likely due to the fact that we have isolated the central peak in our sample, while the second moment values are affected by the presence of high velocity wings and asymmetries in the line profiles.

In Figure~\ref{fig:m2-v-snarrow} we show a comparison between our measured \scentral{} value and $\langle \sigma_\mathrm{m_2} \rangle$ (Equation~\ref{eqn:m2-global}), which provides an estimate of the turbulent width that would have been derived using the \citet{Tamburro2009} methodology. 
Compared to \scentral{}, in all cases the global second moment is larger by $\sim 10 - 50$\%, with a median of 20\%.  
When interpreting the line widths physically, this offset suggests that the second moment leads to higher estimates of the energy necessary to drive \hi{} turbulence.

We can also compare our results to those of \citetalias{Ianj2012}, who have performed the most directly analogous analysis to date.
In this study, the authors generated superprofiles for a number of THINGS galaxies, eight of which overlap with our sample.
However, their approach was somewhat different.
The authors used the naturally-weighted standard maps and worked at the instrumental resolution.
They also chose to model each superprofile as the sum of a narrow and a wide Gaussian profile representing the CNM and the WNM.

The first difference to note is that our sample is smoothed to a fixed physical resolution, as opposed to working at the instrumental resolution of $\sim 10$\arcsec{} for each galaxy. 
While smoothing to a coarser resolution can broaden the intrinsic superprofile because the measured velocity dispersion increases at larger physical scales, matched-resolution cubes allow a more robust comparison from galaxy to galaxy by sampling the \hi{} kinematics on the same physical scale. % due to the increase of velocity dispersion at larger scales.
Since the more massive THINGS galaxies tend to be at larger distances, the varying spatial resolution could lead to spurious trends with any quantity that correlates with galaxy mass.

The superprofiles that we derive are systematically different from \citetalias{Ianj2012} even if we fit them with the double Gaussian function.
For the nine THINGS galaxies in common (DDO 53, DDO 154, Ho I, Ho II, IC 2574, M81 DwB, NGC 2366, NGC 4214, NGC 7793), we typically  measure similar \snarrow{} values, but find systematically smaller values of \sbroad{} by $0.5 - 3$ \kms{}.
This offset is also apparent in the $\snarrow{} / \sbroad{}$ ratio; we find an average $\snarrow{} / \sbroad{} = 0.46$, nearly 25\% higher than measured by \citetalias{Ianj2012}.

We believe that this difference could result from generating our superprofiles from the flux-rescaled cubes instead of the standard cubes; from using robust-weighted cubes instead of natural-weighted cubes; or from choosing different S/N thresholds.

First, the use of flux-rescaled cubes likely lowers the amount of flux in the wings of the profile.
While the central peak of the superprofile is due to the regions of the line-of-sight spectra that are the brightest, the wings arise from low-level flux to either side of the peak.
When we consider an individual line-of-sight \hi{} spectrum, the peak of the spectrum has a higher fraction of flux in the clean components than in the residuals when compared to the lower-level emission to either side of the peak.
Because the flux-rescaling correction effectively lowers the amount of flux in the residuals, the low-level emission on either side of the peak in each line-of-sight spectrum is rescaled to a relatively smaller value compared to the peak regions.
The flux-rescaling correction therefore has the effect of narrowing line-of-sight profiles, which then produces a narrower central peak in the superprofiles and less flux in the wings compared to superprofiles generated from the standard (non-rescaled) data cube.
With smaller wing amplitudes, the broad Gaussian component \sbroad{}, is smaller, which could then make the $\snarrow{} / \sbroad{}$ ratio larger.

Second, the synthesized beam of natural-weighted data cubes, which were used by \citetalias{Ianj2012}, exhibits a positive pedestal that extends to large radii which is not present in the robust-weighted data cubes that we have used to generate the superprofiles. 
Because of the broad pedestals in the synthesized, naturally-weighted beams, each apparently independent beam, as judged by its FWHM, actually includes flux from a much greater area.
These pedestals can therefore lead to additional velocity smearing, beyond what one would expect based on the velocity field of the galaxy and the angular size of the beam.
The neighboring pixels that are included at low-level, however, have offset velocities due to the overall rotation of the galaxy.
Therefore, each line-of-sight spectrum also includes flux from neighboring pixels at offset velocities.
This additional flux at offset velocities may translate to more flux in the wings of the superprofile, and therefore larger measured values of $\sigma_b$ and smaller $\sigma_n / \sigma_b$ ratios when compared to the robust-weighted cubes that lack these positive pedestals.

Finally, the signal-to-noise threshold was different between the two studies. At lower S/N values, determination of the peak is more difficult (see Figure~\ref{fig:mc-vp-offsets}). 
Because the determination of \vp{} is worse for low S/N spectra, additional flux could be incorrectly added into the wings of the superprofile due to these offset \vp{} values. 
This addition of flux into the wings would likely widen the \sbroad{} measurement and lower the \snarrow{} / \sbroad{} value.

\section{Comparison with Physical Properties}
\label{sec:correlations}

In this section we examine how global physical properties of the galaxies correlate with the properties of the global \hi{} superprofiles, as characterized by \scentral{}, \swing{}, \fw{}, and \aw{} (\S~\ref{sec:analysis--parameterization}).

For all tests, we characterize the strength of the correlation compared to a random sample using the Spearman rank correlation coefficient. 
This method makes no assumptions about the functional relationship between the two input data sets, and instead tests only for a monotonic relationship between the two variables. 
The Spearman correlation coefficient $r_s$ varies between -1 (monotonically decreasing) and 1 (monotonically increasing) with 0 implying no relationship. 
The significance of the observed value of $r_s$ is given by $p_s$, where $p_s \leq 0.01$ means that random uncorrelated data produce a $r_s$ value at least as extreme as measured $\leq 1$\% of the time. 
We adopt this threshold to indicate a statistically significant relationship between two quantities.

For all following comparisons, we list $r_s$ and $p_s$ %, as well as $r_p$ and $p_p$ if appropriate,
in Table~\ref{tab:correlations}. 
The coefficients for significant correlations ($p_s \leq 0.01$) are shown in bold.

Throughout this section we also present a number of figures showing the behavior of the superprofiles with a different physical property, starting with Figure~\ref{fig:panels-mbaryon}. 
The four upper panels show \scentral{}, \swing{}, \fw{}, and \aw{}. 
Colors are determined by the physical property itself, with red indicating a low value and blue indicating a high value. 
In the lower panel, we show the scaled superprofile residuals from the HWHM Gaussian fit, as previously seen in the lower panel of Figure~\ref{fig:superprofiles-scaled}. 
In this case, though, we have added color to each line to highlight how the superprofile shape changes with that physical property. 
The colors of each line are the same as the corresponding point in the upper panels, though the lines have a transparency value so that overlapping regions are clearer. 
To better highlight any asymmetry of the superprofiles, we have reversed the velocity axis if necessary such that the wing peak with the higher area is on the left. 
We now discuss these figures for correlations with the physical properties calculated in \S~\ref{sec:sample--other-data}. 
We also remind the reader that many of these properties are physically connected with each other, as shown in Figure~\ref{fig:property-correlations}, %and Table~\ref{tab:phys-correlations},
 so a correlation with one property could be causally due to another.

\subsection{Correlations with Galaxy Mass and Related Quantities}
\label{sec:correlations--mass}

We start by examining trends in the superprofile parameters versus quantities that correlate with galaxy mass. 
While we would not expect the local \hi{} conditions to know much about the overall galaxy potential, galaxy mass is correlated with a host of other properties that are more likely to have direct effects on the \hi{}, such as SFR or \m{HI}. %We show these correlations in Figure~\ref{fig:panels-mbaryon}. 

\subsubsection{Halo Mass}
\label{sec:correlations--mass--mbaryon}

As a proxy for galaxy halo mass, we use \m{baryon,tot} (\S~\ref{sec:data--other-data--mhalo}).
In Figure~\ref{fig:panels-mbaryon} we plot the correlations between galaxy mass and the parameters \scentral{}, \swing{}, \fw{}, and \aw{} derived from the superprofiles.
Neither \scentral{} nor \swing{} shows any significant trend with galaxy mass.
This lack of correlation confirms our expectation that the driver of \hi{} kinematics is not strongly mass-dependent; that is, it is more likely to be dependent on specific phenomena in the ISM.

However, Figure~\ref{fig:panels-mbaryon} does show a strong correlation between galaxy mass and the asymmetry \aw{}, with more massive galaxies exhibiting less asymmetry.
The asymmetry can be seen in the residuals; the red lines of low-mass galaxies show a large variation between one side of the residuals and the other, while the blue lines of higher-mass galaxies have more symmetric structure in the residuals.
However, it is likely that this trend reflects of galaxy properties such as SFR that correlate with mass, instead of mass itself. 
Asymmetries in the line-of-sight spectra induced by star formation may propagate to asymmetries in the superprofiles, as discussed further in \S~\ref{sec:disc--wings--asymmetry}, but are averaged out when more star forming regions are present.

We also find a trend between galaxy mass and \fw{}.
The $p_s$ value is 0.017, only slightly above our cutoff for a significant correlation.
There is a tentative indication that lower mass galaxies with $\m{baryon,tot} \lesssim 5 \times 10^8$ \msun{} tend to have a larger scatter in \fw{}. Galaxies with larger \m{baryon,tot} exhibit higher \fw{} values with less scatter, showing half the standard deviation relative to the median compared with their lower mass counterparts.
If this trend holds true with a larger sample, it would be consistent with higher mass galaxies being able to more consistently drive \hi{} into the wings of the superprofile. 
The presence of high velocity gas in lower mass galaxies could be due to more stochastic processes, with not every galaxy being able to launch high velocity gas at all times, thus leading to more scatter in \fw{}.

We have also looked for correlations with other mass tracers, such as \mhi{}, \mstar{}, and $w_{20}$. 
While the exact values of $r_s$ and $p_s$ change, we find similar trends between superprofile properties and other mass tracers. 
Such behavior is expected  expected based on the strong correlations between \m{baryon,tot} and other mass tracers shown in Figure~\ref{fig:property-correlations}.

\subsubsection{Star Formation Rate}
\label{sec:correlations--mass--sfr}

We plot the behavior of the superprofile parameters with SFR in Figure~\ref{fig:panels-sfr}.
As shown in Figure~\ref{fig:property-correlations}, SFR is strongly correlated with galaxy mass.
We are therefore not surprised to see nearly-identical correlations as those in Figure~\ref{fig:panels-mbaryon}. 
We do find statistically significant correlations between SFR and both \fw{} and \aw{}.
The trends can be seen in the upper panels of Figure~\ref{fig:panels-sfr}, where galaxies with higher SFRs (blue) tend to have higher \fw{} values and lower \aw{} values compared to the galaxies with lower SFRs (red).
This behavior can also be seen in the superprofile residuals plotted in the lower panel of Figure~\ref{fig:panels-sfr}; the blue residuals, with the highest SFR, are also among the largest and most symmetric.
The red lines are typically lower and exhibit varying levels of asymmetry.

Given that the correlations between \fw{} and \aw{} are the same as the correlations with mass, it is possible that the \fw{} and \aw{} correlations are causally connected to any of the other galaxy properties that scale with mass.
However, of all these properties, SFR is the only one that provides a physical mechanism for driving \hi{} gas in the wings.
It may therefore be the actual driver of the correlations with mass. We explore this connection further in \S~\ref{sec:correlations--sf} below.

\subsection{Star Formation}
\label{sec:correlations--sf}

Star formation is typically proposed as the primary driver of \hi{} turbulence in spiral galaxies within $\sim r_\mathrm{25}$ \citeeg{Tamburro2009}. 
If so, then we may expect to find correlations between measures of star formation and the superprofile parameters. 
While the total SFR provides a measure of the overall energy input from star formation, it is strongly dependent on galaxy mass. 
In addition, galaxies with higher SFRs, and thus higher masses, also have larger \hi{} masses to affect with the energy provided by star formation.

We consider two possible measures of star formation other than SFR, which was considered previously in \S~\ref{sec:correlations--mass--sfr}. %These include the global SFR, reflective of the overall energy input from star formation but also strongly dependent on galaxy mass due to the larger gas reservoirs in more massive disk galaxies. 
These include the SFR intensity (i.e., \ave{\sfrsd{}}; \S~\ref{sec:data--other-data--sfr}), which measures the SFR concentration and thus may correlate with the efficiency of locally accelerating \hi{}; and SFR / \mhi{} (\S~\ref{sec:data--other-data--sfr-per-mhi}), which measures the ratio between the available energy of star formation and the mass of gas that the energy must couple to.

\subsubsection{\sfrsd{}}
\label{sec:correlations--sfr--sfr-per-area}

In Figure~\ref{fig:panels-sfr-per-area} we show the relationship between \ave{\sfrsd{}} and the superprofile parameters. 
We find no significant correlations with \scentral{}, \fw{}, or \aw{}. 
However, \swing{} shows a trend with \ave{\sfrsd{}}, such that galaxies with higher $\langle \sfrsd{} \rangle$ values are able to drive \hi{} into the wings with faster average velocities. 

The existence of the correlation between \swing{} and \ave{\sfrsd{}} is not surprising. 
Since energy input from star formation is a local process, more concentrated star formation should be more effective at inducing anomalous motions in the surrounding \hi{}. 
At the same efficiency of converting the star formation energy into kinetic \hi{} energy, a higher concentration of star formation energy imparts a given amount of energy into a smaller mass of \hi{},
thus driving \hi{} in the wings to higher velocities. 
We further explore the connection between star formation energy and kinetic energy in the wings of the superprofile in \S~\ref{sec:disc--energy--wings}.

\subsubsection{SFR / \mhi{}}
\label{sec:correlations--sfr--sfr-per-mhi}

We might also expect SFR per unit M$_\mathrm{HI}$ to affect \hi{} kinematics. 
As discussed in \S~\ref{sec:data--other-data--sfr-per-mhi}, this quantity is best interpreted as the ratio between the rate of energy input from star formation and the mass of gas that can be accelerated by that energy instead of the average SFE.

We show the superprofile parameters as a function of SFR / \mhi{} in Figure~\ref{fig:panels-sfr-per-mhi}. 
We find a correlation between \fw{} and SFR / \mhi{}. 
As a potential explanation for this trend, we invoke the argument that galaxies with higher SFR and lower \mhi{} are better able to accelerate the surrounding \hi{} into the wings of the superprofile compared with their counterparts. 
However, for the most common values of SFR / \mhi{} $\sim 10^{-10}$ \per{yr}, \fw{} values vary by $\sim40$\%, suggesting a large degree of stochasticity in this correlation.
It is also unclear why SFR / \mhi{} affects \fw{} but not \swing{}.

\subsection{Surface Mass Density}
\label{sec:correlations--mass-density}

Disks with higher surface mass density are able to permit more turbulent motions in gas that is still bound to the disk \citeeg{vanDerKruit1981}. 
In addition, a number of gravitational instabilities have been proposed to drive turbulence in galaxy disks \citeeg{Huber2001, Wada2002}, so the amplitude of turbulent motions might be expected to scale with surface mass density.

We find correlations between \ave{\hisd{}} and both \scentral{} and \swing{}, as shown in Figure~\ref{fig:panels-mhi-per-area}.
This correlation may be indicative of turbulence as driven by gravitational instabilities, as explored further in \S~\ref{sec:disc--energy--central}, or of a coupling with \ave{\sfrsd{}}, which tends to scale with \hisd{}.
We note, however, that no correlation was observed between \scentral{} and \sfrsd{}.
A similar correlation with \ave{\Sigma_\mathrm{baryon}} exists with \swing{}, but not with \scentral{}.

\subsection{Inclination}
We end by confirming that the superprofile parameters are not strongly influenced by galaxy inclination, as can be seen in Figure~\ref{fig:panels-i}. 
There are no significant trends in any of the parameters. 
However, we note that uncertainties in the inclination of the dwarfs could potentially mask any underlying systematic effects.

\subsection{Extending the Correlations to Higher Mass Galaxies}
\label{sec:correlations--spirals}

A number of massive spirals were eliminated from our sample based on the selection criteria to ensure high-quality data (\S~\ref{sec:sample--selection}). 
In this section we now include eight of these galaxies to see if the correlations we identify among the dwarfs could hold when extended to higher mass galaxies. 
Since this is a simple check, we do not perform the same rigorous tests as we have for the primary dwarf sample. 
Instead, we merely present our results and assess how the correlation coefficients change with the inclusion of more massive galaxies.

For the higher mass galaxies, we relax selection criteria (1) and (2) as given in \S~\ref{sec:sample--selection} by working at a spatial resolution of 400 pc and by including observations with $\Delta v = 5.2$ \kms{}, with the caveat that we have not characterized the effect of observational properties on these superprofiles. 
We also include galaxies with de Vaucouleurs T-type $\geq 2$. 
Finally, we only use galaxies that are in the clean sample of \citetalias{Ianj2012} to avoid contamination from bulk inflows, outflows, or \hi{} in the Milky Way. 
These relaxed criteria give us eight additional, higher-mass galaxies (NGC~628, NGC~2403, NGC~2903, NGC~2976, NGC~3351, NGC~4736, NGC~5055, and NGC~5236).

For each galaxy, we produce superprofiles in the same manner as we have for the dwarfs in our sample. 
We first smooth the data to a common physical resolution of 400~pc (following \S~\ref{sec:sample--hi-data-prep}). 
We then generate a superprofile for that galaxy after applying the same S/N $ > 5$ threshold as for the primary sample (\S~\ref{sec:superprofiles}). 
Finally, we parameterize each superprofile using the HWHM-scaled Gaussians (\S~\ref{sec:analysis--parameterization}).  
We list the derived superprofile parameters for these galaxies in Table~\ref{tab:measurements-extend400}.

We assess the strength of the correlations with physical properties after including the higher-mass galaxies. % for the significant correlations given in Table~\ref{tab:correlations}.
In Figure~\ref{fig:extend-400} we show relevant correlations between the superprofile parameters and the physical properties with which they are correlated after the inclusion of high-mass galaxies.
The displayed panels are:
(1) \scentral{} versus \ave{\hisd{}};
(2) \scentral{} versus \ave{\sfrsd{}};
(3) \swing{} versus \ave{\hisd{}};
(4) \swing{} versus \ave{\sfrsd{}};
(5) SFR / \m{HI} versus \fw{};
and
(6) SFR versus \aw{}.
The dwarf galaxies from our sample are shown with filled circles, while the higher mass galaxies are shown with open circles.

We find that \scentral{} is no longer correlated with \ave{\hisd{}}, but a trend with \ave{\sfrsd{}} is present with $r_s = 0.404$ and $p_s = 0.024$. 
Although this $p_s$ value is marginally higher than our cutoff, the trend lends credence to the idea that the observed correlation between \scentral{} and \ave{\hisd{}} for the dwarf sample may be tracing a correlation with \sfrsd{}.
%, but with a $p_s$ value higher than our cutoff.

Similarly, \swing{} no longer shows correlations with \ave{\hisd{}} but does with \ave{\sfrsd{}}, implying that \sfrsd{} does indeed affect the gas in the wings. 
The idea that \hi{} gas in the wings of the superprofile is influenced by star formation is also supported because correlations between \fw{} and SFR / \m{HI} and between \aw{} and SFR both remain with the inclusion of higher mass spirals.

%%%%%
%%%%% Discussion
%%%%%
\section{Discussion}
\label{sec:disc}

In this section we discuss the implications for the width of the central peak of the superprofile (\S~\ref{sec:disc--scentral}) as well as the correlations we see with the wings of the superprofile (\S~\ref{sec:disc--wings}).
We then compare the energy available from star formation to the kinetic energy in the \hi{} gas  (\S~\ref{sec:disc--energy}).
Finally, we approximate implied \hi{} scale heights for the sample (\S~\ref{sec:disc--scale-height}) and discuss the similarity of the superprofiles' shapes (\S~\ref{sec:disc--universal-hi-profile}).

\subsection{\hi{} in the Central Peak}
\label{sec:disc--scentral}

The width of the central peak, \scentral{}, has a very small range in our sample, with a median of 7.7 \kms{} and interquartile range of only $\pm 1$ \kms{}.
The small range of observed \scentral{} values may suggest that average turbulence in the ISM is regulated in some way, such as by energy input from physical processes or external heating from the UV background \citeeg{Schaye2004, Tamburro2009}.

The only correlation we have found between \scentral{} and globally-averaged physical properties is with \ave{\hisd{}}, but this correlation disappears with the inclusion of higher mass spirals.
Interestingly, we find no trend between \scentral{} and our measurements of star formation.
In standard lore, star formation is the primary driver of turbulence in the warm ISM.
If star formation were the sole driver of \hi{} line widths in galaxies, we would initially have expected to have seen a connection between some measure of SFR, \ave{\sfrsd{}}, or SFR / \mhi{}.
%However, we caution that the superprofiles for the galaxies are a global average of \hi{} line profiles, so they 
%There may be an explanation for this mismatch.
We now give some potential explanations for this mismatch.

Because regions with higher \ave{\hisd{}} also tend to have higher \ave{\sfrsd{}}, the correlation between \scentral{} and \ave{\hisd{}} could in actuality a correlation between \scentral{} and \ave{\sfrsd{}}.
However, this interpretation is called into question by the lack of correlation between \scentral{} and  \ave{\sfrsd{}} as traced by FUV+24\um{} emission.
The FUV+24\um{} tracer probes SFR averaged over the past $10 - 100$~Myr, with the implicit assumption that the SFR has been constant over that timescale.
It is unlikely that this is the case in our sample galaxies.
Because the timescales over which turbulent gas in the central peak can dissipate energy are $\sim 10$ Myr,
the FUV+24\um{} timescale may be a poor match to the timescales relevant for the \hi{} component.
If we interpret the central peak as turbulent, the \hi{} in the sample galaxies is able to dissipate its energy in $\sim 10^7$ yr. 
Therefore, the star formation that has occurred in the last $\sim 10^7$ yr is the primary influence on the gas.
Because our SFR measurement has been averaged over a longer time, galaxies with similar average SFRs on $\sim 10^8$ yr timescales may in fact may have different SFRs in the past $\sim 10^7$ yr.

We caution that the superprofiles for the galaxies are a global average of individual \hi{} line-of-sight spectra, so they often include regions in a single galaxy with very different star formation properties.
The loss of spatial information in the superprofiles could account for the lack of correlation between star formation and \scentral{}, especially when regions with prominent star formation are mixed with those that lack strong star formation.
We explore the spatial dependence of velocity dispersion in a subsequent paper.
%The galaxies with the highest \ave{\hisd{}} values are often in the low signal-to-noise regime.
%In these cases, only the regions with the highest \hisd{} are included in the superprofile.

It is certainly possible that the correlation with \ave{\hisd{}}, and not with \ave{\sfrsd{}}, indicates that there is indeed no physical connection between the central peak and star formation.
There are hints of this possibility in the \citet{Tamburro2009} results.
In the inner regions of their more massive galaxies, they found a correlation between $\sigma_\mathrm{m2}$ and \sfrsd{}.
However, this correlation broke down at large radii, where $\sigma_\mathrm{m2}$ approaches a nearly constant value but star formation falls off dramatically.
Given the similarity between dwarf galaxies and the outer disks of spirals (in terms of \sfrsd{}, \hisd{}, $\Sigma_\star$, etc.), it is possible that our sample lies primarily in this regime.

If star formation is not setting the velocity dispersion, what else is?
One commonly-adopted mechanism in spirals is the MRI, which works well in the outskirts of massive spiral galaxies where angular velocity declines with radius.
In dwarfs, however, the rotation curves across much of the observable disk are closer to solid body rotation, and therefore lack the strong differential rotation necessary for the MRI.
Most dwarfs in the Local Group also show magnetic field strengths approximately three times smaller than that of spirals \citep{Chyzy2011}.
The combination of these two factors means that the MRI should be less effective in the dwarf galaxies that comprise our sample.
Since more massive galaxies host conditions that are more conducive to MRI-driven turbulence, we might expect to see a correlation between galaxy mass and \scentral{}.
However, no such trend is present.
While it is possible that the range in \scentral{} values is too small to measure such a trend, it would also be surprising if the MRI conspired to produce such similar \scentral{} values across the sample without any external regulation.

Other energy sources for turbulence in galaxies include gravitational instabilities.
Many of these, however, require shear in the rotating gas to function and thus fall prey to the same problems as the MRI.
The most promising of these instabilities is that presented in \citet{Wada2002}, which does not explicitly require shear to drive turbulent velocity dispersions and hearkens back to the observed correlation between \scentral{} and \ave{\hisd{}}.
While the authors have shown that this method can drive turbulence at levels observed in NGC~2915, others have noted that this instability provides energy at levels that are two orders of magnitude smaller than that required to drive the observed turbulence in the ISM \citep{MacLow2004}.
We assess the ability of this gravitational instability to provide enough energy to drive turbulence in our sample galaxies in \S~\ref{sec:disc--energy--central--grav}.

Another proposed method is UV heating, which can drive thermally-broadened line widths to $\sim 6$ \kms{} \citep{Tamburro2009}.
In this case, the widths are due to thermal effects, not turbulence.
However, the measured \scentral{} values of the superprofiles show a much wider temperature range than can be explained by UV heating.
Perhaps, however, UV heating can sets the base velocity dispersion of \hi{} profiles, and any additional dispersion is driven by other physical processes, such as star formation or instabilities.

The mechanism that drives \hi{} velocity dispersions remains an open question and will likely become clearer in future spatially-resolved studies.

\subsection{The Energetic \hi{} in the Wings}
\label{sec:disc--wings}

Compared with \scentral{}, the properties of the superprofile wings are more correlated with galaxy physical properties.
We found that:
(1) the characteristic velocity of the wings, \swing{}, increases with both \sfrsd{} and \hisd{}; 
(2) the fraction of the gas in the wings, \fw{}, increases most strongly with SFR / \mhi{}; 
and 
(3) the asymmetry, \aw{}, is primarily in the wings and decreases with increasing SFR. In this section we discuss potential physical explanations for these trends.

\subsubsection{\sfrsd{}, \hisd{}, and \swing{}}
\label{sec:disc--wings--swing}

We find correlations between \swing{} and both \ave{\sfrsd{}} and \ave{\hisd{}}.
The fact that both correlations exist may be due to the \ave{\sfrsd{}} - \ave{\hisd{}} connection.
The correlation between \swing{} and \hisd{} may also partially be due to the correlation between \scentral{} and \hisd{}; galaxies with wider central peaks, and therefore higher values of \hisd{}, will by definition also have higher \swing{} values, as we consider only gas moving faster than expected compared to the central Gaussian when calculating \swing{}.
We therefore discuss \swing{} properties in relation to star formation, a potential driver for high-velocity gas in galaxies.

The behavior of \swing{} and \sfrsd{} can be explained if the energy from star formation pushes \hi{} to higher velocities than expected from the \scentral{} Gaussian.
Expanding \hi{} structures have been observed in numerous studies down to the instrumental resolution \citeeg{Brinks1986, Bagetakos2011}, including many smaller structures in the Milky Way \citeeg{Ehlerova2005}.
Presumably smaller expanding \hi{} structures exist below the current limits of spatial resolution.
Both Type 2 and Type 3 holes, as defined by \citet{Brinks1986}, show \hi{} offsets in velocity space along a single line-of-sight spectrum, which would contribute \hi{} emission to the wings of a global superprofile.

\hi{} holes are often thought to be due to star formation \citeeg{McCray1987}, although direct spatial correlation with young, massive stars is not always seen \citeeg{Rhode1999}.
Recent studies have found that multiple star formation events over the age of the hole do provide enough energy to drive \hi{} hole formation, though other regions show similar star formation histories without the presence of \hi{} holes \citeeg{Weisz2009, Warren2011, Cannon2011}.
\hi{} gas with anomalous velocities has also been linked to the presence of \hi{} holes and star forming regions in the spiral NGC~6946 \citep{Boomsma2008}.
However, projection effects make these measurements difficult, so such studies are only appropriate in a small number of face-on spirals.

If similar expanding \hi{} structures exist at smaller spatial scales, we can explore their connection with superprofile parameters in more detail using the canonical Chevalier equation \citep{Chevalier1974}, which relates star formation energy and \hi{} hole properties:
\begin{equation}
E \left( \mathrm{erg} \right) = 5.3 \times 10^{43} \left( \frac{n_0}{ \mathrm{cm}^{-3} } \right)^{1.12} \left( \frac{r_\mathrm{hole}}{\mathrm{pc}} \right)^{3.12} \left( \frac{v_\mathrm{exp}}{1 \; \mathrm{km \; s}^{-1}} \right)^{1.4}.
\end{equation}
Here, $E$ is the enclosed, single-burst energy required to drive the expansion, $n_0$ is the density of the surrounding ambient medium, $r_\mathrm{hole}$ is the radius of the hole, and $v_\mathrm{exp}$ is the expansion velocity. 
% chevalier caveats
This equation provides a description of the energy of an expanding shell in an idealized ISM, with the assumption of uniform density which does not hold true on global scales.
This equation is also complicated by the fact that multiple bursts can provide energy to drive \hi{} hole expansion.
However, it still allows us to explore the relation between the physical quantities involved in the expansion. 
The energy density of the hole, $\sim E_\mathrm{hole} / r^3$, scales most strongly with $v_\mathrm{exp}$, with weaker dependencies on $n_0$ and $r_\mathrm{hole}$.
If we assume that the energy source for the hole is due to star formation, a higher concentration of star formation energy within the hole should lead to faster expansion velocities.
Because there is no reason not to expect smaller expanding structures below the 200 pc resolution of our data, these structures may manifest as the high-velocity wings observed in the superprofiles.
Thus, for a given \hi{} mass, higher concentrations of energy due to larger \sfrsd{} should lead to faster expansion velocities and therefore to higher measured \swing{} values.
It is unclear, however, why \sfrsd{} would influence \swing{} without also driving more gas into the wings of the profile, as measured by \fw{}.

\subsubsection{SFR / \mhi{} and \fw{}}
\label{sec:disc--wings--fw}

The fraction of gas in the wings is correlated not with \sfrsd{} but with SFR / \mhi{} for dwarf galaxies. 
Such behavior can again be ascribed to expanding \hi{} structures. 
At a fixed SFR, and therefore a fixed energy input into the ISM, galaxies with smaller \hi{} masses should show more pronounced effects on the \hi{} kinematics. 
Therefore, galaxies with relatively high SFRs compared to their \hi{} masses should be able to perturb the \hi{} content more easily, as seen in the correlation between SFR / \mhi{} and \fw{}. 
More \hi{} is pushed into the wings of the superprofile if the galaxy has a high SFR relative to its \hi{} content, or if it has a smaller amount of \hi{} to move around with the energy available from star formation.

The correlations between \fw{} and \swing{} indicate that star formation does indeed play a role in driving \hi{} to anomalous velocities
seen in the wings of the superprofile. 
It is unclear why \fw{} and not \swing{} would scale with SFR / \mhi{}. However, the correlation coefficient between SFR / \mhi{} and \swing{} implies $p_s = 0.066$, which indicates only a 7\% probability of finding a correlation this extreme from a random sample.

\subsubsection{Star Formation as a Driver of Asymmetry}
\label{sec:disc--wings--asymmetry}

Since the properties of the superprofile wings appear to be connected to star formation, it is not surprising that asymmetries in the wings can also be attributed to star formation. 
Due to the inhomogeneity of the ISM, individual star forming regions can affect the local \hi{} gas asymmetrically.
This effect can then result in asymmetric line profiles near star forming regions, which contributes to asymmetry in the wings of the global superprofile.
The average of a large number of asymmetric \hi{} regions should average out to produce a symmetric superprofile, while the average of only a few asymmetric regions is more likely to retain net asymmetry in the superprofile.
If galaxies with larger SFR have more individual star forming events compared to their counterparts with smaller SFR, we would expect an anti-correlation between SFR and \aw{}, as is observed.
We note that it may also be easier for more massive galaxies to remove the signatures of asymmetric \hi{} motions, regardless of their origin, due to their deeper gravitational potential wells.

Asymmetric \hi{} motions have already been observed near star forming regions.
\citet{Young2003} found that \hi{} line-of-sight spectra exhibit asymmetry near regions of star formation on 200 pc scales, though this behavior was not seen in the sample observed by \citet{Begum2008}, with somewhat more coarse resolutions of 300 - 700 pc. 
We also find that some galaxies show asymmetric profiles near regions of star formation. 
To measure this, we use the difference between \vp{} from \gh{} fits and the intensity-weighted mean velocity from the first moment map ($v_\mathrm{IWM}$). 
As previously seen in Figure~\ref{fig:peak-fit-methods}, $v_\mathrm{IWM}$ and \vp{} from \gh{} fits are offset for asymmetric profiles.  
In Figure~\ref{fig:sfr-asymmetry}, we plot \sfrsd{} compared to the difference between \vp{} and $v_\mathrm{IWM}$/
Red indicates regions where $\vp{} < v_\mathrm{IWM}$, while blue shows regions where $\vp{} > v_\mathrm{IWM}$ \kms{}.
Transparency has been added to show the underlying \sfrsd{} in grey, and regions where $\vp = v_\mathrm{IWM}$ are more transparent than regions where they are different.
In these galaxies, some star formation regions also show asymmetric profiles. 
The spatial overlap between star formation and asymmetric profiles is not proof that star formation is the cause of asymmetry, as there are regions with no apparent star formation that also show asymmetric profiles, but it indicates that star formation may be one cause of asymmetry in \hi{} line-of-sight spectra. 
If this is the case, the correlation between \aw{} and SFR is not surprising.

%\textbf{Leave the rest of this section in or remove it?}
We next estimate the number of star forming events, $N_\mathrm{SF}$, in a galaxy based on its SFR plus a fiducial star formation timescale and mass. We first assume that star formation is linearly proportional to \htwo{} mass, with a timescale $\tau_\mathrm{dep} \sim \Sigma_\mathrm{H_2} / \Sigma_\mathrm{SFR}$. This timescale estimates approximately how long it will take a galaxy to use up its entire \htwo{} reservoir, and has been found to remarkably independent to environment \citep{Bigiel2011, Bolatto2011}. We can then estimate $N_\mathrm{SF}$:
\begin{equation}
N_\mathrm{SF} = \frac{ \mathrm{SFR} \times \tau_\mathrm{dep} }{ \m{SF} }
\end{equation}
where \m{SF} is the typical mass of a star forming region. The numerator is an estimate of \htwo{} mass in our galaxies, while the denominator is the average mass of a star formation clump.

It is becoming clear that $\tau_\mathrm{dep}$ may be universal as it does not appear to vary much from galaxy to galaxy \citeeg{Bigiel2011}.
\citet{Bolatto2011} measure $\tau_\mathrm{dep} \sim 1.6$ Gyr in the Small Magellanic Cloud for the same spatial scales of 200~pc as our data, a value very similar to the $\tau_\mathrm{dep} \sim 2.35$ Gyr measured by \citet{Bigiel2011}.
The SMC value is likely the best comparison, since it has an \hi{} mass and SFR similar to our sample \citep{Stanimirovic1999, Harris2004}.
If we assume that star formation arises from Giant Molecular Clouds (GMCs) with average masses of \m{MC}, whose sizes are well-matched to our 200~pc scale and therefore provide an estimate of the number of 200~pc resolution elements with star formation, we can calculate the number of spatially-resolved elements with star formation:
\begin{equation}
N_\mathrm{SF} = 1.6 
\left(  \frac{ \mathrm{SFR} }{ 1 \times 10^{-3} \; \msun{} \; \mathrm{yr}^{-1}} \right) 
\left( \frac{ \tau_\mathrm{dep} } { 1.6 \times 10^9 \; \mathrm{yr}} \right) 
\left( \frac{ \mathrm{\m{MC}} }  { 1 \times 10^6 \; \msun{} } \right)^{-1} .
\end{equation}
This equation provides an approximation of the number of resolution elements in our sample that have star formation.

At the low end of our sample, the observed SFR $\sim 1 \times 10^{-3}$ \msun{} \per{yr} \per[2]{kpc} yields only a few regions of active star formation. 
Indeed, many of the low-mass galaxies in our sample show only a few clumps of star formation as traced by FUV + 24\um{} emission, while the higher mass galaxies have a more widespread, smooth star formation distribution across their disks. 
With our above assumption that each SF event has a chance to drive asymmetric \hi{} outflows, we would therefore expect that these potential asymmetries do not always average out at the low star formation rates characteristic of our sample. 
At the higher end of our sample, where SFR $\sim 1 \times 10^{-1}$, we would expect a few hundred regions of active star formation. 
While each individual region may produce asymmetric \hi{} motions, the average over the large number of SF events in the entire galaxy produces an overall symmetric distribution. 
However, very large \hi{} holes or extreme star formation events could still produce observable asymmetric \hi{} outflows even in galaxies with relatively high star formation rates. 
This behavior has been seen in NGC~2366 \citep{vanEymeren2009}, which also has a high $\aw{} = 0.22$ value compared to its relatively high SFR, though it is unclear if this asymmetry is due to star formation or the high degree of non-circular motions in the northwestern region \citep{Oh2011}.

\subsection{Comparison of Energy in the \hi{} gas to Energy Sources}
\label{sec:disc--energy}

In this section we estimate the kinetic energy of the \hi{} gas using the superprofile and compare it to the energy available from physical processes.
In \S~\ref{sec:disc--energy--central} we assess the ability of both the \citet{Wada2002} gravitational instability (\S~\ref{sec:disc--energy--central--grav}) and star formation (\S~\ref{sec:disc--energy--central--sf}) to provide enough energy to drive turbulence at levels indicated by the central peak.
We then examine the efficiencies required for star formation to move \hi{} into the wings of the superprofiles (\S~\ref{sec:disc--energy--wings}).
Finally, we discuss whether star formation can drive the full \hi{} kinematics measured by the superprofile (\S~\ref{sec:disc--energy--full}).

\subsubsection{Energy in the central \hi{} peak}
\label{sec:disc--energy--central}

We first estimate the kinetic energy contained in the central peak of the superprofile.
We assume that the majority of gas in the central peak can be reasonably approximated by a Gaussian profile with a width \scentral{} and that the velocity dispersion is isotropic in three dimensions.
The energy in the central peak is therefore:
\begin{equation}
E_\mathrm{HI,central} = \frac{3}{2} \left( 1 - \fw{} \right) \, (1 - f_\mathrm{cold}) \, \m{HI} \, \scentral{} ^ 2
\end{equation}
where \m{HI} is the measured \hi{} mass of the superprofile.
The $(1 - \fw{})$ factor accounts for the fraction of \hi{} in the wings, so $(1 - \fw{}) \m{HI}$ represents the \hi{} mass contained in the central peak of the superprofile.
The $f_\mathrm{cold}$ variable represents the fraction of \hi{} gas that is in the cold phase, which has a narrower velocity dispersion compared to the average turbulent component.
The velocity dispersion of this gas is likely thermal, and therefore cold \hi{} should not be included in our calculation of turbulent energy.
Typical measured fractions are $f_\mathrm{cold} \lesssim 10 - 20$\% \citeeg{Young2003, Warren2012}.
Without strong constraints on individual values for cold \hi{} fractions, we choose $f_\mathrm{cold} = 0.15$ for our analysis.

\subsubsection{Energy from Gravitational Instabilities}
\label{sec:disc--energy--central--grav}

The energy provided by gravitational instabilities over a timescale $\tau$ is:
\begin{equation}
E_\mathrm{grav} = \dot \varepsilon_\mathrm{grav} \, (1 - \fw{}) (1 - f_\mathrm{cold}) \m{HI} \, \tau.
\end{equation}
As before, the $(1 - \fw{}) (1 - f_\mathrm{cold}) \m{HI}$ factor represents the approximate \hi{} mass contained in the turbulent central peak.

We now calculate the amount of energy released into the ISM based on the gravitational instability proposed by \citet{Wada2002}.
This instability allows the gas to extract energy from rotation instead of from shear, so it is a potential source of energy for dwarf galaxies that lie primarily in the regime of solid body rotation.
\citet{Wada2002} approximate the energy supply rate per unit mass as:
\begin{equation}
\frac{\dot \varepsilon_\mathrm{grav}}{ \mathrm{km^2 \, s^{-3}}} \sim
5 \left( \frac{\Omega}{\mathrm{s}^{-1}} \right)
\left( \frac{ \Sigma_\mathrm{gas} }{ 10 \, \msun \, \mathrm{pc}^2} \right)
\left( \frac{ \lambda }{100 \, \mathrm{pc}} \right)^2
\left( \frac{h_z}{100 \, \mathrm{pc}} \right)^{-1}.
\end{equation}
Here, $\Omega$ is the angular velocity, $\Sigma_\mathrm{gas}$ is the gas surface density, $\lambda$ is the scale length of turbulence, and $h_z$ is the scale height of the disk.
We approximate $\Omega \sim (w_{20} / 2) / (1.5 r_{25})$, where $w_\mathrm{20}$ has been corrected for inclination and $1.5 r_{25}$ is the approximate extent of \hi{} in galaxies.
This approximation provides an order-of-magnitude estimate for the angular velocity of gas in our sample \citep{Giovanelli1988, Swaters2009}.
We use the \ave{\Sigma_\mathrm{gas}} measurement for the gas surface density and choose $\lambda = 100$ pc following \citet{Tamburro2009}.
Based on studies of the scale height in dwarf galaxies by \citet{Banerjee2011}, we approximate the scale height of our sample as 500 pc.

We must then find the timescale over which the \hi{} can dissipate its energy.
If the \hi{} in the central peak is turbulent, this timescale is the turbulent timescale as given by \citet{MacLow1999}:
\begin{equation}
\tau_D \simeq 9.8 \; \mathrm{Myr} \; \left( \frac{\lambda}{ 100 \mathrm{pc} } \right)
  \left( \frac{ \sigma }{ 10 \, \mathrm{km \, s^{-1}} } \right)^{-1},
\label{eqn:tau_d}
\end{equation}
where $\lambda$ is the turbulent driving scale  and $\sigma$ is the \hi{} velocity dispersion.
Following \citet{Tamburro2009}, we estimate $\lambda \sim 100$ pc. 
We also set $\sigma = \scentral{}$ as measured from the superprofile for each galaxy. For our measured \scentral{} values, the turbulent timescale ranges from $9 - 16$ Myr.

To convert the available energy to \hi{} kinetic energy, the conversion efficiency, $\epsilon_\mathrm{grav } \equiv E_\mathrm{HI} / E_\mathrm{grav}$ must be taken into account.
We do not adopt a single value for $\epsilon_\mathrm{grav}$, and instead measure the range of the range of $\epsilon_\mathrm{grav}$ that is compatible with our data.
In general, implied efficiencies of $\epsilon_\mathrm{grav} > 1$ are unphysical in that the \hi{} component has more turbulent kinetic energy than can be provided over the timescale.

In Figure~\ref{fig:snarrow-e-grav}, we compare $E_\mathrm{grav}$ to the kinetic energy contained in the central peak.
In the left panel, we plot $E_\mathrm{grav}$ versus the energy in the central peak. 
We have shown the background in grey to represent the fact that the required efficiencies are $> 1$ and are therefore unphysical.
Dashed lines represent constant $\epsilon_\mathrm{grav}$.
The left panel shows the inferred value of $\epsilon_\mathrm{grav}$ necessary to drive turbulence.

In all cases, the gravitational instability cannot provide enough energy to drive the observed levels of \hi{} turbulence, falling short by a factor of $10 - 10^3$ in spite of the fact that the correlation between \ave{\hisd{}} and \scentral{} may have initially pointed at this driver.
The discrepancy between $E_\mathrm{HI, central}$ and $E_\mathrm{grav}$ appears to be more extreme in small galaxies with low values of $E_\mathrm{grav}$.
We note that in nearly all cases, changing any of our assumptions by a factor of two does not alter the result that the \hi{} harbors far more energy on average across the disk than the gravitational instability can provide.
The inability of this instability to drive turbulence has been noted before \citeeg{MacLow2004}, so our results confirm this idea.
It is possible that other gravitational instabilities are operating to produce the observed \scentral{} values, but any candidate instability must be able to function efficiently in galaxies with low internal shear.

\subsubsection{Energy from Star Formation}
\label{sec:disc--energy--central--sf}

We next turn to star formation as a driver of \hi{} velocity dispersion.
Even though there is no straightforward correlation between measures of star formation and \scentral{}, this assessment provides a limit on the efficiencies necessary to couple energy from star formation to the \hi{} gas if it is indeed, as widely regarded, the driver of \hi{} turbulence for the bulk of the gas.

From the measured star formation rate, we can estimate the energy released into the ISM by SNe over the turbulent dissipation timescale $\tau_D$ as:
\begin{equation}
E_\mathrm{SF} \sim \dot{E}_\mathrm{SF} \, \tau_D
\label{eqn:e-sf-hi}
\end{equation}
where $\dot{E}_\mathrm{SF} \tau_D$ is the total amount of energy released from the stellar population over one turbulent timescale.
This equation implicitly assumes that the rate of energy input from the stellar population has been constant over $\tau_D$.

To estimate $\dot{E}_\mathrm{SF}$, we use the formalism proposed by \citet{Tamburro2009}, assuming that the majority of the star formation energy is released by SNe explosions.
On average, each explosion provides $10^{51}$ ergs of mechanical energy.
We take the number of SN per unit stellar mass formed to be $\eta_\mathrm{SN} = 1.3 \times 10^{-2} $ SN \per{\msun{}}, assuming a \citet{Kroupa2001} IMF with an upper mass limit of 120 \msun{}.
We next assume that the SFR measured by FUV + 24\um{} observations has been constant over the $\tau_D$ timescale.
The total energy available to the \hi{} due to SNe is then:
\begin{equation}
E_\mathrm{SF} = \eta_\mathrm{SN} \left( \mathrm{SFR} \times \tau_D \right) 10^{51} \mathrm{ergs}.
\label{eqn:e-sn}
\end{equation}
Since the measured SFRs of our sample are averaged over $10 - 100$ Myr, this equation assumes that this average measurement is representative of the SFR over $\tau_D$ for our galaxies, which is $\sim 10$ Myr (e.g., Equation~\ref{eqn:tau_d}).
This assumption may fail in the case of galaxies with bursty star formation histories, since a large recent burst may be able to affect the \hi{} gas while an older burst may not.
However, the energy from supernovae is smoothed out over $\sim 50$ Myr after a burst, so minor fluctuations in the star formation history may not be too important.
It also assumes that the initial mass function (IMF) is well-sampled, which may be an issue for galaxies with very low SFRs.
This formalism indicates that the galaxies with the lowest SFRs ($\sim0.5 - 1 \times 10^{-3}$ \msun{} \per{yr}) would have approximate 1 SNe over $1 - 2 \times 10^5$ yr. 
We note that stellar winds can provide additional mechanical energy into the ISM.

As with $E_\mathrm{grav}$, the conversion efficiency between energy available from star formation and \hi{} kinetic energy, $\epsilon_\mathrm{SF} \equiv E_\mathrm{HI} / E_\mathrm{SF}$, must be taken into account.
Values of $\epsilon_\mathrm{SF} > 1$ are unphysical, as in these cases star formation provides less energy than is contained in the central \hi{} peak.
Additional limits have been placed on $\epsilon_\mathrm{SF}$ by simulations. \citet{Thornton1998} found that the average efficiency $\ave{\epsilon_\mathrm{SF}} \sim 0.1$, while other simulations measure efficiencies that can be as high as 0.5 \citep{TenorioTagle1991}. 
As with $\epsilon_\mathrm{grav}$, we measure the range of $\epsilon_\mathrm{SF}$ that is compatible with our data.

In Figure~\ref{fig:snarrow-e-sf}, we compare the energy provided by SNe over one turbulent timescale to the energy in the \hi{} gas.
The format is the same as Figure~\ref{fig:snarrow-e-grav}.
In the left panel, we plot the star formation energy versus the kinetic energy in the central \hi{} peak.
The dashed grey lines show constant $\epsilon_\mathrm{SF}$. 
The unphysical region where $\epsilon_\mathrm{SF} > 1$ is shown in dark grey.
The simulations by \citet{Thornton1998} suggest that $\epsilon_\mathrm{SF}$ is never higher than 0.1; we shade regions above this threshold in light grey.
The left panel shows the inferred values of $\epsilon_\mathrm{SF}$ necessary to drive turbulence with star formation energy.
Again, we have shown efficiencies above the more stringent theoretical maximum of 0.1 from \citet{Thornton1998} in light grey.

For all galaxies, we find that recent star formation provides enough energy to drive the observed turbulence over a single turbulent timescale with efficiencies $0.01 \lesssim \epsilon_\mathrm{SF} \lesssim 0.1$, well in line with the limits from both \citet{TenorioTagle1991} and \citet{Thornton1998}.
However, it is likely that some fraction of this energy \citep[$\sim 35$\%, e.g.,][]{Joung2009} goes into accelerating \hi{} from its undisturbed state into the wings of the superprofile, as explored in the following section.
In this sense, these efficiency estimates are lower limits.
On the other hand, some of the mass in the central peak is likely kinematically associated with the wings, thus lowering $\epsilon_\mathrm{SF}$ values.

\subsubsection{The Superprofile Wings}
\label{sec:disc--energy--wings}

Next we compare the kinetic energy contained in the wings of the superprofiles to that provided by star formation. 
Although the wings contain only a small fraction of the \hi{} mass ($\langle{} \fw{} \rangle \sim 0.11$), the velocities are very high ($\langle \swing{} / \scentral{} \rangle \sim 3$), suggesting that they may harbor a significant fraction of the kinetic energy.

We calculate the kinetic energy in the wings as follows.
If all \hi{} in the galaxy started in the central turbulent component, some extra energy is necessary to accelerate the gas from $v \sim \scentral{}$ to $v \sim \swing{}$.
The excess kinetic energy in the wings is then:
\begin{equation}
E_\mathrm{wings} = \frac{3}{2} \fw{} \, \mhi{} \, \left( \sigma_\mathrm{wings} ^ 2 - \sigma_\mathrm{central}^2 \right).
\end{equation}
Based on our definition of \swing{} (Equation \ref{eqn:swing}), this gives the total energy necessary to accelerate a mass $\fw{} \mhi{}$ from a Gaussian velocity distribution into the observed wings of the superprofile.
Because some of the \hi{} in the central peak may be kinematically associated with the wings, this assumption provides a lower limit on the energy contained in the wings.

We next must choose a timescale over which to consider energy input from star formation for the wings.
The relevant timescale to consider is not necessarily straightforward because the source of kinematics in the wings is unclear.
If the wings are representative of bulk motions, such as away from star forming regions, the relevant timescale should be related to how long these bulk motions are expected to persist.
On the other hand, if the wings are turbulent, the best timescale may be the turbulent timescale given in Equation~\ref{eqn:tau_d}.
Because the source of the wings is not necessarily clear and because neither case provides a definitive timescale, we assess the ability of star formation to provide enough energy over both timescales in turn.

First, we consider the scenario where the wings represent bulk gas motion away from star forming regions.
To estimate the relevant timescale associated with this component, we capitalize on recent studies of \hi{} holes, as these structures often exhibit velocity structures similar to what we expect to find in the wings. 
% and are expected to be caused by star formation.
The kinematic age of \hi{} holes can be estimated based on their size and expansion velocities, if observable.
This calculation is very uncertain and typically provides at most an upper limit to the true age as expansion velocities are expected to slow over time.
However, they provide an order-of-magnitude estimate of the timescale over which these structures are observable.
Seven galaxies overlap between our sample and that of \cite{Bagetakos2011} (NGC 2366, Holmberg II,  IC 2574, Holmberg I, NGC 4214, DDO 154, and NGC 7793), who find a mean kinematic age in dwarfs of $\sim 32.5$ Myr. 
We therefore adopt this value for $\tau_\mathrm{holes}$ with the caveat that it is uncertain, and only consider star formation energy input over the approximate timescale on which \hi{} signatures of expansions are expected to decay.
We use the same formalism as in Equation~\ref{eqn:e-sn} and \S~\ref{sec:disc--energy--central--sf} to estimate star formation energy, substituting 32.5~Myr for $\tau_D$.

In Figure~\ref{fig:swings-e-sf-holes}, we compare the energy available from star formation to the energy in the superprofile wings.
In the left panel, we plot the star formation energy versus the wing kinetic energy for each galaxy.
Dashed grey lines indicate a constant efficiency of transferring star formation energy to kinetic \hi{} energy, where $\epsilon_\mathrm{SF,wings} \equiv E_\mathrm{wings} / E_\mathrm{SF}$.
As in Figures~\ref{fig:snarrow-e-grav} and \ref{fig:snarrow-e-sf}, grey regions of the plot show unphysical or theoretically prohibited efficiencies, i.e., $\epsilon_\mathrm{SF,wings} > 1$ and $\epsilon_\mathrm{SF,wings} > 0.1$.
The right panel shows the inferred efficiency $\epsilon_\mathrm{SF,wings}$ versus the star formation energy available.

We find that all of galaxies require efficiencies of only $< 0.05$ to produce enough energy to drive \hi{} gas into the wings over a timescale of $32.5$ Myr.
The distribution has a median $\epsilon_\mathrm{SF,wings} = 0.013$ with a standard deviation of $0.007$.
These estimates are well below with the theoretical maximum of $0.1 - 0.5$ found by \citet{TenorioTagle1991} and \citet{Thornton1998}, and are in line with simulations by \citet{Joung2009} that show that $\sim 35$\% of star formation energy goes into driving large-scale bulk motions.
The required efficiencies are also much lower than estimates of efficiencies necessary to drive larger \hi{} holes, which range between 1 - 40\% at their kinematic ages \citeeg{Weisz2009, Warren2011, Cannon2011, Bagetakos2011}.
However, we have derived these estimates from global properties.
A more precise determination is necessary using spatially-resolved data scales, as previous studies have shown that the value of the second moment declines with radius \cite{Tamburro2009}, therefore changing the energy in the wings based on location in the galaxy.

We next consider the possibility that the gas in the wings is instead representative of a turbulent component. In this case, the relevant timescale to consider is the turbulent dissipation timescale, as given by Equation~\ref{eqn:tau_d}.
Even though the velocity profile of \hi{} in the wings is not Gaussian in our parameterization, it again provides an order-of-magnitude estimate of how long this component can dissipate its energy if it is indeed turbulent.
In this case, it may be more relevant to use measured \sbroad{} values for double Gaussian fits instead of \swing{}, because our parameterization explicitly removes gas with small velocities from the \swing{} calculation.
Because the double Gaussian fits appear to be determined primarily by the wings (see Appendix~\ref{appendix:double-gaussian}), we choose to use \sbroad{} to calculate $\tau_D$ as the second timescale to consider.
If we substitute \sbroad{} for $\sigma$ in Equation~\ref{eqn:tau_d}, we obtain values for $\tau_D \sim 6 - 10$ Myr, or $\sim 70$\% of those determined for the central component.
As before, we use Equation~\ref{eqn:e-sn} to calculate the input energy over this turbulent dissipation timescale determined for the wings.

Figure~\ref{fig:swings-e-sf-turb} shows the comparison of energy in the wings to energy provided by star formation over one turbulent timescale.
The format is the same as Figure~\ref{fig:snarrow-e-sf}.
In this case, implied efficiencies are higher by a factor of $\sim 3-5$ due to the difference in timescales, and are much closer to the theoretical maximum of 0.1 found by \citet{Thornton1998}.

\subsubsection{The Entire Superprofile}
\label{sec:disc--energy--full}

In this section we assess whether star formation provides enough energy to produce the \hi{} velocity distribution seen in the superprofiles. In this case, the kinetic energy of the entire superprofile is simply:
\begin{equation}
E_\mathrm{SP} = \frac{3}{2}\sumlim{v} M(v) \, v^2
\end{equation}
where $M(v)$ is the \hi{} mass at velocity $v$. 

As with the central peak and the wings, we must determine the timescale over which to calculate energy input from star formation. 
Since the relevant timescales are unclear, we use both the turbulent timescale for the central peak and a fixed timescale of $\tau = 32.5$ Myr in Equation~\ref{eqn:e-sn} to calculate energy input.

We show the comparison between energies over the turbulent timescale in Figure~\ref{fig:full-e-sf-turb}. 
Over this timescale, star formation provides enough energy to drive the full shape of the superprofile, but many galaxies lie in the $\epsilon > 0.1$ region unfavored by simulations \citep{Thornton1998}.

Figure~\ref{fig:full-e-sf-fixed} again shows the comparison between star formation and \hi{} energies, but over the the fixed timescale of $\tau = 32.5$ Myr.
Over this timescale, star formation provides enough energy to drive kinematics in the entire \hi{} superprofile at with $\epsilon < 0.1$.
We note, however, that it is likely that the central peak and the wings of the superprofile have different associated timescales and efficiencies, so choosing single values to represent both components may not be physically appropriate.

\subsection{Estimating the Scale Height of \hi{} in the Sample}
\label{sec:disc--scale-height}

The scale height of the \hi{} layer perpendicular to an isothermal, self-gravitating disk can be determined based on its velocity dispersion and disk surface mass density \citep{vanDerKruit1981}. 
We use a method similar to that presented in \citet{Ott2001} and \citet{Warren2011} to approximate \hi{} scale heights for our sample.
The scale height $h_z$ is given by \citet{vanDerKruit1981} as:
\begin{equation}
h_z = \frac{ \sigma_\mathrm{gas} } {\sqrt{4 \pi G \rho_\mathrm{t}}},
\end{equation}
where $\sigma_\mathrm{gas}$ is the velocity dispersion perpendicular to the disk, $G$ is the gravitational constant, and $\rho_\mathrm{t}$ is the stellar mass density of the disk. 
If we assume that \hi{} has a Gaussian distribution in the $z$-direction, we find that:
\begin{equation}
N_\mathrm{HI} = \sqrt{2 \pi} h_z n_\mathrm{HI,0},
\end{equation}
where $N_\mathrm{HI}$ is the \hi{} column density and $n_\mathrm{HI,0}$ is the number density at the midplane of the disk.
These two equations were combined by \citet{Ott2001} to find an expression for \hi{} scale height in terms of observables:
\begin{equation}
h_z = 579
\left( \frac{ \sigma_\mathrm{gas}}{ 10 \, \mathrm{km \, s}^{-1} } \right)^2 \left( \frac{ N_\mathrm{HI}}{ 10^{21} \; \mathrm{cm}^{-2}} \right)^{-1}
\left( \frac{ \rho_\mathrm{HI} }{ \rho_t } \right) \; \mathrm{pc},
\end{equation}
where $\rho_\mathrm{HI} / \rho_t$ is the ratio between \hi{} density to total disk density, and can be approximated as $( \rho_\mathrm{HI} / \rho_t ) = ( \m{hi} / \m{t} )$ where $\m{t} = 1.36 \m{HI} + \m{\star}$. %This approach neglects the dark matter contribution to the disk, which may be important for dwarf galaxies \citeeg{}

Using our superprofiles and measured galaxy properties, we can estimate the average \hi{} scale height of the disk. We first assume that the velocity dispersion is isotropic, such that $\sigma_z = \scentral{}$.
% discuss validity of assumption?.
Second, we convert \ave{\hisd{}} to $N_\mathrm{HI}$ units with the caveat that we have averaged these quantities over the disk and have not included the contribution of dark matter.
We note that since these values are averaged over the disk of the galaxy, any spatial variation in these parameters is no longer distinct.
\hi{} scale heights in galaxies are expected to flare at large radii, as gas velocity dispersions remain relatively constant but disk surface density declines with radius.
This method therefore gives an estimate for \hi{} scale height that is weighted toward the regions with the highest \hi{} surface densities.
The average \hi{} scale heights derived from these values are listed in Table~\ref{tab:scale-height}. The majority of galaxies have implied scale heights between $ 100 < \ave{h_z} < 700$ pc, with a median of $320$~pc and interquartile range of $210 - 480$~pc.

Scale heights for some of the galaxies in our sample have been determined using other methods. \citet{Banerjee2011} modeled the dark matter and baryonic components of halos for DDO~154, Ho~II, IC~2574, and NGC~2366.
They obtained scale heights 
between $\sim 130$ pc at $r = 0$ kpc to $\sim 1$ kpc at $r = 6$ kpc for DDO~154; 
between $\sim 180$ pc at $r = 1$ kpc to  $\sim 1$ kpc at $r = 7$ kpc for NGC 2366; 
between $\sim 350$ pc at $r = 1.5$ kpc to $\sim 700$ pc at $r = 9$ kpc for IC 2574;
 and a fixed scale height of $\sim 400$ pc at all radii for Ho~II.
Compared to the \citet{Banerjee2011} scale heights, our method yields scale heights that are within the same range for DDO~154 and NGC~2366 but smaller by 35\% and 25\% for IC~2574 and Ho~II, respectively. %The majority of galaxies have implied scale heights between $ 100 < \ave{h_z} < 700$ pc.

\subsection{A Universal \hi{} Profile Shape?}
\label{sec:disc--universal-hi-profile}

As seen in Figure~\ref{fig:superprofiles-scaled}, the superprofiles show a distinct velocity distribution: a central peak with strong contributions from non-Gaussian wings.
After normalization to the same HWHM, the uniformity of the profiles is striking, especially considering that the low-mass dwarf galaxies in our sample are typically characterized by irregular velocity fields and morphologies, and stochastic, varied star formation histories \citep{Weisz2011}.
The residuals also show a surprisingly similar shape, with the peak often occurring at $2 \times \mathrm{HWHM}$ across the observed range of \fw{}.
This global regularity exists in spite of the fact that the individual line-of-sight \hi{} profiles have a much more varied shape, with some showing asymmetry \citeeg{Young2003,Warren2012} or double peaks indicative of expanding structures \citeeg{Bagetakos2011}.
Statistically, however, the sum of these profiles generates the same kinematic distribution from galaxy to galaxy.

The shape of the superprofiles on global scales is qualitatively similar to those found by other studies of average \hi{} line profiles, which also show a mostly Gaussian central peak with broader wings \citeeg{Dickey1990, Boulanger1992, Kamphuis1993, Braun1997, Petric2007, Ianj2012}.
It also matches the shape of simulated \hi{} profiles found by \citet{Joung2009} for a supernova-driven turbulent medium, though at lower star formation intensities.
However, quantitative comparisons among the surveys are hindered by the vast differences both in observational parameters such as spatial and spectral resolution as well as in techniques for removing the rotational velocity.
Nonetheless, many of the studies, including our own, find that the broad central peaks can be characterized reasonably well by Gaussians with widths of 5-10 \kms{}, with additional wings to either side.
The similarities indicate that the shape of the \hi{} line profile, and therefore the general kinematic structure of \hi{} line profiles, are relatively independent of galaxy properties.
However, uniform studies of larger spirals with better velocity resolution, now possible with the larger bandwidth of the newly-updated JVLA, are necessary to confirm this idea.

%It may not be surprising to see such similar superprofile shapes, however.

%The similarity also implies that the superprofile shape must reflect the origin of \hi{} velocity dispersion and kinematics.
%This holds across the wide range of observed SFR, $\langle \Sigma_\mathrm{HI} \rangle$, and other physical properties in our sample.
%Therefore, the ability of physical processes to drive overall \hi{} kinematics and the magnitude of \hi{} turbulence are similar across a wide range of galaxy types.

%%%%%%
%%%%%% CONCLUSIONS
%%%%%%
\section{Conclusions}
\label{sec:conclusions}

We have generated a measure of global \hi{} kinematics in a sample of nearby dwarf galaxies from VLA-ANGST and THINGS by summing the contribution to a global line profile for each line-of-sight spectrum after removing rotation from each spectrum. 
The resulting superprofile for an individual galaxy provides an intensity-weighted average of its individual \hi{} line profiles.

We interpret the superprofiles as composed of a central peak indicating average turbulence with higher-velocity wings to either side. 
We parameterized them with four parameters describing the width of the central peak (\scentral{}), the characteristic velocity of the wings (\swing{}), the fraction of gas in the wings (\fw{}), and the asymmetry (\aw{}). 
We have compared these parameters to various global galaxy properties in order to determine what, if any, physical causes are behind \hi{} kinematics.

\begin{itemize}

\item  The dynamic range of \scentral{} is quite small, varying only between $\sim6 - 10$ \kms{} across our sample. 
We find a correlation between \scentral{} and \ave{\hisd{}} in the dwarf sample which is not significant once higher mass galaxies are added. 
The measured \scentral{} values are close to but slightly higher than line widths that can be driven by background UV heating. 
It is possible that base \hi{} velocity dispersions are set by this heating, with star formation imparting only additional energy.

\item The characteristic velocity of gas in the wings, \swing{}, increases with \ave{\sfrsd{}}, \ave{\Sigma_\mathrm{baryon}}, and \ave{\hisd{}}, implying that star formation could be one way to accelerate \hi{} to velocities faster than expected compared to the surrounding turbulent medium.

\item The fraction of gas in the wings, \fw{}, increases with galaxy mass and with SFR / \mhi{}, so galaxies with relatively high SFR or low \mhi{} could be better able to accelerate \hi{} to higher velocities.

\item The asymmetry, \aw{}, decreases with both SFR  and with galaxy mass, and is primarily in the wing regions. 
This supports the idea that star formation can accelerate \hi{} away asymmetrically, so galaxies with smaller SFR likely have fewer star-forming regions and thus show more asymmetry.

\end{itemize}

In all cases, our trends exhibit large scatter. Since many of the physical properties we examined vary on both radial and spatial scales, future analyses must incorporate this information to disentangle any causal connection with \hi{} gas kinematics.

We have also compared the energy contained in the \hi{} superprofiles with the energy provided by the gravitational instability from \citet{Wada2002} and by star formation. 
We find that this gravitational instability cannot provide enough energy to drive turbulent line widths on timescales of $\sim 10$ Myr, while star formation can. Star formation also imparts enough energy to accelerate gas into the wings of the profile over timescales of $\sim 32.5$ Myr, with implied efficiencies below the theoretical maximum of 0.1 - 0.5.

We derived average \hi{} scale heights for the sample, with most galaxies exhibiting scale heights of a few hundred pc.

Finally, we found that the average \hi{} superprofile shape, when scaled to the same HWHM, has a remarkably similar shape from galaxy to galaxy, with variations primarily in the wings at low levels. 
The shape of the central component differs from a Gaussian by only $\sim 0.05$\, with more varied wings showing additional emission 5-10\% above the scaled Gaussian fit. 
This similarity implies that the physical processes setting the kinematics of \hi{} in galaxies function similarly in all dwarf galaxies.

\begin{acknowledgments}

We thank the anonymous referee for comments that improved the quality of this work.
We also thank Fabian Walter and the THINGS team for providing additional data sets used in this paper, as well as Cliff Johnson and Daniel Dale for generously allowing us to use their 3.6\um{} point-subtracted maps. % THINGS, LVL
The National Radio Astronomy Observatory is a facility of the National Science foundation operated under cooperative agreement by Associated Universities, Inc.  % NRAO
Support for this work was provided by the National Science Foundation collaborative research grant ``Star Formation, Feedback, and the ISM: Time Resolved Constraints from a Large VLA Survey of Nearby Galaxies,'' grant number AST-0807710.  % NSF grant
This material is based on work supported by the National Science Foundation under grant No. DGE-0718124 as awarded to A.M.S.  %GRFP

\end{acknowledgments}

%%%%
%%%% APPENDIX
%%%%
\begin{appendix}

\section{Noise Estimates for the Superprofiles}
\label{appendix:noise}
The flux rescaling process is important for interferometric data because it provides an accurate measurement of the true \hi{} flux of each galaxy.
Since this process only rescales the residuals, the noise properties of the rescaled data cube are complicated.
While the highest noise peaks can show up in the clean components of the rescaled data cube, the majority of the noise is still in the residuals. Traditional estimates of noise therefore provide an overestimate of the noise for superprofiles generated from rescaled data cubes.

The noise on a single point of a superprofile generated from a \emph{standard} cube can be approximated as:
\begin{equation}
\sigma_\mathrm{SP} = \sigma_\mathrm{chan} \times \sqrt{ N_\mathrm{pix} / N_\mathrm{pix/beam}},
\label{eqn:noise-std}
\end{equation}
where $\sigma_\mathrm{chan}$ is the \emph{rms} noise level in a single channel of the data cube, $N_\mathrm{pix}$ is the number of pixels contributing to each superprofile point, and $N_\mathrm{pix/beam}$ is the number of pixels per beam \citepalias{Ianj2012}. 

In the left panel of Figure~\ref{fig:noise-jvm-std}, we apply this formula to the superprofile generated from the standard cube and to that from the rescaled cube. 
The upper panel shows the absolute flux measured in each superprofile. 
In the lower panel we have normalized the superprofiles so that the maxima are the same. 
It is clear that the fractional noise is much larger in the rescaled cube.

In the right panel, we have rescaled the noise for the superprofile generated from the rescaled cube by the ratio of total fluxes, $F_\mathrm{rescaled} / F_\mathrm{standard}$ such that the noise is given by Equation~\ref{eqn:noise-jvm}. 
This rescaling produces a noise estimate with a similar fractional uncertainty compared to the standard cube. 
This estimate is not exact, because the highest noise spikes are contained in the clean components. 
However, the majority of the noise is in the residuals, which are rescaled.
This method also provides a noise estimate that matches the fractional uncertainty on each point compared to superprofiles generated from the standard cubes.
Since it is a better representation of the fractional uncertainty on each superprofile point, we adopt Equation~\ref{eqn:noise-jvm} when calculating the effects of noise on the superprofiles.

\section{Effects of Observational Settings on Superprofile Shapes and Measured Parameters}
\label{appendix:uncertainties}

We have performed a number of tests to ensure the validity of our results and to estimate uncertainties on the measured parameters,  \scentral{}, \swing{}, \fw{}, and \aw{}. 
In particular, we have examined the effects of \vp{} uncertainties (\ref{appendix:uncertainties--vp}); finite spatial resolution (\ref{appendix:uncertainties--spatial-resolution}); finite velocity resolution (\ref{appendix:uncertainties--velocity-resolution}); and noise on each superprofile point (\ref{appendix:uncertainties--noise}). 
Finally, we review the final uncertainties on the measured parameters (\ref{appendix:uncertainties--overview}).

\subsection{Uncertainties in \vp{}}
\label{appendix:uncertainties--vp}

We explore how our superprofile parameters are affected by the uncertainties in determining \vp{}, which could possibly generate broader superprofiles or more flux in the wings.

For each of our four test galaxies (GR~8, Sextans~A, UGC~4483, and NGC~2366), we start with the pixel-by-pixel uncertainty in determining \vp{} as a function of S/N ratio, determined from our Monte Carlo tests for the four test galaxies (\S~\ref{sec:superprofiles--signal-to-noise}; Figure~\ref{fig:mc-vp-offsets}). 
We first assume that all pixels would contribute a Gaussian with a width of \scentral{} in the absence of any uncertainties on \vp{}. 
For each pixel above our S/N $> 5$ threshold, we generate a random offset drawn from a Gaussian with that pixel's standard deviation of determining \vp{} and with the S/N as the amplitude. 
We generate a fake superprofile by summing all of the \scentral{} Gaussians with their respective velocity offsets.

The results of this test are shown in Figure~\ref{fig:mc-vp-offset-gaussians} for our four test galaxies. 
For each of the four galaxies, the upper panel shows the ``observed'' fake superprofile as the black solid line; the input superprofile we would have expected in the absence of any uncertainties on \vp{} as the blue dash-dot line; and the HWHM-scaled Gaussian fit as the dashed red line. 
In all cases, the differences are smaller than the line widths. The lower panel shows the residuals (i.e., ``observed'' - fit and ``observed'' - input).

The differences between the input Gaussian, the ``observed'' fake superprofile, and the fit are $< 0.005$ in all cases. 
We also find that the width of the superprofile is increased by $< 0.5$\%. The uncertainties in \vp{} therefore have a negligible effect on the superprofile shapes and parameters.

\subsection{Finite Spatial Resolution}
\label{appendix:uncertainties--spatial-resolution}

The combination of finite spatial resolution and rising rotation curves at the centers of galaxies can increase the width of observed \hi{} line-of-sight spectra in the central regions, which could then either increase the width or mimic \hi{} in the wings of the observed superprofile. 
Our sample of dwarf galaxies likely have either slowly-rising rotation curves \citep[DDO 154 and NGC 2366; e.g.,][]{deBlok2008} or primarily display solid body rotation typical of dwarfs \citep{Oh2011}. 
Because these rotation curves have a smaller gradient with increasing radius, we expect to see less of an effect from beam smearing in the central regions compared to larger spiral galaxies, but we must still to understand its effects.

To quantify the effects of beam smearing, we have developed a Python module to generate a suite of model galaxy observations using NGC 2366 as our test galaxy. 
This galaxy has the steepest rotation curve of our sample, and thus would be the most affected by this particular bias. 
We use the observed \hi{} surface brightness distribution plus the observed inclination and position angle from \citet{deBlok2008} to generate the model \hi{} distribution. 
We also assume that the disk has an exponential distribution in the z-direction with a scale height $h_z = 500$ pc, a typical observed value for \hi{} scale heights in dwarfs \citeeg{Banerjee2011, Warren2011}. 
Changing this assumption to either 100 pc or 1 kpc does not strongly influence our results.

We next impose a rotation curve that can be modeled as a linear rise for radii smaller than $r_\mathrm{flat}$, with a flat regime at larger radii with circular velocity $v_\mathrm{flat}$. 

Finally, we assume that all line-of-sight \hi{} spectra have Gaussian velocity distributions with a dispersion of 6 \kms{}. 
While this assumption is not necessarily indicative of the true dispersion as a function of radius \citeeg{Tamburro2009}, it does allow us to quantify the effects of beam smearing on a uniform \hi{} profile. 
In order to estimate the effects of declining velocity dispersion, we also generate a model where the intrinsic velocity dispersion is chosen by an exponential fit to the  radial average of the second moment map.

For this test, we use three models: one with the observed $r_\mathrm{flat} = 1.9$ kpc, $v_\mathrm{flat} = 60$ \kms{} \citep{deBlok2008}, and $\sigma_\mathrm{HI} = 6$ \kms{}; a second with an extreme $r_\mathrm{flat} = 0.5$, $v_\mathrm{flat} = 60$, and, $\sigma_\mathrm{HI} = 6$ \kms{} ; and a third with $r_\mathrm{flat} = 1.9$ $v_\mathrm{flat} = 60$, and $\sigma_\mathrm{HI} \left( r \right)$ set by the exponential fit to the second moment map.

To place \hi{} clouds in the model cube, we draw a sample of points from the \hi{} surface brightness distribution, assuming each cloud represents a gas cloud at the observed spatial position. 
We then distribute these points randomly in the z-direction using our assumed exponential $z$ distribution. 
Each point is smoothed in velocity space into a Gaussian with the central velocity determined by the rotation curve at that radius and the width dependent on the model. 
Finally, we scale the cube so that the total \hi{} mass is the same as measured in the galaxy. This yields a cube that has \hi{} line-of-sight spectra unaffected by spatial resolution with a velocity resolution of 2.6 \kms{}.

To reproduce the effects of finite spatial resolution, we smooth the cube to our working 200 pc resolution using a circular Gaussian beam with a FWHM of 200~pc.  
For both the true and convolved cube, we find \vp{} for each pixel and then generate a superprofile using the same method as described in \S \ref{sec:superprofiles}. 
We fit and parameterize the superprofile using a single Gaussian scaled by the superprofile's HWHM (\S \ref{sec:analysis--parameterization}).

We show the resulting true and smoothed model superprofiles compared to the observed superprofile in Figure~\ref{fig:n2366-models}. 
In all cases, the superprofile from the smoothed cube shows very small differences compared to the true cube. 
The measured Gaussian dispersion is slightly wider by $\Delta \scentral{}_\mathrm{,spatial} \lesssim 0.5$ \kms{}. 
Additionally, the convolved cubes have a negligible fraction in the wings ($\Delta \fw{}_\mathrm{,spatial} \lesssim 0.01$), a value much lower than the typical range observed in our sample ($0.05 < \fw{} < 0.15$). 
With such a small contribution to the wings, \swing{} should not be strongly affected by beam smearing.  
We note that the superprofile generated from the model with exponentially declining $\sigma_\mathrm{HI}$ is well-fit by a Gaussian with a width $\sim 10$ \kms{}, similar to the \citet{Tamburro2009} results. 
None of these models is able to reproduce wings at the observed magnitude.

In summary, while finite spatial resolution does contribute a small amount of broadening, it is at a low level compared to the observed widths and is not strong enough to generate spurious flux in the wings.

\subsection{Finite Velocity Resolution}
\label{appendix:uncertainties--velocity-resolution}

We also examine the effects that finite velocity resolution has on our results. 
Since our sample is composed of galaxies observed with a variety of velocity resolutions (0.6, 1.3, and 2.6 \kms{}), we must quantify any effects that arise from these differences. 
For each galaxy, we bin the observed standard and flux-rescaled cubes to the coarser velocity resolutions of our sample. 
We then find \vp{}, generate a superprofile, and measure parameters for each new resolution.

In Figure~\ref{fig:vel-res-sps} we show the superprofiles generated from the binned cubes for DDO~125. 
The upper panels shows the superprofiles themselves, and the lower panels show the difference between superprofiles generated from the binned cube and the original cube. 
The superprofiles behave similarly for $\Delta v = 0.6, 1.3$ and 2.6 \kms{}, but the superprofile for the $\Delta v = 5.2$ \kms{} cube shows $> 5$ \% differences compared to the original. 
Similar behavior is evident in all the superprofiles generated from cubes binned to $\Delta v = 5.2$ \kms{}, and thus galaxies whose original data cubes have $\Delta v = 5.2$ \kms{} have been excluded from our sample.

We show the resulting parameters as a function of bin size for all galaxies in Figure~\ref{fig:vel-res-params}. 
The four panels in the plot are \scentral{}, \swing{}, \fw{}, and \aw{}. 
Each color represents a different galaxy. 
We find that three of our four parameters (\scentral{}, \swing{} and \fw{}) are relatively well-behaved with increasing velocity resolution, showing minor variations relative to the range seen in the sample. 
However, the standard deviation of variations in \aw{} is $\sim 0.06$; we account for this variation in the final uncertainties (\S \ref{appendix:uncertainties--overview}). 

In general, coarser velocity resolution slightly increases the width of the central component and the  velocity of the wings, which then places slightly less flux in the wings. 
Comparing the same galaxy at 1.3 \kms{} to 0.6 \kms{}, we find median differences of $\Delta \scentral{}_\mathrm{,vel} = 0.08$, $\Delta \swing{}_\mathrm{,vel}  = 0.13$, and $\Delta \fw{}_\mathrm{,vel}  = -0.005$. Comparing between 2.6 and 1.3 \kms{}, we find median differences of $\Delta \scentral{}_\mathrm{,vel}  = 0.17$, $\Delta \swing{}_\mathrm{,vel}  = 0.32$, and $\Delta \fw{}_\mathrm{,vel}  = -0.006$.

\subsection{Noise}
\label{appendix:uncertainties--noise}

The final influence on the measured parameters is the noise of the superprofiles and is especially important for the lowest-mass dwarfs. 
To quantify the effects of noise on the measured parameters, we start with the noise estimate for each point (Equation~\ref{eqn:noise-jvm}; \S~\ref{sec:superprofiles--uncertainties}). 
For each sample galaxy, we assume that the measured superprofile is true. 
We then add noise to each point drawn from a Gaussian distribution with a width $\sigma_\mathrm{SP}$ calculated using Equation~\ref{eqn:noise-jvm}. 
Finally, we measure the parameters from this ``noisy'' superprofile.

After repeating the above process 10,000 times, we have obtained an estimate of the typical range of parameters allowed by the noise on the superprofiles. 
For each parameter, we fit a Gaussian to the histogram of ``noisy'' parameter values and adopt its width as the uncertainty due to superprofile noise on each parameter.

The median uncertainties and interquartile ranges on the parameters due to noise are: 
$\Delta \scentral{}_\mathrm{,noise} = 0.05_{0.03}^{0.13}$ \kms{}; 
$\Delta \swing{}_\mathrm{,noise} = 1.9_{0.8}^{2.5}$ \kms{}; 
$\Delta \fw{}_\mathrm{,noise} = 0.014_{0.006}^{0.021}$; and 
$\Delta \aw{}_\mathrm{,noise} = 0.030_{0.009}^{0.051}$

\subsection{Overview of Parameter Uncertainties}
\label{appendix:uncertainties--overview}

We use the above tests to estimate the total uncertainty of the measured superprofile parameters. 
We include only uncertainties that have a non-negligible effect on each parameter in its total uncertainty estimate by adding the uncertainties due to various observational effects in quadrature.

For each galaxy, the final uncertainty for \scentral{} is: 
\begin{equation}
\Delta \scentral{} = \sqrt{ 
(0.5 \; \mathrm{km \; s}^{-1}) ^2 + 
(0.17 \; \mathrm{km \; s}^{-1})^2 + 
(\Delta \sigma_\mathrm{central,noise})^2 }.
\end{equation}
The $0.5$ \kms{} and $0.17$ \kms{} errors are due to the effects of finite spatial resolution and finite velocity resolution. 
The value for $\Delta \scentral{}_\mathrm{,noise}$ is different for each galaxy. 
We do not include errors from \vp{} uncertainties, as they are an order of magnitude lower than the other uncertainties.
The $\Delta \scentral{}$ values are dominated by errors due to finite spatial resolution.

The final uncertainty of \swing{} is:
\begin{equation}
\Delta \swing{} = \sqrt{ 
(0.13 \; \mathrm{km \; s}^{-1})^2 + 
(\Delta \sigma_\mathrm{wings,noise})^2 }.
\end{equation}
The $0.13$ \kms{} error is due to the effects of finite velocity resolution, and $\Delta \sigma_\mathrm{wings,noise}$ is different for every galaxy.
We do not include errors from \vp{} uncertainties or finite spatial resolution, as the small amounts of flux they contribute to the wings do  not strongly change \swing{} values.
The $\Delta \swing{}$ values are dominated by uncertainties due to noise on the superprofiles.

The final uncertainties for \fw{} is:
\begin{equation}
\Delta \fw{} = \sqrt{
(0.01) ^ 2 +
(\Delta f_\mathrm{wings,noise})^2
}.
\end{equation}
The 0.01 uncertainty is due to finite spatial resolution. 
As before, $\Delta f_\mathrm{wings,noise}$ is different for each galaxy. 
We neglect uncertainties due to finite velocity resolution and \vp{} uncertainties, as they are over an order of magnitude smaller than the included uncertainties. 
Both values contribute approximately equally to the final uncertainty $\Delta \fw{}$.

The final uncertainty for \aw{} is:
\begin{equation}
\Delta \aw{} = \sqrt{
(0.06)^2 +
(\Delta a_\mathrm{noise})^2
}.
\end{equation}
The $0.06$ uncertainty is due to finite velocity resolution, as the peak determination is less precise in cubes that have larger velocity resolutions. 
We do not include uncertainties from finite spatial resolution in this estimate. 
The included uncertainties contribute fractionally different amounts in different galaxies, so $\Delta \aw{}$ is not typically dominated by either uncertainty.

\section{The Interpretation of Double Gaussian Fits}
\label{appendix:double-gaussian}

In a recent study, \citetalias{Ianj2012} parameterized superprofiles for THINGS galaxies using double Gaussian fits and then argued that the two components were representative of cold and warm \hi{} in the galaxies. 
We had independently pursued this approach for our study based on the low reduced $\chi^2$ value of double Gaussian fits compared to single Gaussian fits. Unlike \citetalias{Ianj2012}, we chose to abandon this parameterization in favor of the simpler HWHM parameterization for reasons described in this appendix.

We have previously given results for our double Gaussian fits in \S~\ref{sec:analysis--interpretation} and Table~\ref{tab:parameters--double-gaussian}. However, a number of potential concerns about the physical meaning of the double Gaussian fits arose as we explored them in more detail.

We first refer to the strong similarity in shape seen in Figure~\ref{fig:superprofiles-scaled}, when all superprofiles are scaled by their HWHM value. 
We now question whether the same similarity is seen when scaling by width of the narrow or broad Gaussian components, which might be expected if \snarrow{} or \sbroad{} were physically relevant quantities. 
In Figure~\ref{fig:sps-scaled-2gauss-scale} we again show the scaled superprofiles for all the galaxies, but we now scale the velocity axis by a different measured parameter in each panel: HWHM (upper panel); the width of the narrow component of the double Gaussian fit, \snarrow{} (middle panel); and the width of the broad component of the double Gaussian fit, \sbroad{} (lower panel). 
The left column shows the superprofiles after normalizing the scaled velocity axes so that the median width of the superprofiles is aligned, to better show the variation in shape. 
We then plot the median superprofile in red. The right column of Figure~\ref{fig:sps-scaled-2gauss-scale} shows each superprofile minus the median superprofile for that scaling over the same velocity range.

The HWHM scaling removes most of the variations in shape among the profiles. 
In contrast, the specific values of \snarrow{} and \sbroad{} have little direct bearing on the overall profile shape. 
Quantitatively, the \emph{rms} residuals around the median scaled superprofile are 5 times larger for the superprofiles scaled by either \snarrow{} or \sbroad{} compared to those for the HWHM-scaled superprofile. 
The HWHM-scaling appears to provide the best characterization of the superprofile shape.

We next ask if the Gaussian components behave similarly when scaled to the superprofile HWHM. 
Since the HWHM-scaled superprofile shapes are very similar, we would expect the narrow and the broad components of the double Gaussian fits to have similar properties when scaled in the same way. 
We plot the results of this test in Figure~\ref{fig:superprofiles-scaled-2gauss-comp}. 
The left panel shows the superprofile for each galaxy scaled to its HWHM. We then overplot the two Gaussian components, which have also been scaled to the same HWHM as the corresponding superprofile. 
The upper panel shows the narrow component, while the lower panel shows the broad component. 
At first glance, the shapes of the scaled Gaussian components are strikingly different compared to the similarity in the global superprofiles' shapes. 
In other words, the properties of the two Gaussian components vary wildly, while somehow conspiring to preserve the same overall shape.

The right panel of Figure~\ref{fig:superprofiles-scaled-2gauss-comp} is the same as the left panel, but for clarity only Sextans B and NGC 7793 are shown.  Although the upper $\sim50$\% of the superprofiles are nearly identical, the best-fit double Gaussians are very different.
Since the superprofile for NGC 7793 has broader wings relative to the superprofile HWHM, the broad Gaussian is forced to a larger $\sbroad{}$ value and a smaller amplitude to fit the wings. The narrow component is then left to make up the remainder of the central peak.
Because the wings are broad, which leads to a low-amplitude broad Gaussian, the narrower component is forced to a high amplitude to match the center of the profile.
In contrast, the superprofile for Sextans B has lower level wings. 
The broad Gaussian component therefore has a higher amplitude, leaving less of the central peak for the narrow component. 
This behavior indicates that the relative amplitude and fractional area of the broad component are strongly driven by the shape of the wings. 
Even though the wings of NGC 7793 have only $\sim9$\% more flux than those of Sextans B, the relative amplitudes of the broad components differ by nearly a factor of two, and the fractional areas of the broad components differ by more than a factor of three. 
The large differences in double Gaussian parameters for superprofiles with remarkably similar shapes is the first indication that the double Gaussian fits may not be tracing physically-meaningful properties.

We see further evidence for this behavior when we quantitatively explore the correlations among the double Gaussian parameters. 
To provide a relative comparison, we have generated superprofiles for galaxies in \citetalias{Ianj2012} following their methodology for their well-behaved ``clean'' subsample. 
In particular, we have used the natural weighted cubes at their instrumental resolution and included all pixels with $\mathrm{S/N} > 3$. 
We do not apply any masking. We can then calculate the HWHM of each superprofile.

Based on Figure~\ref{fig:superprofiles-scaled-2gauss-comp}, we have speculated that the relative strength of the narrow and broad components are being set primarily by the relative velocity of the wings compared to the central profile (i.e., $\sbroad{} / \mathrm{HWHM}$). 
We explore this idea in Figure~\ref{fig:comp-sb-amp}, where we plot the amplitude ratio between the narrow and broad Gaussian components versus $\sbroad{} / \mathrm{HWHM}$. 
The left panel of the plot shows the galaxies in our sample, using the double Gaussian fits to the superprofiles discussed in this paper (Table~\ref{tab:parameters--double-gaussian}). 
The right panel shows the equivalent data for the clean subsample of \citetalias{Ianj2012}.
For this comparison, we calculate the Narrow / Broad amplitude ratio, based on the formula for the area under a Gaussian, as follows:
\begin{equation}
  \mathrm{Narrow \; / \;  Broad \; Amplitude} = \frac{A_n}{A_b} \frac{\sigma_b}{\sigma_n},
\end{equation}
where $A_n / A_b$ is the ratio of the areas under the narrow and broad components, and $\sigma_n$ and $\sigma_b$ are the widths of the narrow and broad Gaussian components. 
All numbers are taken directly from \citetalias{Ianj2012}.
In this case, we have taken $\sigma_b$ directly from \citetalias{Ianj2012} and calculated the HWHM from our I2012-like superprofiles.

It is immediately clear that profiles with broader wings relative to their characteristic HWHM width have narrow components with much higher amplitude ratios compared to the broad component. 
This correlation is a quantitative representation of the behavior shown in Figure~\ref{fig:superprofiles-scaled-2gauss-comp}; the broad Gaussian component primarily fits the wings, leaving the narrow Gaussian to fit the remainder of the superprofile as best it can. 
This behavior is exacerbated by the $1 / \sqrt{N}$ weighting scheme. 
Because this scheme gives pixels with fewer contributing points more weight, the superprofile wings, which are produced by only a small fraction of the \hi{}, are weighted most strongly. 
Since the relative broadness of the wings strongly affects the relative amplitudes of the two Gaussians, it is unclear if the two components are tracing the same type of gas in superprofiles with varying levels of wing importance.

We can also look at the ratio of narrow to broad component areas compared to the fraction of gas in the wings of the profile (\fw{}) as measured in \S~\ref{sec:analysis--parameterization}.
In Figure~\ref{fig:comp-fw-anab} we plot $A_n / A_b$ versus \fw{}. Again, the left panel shows data from our sample, while the right panel shows that from \citetalias{Ianj2012}. 
%The measurements in the left panel are from the 200-pc superprofiles discussed in the body of this paper. 
For the latter, we have taken $A_n / A_b$ directly from \citetalias{Ianj2012}, and we have used our I2012-like superprofiles to measure \fw{} with the caveat that we have not brought the \citetalias{Ianj2012} sample to a common physical resolution.
 If the broad Gaussian fit were indeed tracing a warm component, then on average we would expect galaxies with more flux in the wings to have a higher fraction of gas in the broad component, leading to lower \anab{} values. 
 For the \citetalias{Ianj2012} numbers, we find an unexpected positive correlation, where galaxies with \emph{more} flux in the wings, as measured with \fw{}, have a higher \anab{} values and therefore a \emph{higher} fraction of gas in the narrow component.
 We do not see a similar trend in our data, but it is possibly due to the fact that we are measuring smaller galaxies with lower overall S/N ratios, higher asymmetries, and a smaller range in \fw{}.

These two figures call the physical interpretation of the double Gaussian fits into question. 
It is unclear if the parameters are measuring physical quantities in the galaxies, or if they simply reflect the strength of the wings. 
The correlations between physical properties and both $\snarrow{} / \sbroad{}$ and \anab{} found by \citetalias{Ianj2012} were attributed to star formation, which could be responsible for driving gas into the wings of the profile.
We have discussed a similar idea in this paper using a different superprofile parameterization. 
Surprisingly, though, \anab{} and $\snarrow{} / \sbroad{}$ show, if anything, only weak trends with the direct measure of star formation represented by H$\alpha$ luminosities.
Additionally, since the wings of the profile appear to set the relationship between a number of the other measured parameters, it not clear if the narrow component is truly tracing cold \hi{} or if it is just another reflection of the superprofile wings. 
%If $\sigma_b$ truly does set the $A_n / A_b$ and $\sigma_n / \sigma_b$ ratios, a number of correlations with physical properties can be explained by the relationship between $\sigma_b$ and star formation. 

We can also turn to constraints provided by previous studies of individual \hi{} line-of-sight spectra. 
Many previous studies of spatially-resolved \hi{} line profiles in external galaxies have found that the line-of-sight profiles are often well-fit by single Gaussians, with only $\lesssim 20$\% of profiles exhibiting non-Gaussian structures such as broad wings or asymmetries \citeeg{Young2003, Warren2012}. 
The narrow and broad Gaussians fit to these small number of line profiles are often interpreted as the emission from CNM and WNM. 
However, it is unclear if this interpretation extends to the statistical measurement of \hi{} line shapes measured by the superprofiles, since information from each individual line profile is no longer distinct. 
For example, the asymmetric line profiles, when added together into a superprofile, can combine to form broader wings than what would be measured on a spatially-resolved basis.
The spatially-resolved studies of \hi{} line profiles have estimated the fraction of \hi{} mass in the cold component to be only $\sim 20$\% of the total \hi{} mass, which implies \anab{} values of $\sim 25$\% -- much less than those measured by double Gaussian fits.

As an additional test of the interpretation of the Gaussian fit parameters, we can compare their values to limits placed on cold \hi{} fractions based on double Gaussian fitting to individual line-of-sight profiles. 
\citet{Warren2012} estimated the \emph{spatially-resolved} minimum and maximum fraction of cold \hi{}, $\mathcal{F}_{cold}$, characterized by individual line of sight profiles that are best fit by a double Gaussian whose narrow component width is $\sigma < 6$ \kms{}, in a number of our sample galaxies. 
In Figure~\ref{fig:anab-warren} we plot the limits on \anab{} from \citet{Warren2012} versus \anab{} values measured from double Gaussian fits to the global superprofile. 
Each box represents a single galaxy. 
The position and width of the box on the $x$-axis is determined by the \anab{} value and associated error from double Gaussian fits. 
The size of the box on the $y$-axis is determined by the allowed range of $\anab{} = \mathcal{F}_{cold} / (1 - \mathcal{F}_{cold})$ values for that galaxy as given by \citet{Warren2012}. 
Grey boxes represent measurements from galaxies presented in this paper, while red boxes are those from \citetalias{Ianj2012}. 
For simplicity, we have assumed that our $\anab{}$ uncertainty is 0.05.

The dashed line in Figure~\ref{fig:anab-warren} is a line of equality, where the independent limits on \anab{} match double Gaussian \anab{} values. 
That is, the boxes would roughly lie along the dashed line if double Gaussian \anab{} values matched the independent limits. 
However, this behavior is not seen. 
Instead, the independent limits on \anab{} tend to be smaller than those measured by double Gaussian fits. 
While we might expect the independent limits on \anab{} to be smaller than the double Gaussian \anab{} values because the superprofiles are a higher S/N representation of \hi{} spectra, we would at least expect to see the a positive correlation between independent limits and double Gaussian \anab{} values. 
However, the data, especially the red boxes, show, if anything, a negative correlation. 
The fact that double Gaussian \anab{} values do not match independent limits indicates that the cold and warm \hi{} interpretation may not be valid.

Finally, we can examine limits on the \hi{} velocity dispersions from observational studies. 
\citet{Petric2007} found that double Gaussian fits to median \hi{} profile shapes at different radii exhibited the same ratio of narrow to broad components. 
This finding complicates the CNM/WNM interpretation, as \hi{} in the outskirts of disks would not be expected to have the same fraction of gas in the warm component as regions inside $r_{25}$. 
Second, the CNM/WNM interpretation of \anab{} values implicitly assumes that there are two well-thermalized \hi{} components in the ISM. 
If this were the case, we might expect to see two characteristic velocity dispersions in the statistical ensemble of individual \hi{} spectra widths. \citet{Braun2009} measured the non-thermal velocity dispersion of \hi{} in M31 and found no evidence of a bimodal distribution. 
Instead, they find a range of non-thermal velocity dispersions between $3 - 25$ \kms{}, with most line-of-sight profiles exhibiting widths $\sim 8$ \kms{}.  Based on this evidence, it therefore is more physically meaningful to interpret the central peak of the superprofiles, which show similar velocity dispersions of $\sim 8$ \kms{}, as turbulence in the ISM instead of as the composite of emission from distinct cold and warm \hi{} components. 
However, this last constraint may not apply if the CNM and WNM components of the ISM are well-mixed on scales smaller than the spatial resolution of the observations.

With the strength of the evidence presented above, we have chosen to parameterize the superprofiles as composed primarily of a central turbulent peak with wings to either side, instead of as two Gaussian components. 
The parameters of double Gaussians, when interpreted as representative of the CNM and WNM, do not behave as expected when exploring the superprofiles in more detail, as the fit is dominated by the small amount of \hi{} in the wings. 
Additionally, evidence from other studies does not support the cold and warm gas interpretation of the double Gaussian fits. 
Previously-derived limits on \anab{} do not match \anab{} values from the CNM/WNM interpretation of double Gaussian fits, and there is no evidence for a bimodal temperature distribution in individual \hi{} line of sight spectra even though such an interpretation is relevant in localized regions. 
We therefore believe the superprofiles are better interpreted not as two Gaussian components representing the CNM and WNM, but instead as a central turbulent peak with wings generated by kinematically disturbed gas.

\end{appendix}

% References
\bibliographystyle{apj}
\bibliography{global_energy_astroph}

%
% Figures!
%

\clearpage

\clearpage{}

\clearpage{}

\clearpage{}

\clearpage{}

\clearpage{}
\clearpage{}
\addtocounter{figure}{-1}
\begin{figure*}
  \centering
  \ifimage
  \includegraphics[width=5cm]{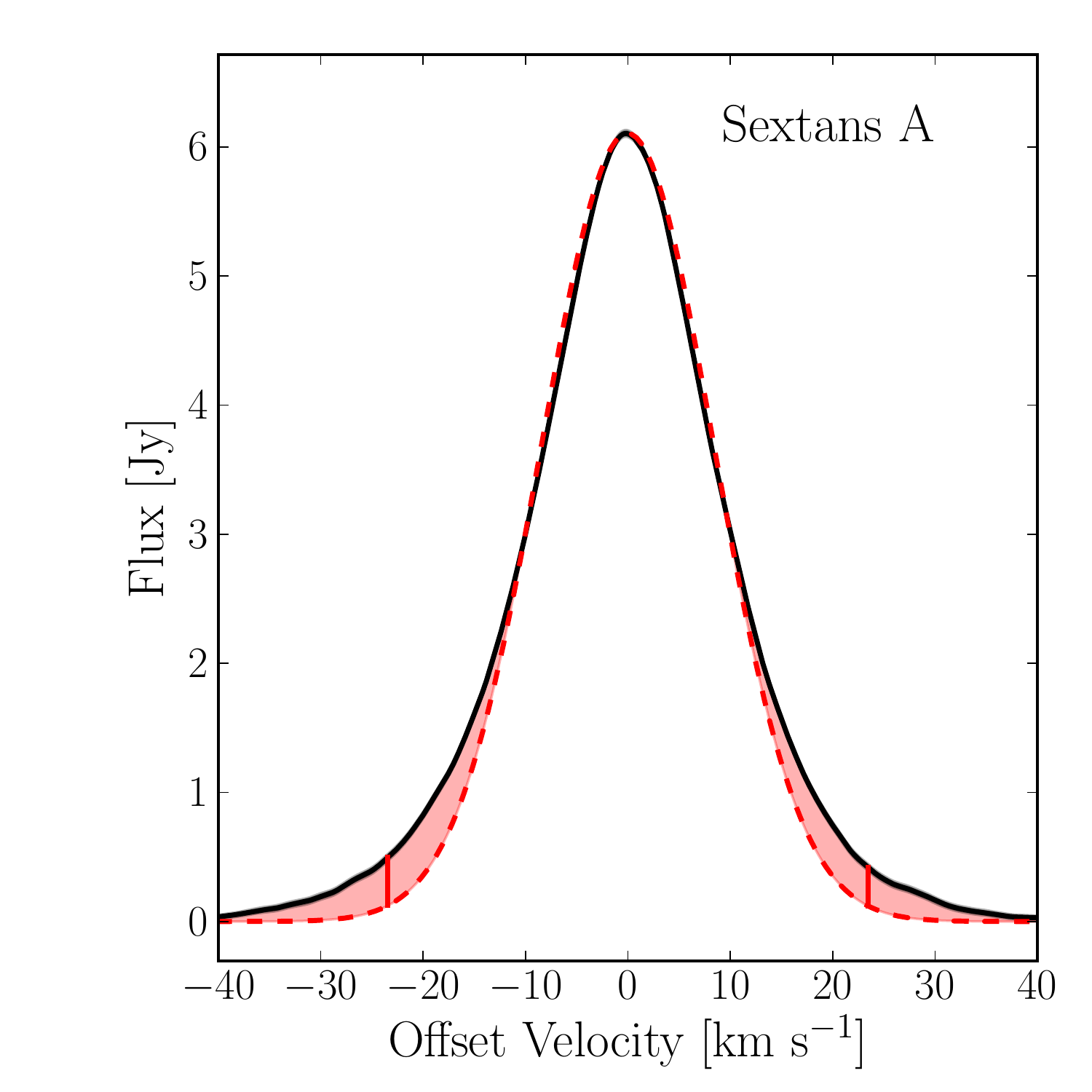}%{{figures/sexa.ro.phys200.superprofile.hwhm}.pdf}
  \includegraphics[width=5cm]{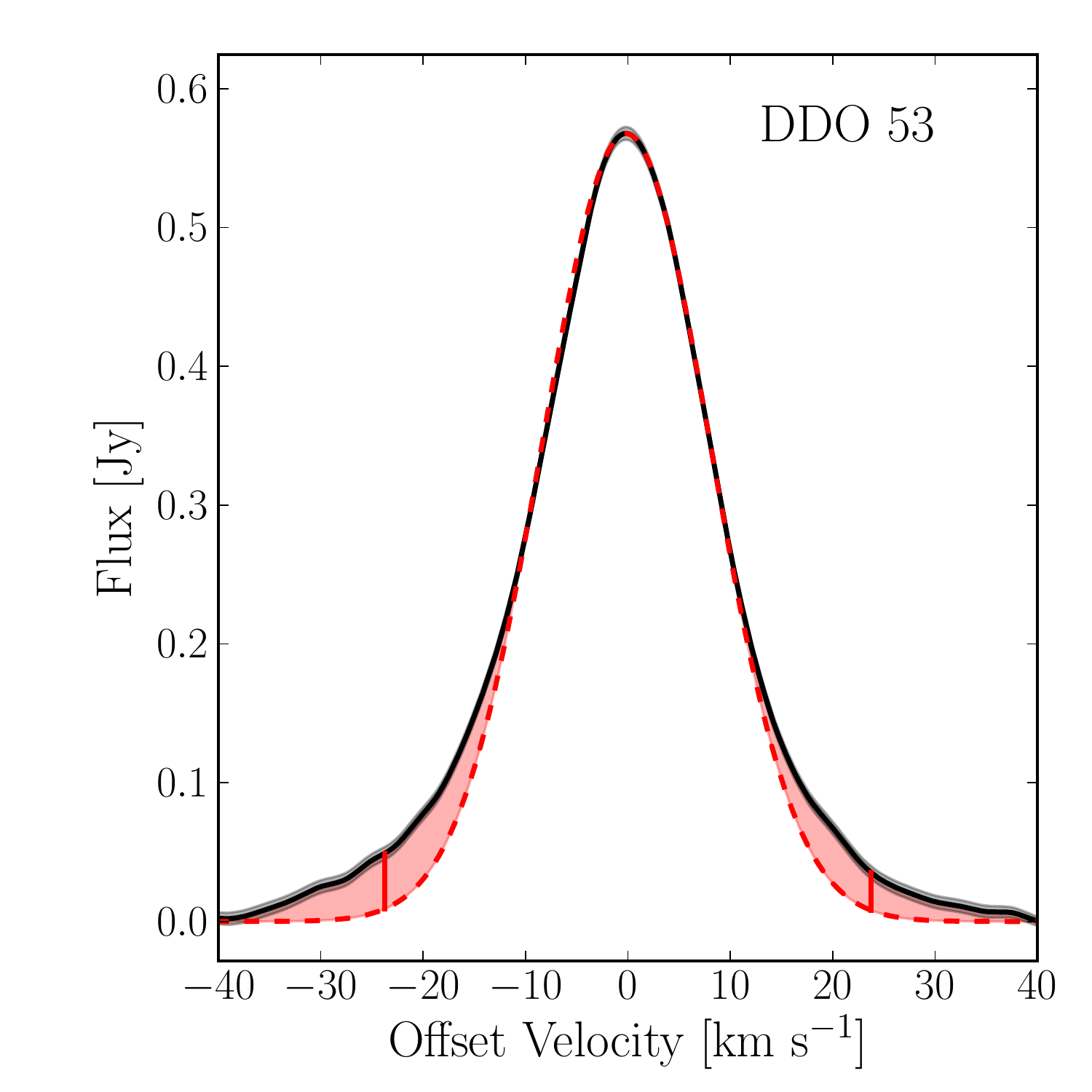}%{{figures/ddo53.ro.phys200.superprofile.hwhm}.pdf}
  \includegraphics[width=5cm]{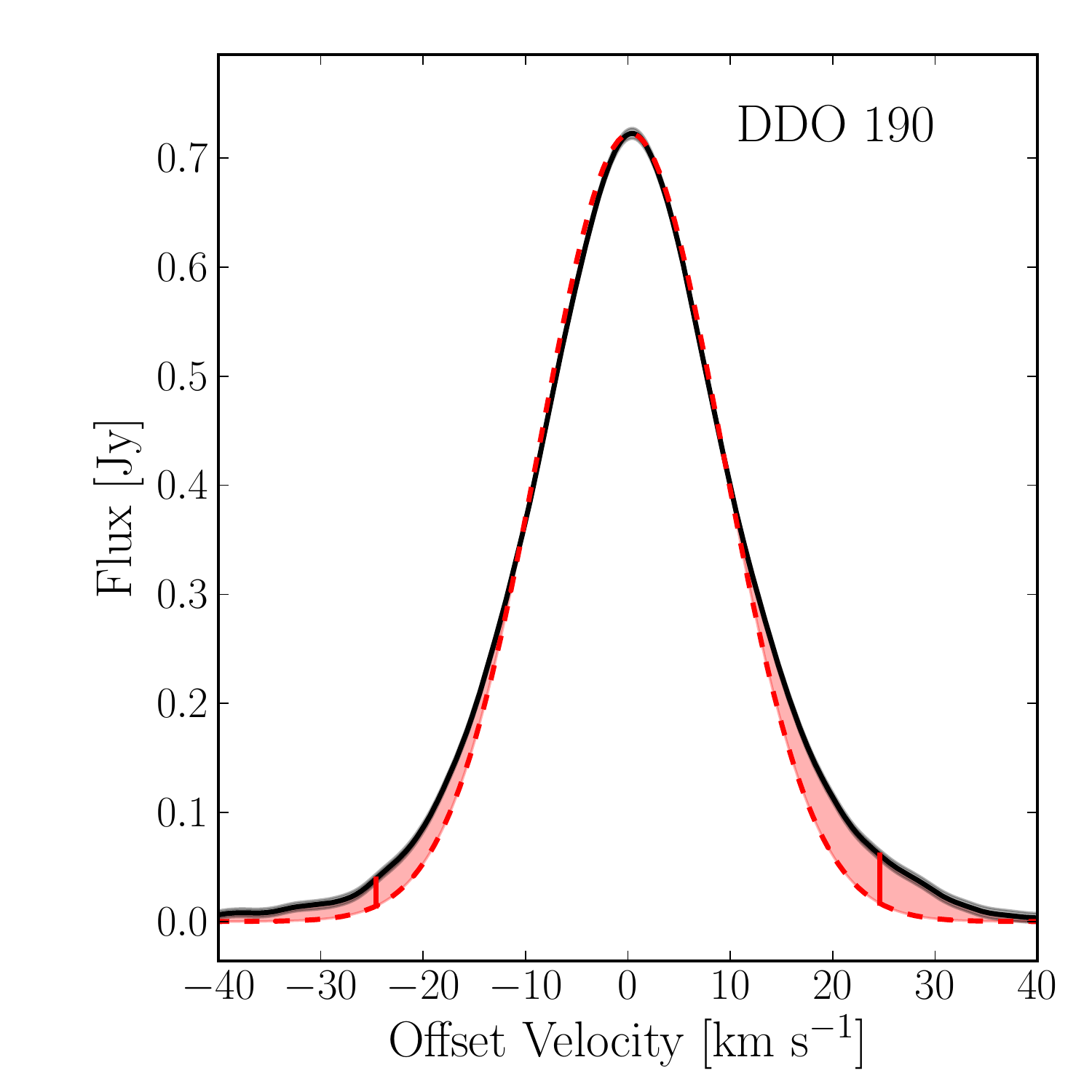}%{{figures/ddo190.ro.phys200.superprofile.hwhm}.pdf}
  \\
  \includegraphics[width=5cm]{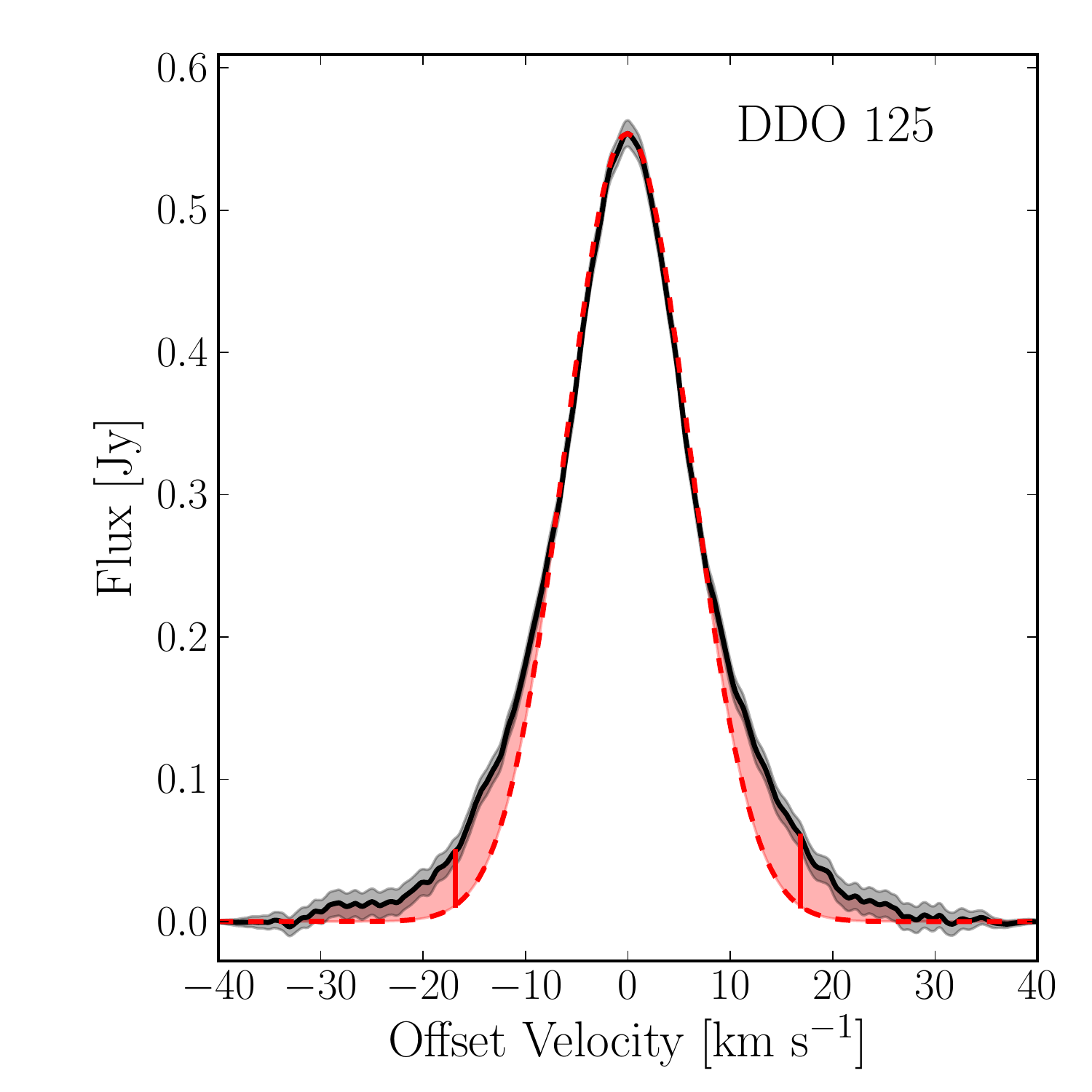}%{{figures/ddo125.ro.phys200.superprofile.hwhm}.pdf}
  \includegraphics[width=5cm]{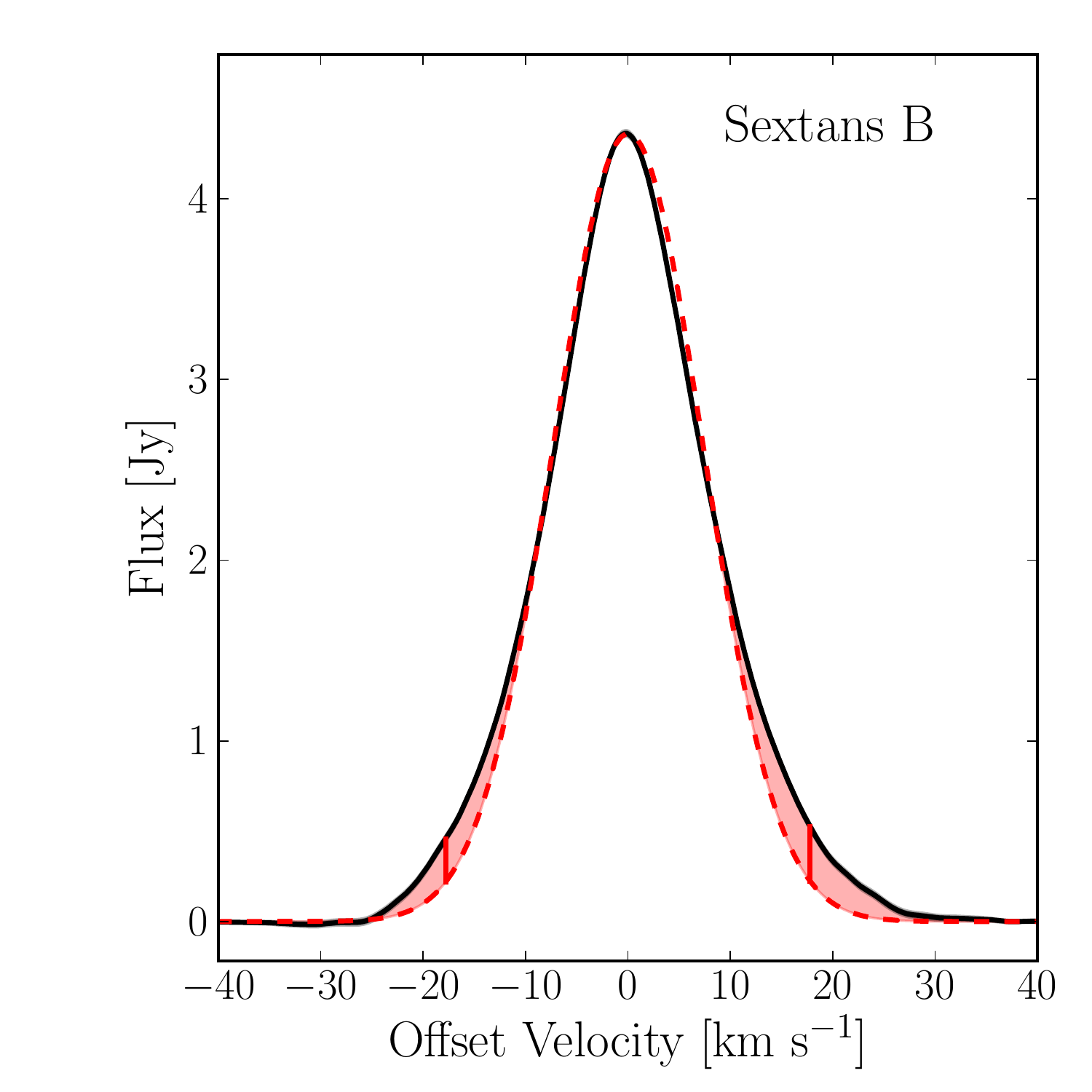}%{{figures/sexb.ro.phys200.superprofile.hwhm}.pdf}
  \includegraphics[width=5cm]{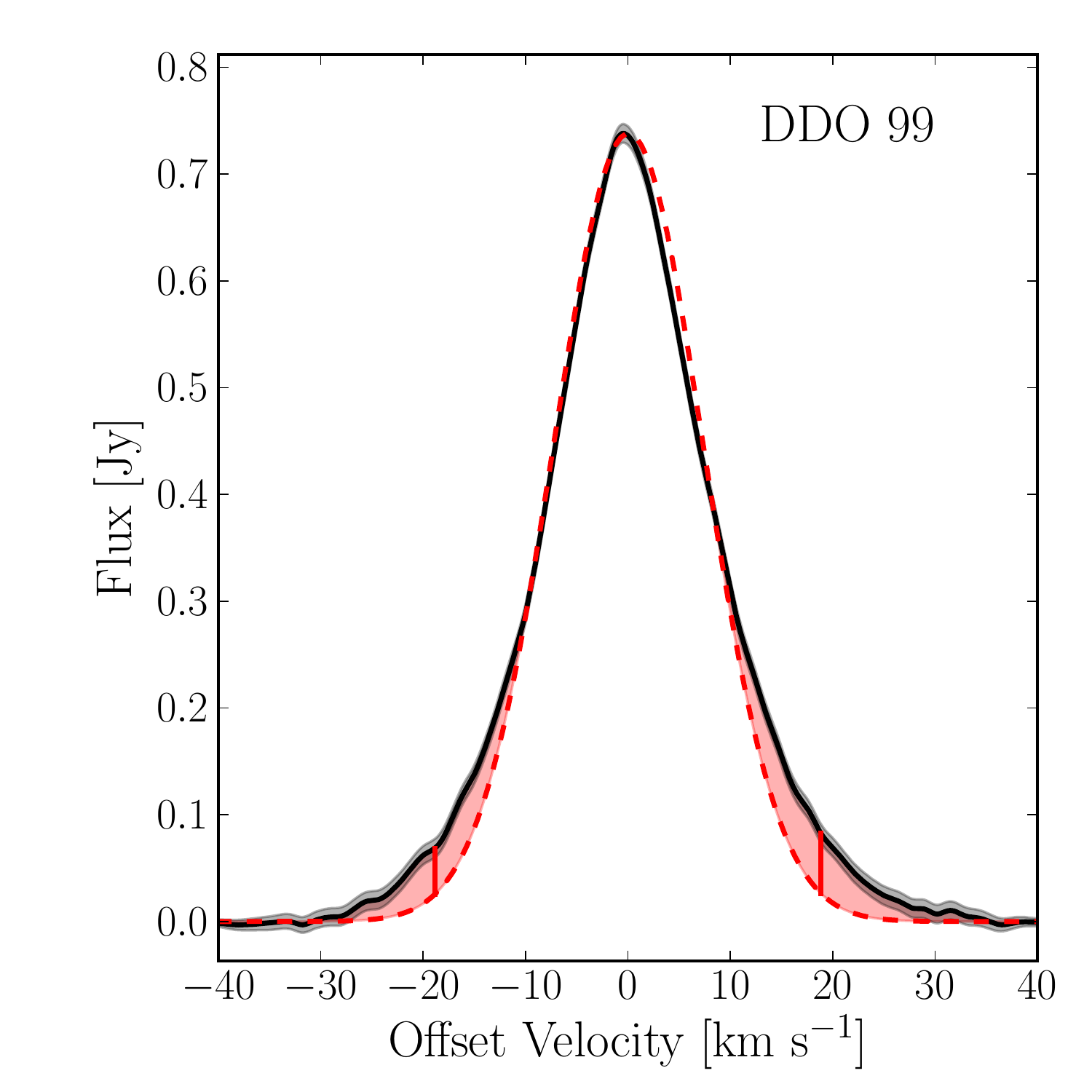}%{{figures/ddo99.ro.phys200.superprofile.hwhm}.pdf}
  \\
  \includegraphics[width=5cm]{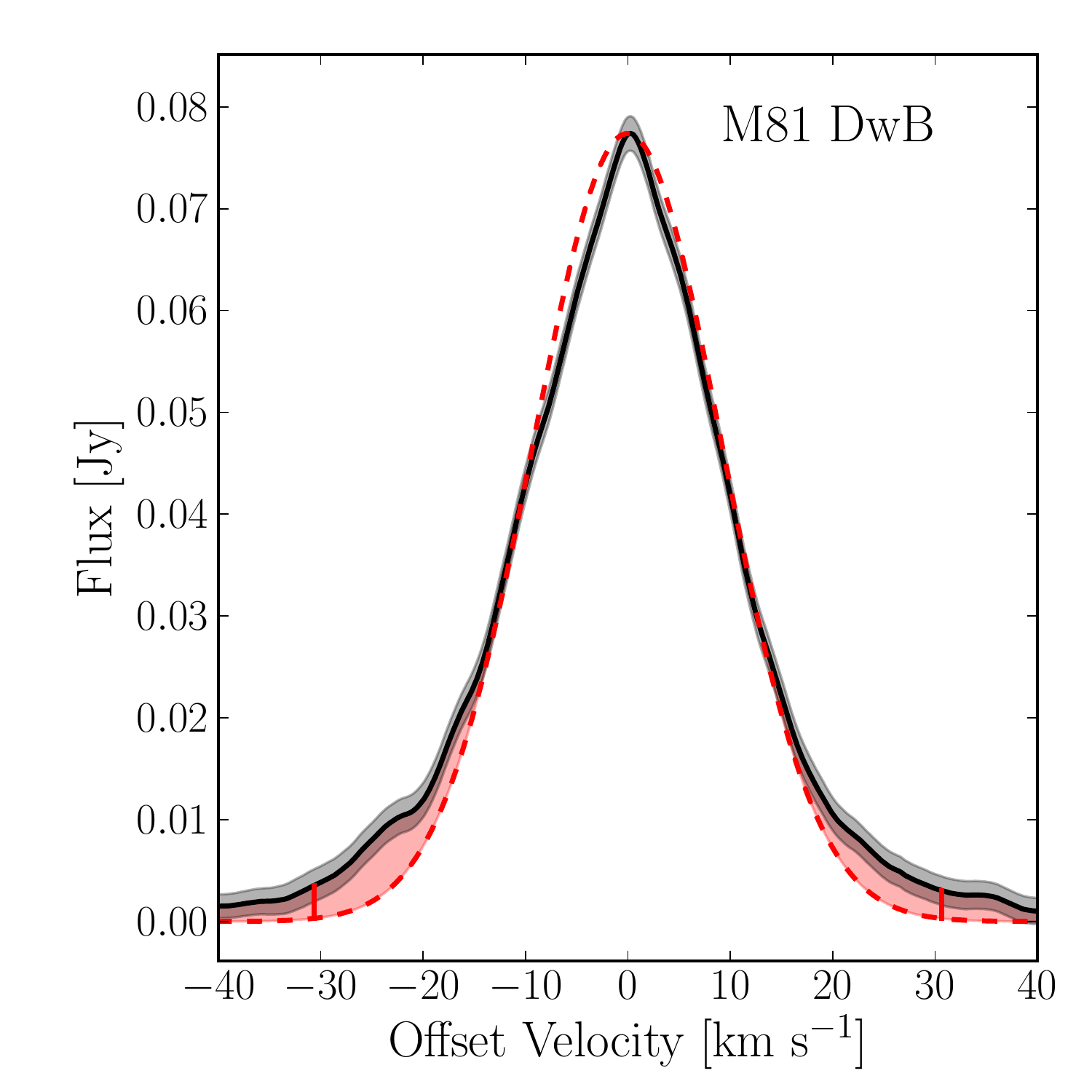}%{{figures/m81dwb.ro.phys200.superprofile.hwhm}.pdf}
  \includegraphics[width=5cm]{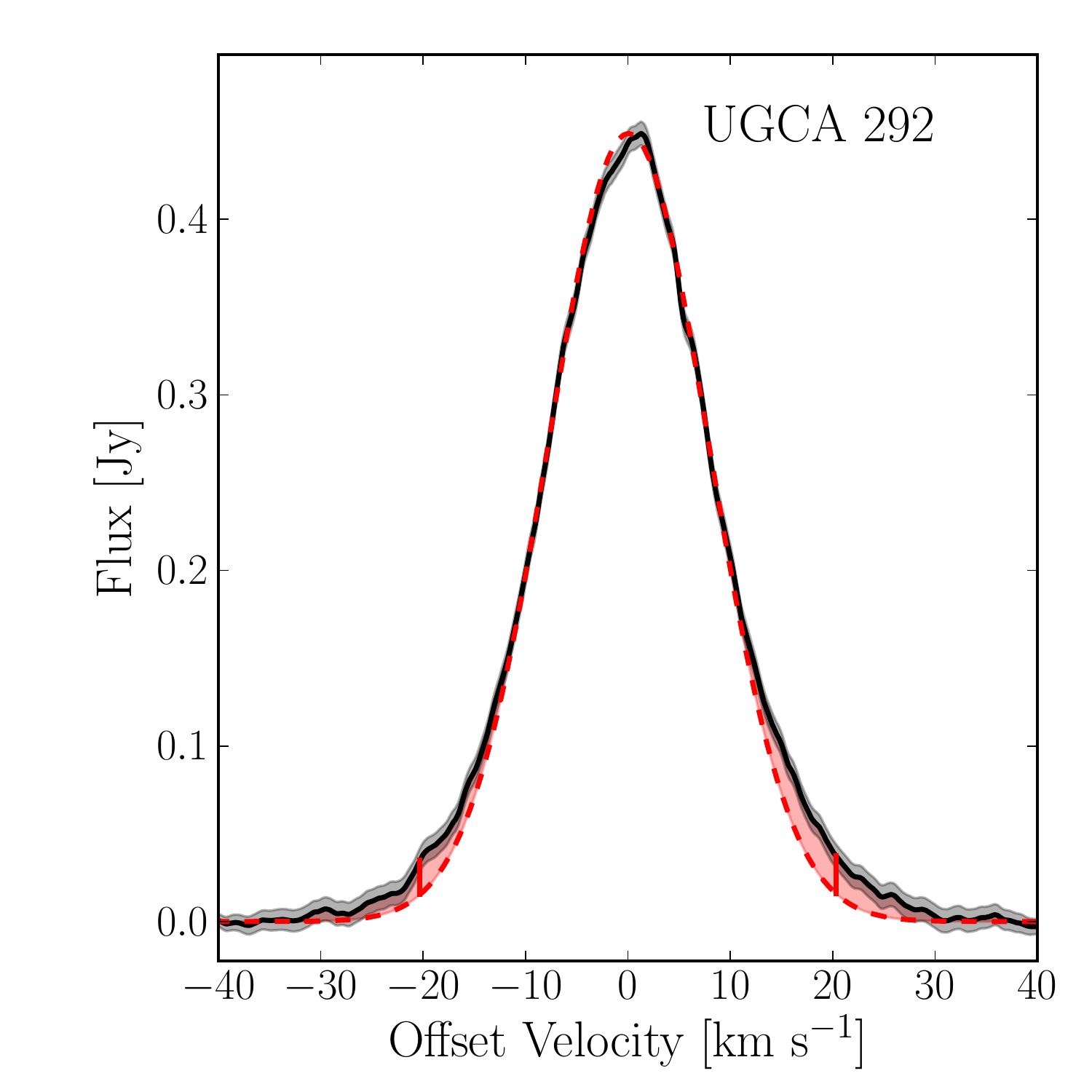}%{{figures/ua292.ro.phys200.superprofile.hwhm}.pdf}
  \includegraphics[width=5cm]{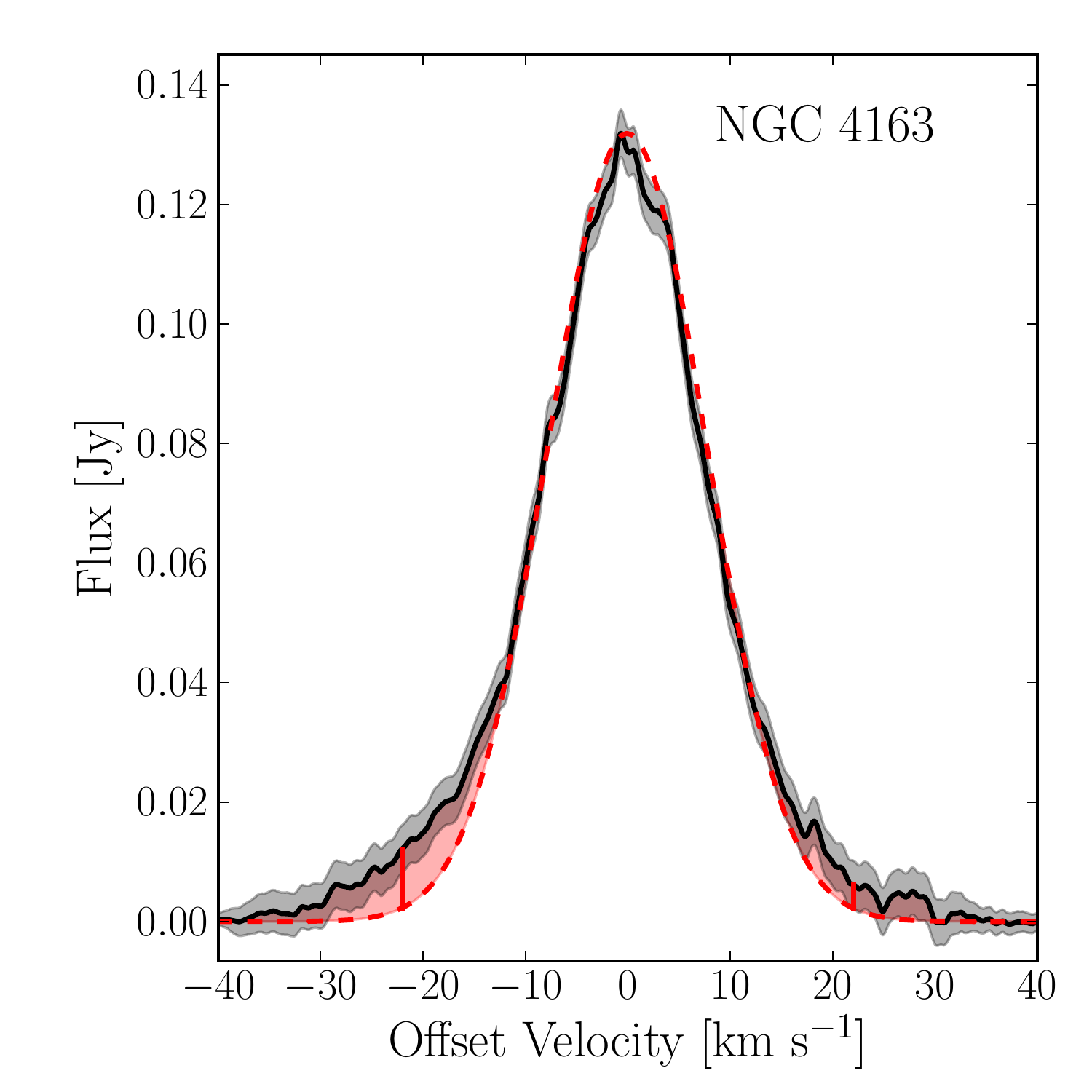}%{{figures/n4163.ro.phys200.superprofile.hwhm}.pdf}

  \fi
  \caption{\hi{} superprofiles (continued). Galaxies are ordered by decreasing \m{baryon,tot}.}
\end{figure*}
\clearpage{}
\addtocounter{figure}{-1}
\begin{figure*}
  \centering
  \ifimage
  \includegraphics[width=5cm]{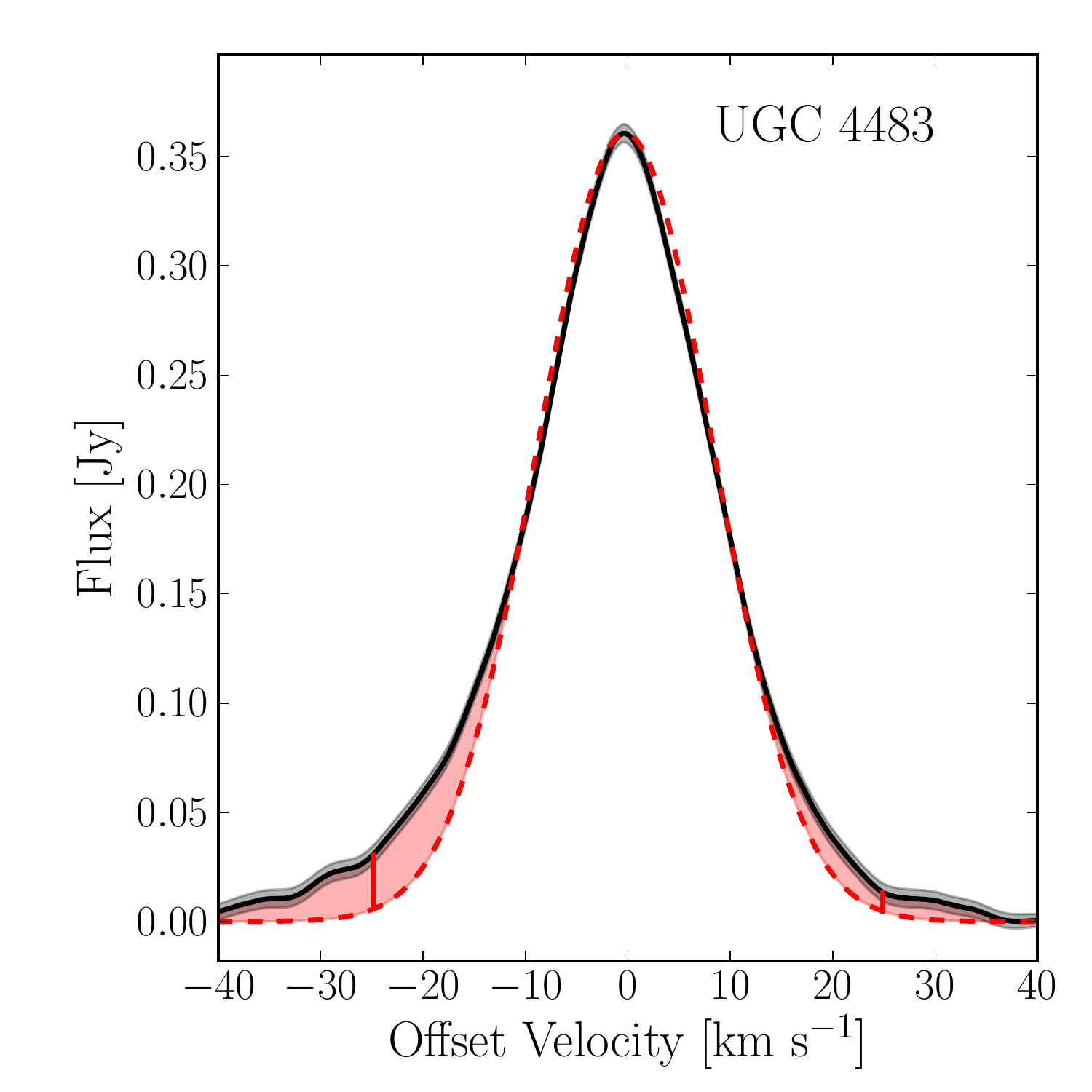}%{{figures/u4483.ro.phys200.superprofile.hwhm}.pdf}
  \includegraphics[width=5cm]{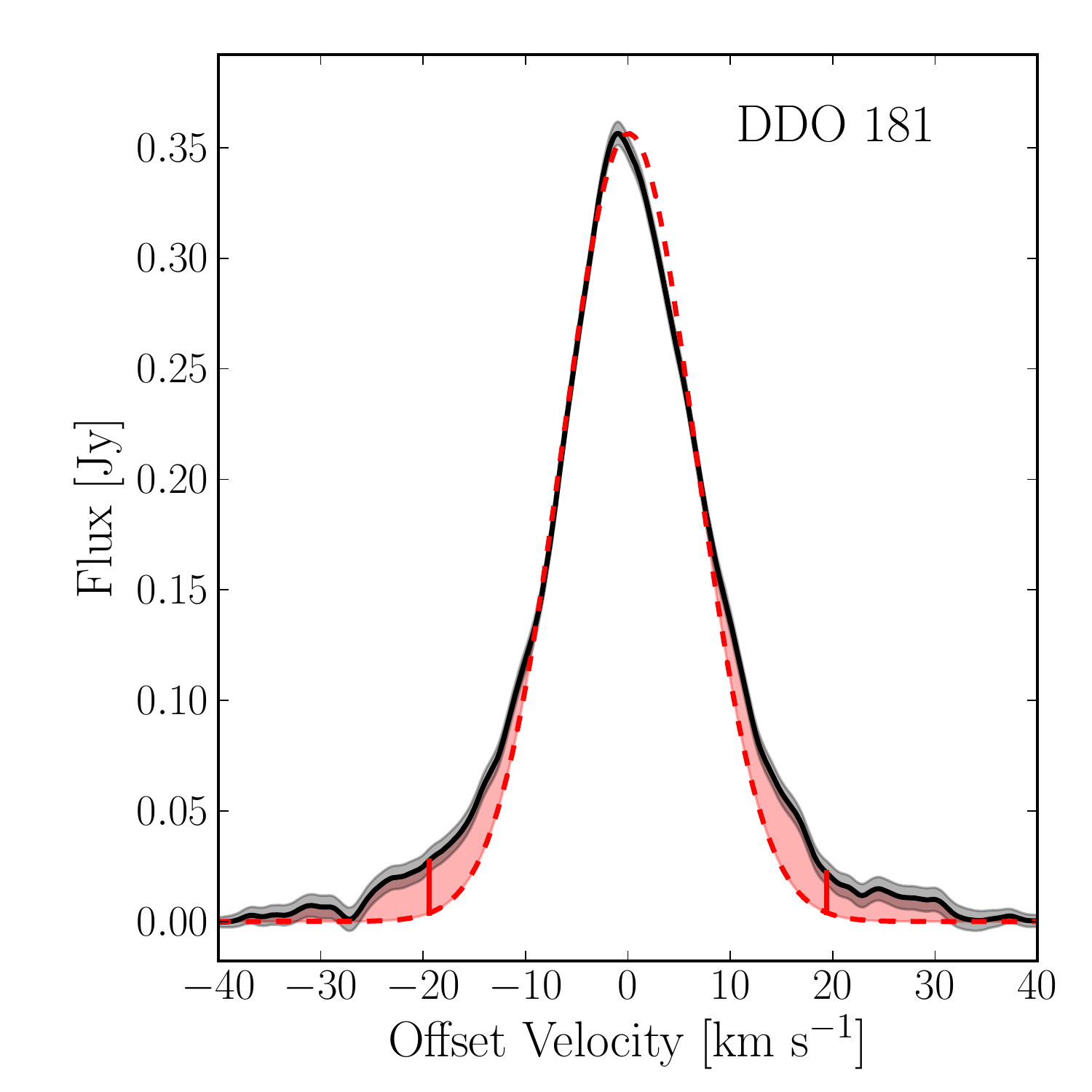}%{{figures/ddo181.ro.phys200.superprofile.hwhm}.pdf}
  \includegraphics[width=5cm]{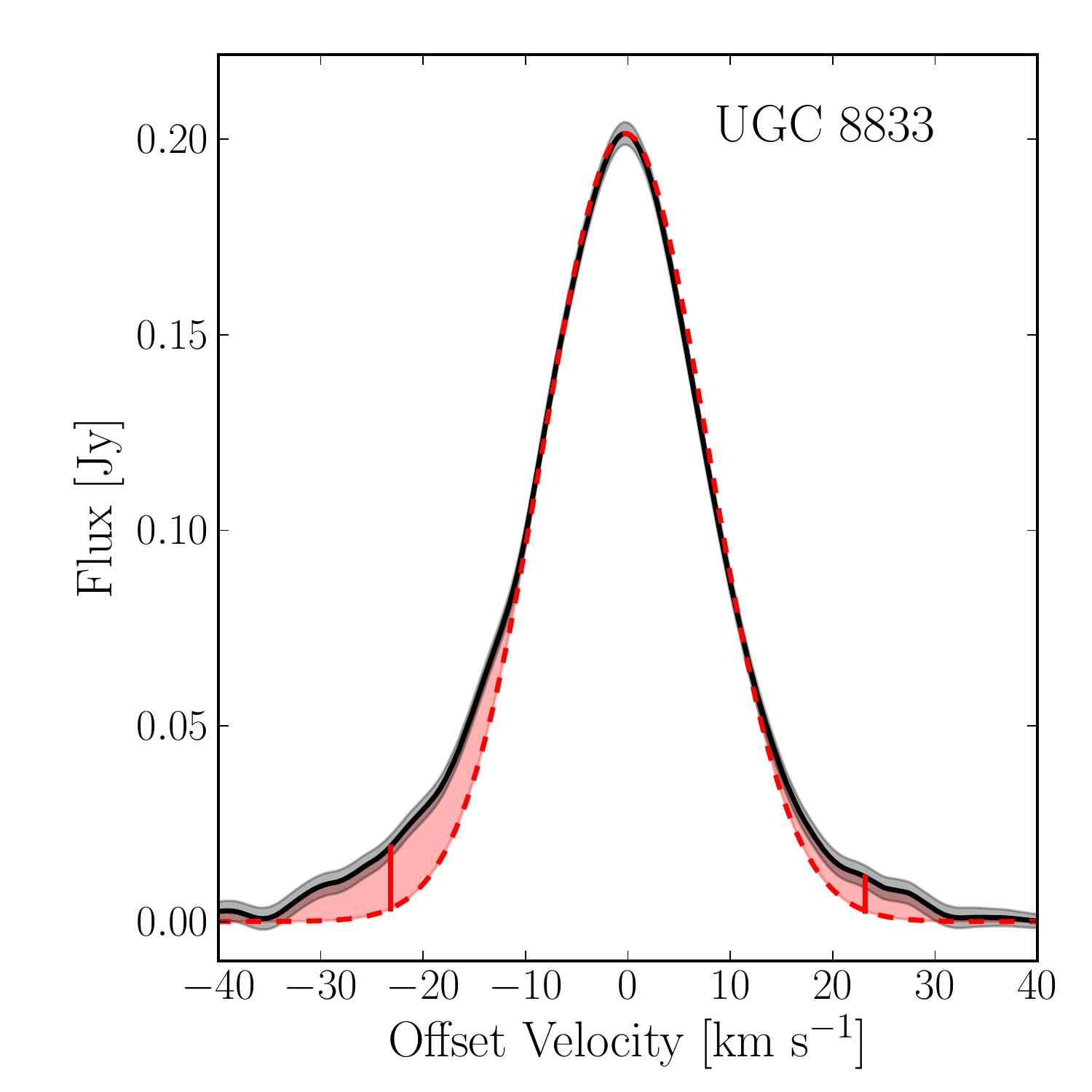}%{{figures/u8833.ro.phys200.superprofile.hwhm}.pdf}
  \\
  \includegraphics[width=5cm]{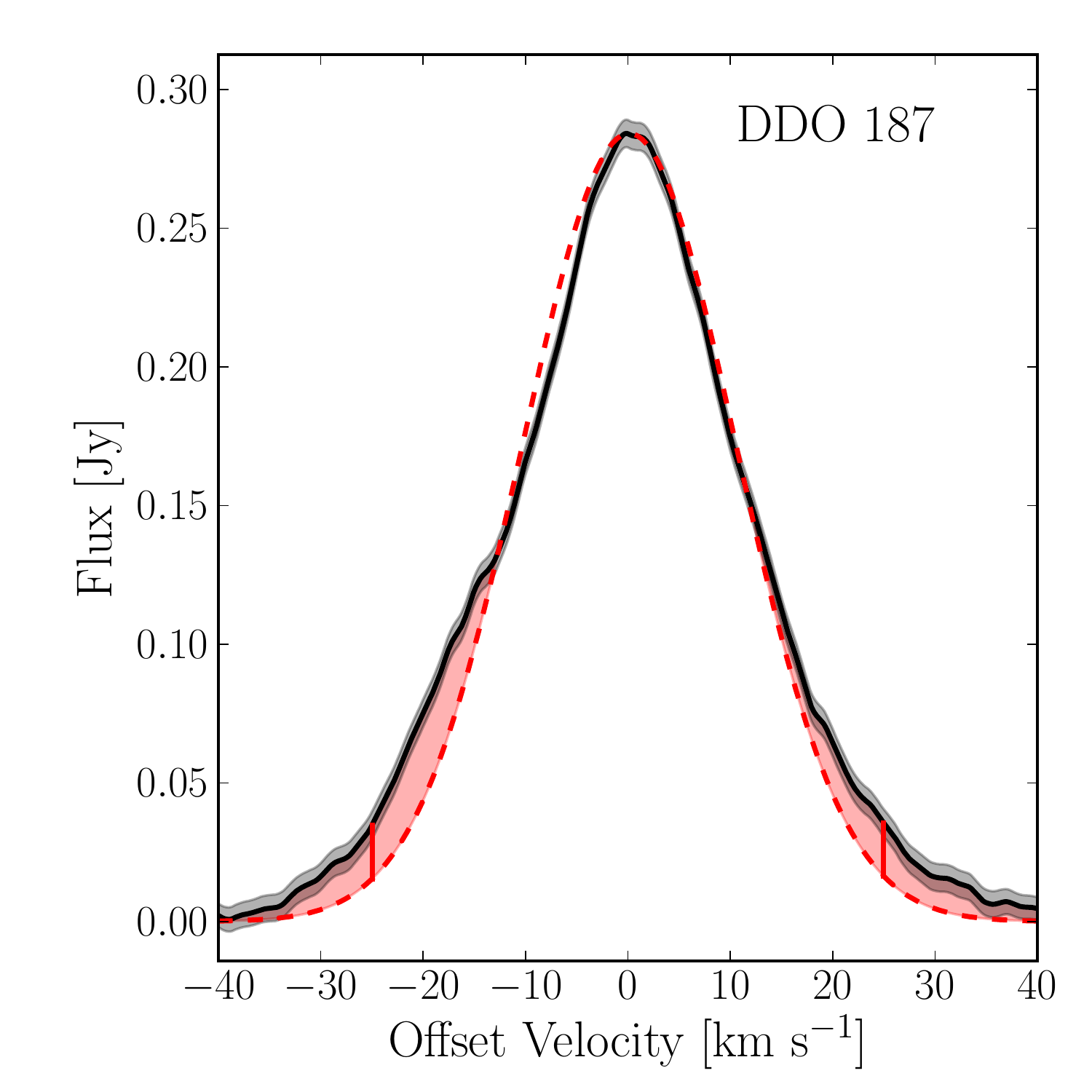}%{{figures/ddo187.ro.phys200.superprofile.hwhm}.pdf}
  \includegraphics[width=5cm]{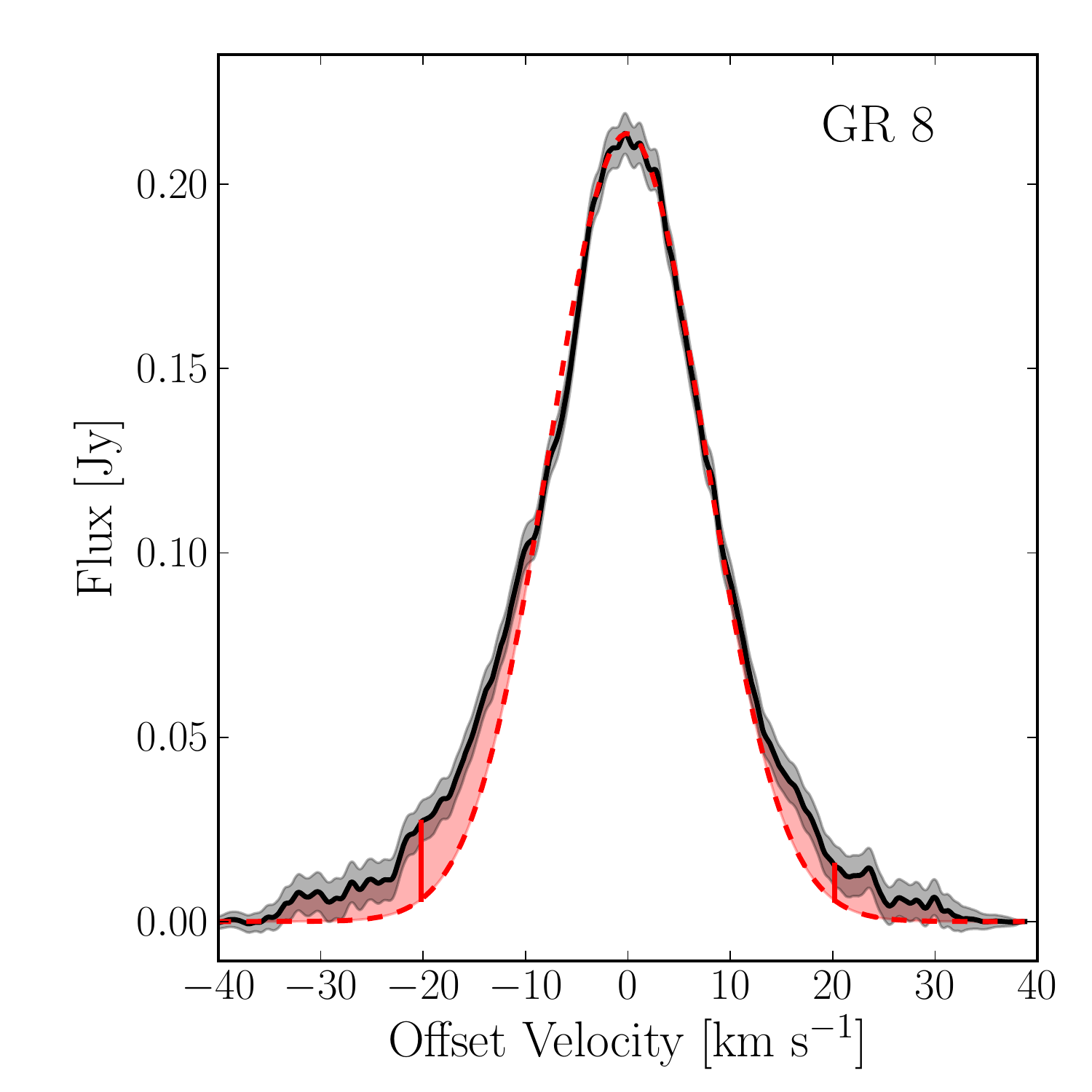}%{{figures/gr8.ro.phys200.superprofile.hwhm}.pdf}
  \fi
  \caption{\hi{} superprofiles (continued). Galaxies are ordered by decreasing \m{baryon,tot}.}
\end{figure*}

\clearpage{}
\begin{figure}
  \centering
  \ifimage
  \includegraphics{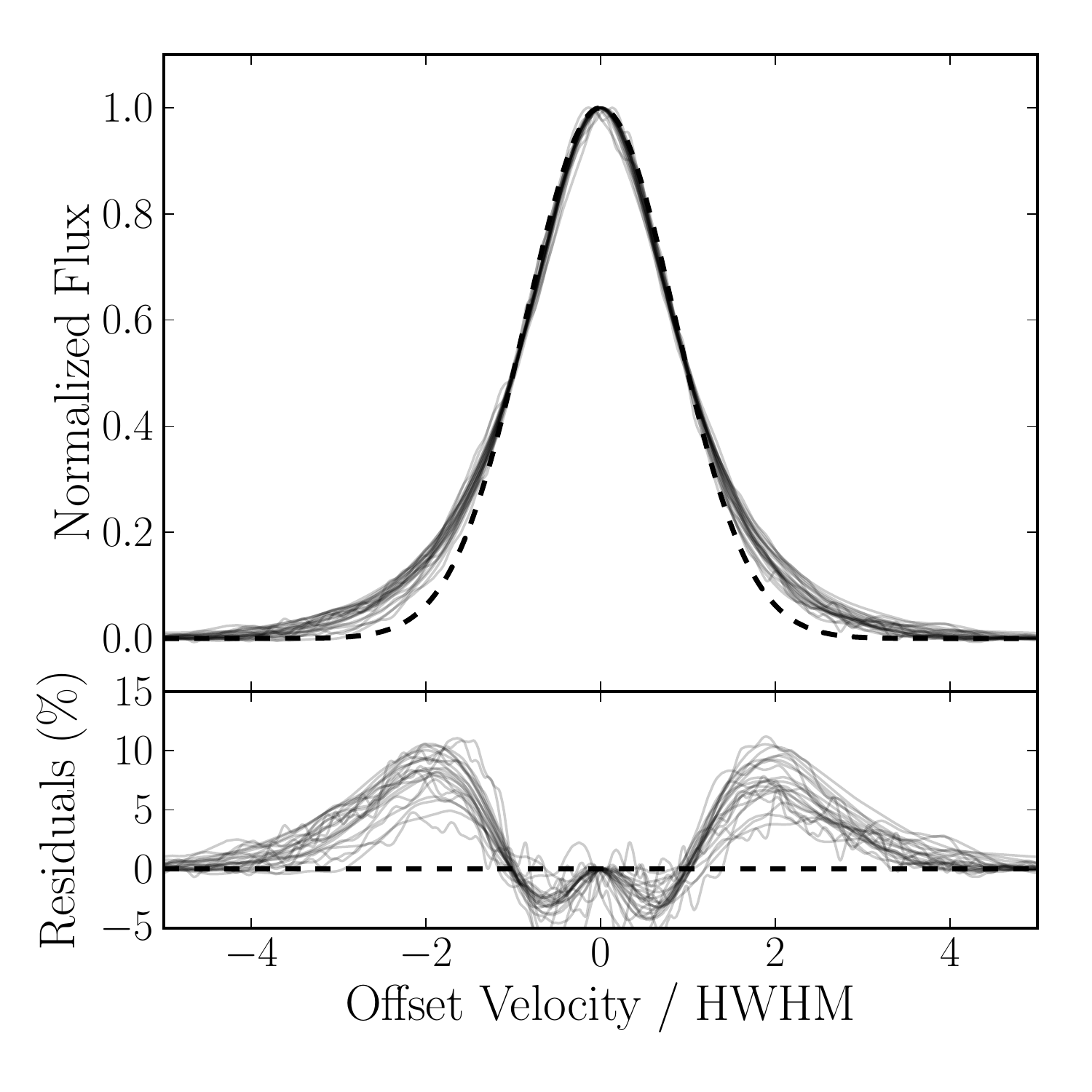}%{{figures/superprofiles-scaled.bw}.pdf}
  \fi
  \caption{Each solid black line is the superprofile for a single galaxy. They have been normalized to their maximum height on the y-axis. The velocity axis has been scaled by the superprofile HWHM value. The superprofile lines have been plotted with transparency; overlap regions are darker. The thick dashed line is a Gaussian with amplitude $= 1$ and HWHM $= 1$. The lower panel shows the residuals from the Gaussian overlay. Compared to the Gaussian, the superprofiles are slightly more narrowly-peaked and have more flux in the wings. The superprofiles show a remarkably uniform shape, with the primary variations in the amount of gas moving faster than expected from the Gaussian overlay.  \label{fig:superprofiles-scaled}}
\end{figure}

\clearpage
\begin{figure}
  \centering
  \ifimage
  \includegraphics[width=6in]{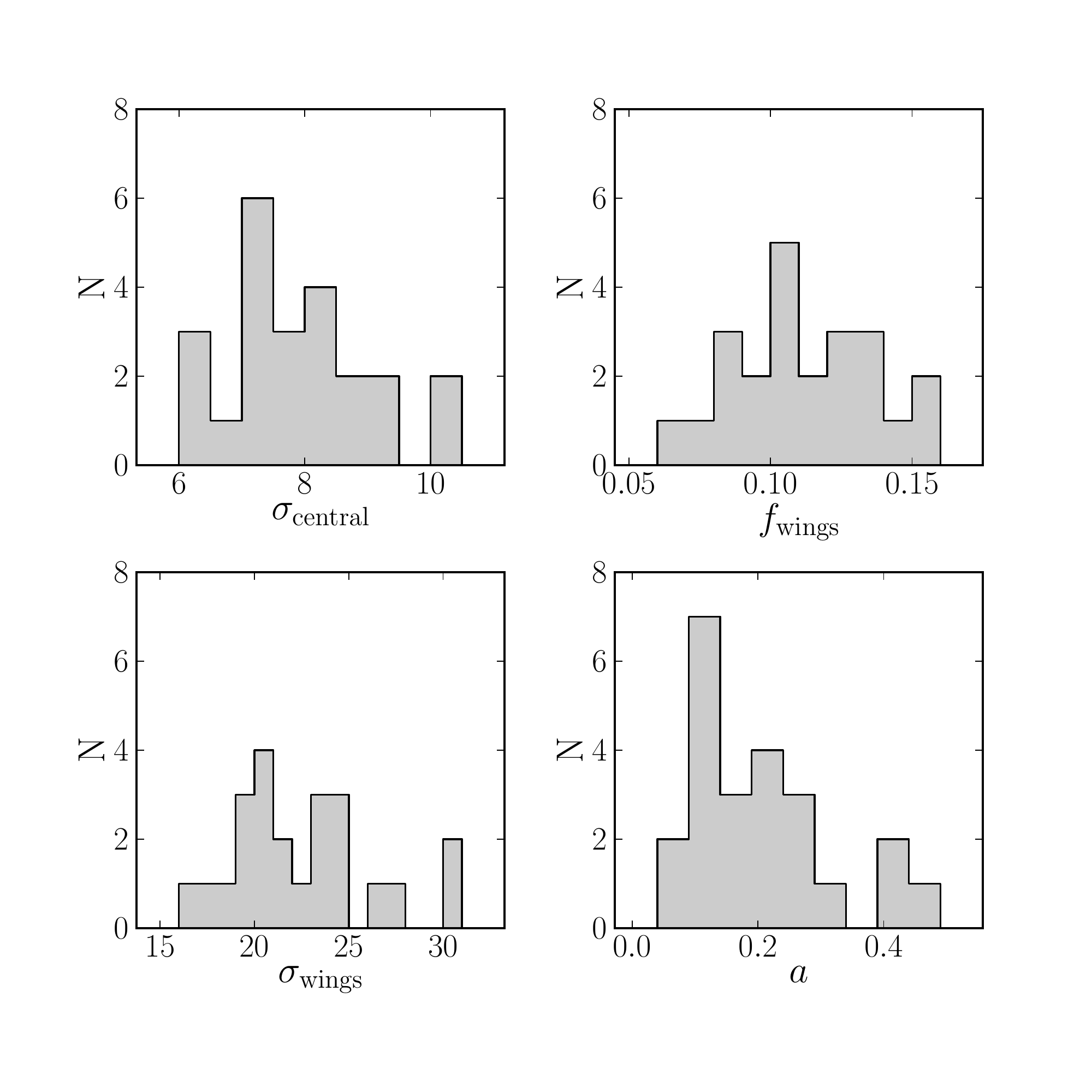}%{{figures/param-histograms.hwhm}.pdf}
  \fi
  \caption{The distribution of measured parameters \scentral{}, \swing{}, \fw{}, and \aw{} in our sample galaxies.  \label{fig:parameter-histograms}}
\end{figure}

\clearpage{}
\begin{figure}
  \centering
  \ifimage
  \includegraphics[height=6in]{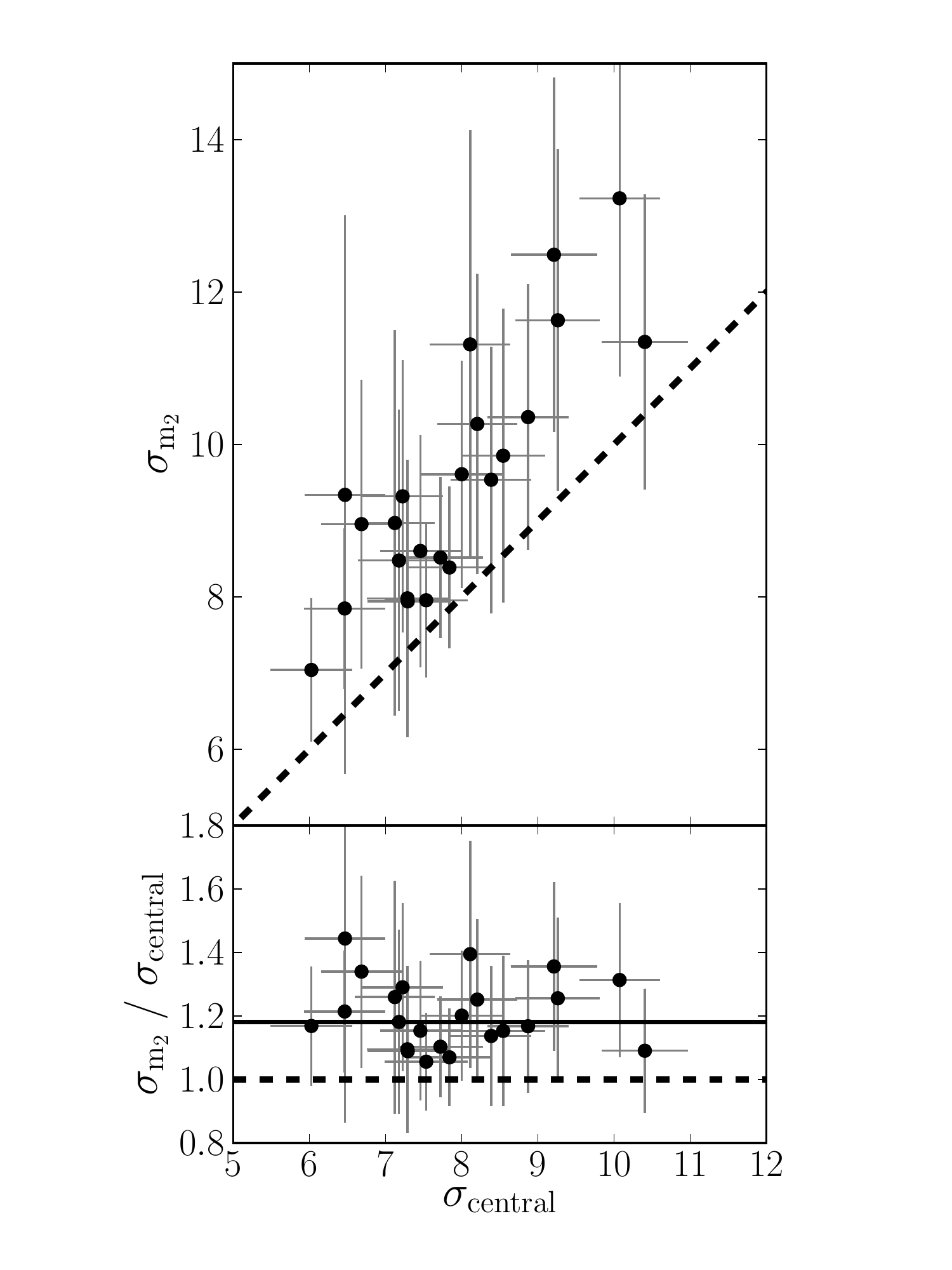}%{{figures/comp-superprofile-m2.s1.ro.phys200}.pdf}
  \fi
  \caption{A comparison of the global second moment versus the dispersion measured by the superprofile HWHM-scaled Gaussian. The dashed line shows where the $\sigma_\mathrm{m_2} = \scentral{}$. The error bars on the x-axis are the approximate uncertainties on $\scentral{}$, while those on y-axis represent the weighted standard deviation of second moment values in the pixels considered. The solid black line at $\sigma_\mathrm{m_2} / \scentral{} \sim 1.2$ on the lower panel indicates the median ratio between $\sigma_\mathrm{m2}$ and \scentral{}. The second moment overestimates the width by $\sim10 - 50$\%.  \label{fig:m2-v-snarrow}}
\end{figure}

\clearpage{}
\begin{figure}
  \centering
  \ifimage
    \includegraphics[height=7in]{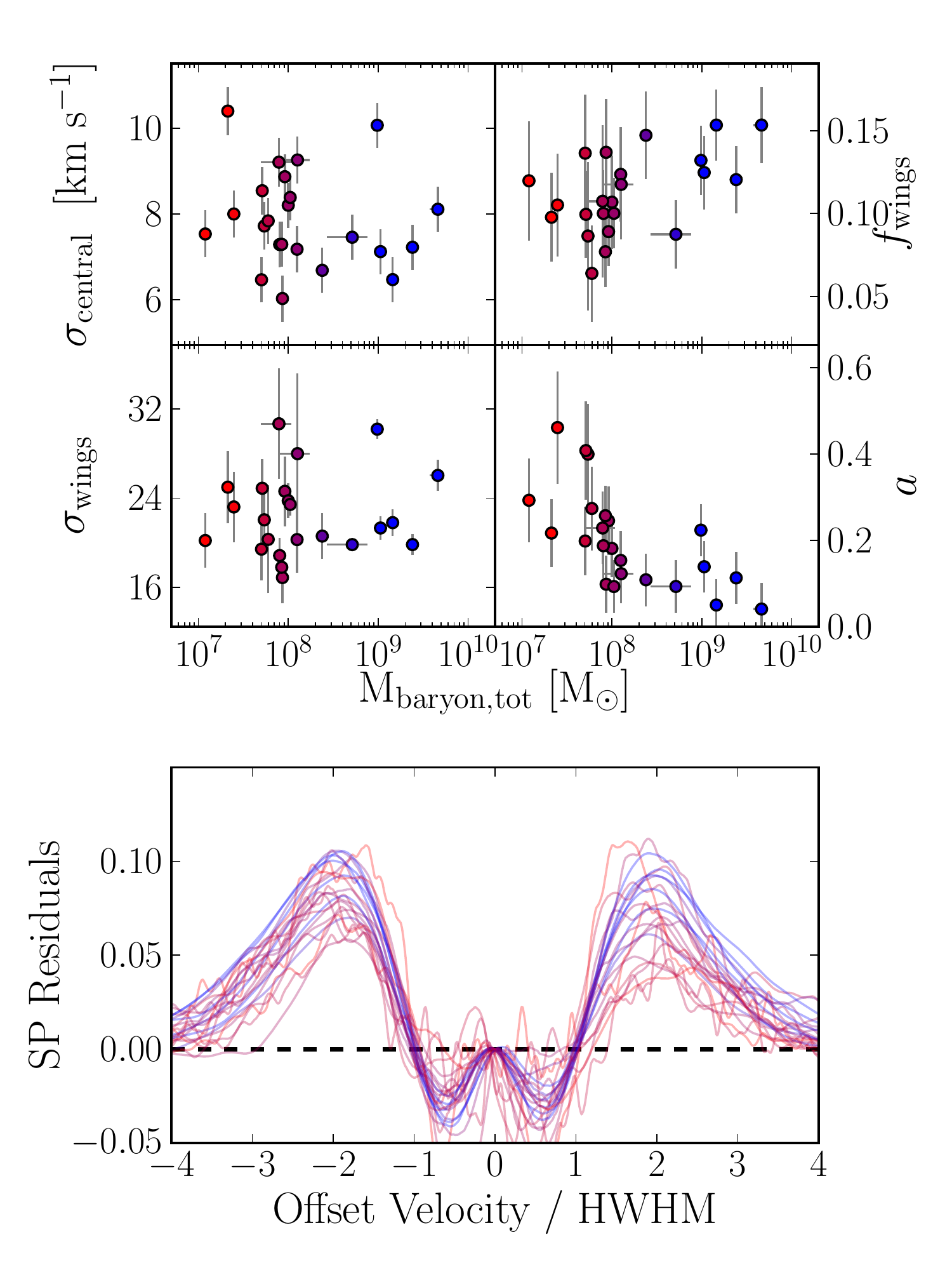}%{{figures/four-panel.hwhm.m_baryon}.pdf}
  \fi
  \caption{The upper panel shows the measured parameters versus galaxy \m{baryon,tot}. Each point is colored from blue to red based on increasing \m{baryon,tot}. The bottom panel shows the residuals of the normalized superprofiles. The color of the line corresponds to that on the upper panel. The lines are plotted with a transparency value, so overlap regions are more saturated. Both \fw{} and \aw{} show trends with increasing star formation.  \label{fig:panels-mbaryon}}
\end{figure}

\clearpage{}
\begin{figure}
  \centering
  \ifimage
  \includegraphics[height=7in]{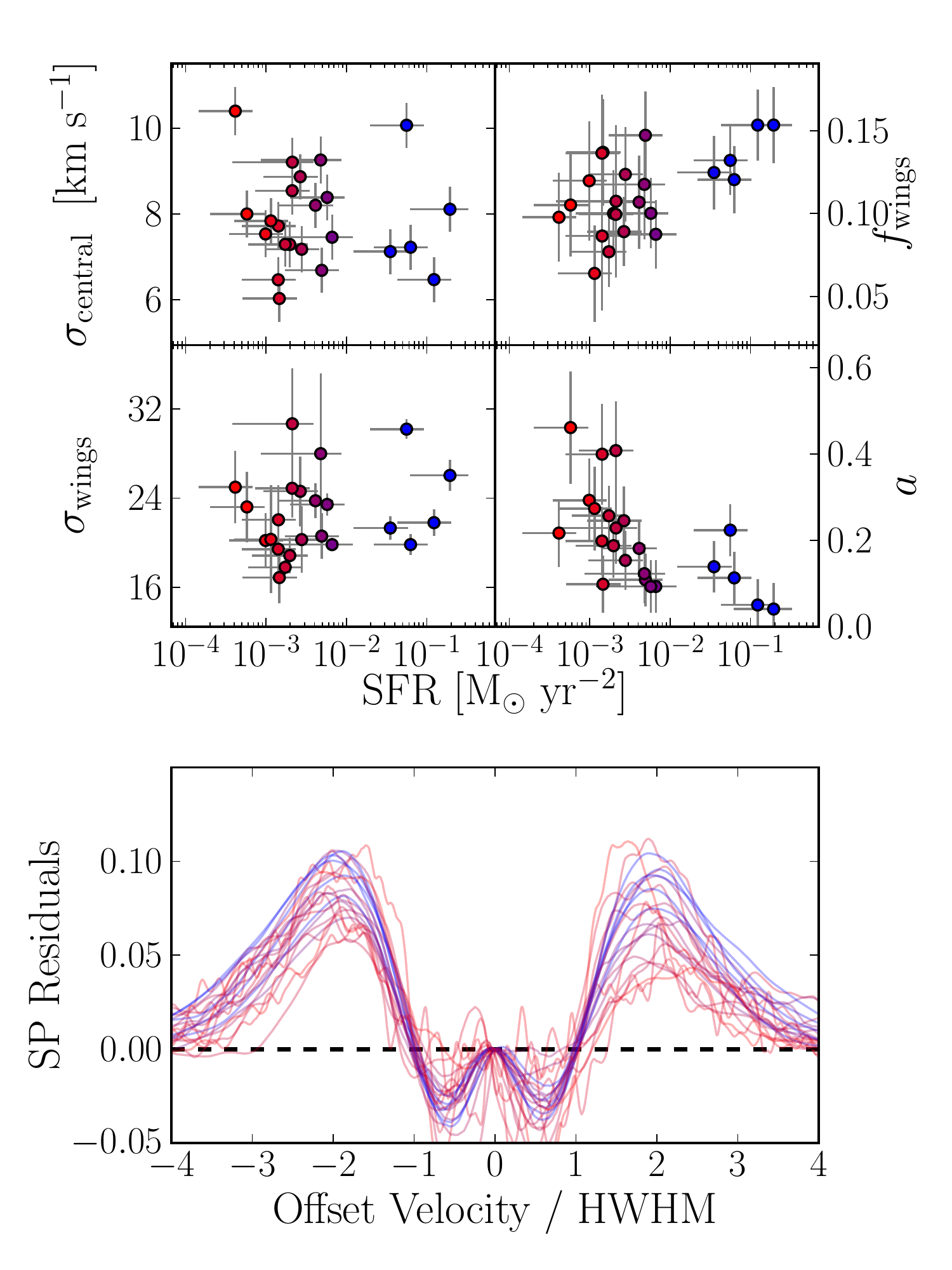}%{{figures/four-panel.hwhm.sfr_pix}.pdf}
  \fi
  \caption{Same as Figure~\ref{fig:panels-mbaryon}, but for SFR. We find that \aw{} decreases with increasing SFR. \label{fig:panels-sfr}}
\end{figure}

\clearpage{}
\begin{figure}
  \centering
  \ifimage
  \includegraphics[height=7in]{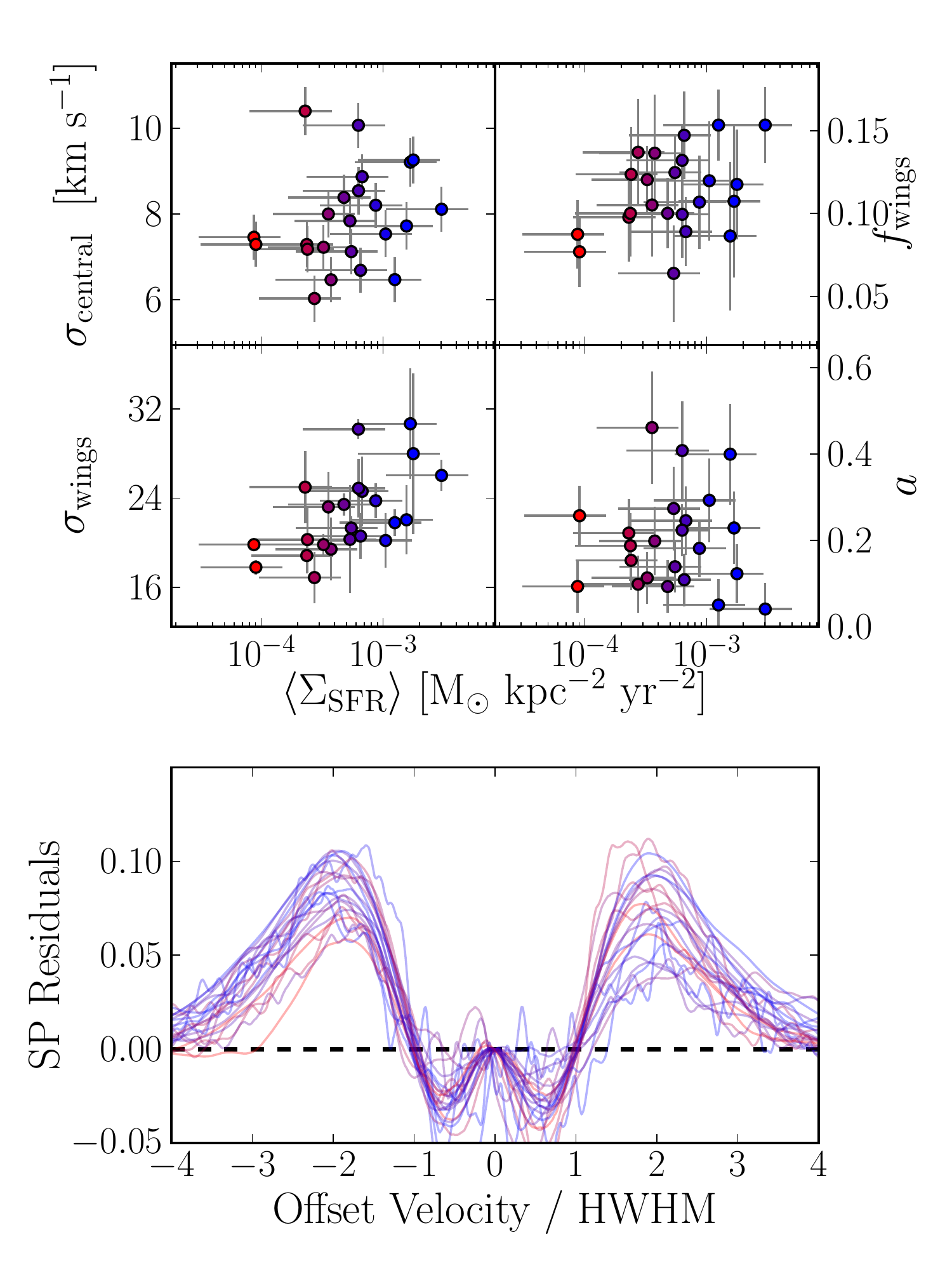}%{{figures/four-panel.hwhm.sfr_pix-per-area_pix}.pdf}
  \fi
  \caption{Same as Figure~\ref{fig:panels-mbaryon}, but for \ave{\sfrsd{}}. We find that galaxies with higher \ave{\sfrsd{}} have higher characteristic wing velocities. \label{fig:panels-sfr-per-area}}
\end{figure}

\clearpage{}
\begin{figure}
  \centering
  \ifimage
  \includegraphics[height=7in]{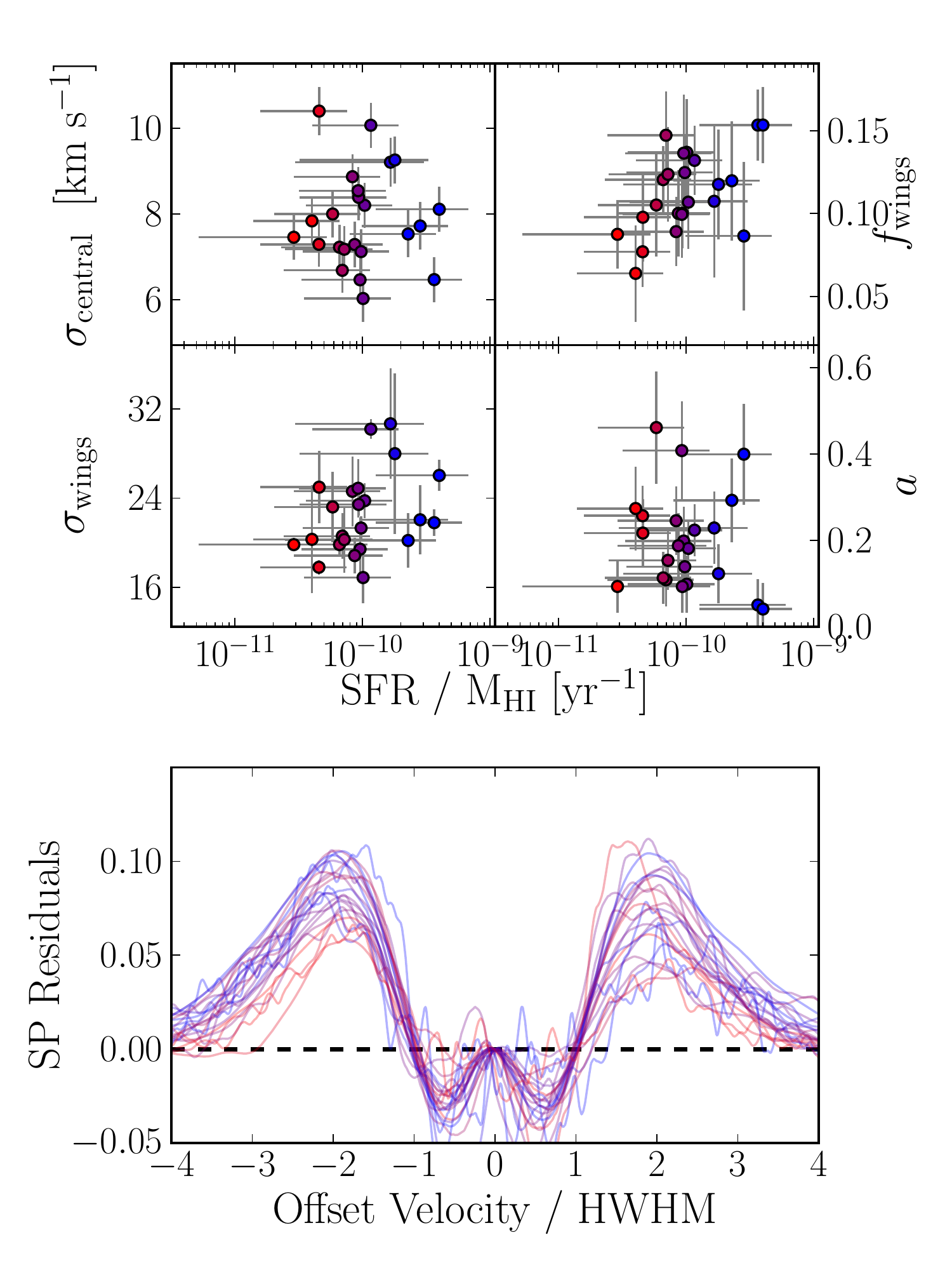}%{{figures/four-panel.hwhm.sfr_pix-per-m_hi_pix}.pdf}
  \fi
  \caption{Same as Figure~\ref{fig:panels-mbaryon}, but for SFR / \mhi{} . Galaxies with higher SFR / \mhi{} also have a higher fraction of gas in the wings of their superprofiles. \label{fig:panels-sfr-per-mhi}}
\end{figure}

\clearpage
\begin{figure}
  \centering
  \ifimage
    \includegraphics[height=7in]{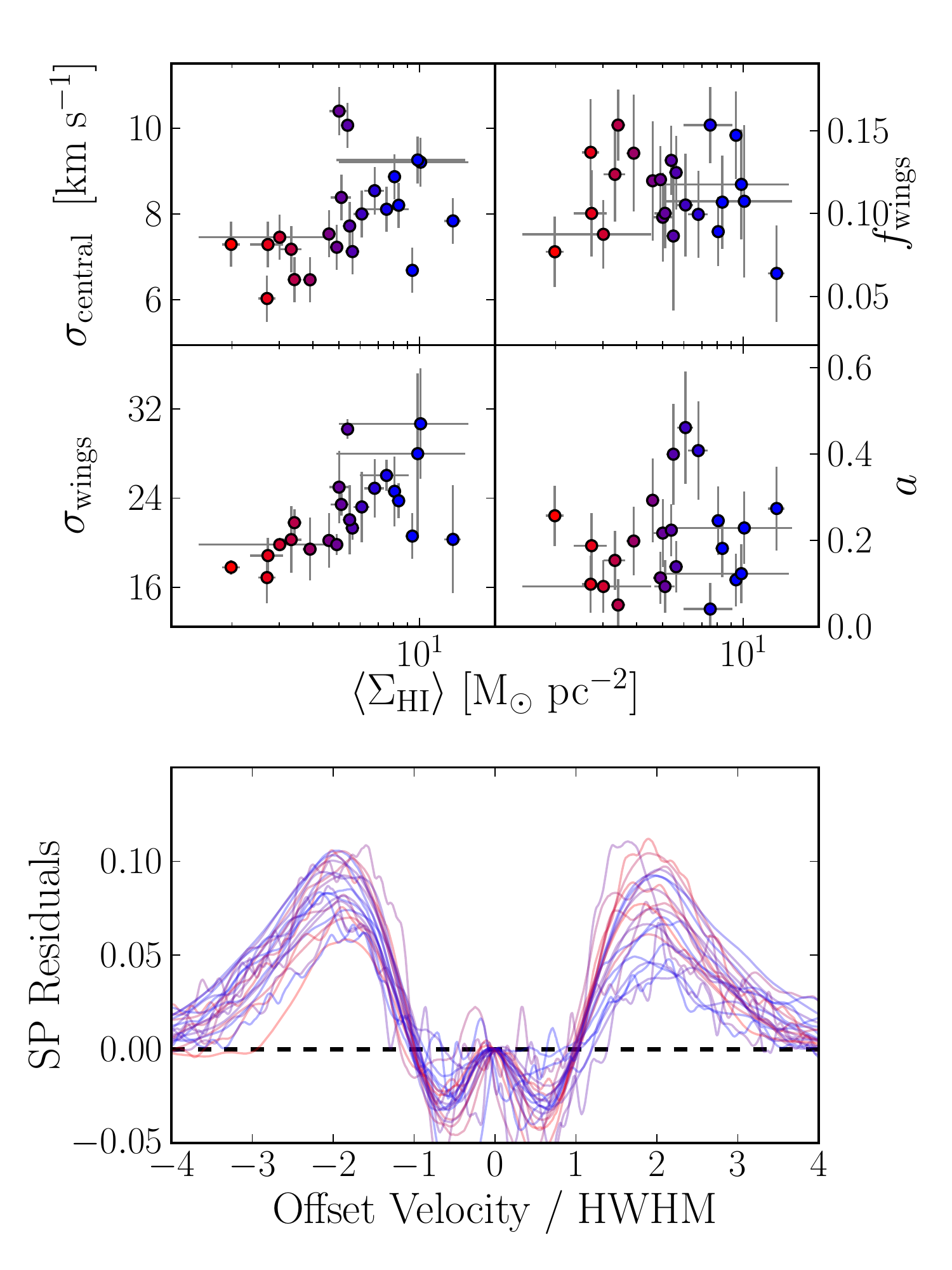}%{{figures/four-panel.hwhm.m_hi_pix-per-area_pix}.pdf}
  \fi
  \caption{The superprofile parameters as a function of \ave{\hisd{}}. We find that both \scentral{} and \swing{} increase with increasing \ave{\hisd{}}.
\label{fig:panels-mhi-per-area}}
\end{figure}

\clearpage
\begin{figure}
  \centering
  \ifimage
    \includegraphics[height=7in]{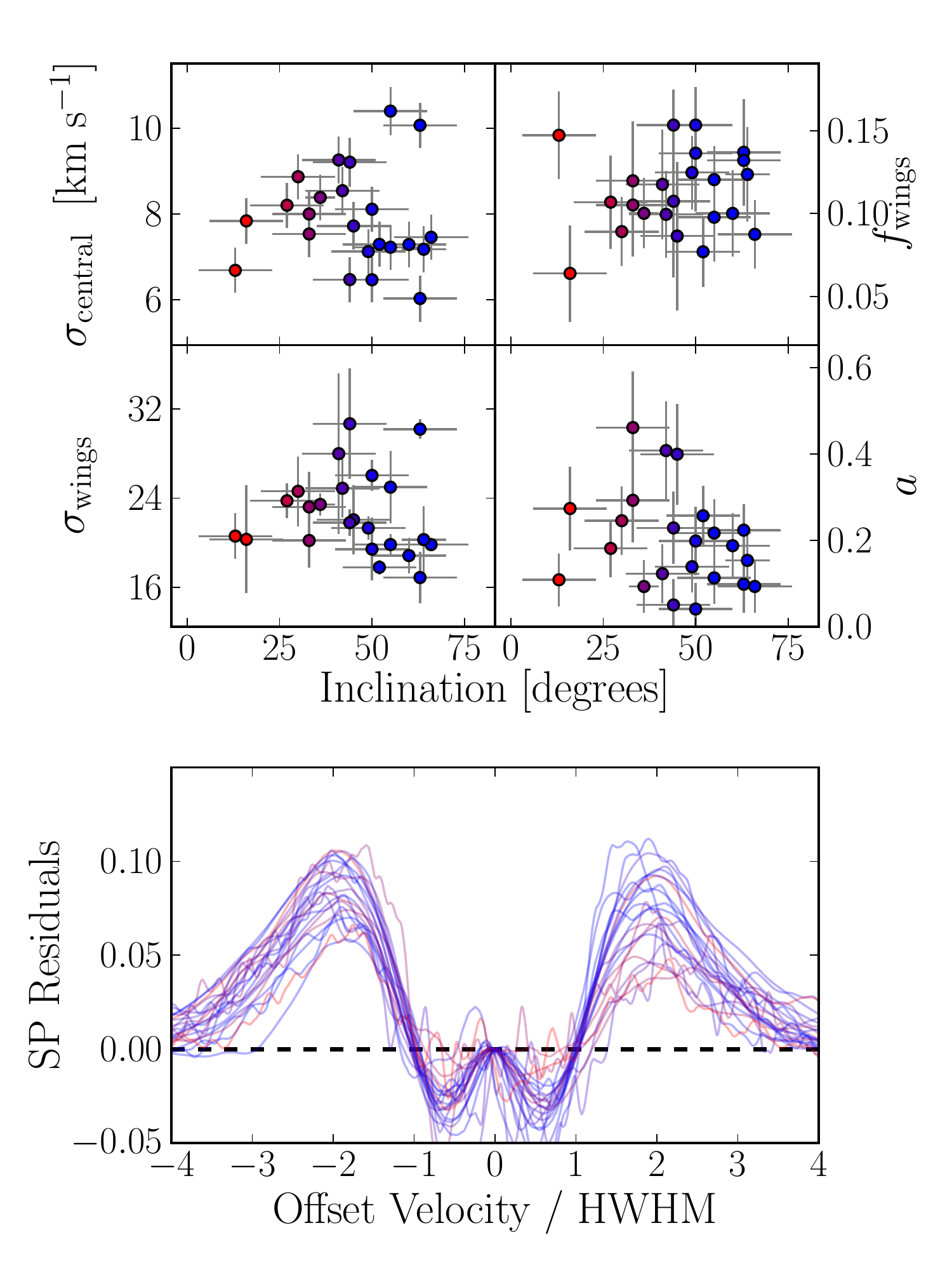}%{{figures/four-panel.hwhm.i}.pdf}
  \fi
  \caption{The superprofile parameters as a function of galaxy inclination. The measured parameters do not change systematically as a function of inclination, but we note that the inclinations for many of the dwarfs in our sample are very uncertain.  \label{fig:panels-i}}
\end{figure}

\clearpage
\begin{figure}
  \centering
  \ifimage
  	\includegraphics[height=5cm]{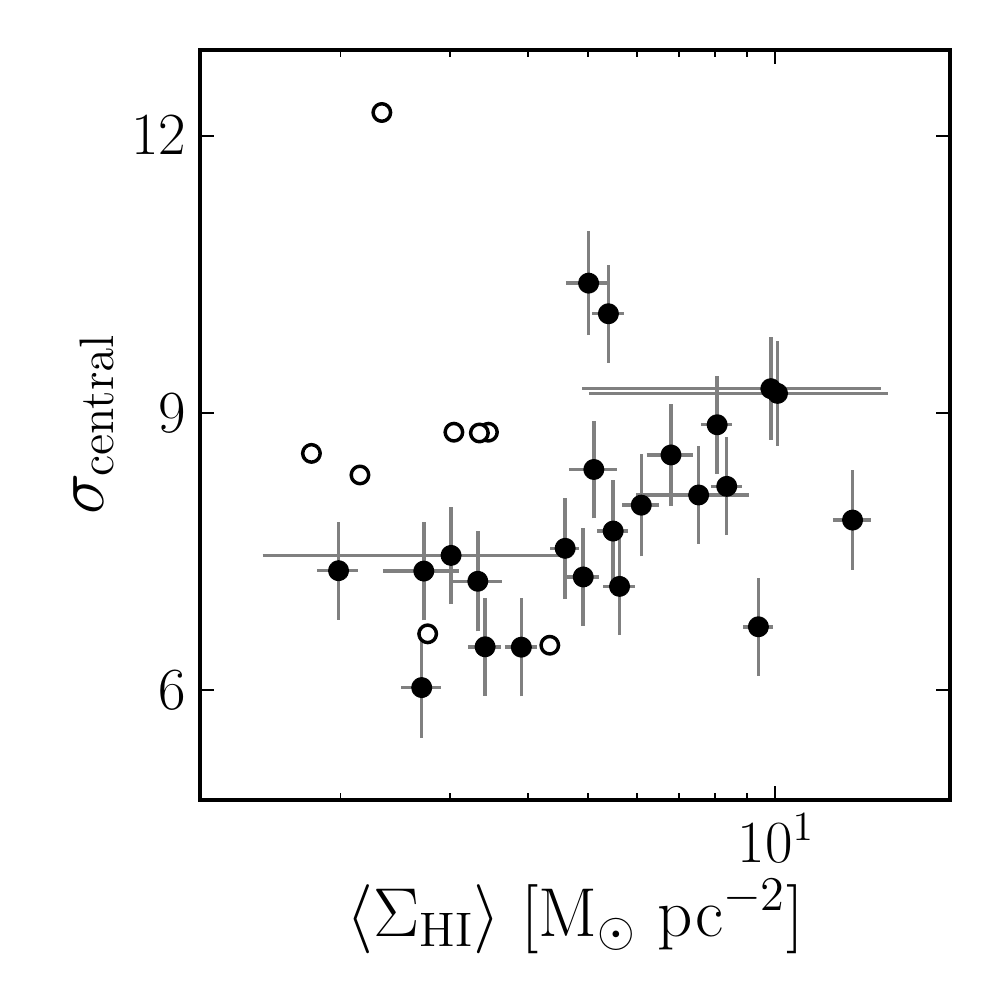}%{{figures/extend.s1.m_hi_pix-per-area_pix}.pdf}
  	\includegraphics[height=5cm]{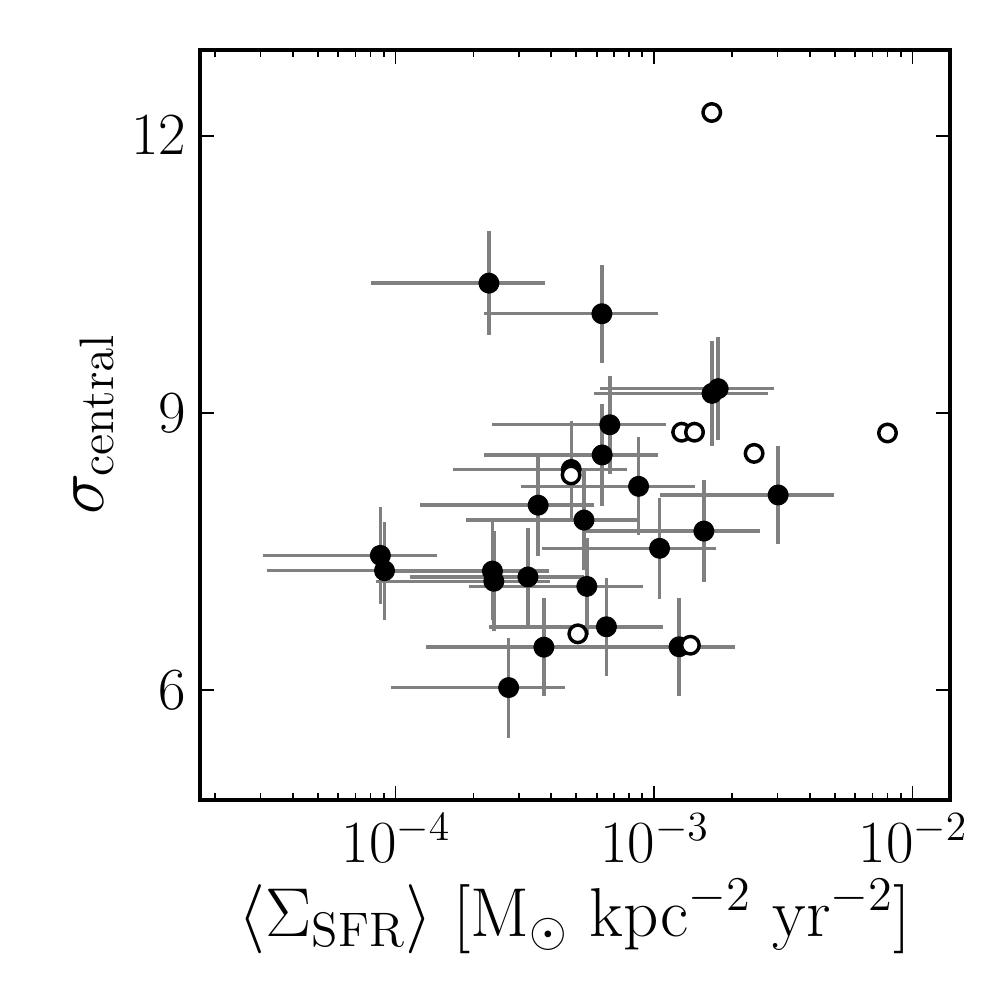}%{{figures/extend.s1.sfr_pix-per-area_pix}.pdf}
  	\includegraphics[height=5cm]{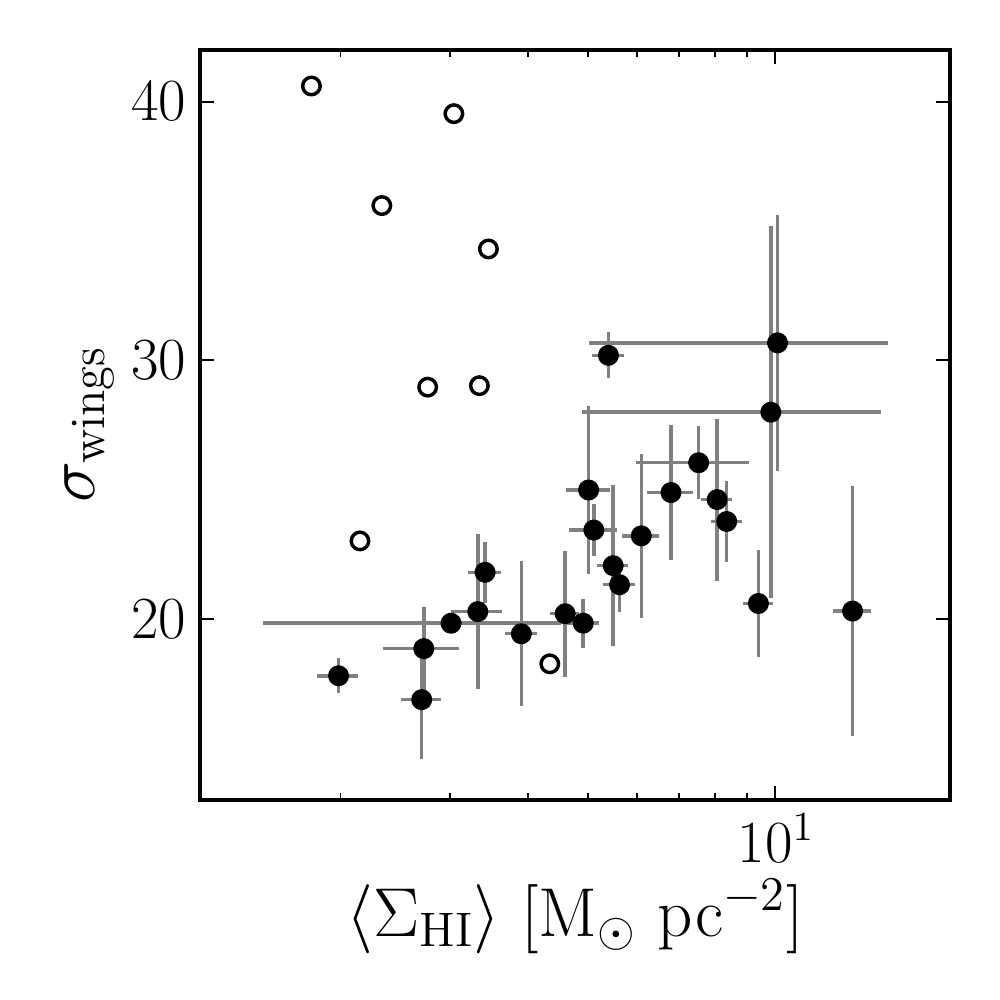}%{{figures/extend.s2.m_hi_pix-per-area_pix}.pdf}
  	\\
  	\includegraphics[height=5cm]{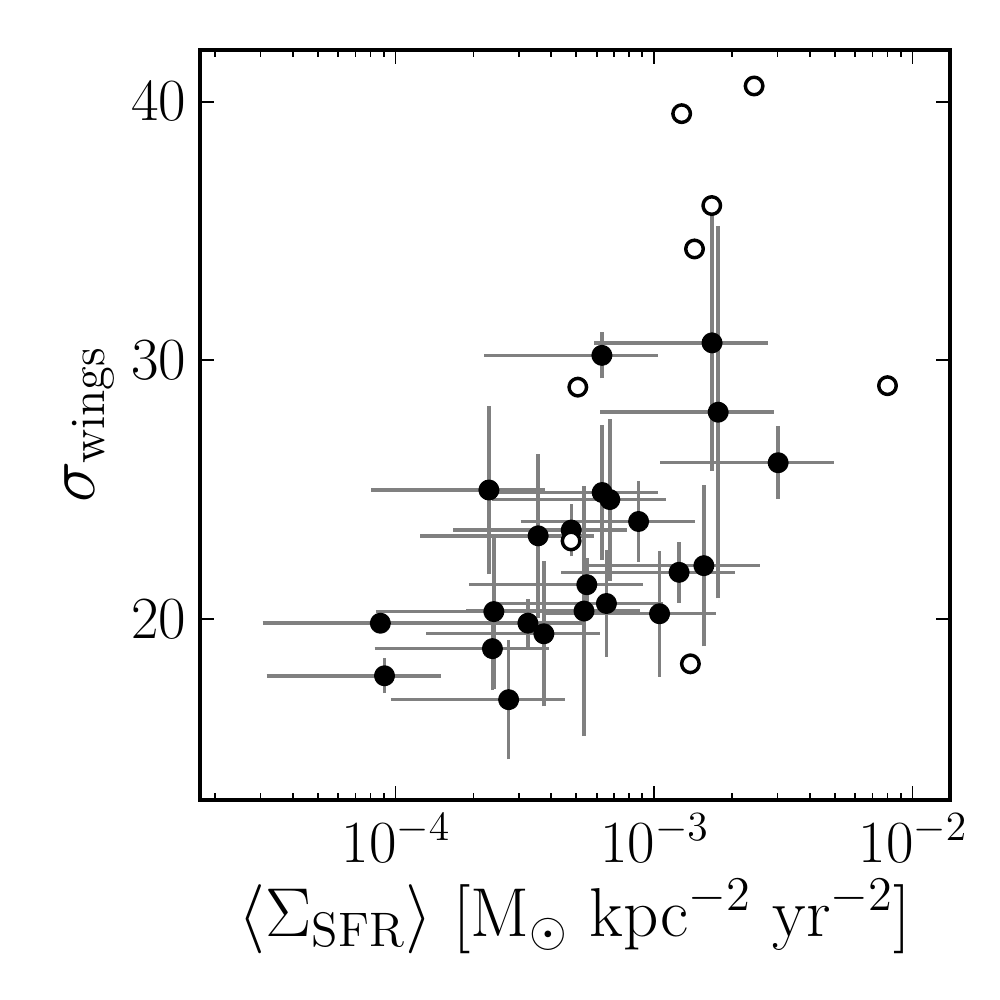}%{{figures/extend.s2.sfr_pix-per-area_pix}.pdf}
    \includegraphics[height=5cm]{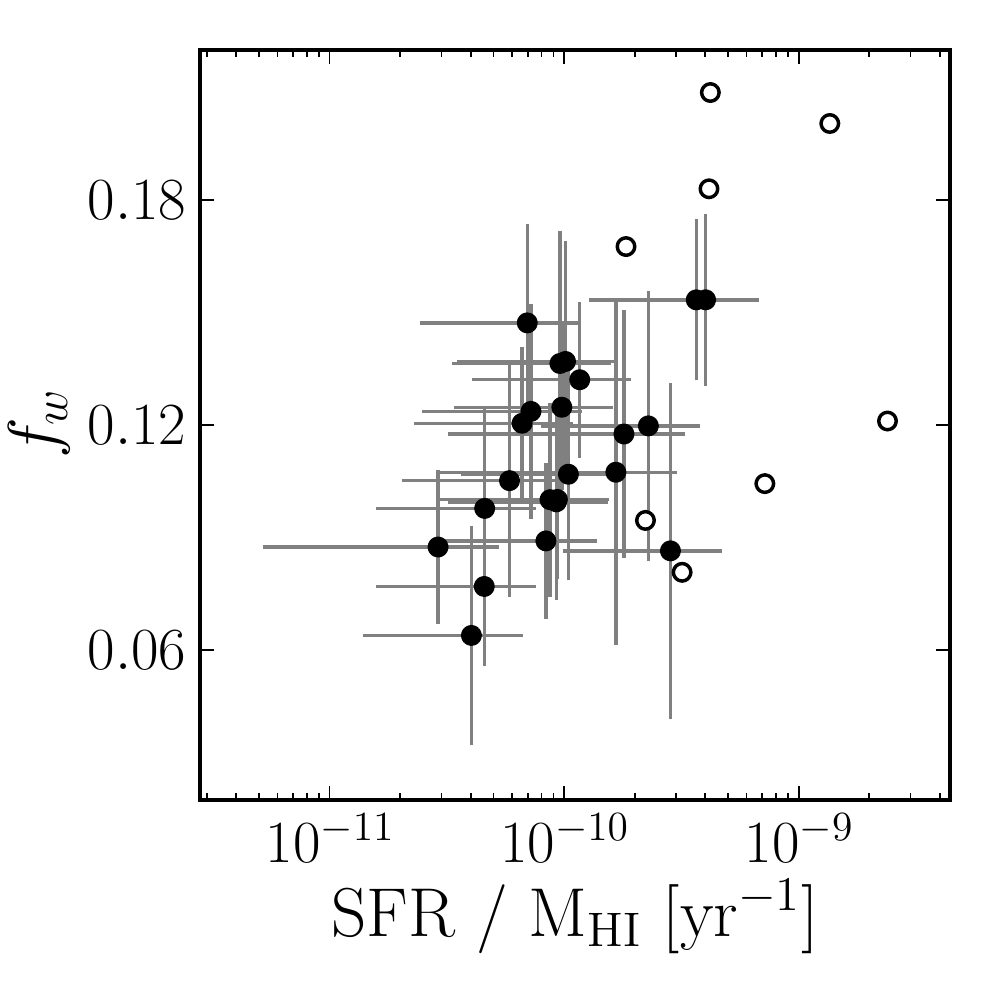}%{{figures/extend.fw.sfr_pix-per-m_hi_pix}.pdf}
    \includegraphics[height=5cm]{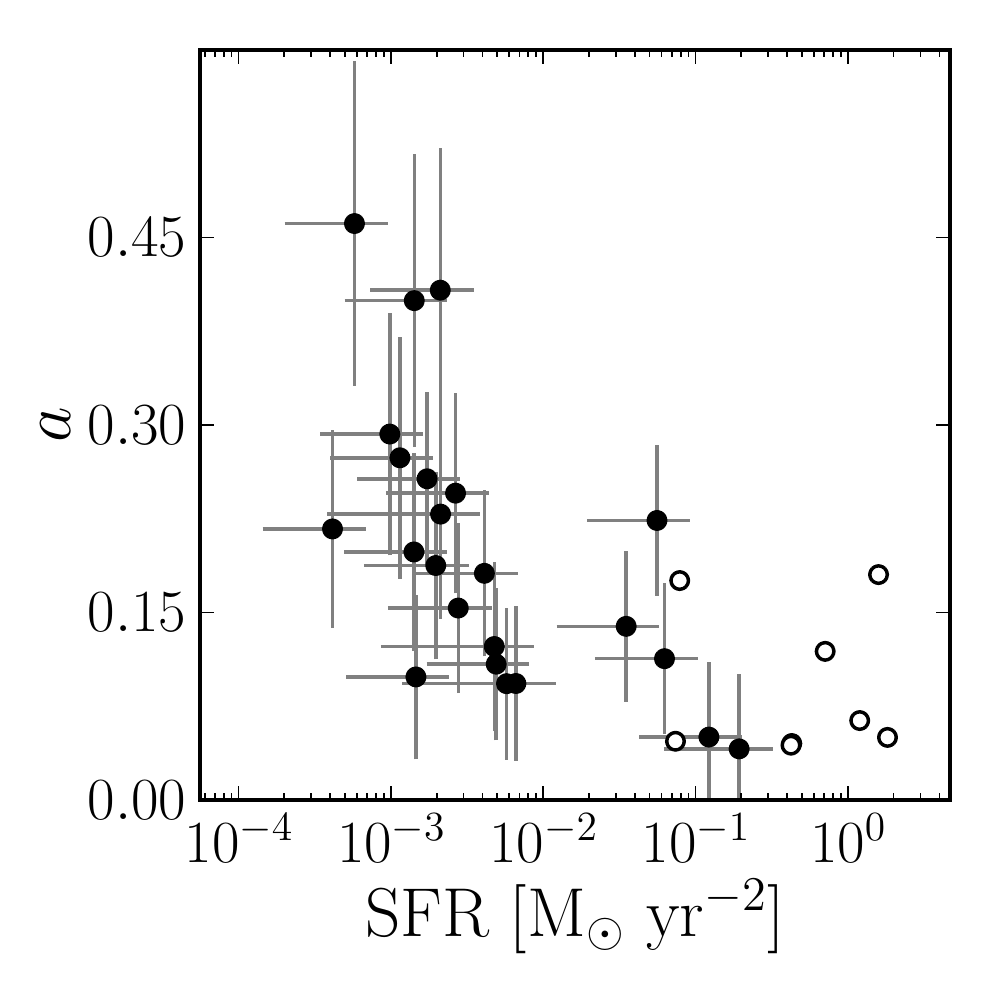}%{{figures/extend.aw.sfr_pix}.pdf}
  \fi
  \caption{Observed superprofile properties versus physical properties when higher mass galaxies are included. The main dwarf sample is shown with filled black circles, while the higher mass galaxies are shown with open black circles. In most cases, the higher mass galaxies fit the correlations, but in a few, they do not (i.e., \scentral{} versus \ave{\hisd{}} and \swing{} versus \ave{\hisd{}}).
  \label{fig:extend-400}}
\end{figure}

\clearpage
\begin{figure}
  \centering
  \ifimage
  \includegraphics[width=8cm]{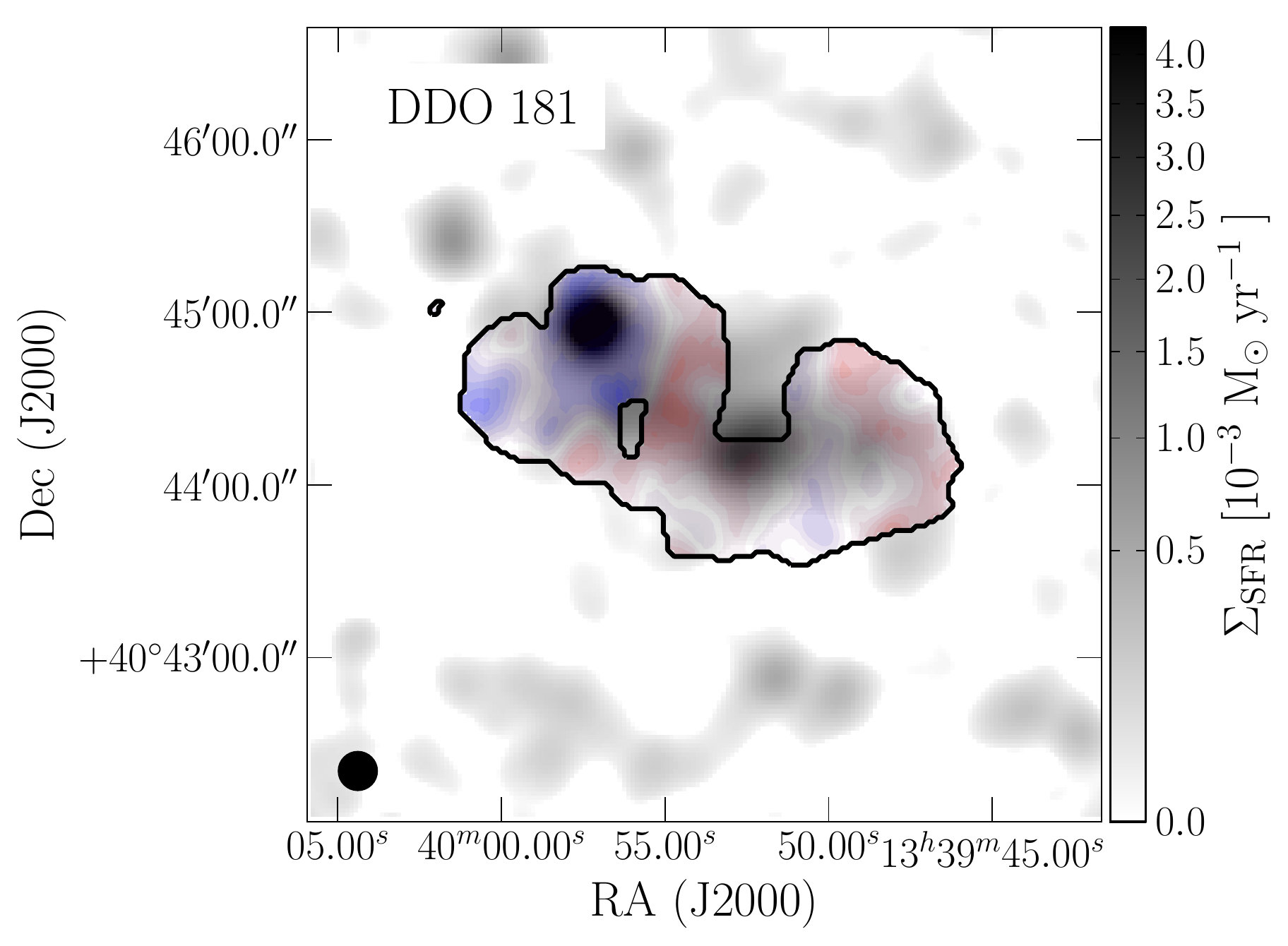}%{{figures/ddo181.ro.phys200.sfr.asymmetry}.pdf}
  \includegraphics[width=8cm]{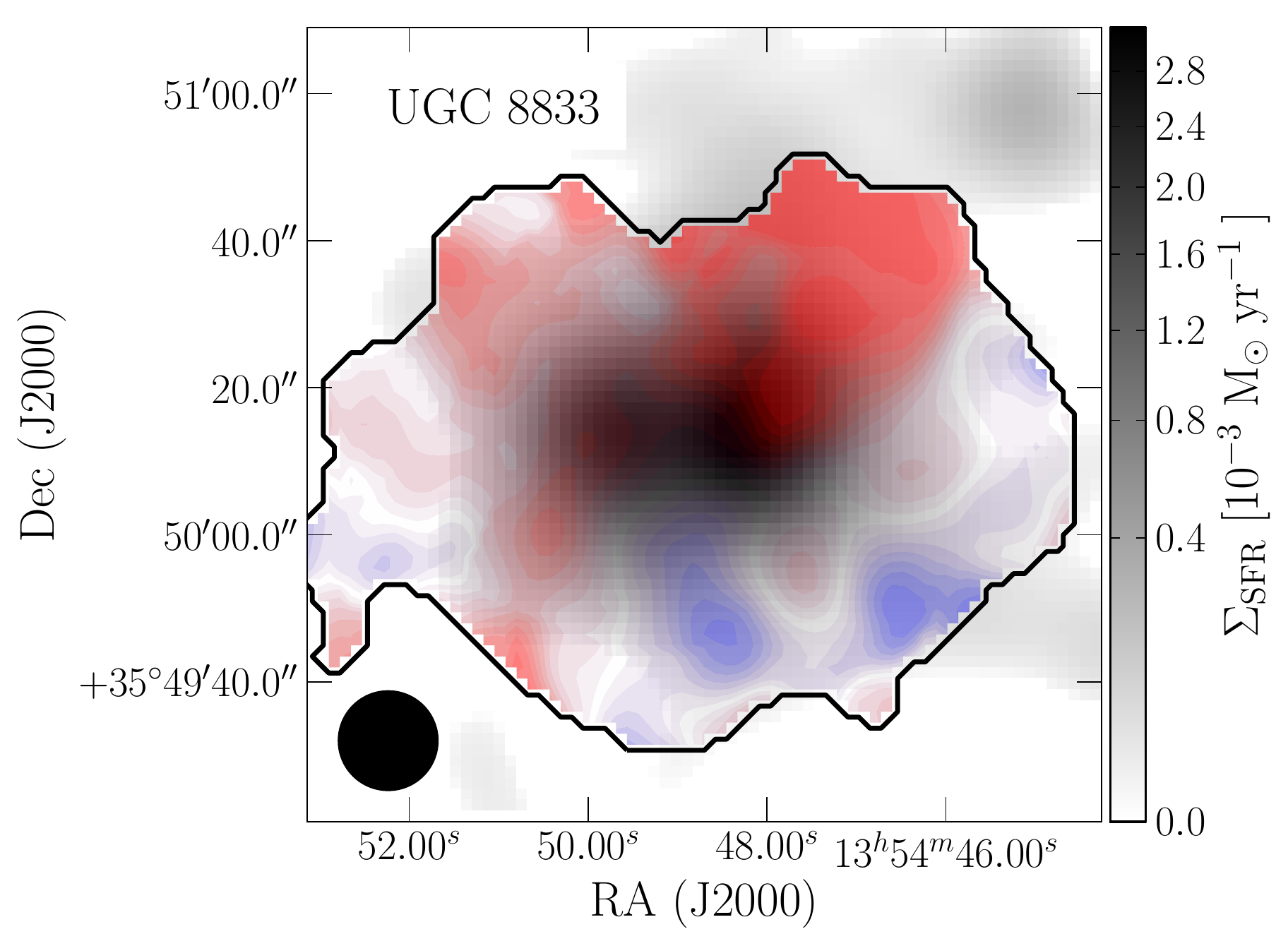}%{{figures/u8833.ro.phys200.sfr.asymmetry}.pdf}
  \\
  \includegraphics[width=8cm]{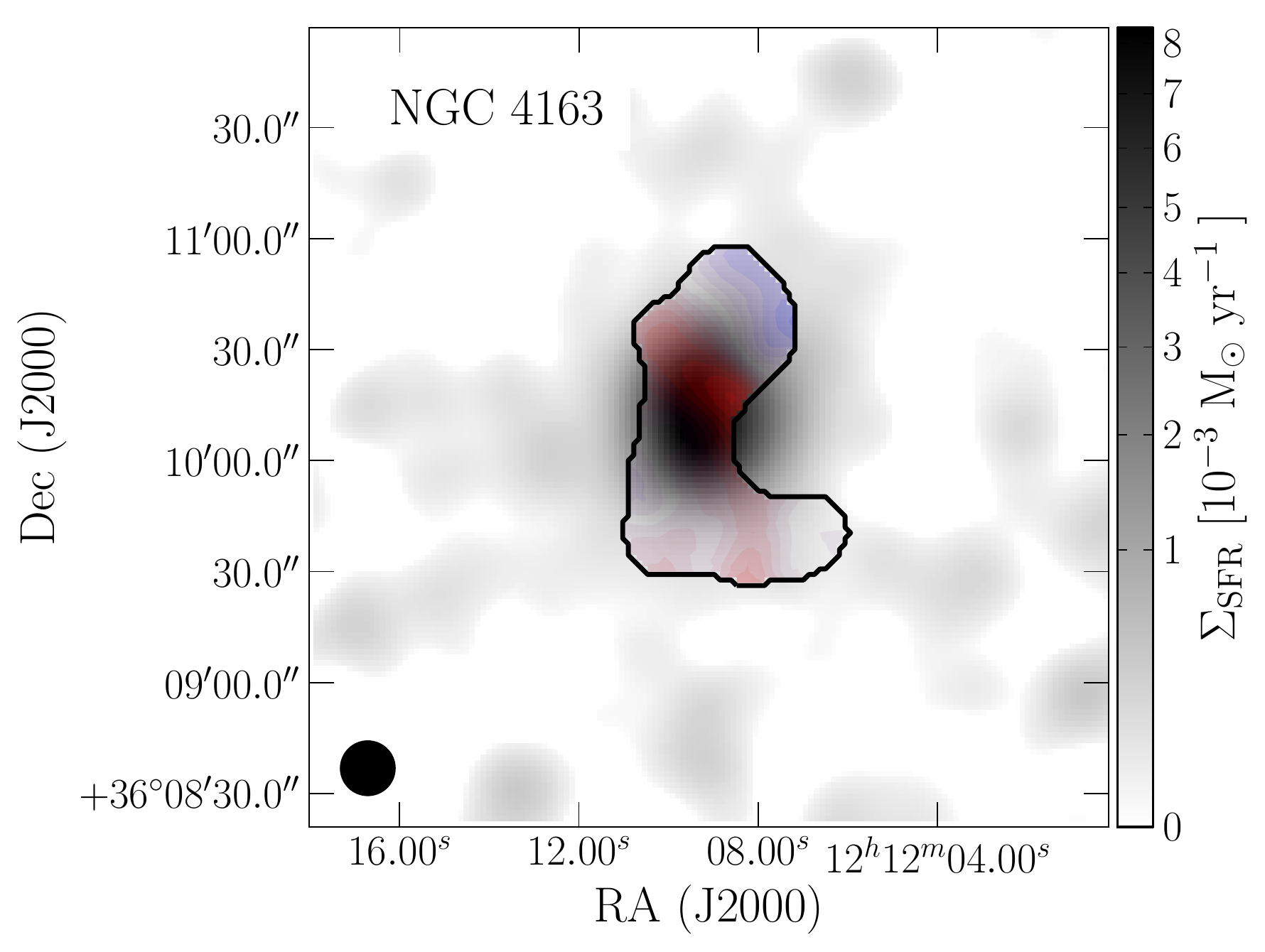}%{{figures/n4163.ro.phys200.sfr.asymmetry}.pdf}
  \includegraphics[width=8cm]{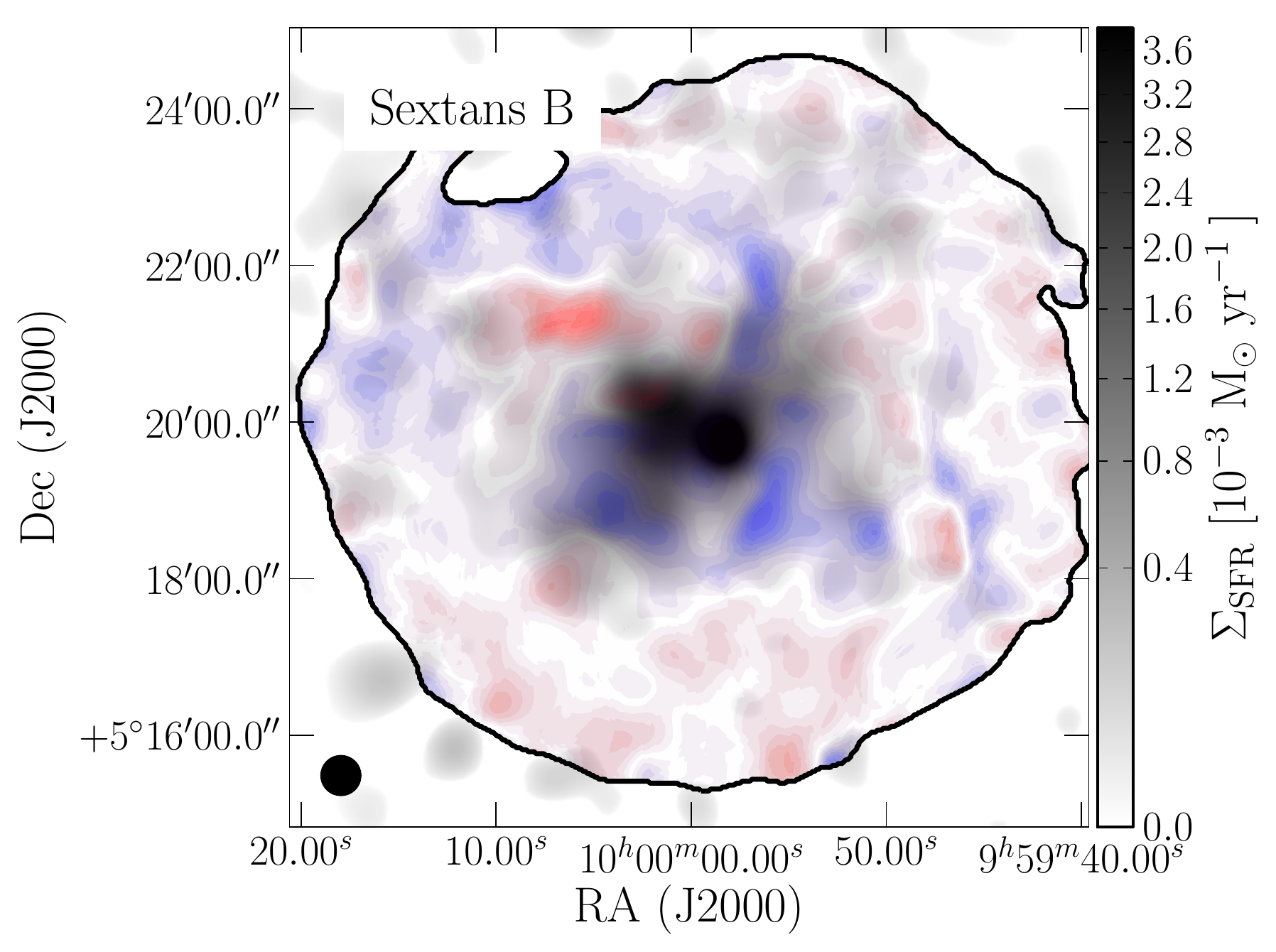}%{{figures/sexb.ro.phys200.sfr.asymmetry}.pdf}
  \fi
  \caption{  Example galaxies in our sample that show asymmetric \hi{} line-of-sight profiles near star forming regions. 
The background image is \sfrsd{}, with the black line representing the $\mathrm{S/N} > 5$ threshold.
The color overlays represent asymmetric line-of-sight profiles, with blue indicating line-of-sight spectra where the first moment is smaller than \vp{} and red showing where the first moment is larger than \vp{}.
The color ranges from -5 to 5 \kms{} differences.
Regions with small absolute differences between \vp{} and the first moment are also shown with more transparency, and those with large absolute differences are less transparent.
  We note that not all star forming regions are associated with asymmetrical line-of-sight profiles, and not all asymmetric line-of-sight profiles are near star forming regions.
However, the observed overlap between some star forming regions and the strongest \hi{} line-of-sight asymmetries may indicate that star formation can be one driver of asymmetry in \hi{} line profiles.
  \label{fig:sfr-asymmetry}}
\end{figure}

\clearpage{}
\begin{figure}
  \centering
  \ifimage
    \includegraphics[width=3in]{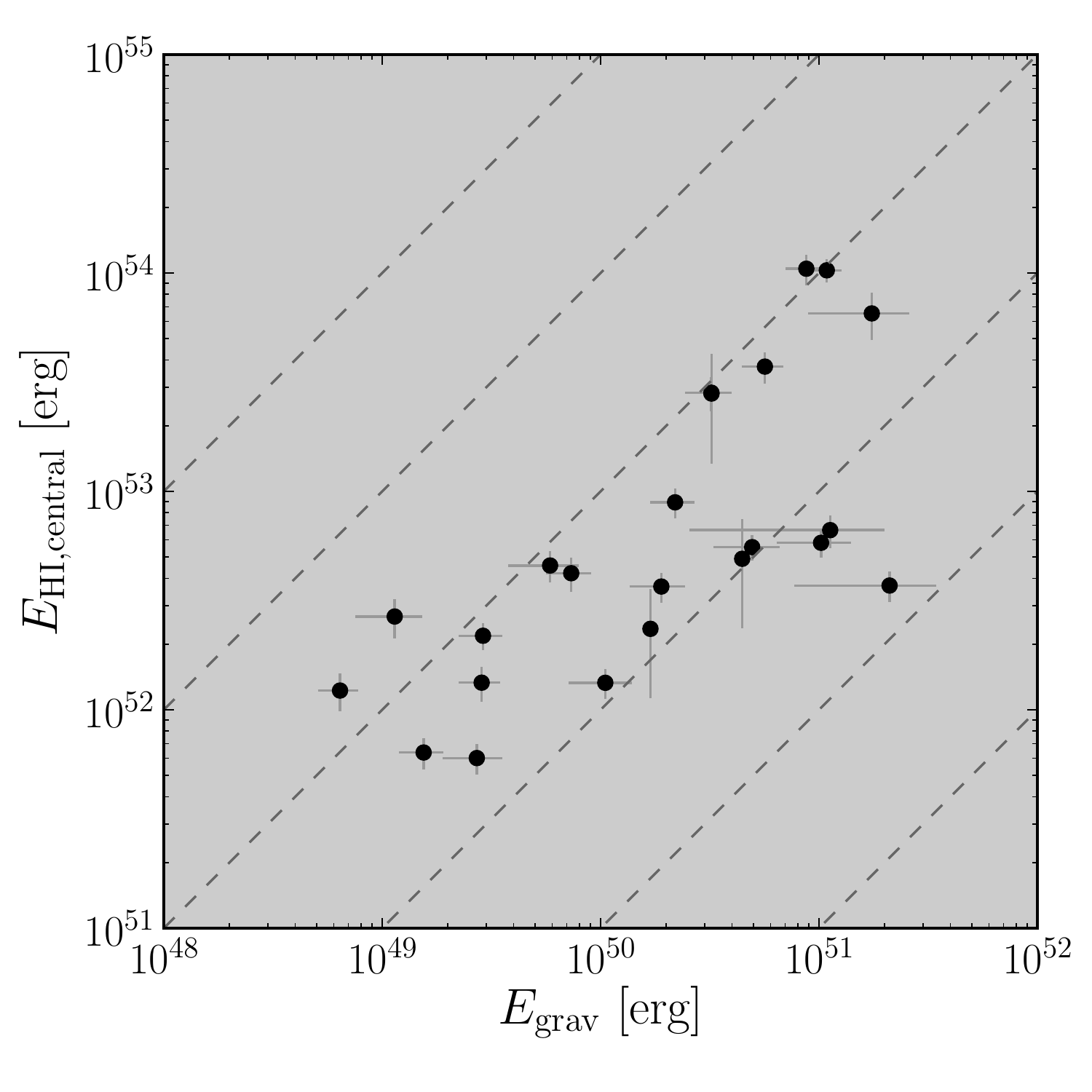}%{{figures/vsub.ro.phys200.jvm.e_grav.e_central.turbulent}.pdf}
    \includegraphics[width=3in]{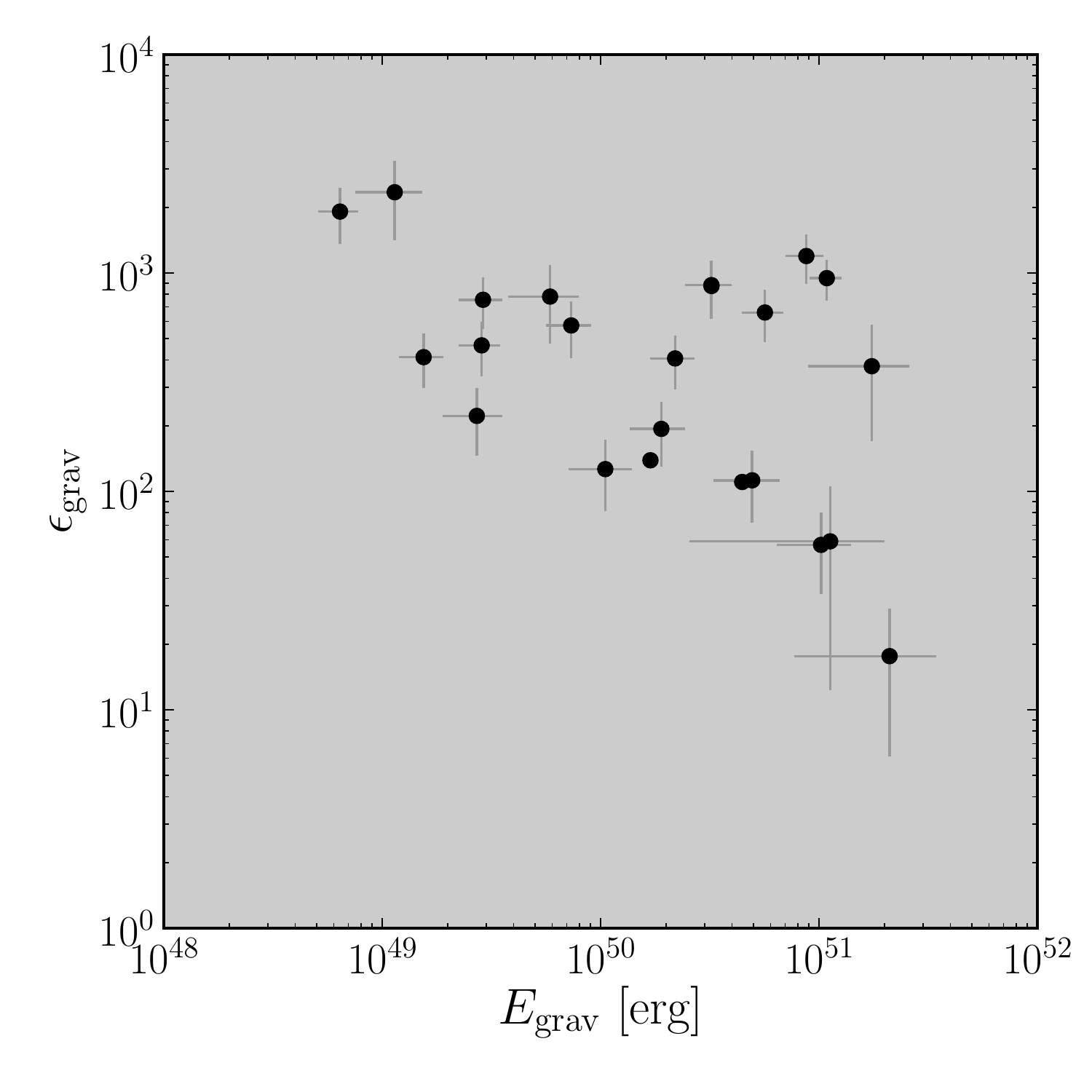}%{{figures/vsub.ro.phys200.jvm.e_grav.eff_central.turbulent}.pdf}
  \fi
  \caption{The left panel shows the energy available from the gravitational instability in \citet{Wada2002} over one turbulent timescale versus the turbulent energy in the central \hi{} component as determined by the HWHM Gaussian fit (left). The entire background has been shaded in grey to represent the fact that all required efficiencies are unphysical, i.e., $> 1$. The right panel shows the associated implied efficiencies ($\epsilon \equiv E_\mathrm{HI, central} / E_\mathrm{grav}$) versus input gravitational energy. It is clear that this instability cannot provide enough energy to drive the central peak in these galaxies over a single turbulent dissipation timescale. \label{fig:snarrow-e-grav}}
\end{figure}

\clearpage{}
\begin{figure}
  \centering
  \ifimage
    \includegraphics[width=3in]{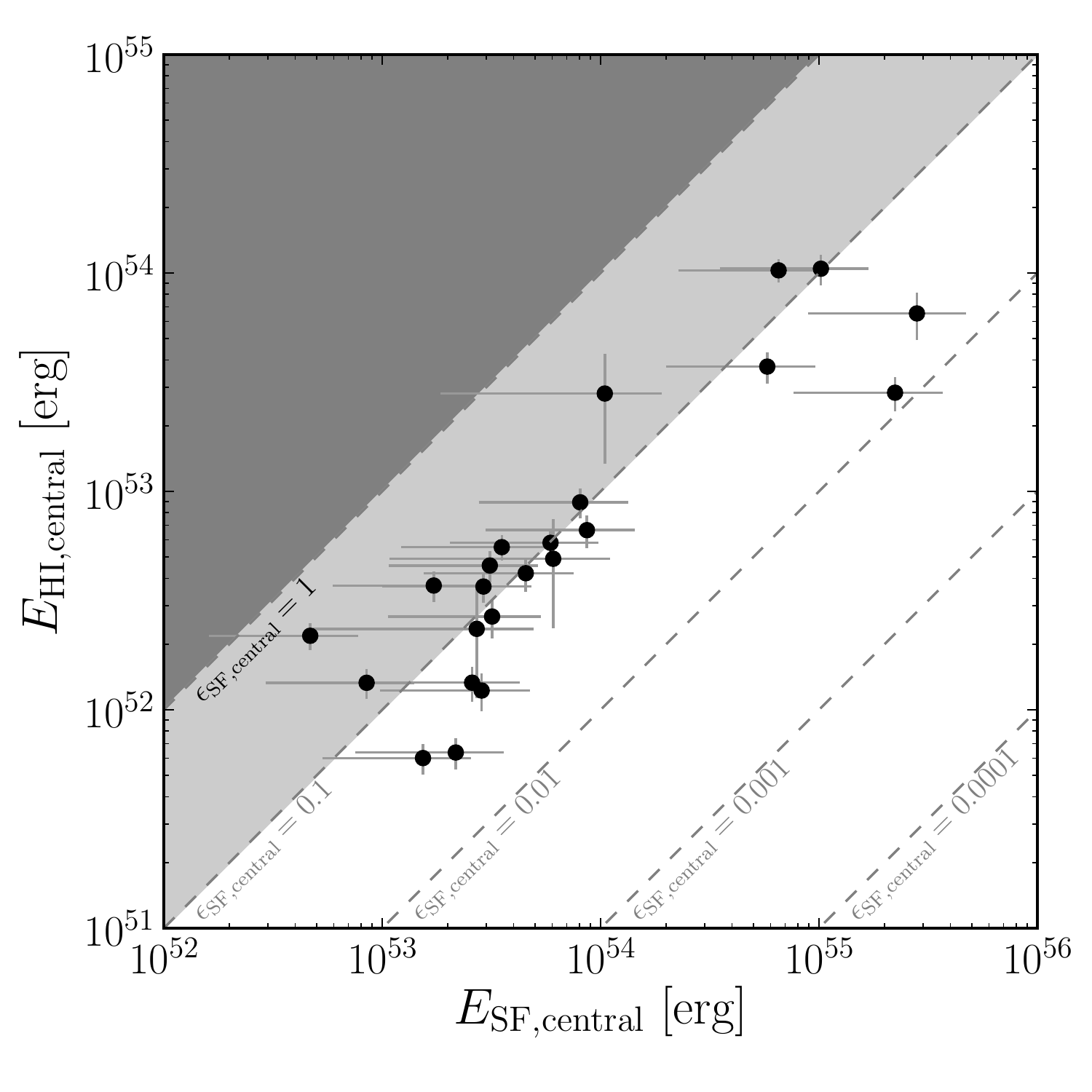}%{{figures/vsub.ro.phys200.jvm.e_sfr.e_central.turbulent}.pdf}
    \includegraphics[width=3in]{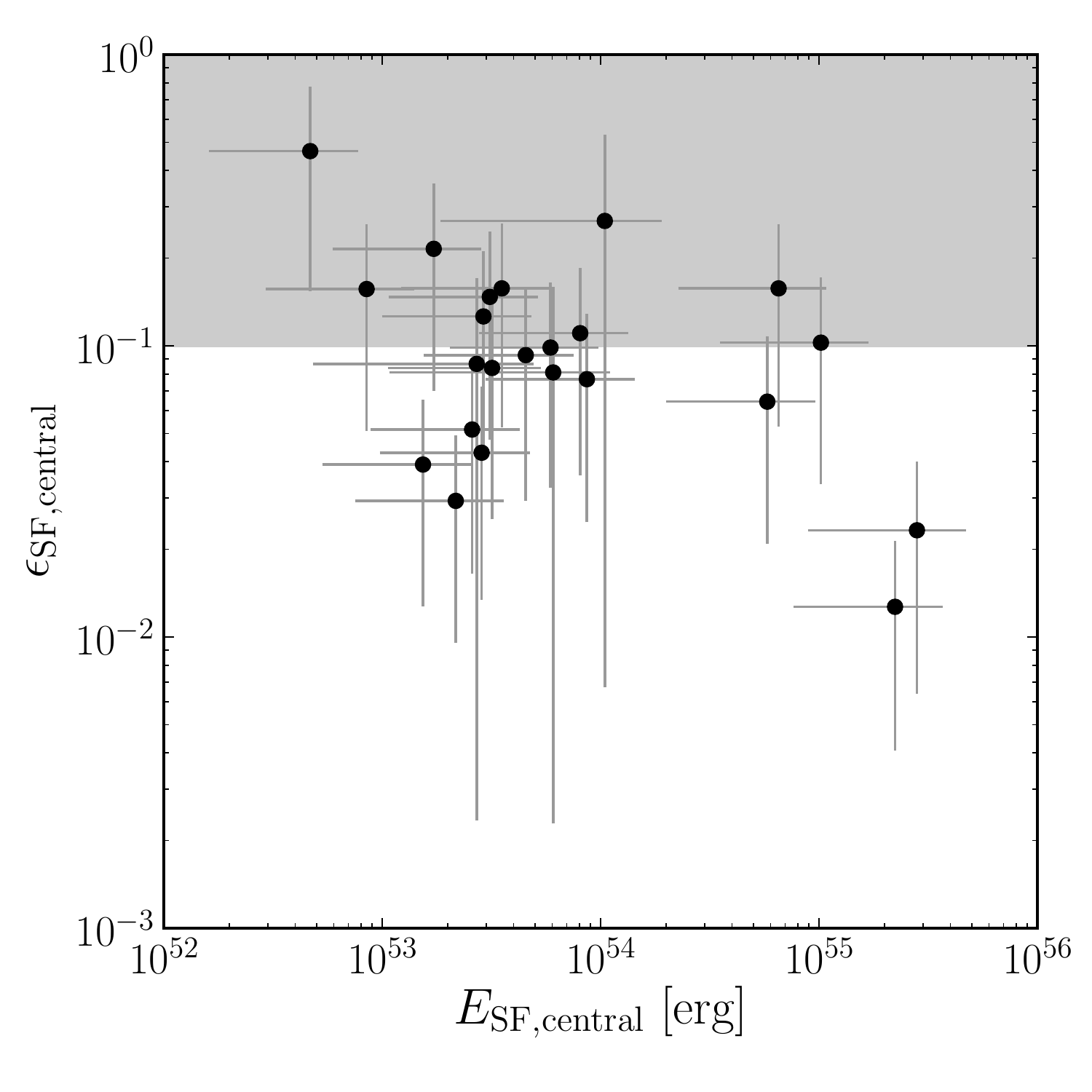}%{{figures/vsub.ro.phys200.jvm.e_sfr.eff_central.turbulent}.pdf}
  \fi
  \caption{The left panel shows the energy available from star formation over one turbulent timescale versus the central \hi{} component as determined by the HWHM Gaussian fit (left).
  The dashed grey lines indicate constant efficiency, and the shaded gray regions represent $\epsilon_\mathrm{SF, turb} > 0.1$ (light grey) and $\epsilon_\mathrm{SF, turb} > 1$ (dark grey).
The right panel shows $\epsilon_\mathrm{SF,turb}$ versus $E_\mathrm{SF}$.
 The mean efficiency required is $\epsilon_\mathrm{SF,turb} = 0.117 \pm 0.096$.
  \label{fig:snarrow-e-sf}}
\end{figure}

\clearpage{}
\begin{figure}
  \centering
  \ifimage
    \includegraphics[width=3in]{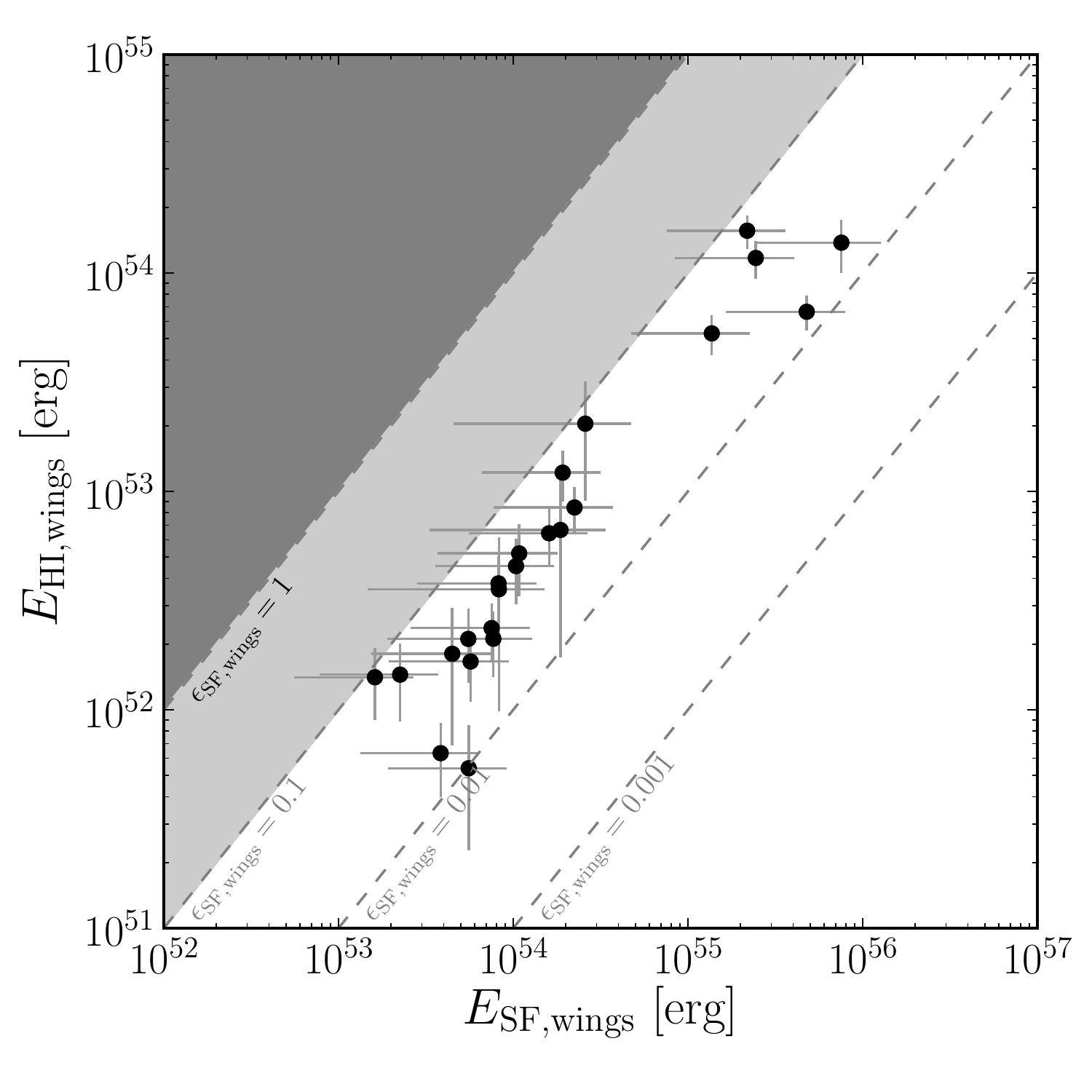}%{{figures/vsub.ro.phys200.jvm.e_sfr.e_wings.holes}.pdf}
    \includegraphics[width=3in]{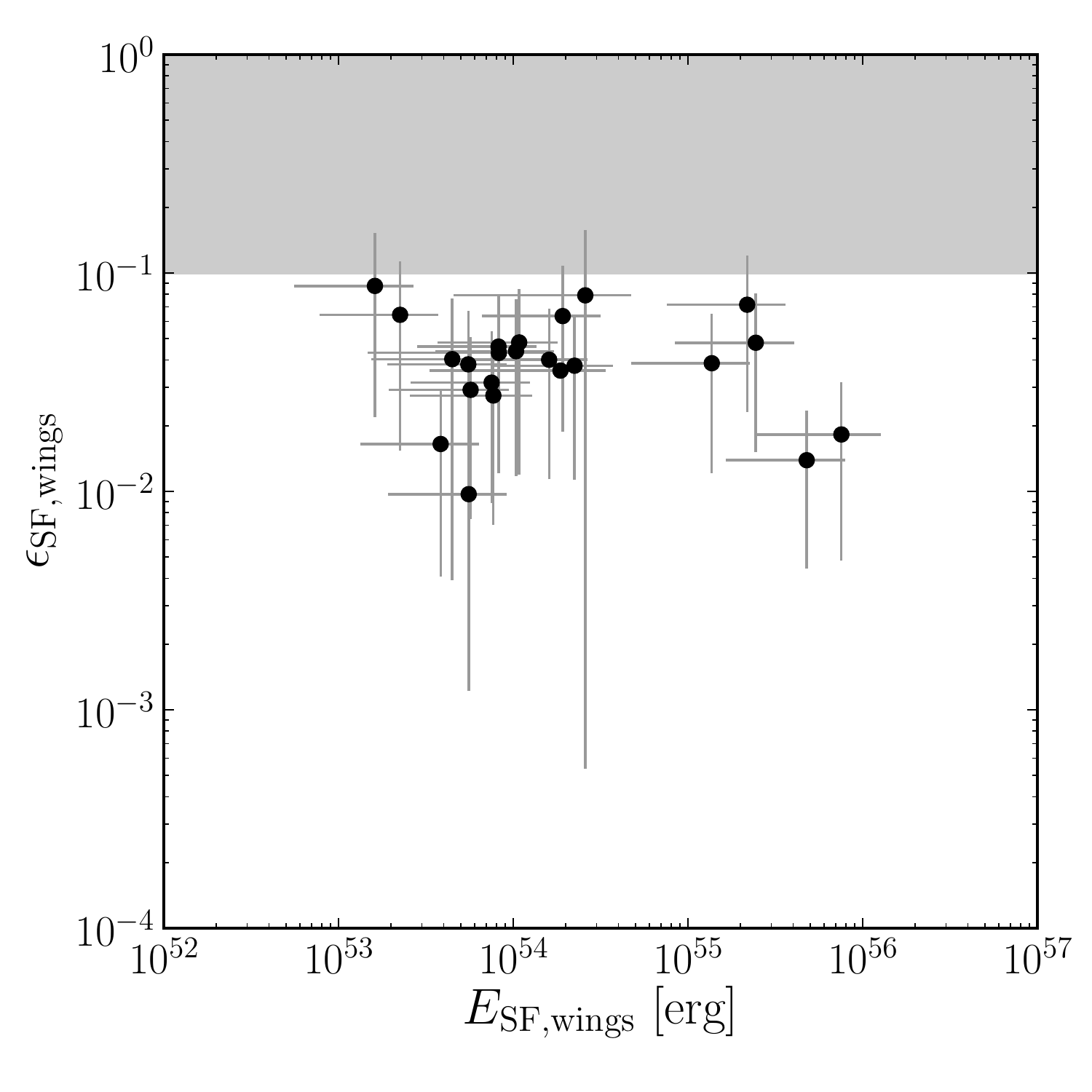}%{{figures/vsub.ro.phys200.jvm.e_sfr.eff_wings.holes}.pdf}
  \fi
  \caption{\label{fig:swings-e-sf-holes} The left panel shows the energy available from star formation over 32.5 Myr versus \hi{} energy in the wings of the superprofile. 
    The dashed grey lines indicate constant efficiency, and the shaded gray regions represent $\epsilon_\mathrm{SF, wings} > 0.1$ (light grey) and $\epsilon_\mathrm{SF, wings} > 1$ (dark grey).
The right panel shows $\epsilon_\mathrm{SF,wings}$ versus $E_\mathrm{SF}$.
 The mean efficiency required is $\epsilon_\mathrm{SF,wings} = 0.042 \pm 0.020$.
 }
\end{figure}

\clearpage{}
\begin{figure}
  \centering
  \ifimage
    \includegraphics[width=3in]{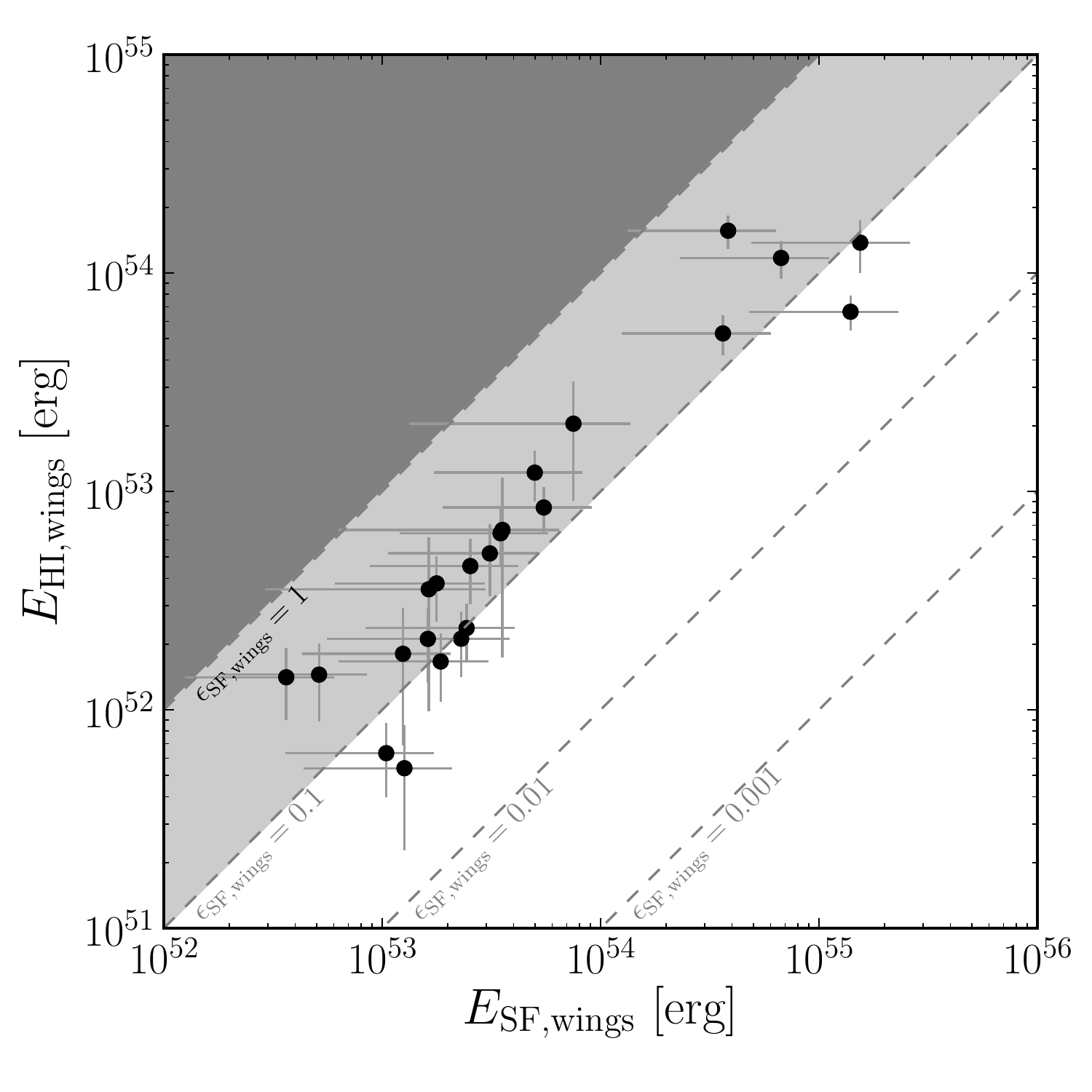}%{{figures/vsub.ro.phys200.jvm.e_sfr.e_wings.turbulent}.pdf}
    \includegraphics[width=3in]{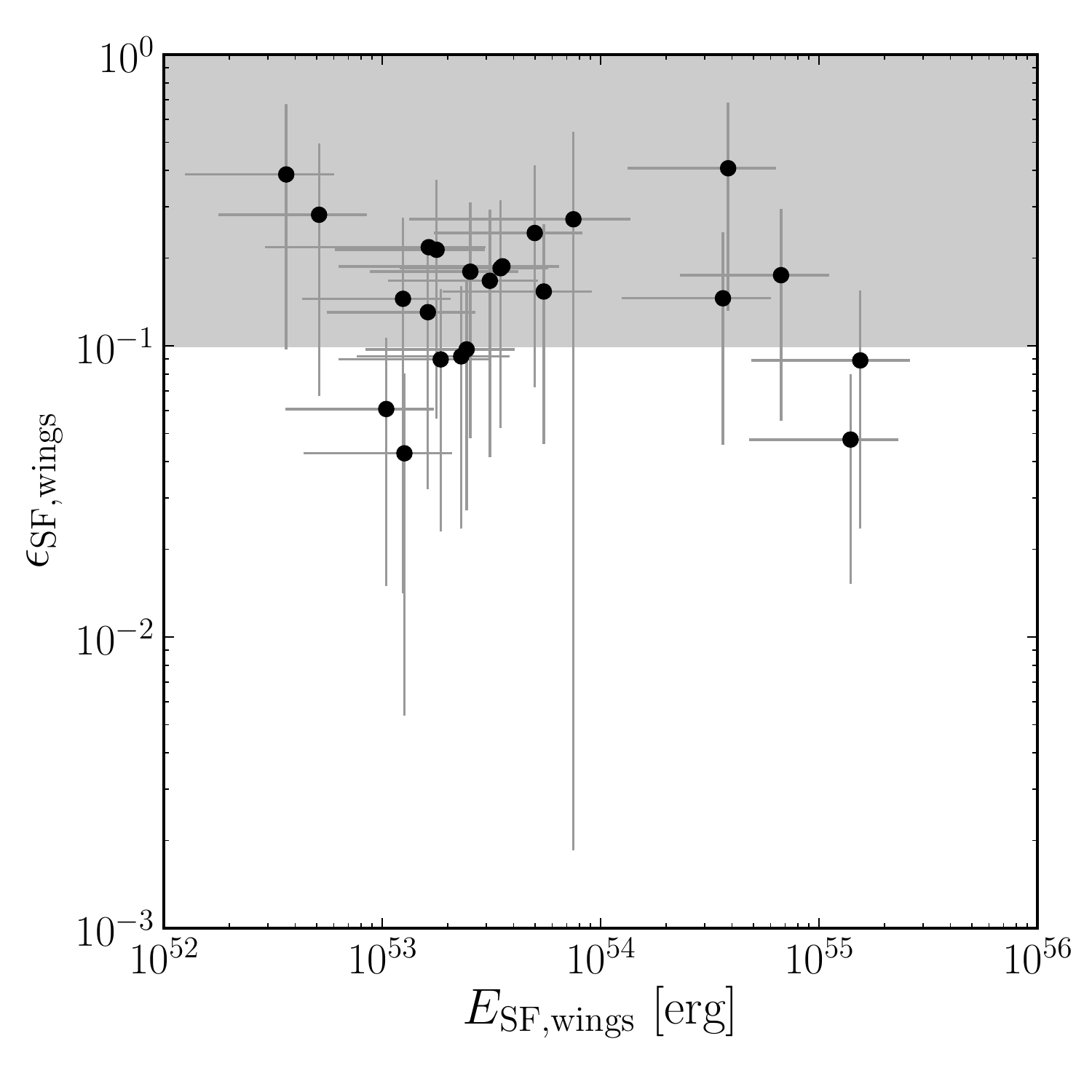}%{{figures/vsub.ro.phys200.jvm.e_sfr.eff_wings.turbulent}.pdf}
  \fi
  \caption{\label{fig:swings-e-sf-turb} The left panel shows the energy available from star formation over one turbulent timescale versus \hi{} energy in the wings of the superprofile. 
      The dashed grey lines indicate constant efficiency, and the shaded gray regions represent $\epsilon_\mathrm{SF, wings} > 0.1$ (light grey) and $\epsilon_\mathrm{SF, wings} > 1$ (dark grey).
The right panel shows $\epsilon_\mathrm{SF,wings}$ versus $E_\mathrm{SF}$.
 The mean efficiency required is $\epsilon_\mathrm{SF,wings} = 0.174 \pm 0.095$.
}
\end{figure}

\clearpage
\begin{figure}
  \centering
  \ifimage
    \includegraphics[width=3in]{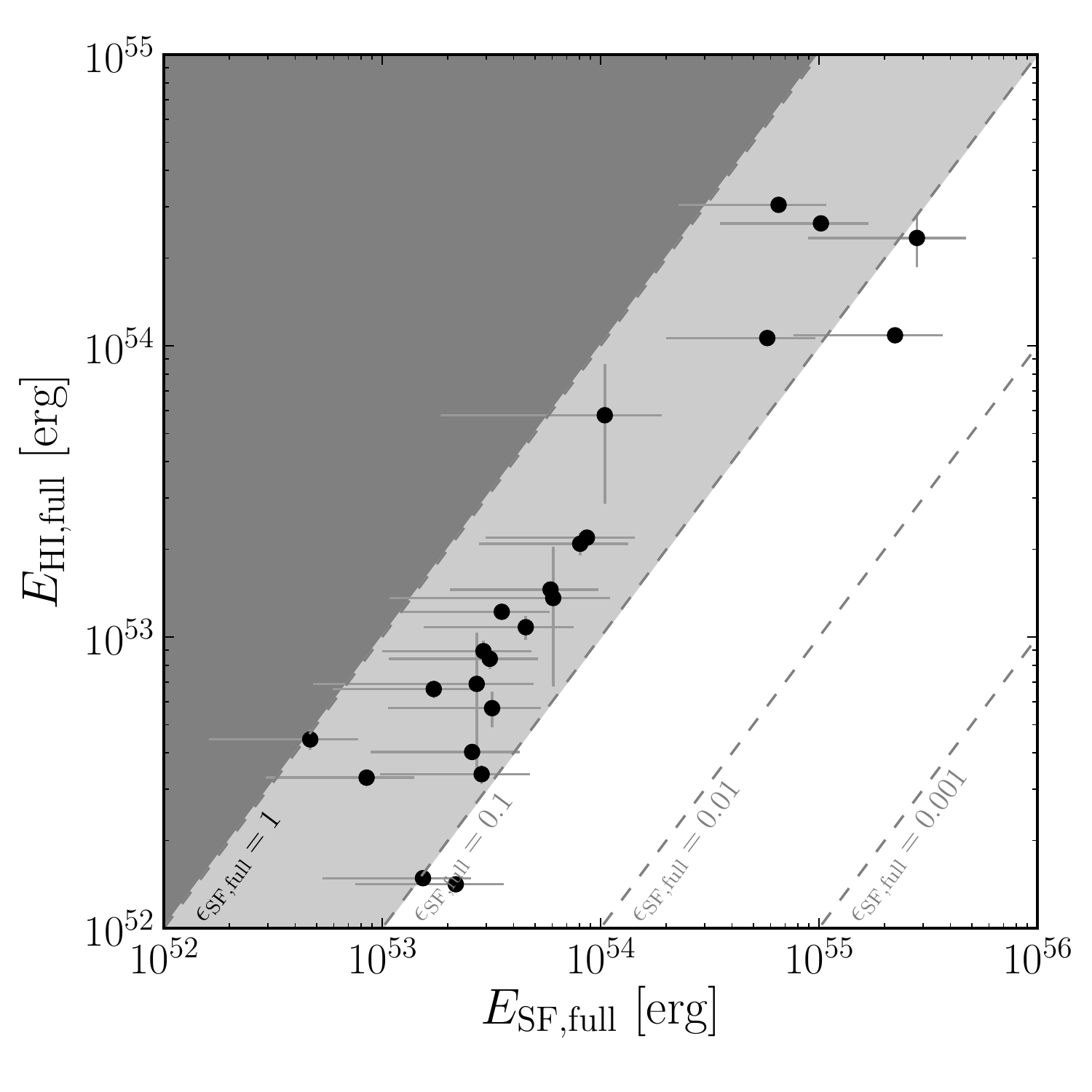}%{{figures/vsub.ro.phys200.jvm.e_sfr.e_full.turbulent}.pdf}
    \includegraphics[width=3in]{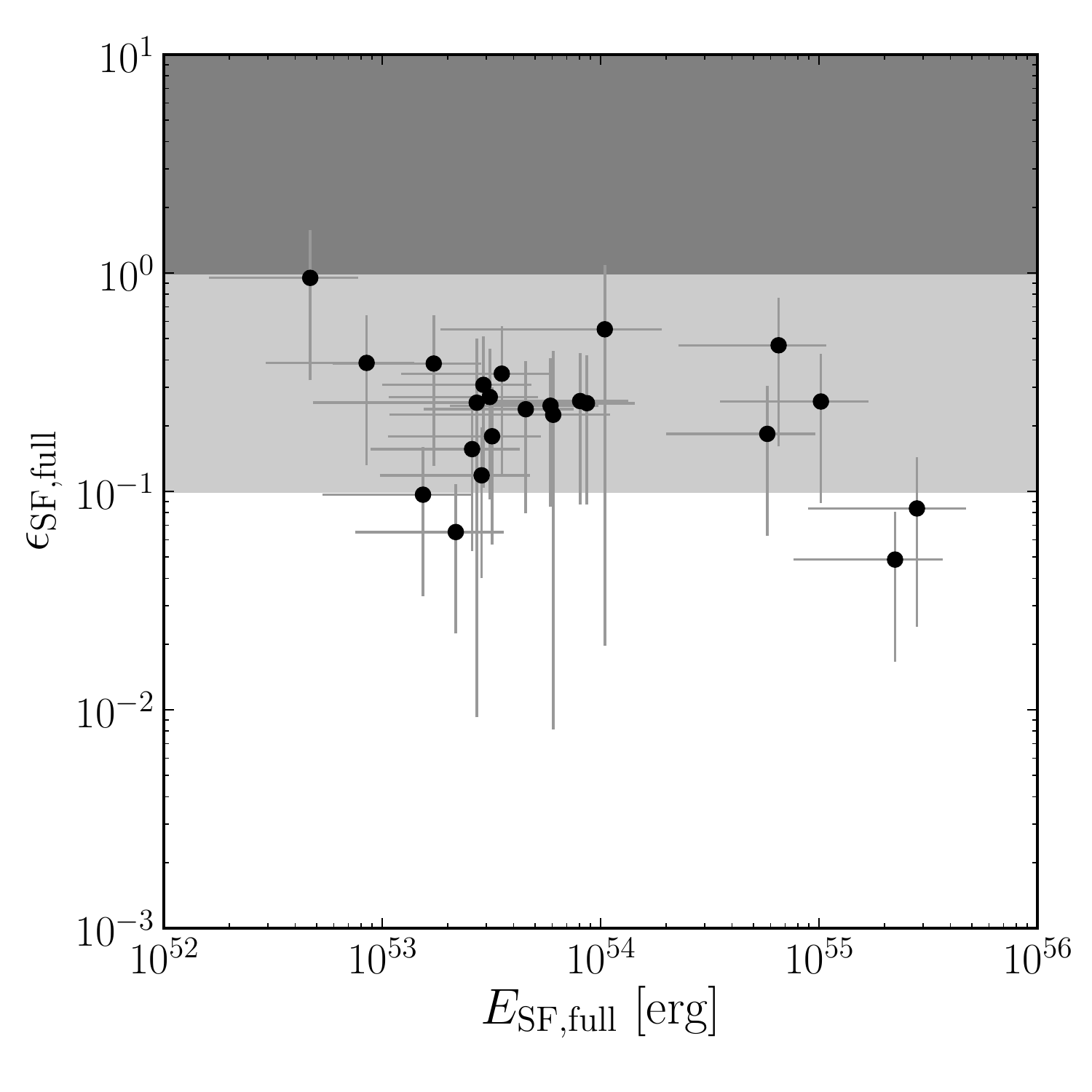}%{{figures/vsub.ro.phys200.jvm.e_sfr.eff_full.turbulent}.pdf}
  \fi
  \caption{\label{fig:full-e-sf-turb} The left panel shows the energy available from star formation over a single turbulent timescale versus \hi{} energy in the full superprofile. 
    The dashed grey lines indicate constant efficiency, and the shaded gray regions represent $\epsilon_\mathrm{SF, full} > 0.1$ (light grey) and $\epsilon_\mathrm{SF, full} > 1$ (dark grey).
The right panel shows $\epsilon_\mathrm{SF,full}$ versus $E_\mathrm{SF}$.
 The mean efficiency required is $\epsilon_\mathrm{SF,full} = 0.275 \pm 0.190$.
}
\end{figure}

\clearpage{}
\begin{figure}
  \centering
  \ifimage
    \includegraphics[width=3in]{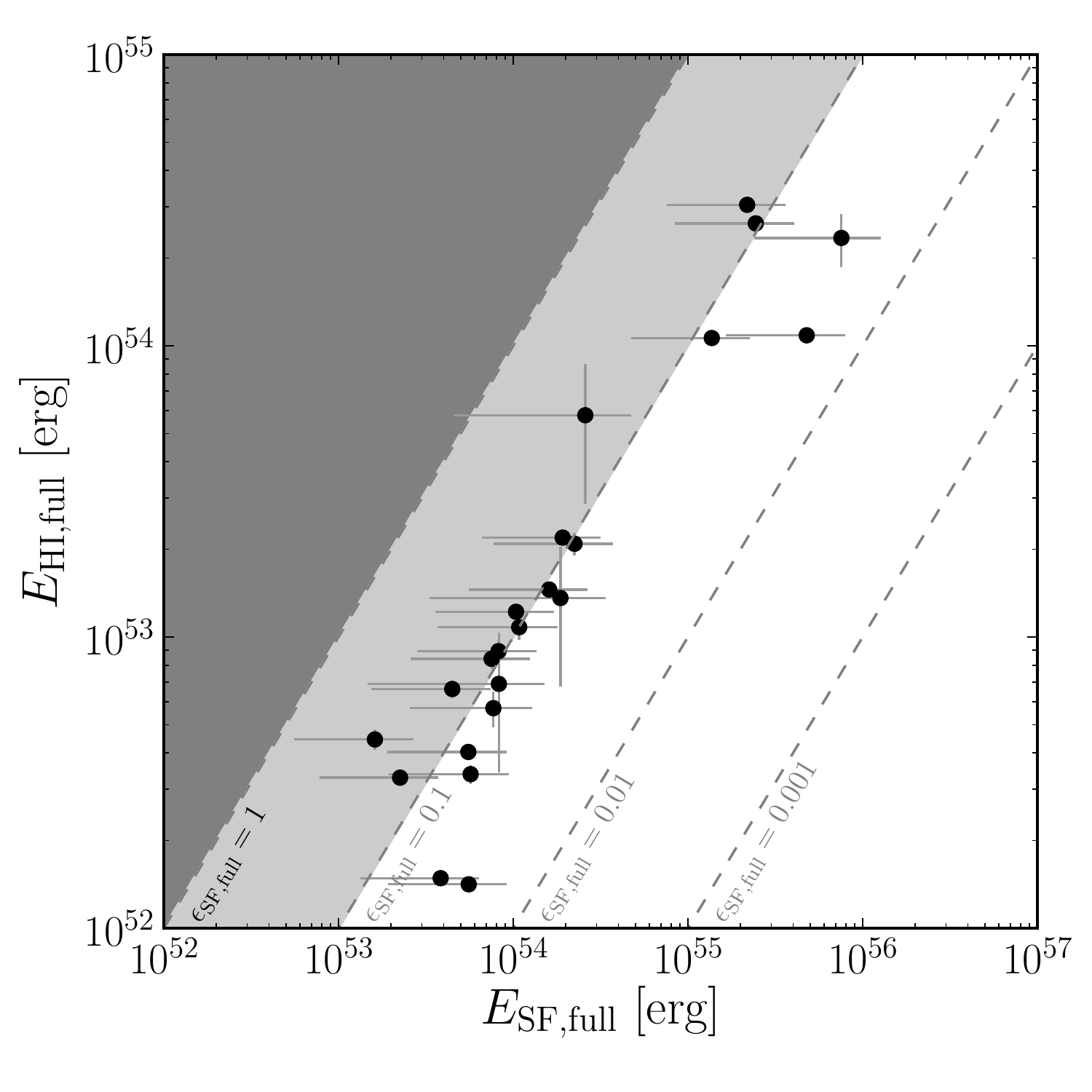}%{{figures/vsub.ro.phys200.jvm.e_sfr.e_full.holes}.pdf}
    \includegraphics[width=3in]{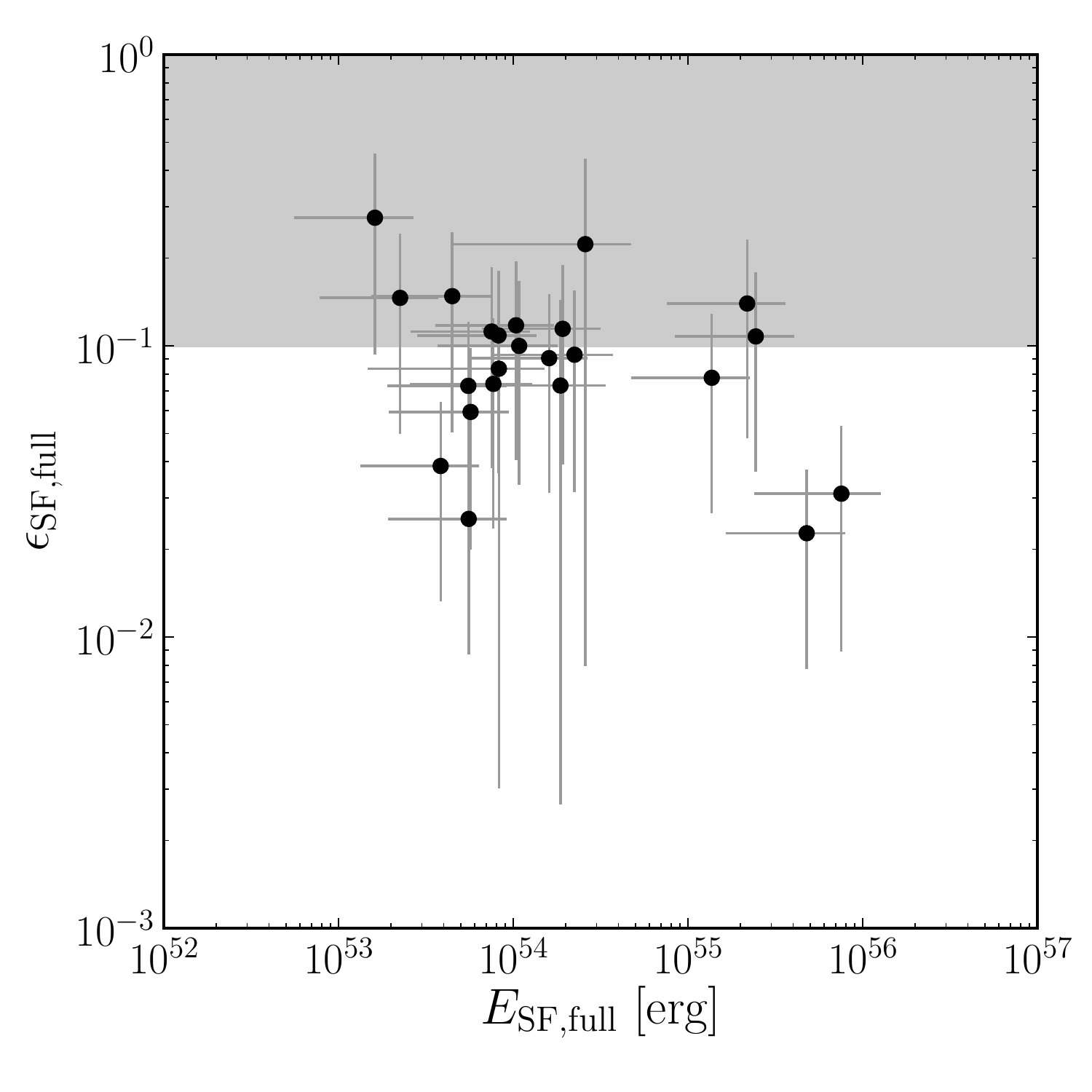}%{{figures/vsub.ro.phys200.jvm.e_sfr.eff_full.holes}.pdf}
  \fi
  \caption{\label{fig:full-e-sf-fixed} The left panel shows the energy available from star formation over 32.5 Myr versus \hi{} energy in the full superprofile component as determined by the HWHM Gaussian fit. 
    The dashed grey lines indicate constant efficiency, and the shaded gray regions represent $\epsilon_\mathrm{SF, full} > 0.1$ (light grey) and $\epsilon_\mathrm{SF, full} > 1$ (dark grey).
The right panel shows $\epsilon_\mathrm{SF,full}$ versus $E_\mathrm{SF}$.
 The mean efficiency required is $\epsilon_\mathrm{SF,full} = 0.102 \pm 0.058$.
 }
\end{figure}

\clearpage
\begin{figure}
  \centering
  \ifimage
    \includegraphics[width=5in]{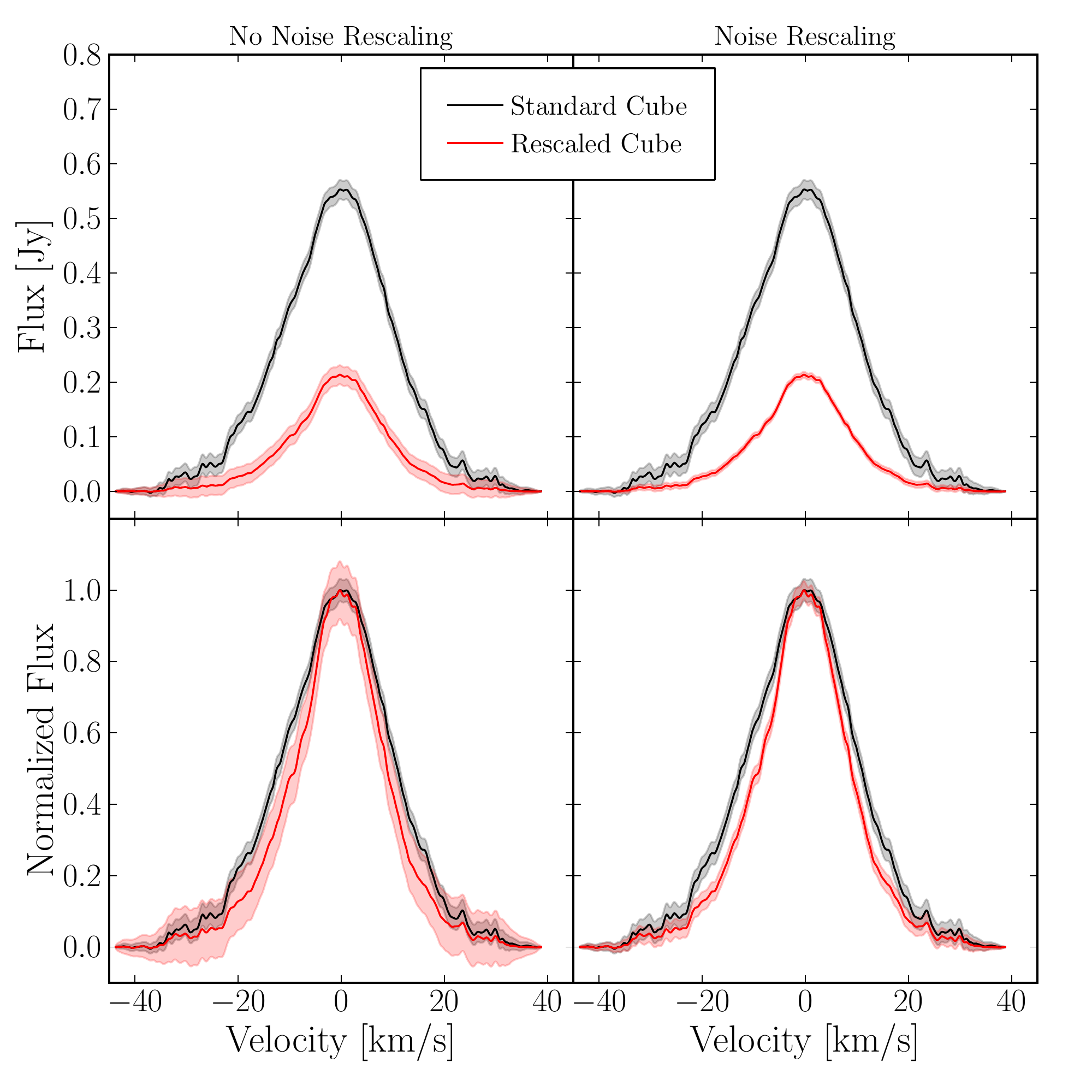}%{{figures/gr8.noise-jvm-std}.pdf}
  \fi
  \caption{A comparison of noise estimates for the superprofiles generated from the standard cube (black) and from the rescaled cube (red) for the galaxy GR 8. The solid lines represent the superprofiles, while the shaded regions show the 1-$\sigma$ noise estimate. The upper panels show the absolute flux measured in each superprofile, while the lower panels show the superprofiles after normalizing to the same maximum amplitude.
The left panels shows the noise estimate given by Equation~\ref{eqn:noise-std} for both the standard and rescaled superprofiles. The right panels shows the noise estimate given by Equation~\ref{eqn:noise-std} for the standard superprofile and that given by Equation~\ref{eqn:noise-jvm} for the rescaled superprofile, where the noise is scaled by the ratio of fluxes between the rescaled and standard superprofile. The rescaled noise approximately matches the fractional uncertainty of noise in the standard superprofile.
 \label{fig:noise-jvm-std}}
\end{figure}

\clearpage{}
\begin{figure}
  \centering
  \ifimage
  \includegraphics[width=3in]{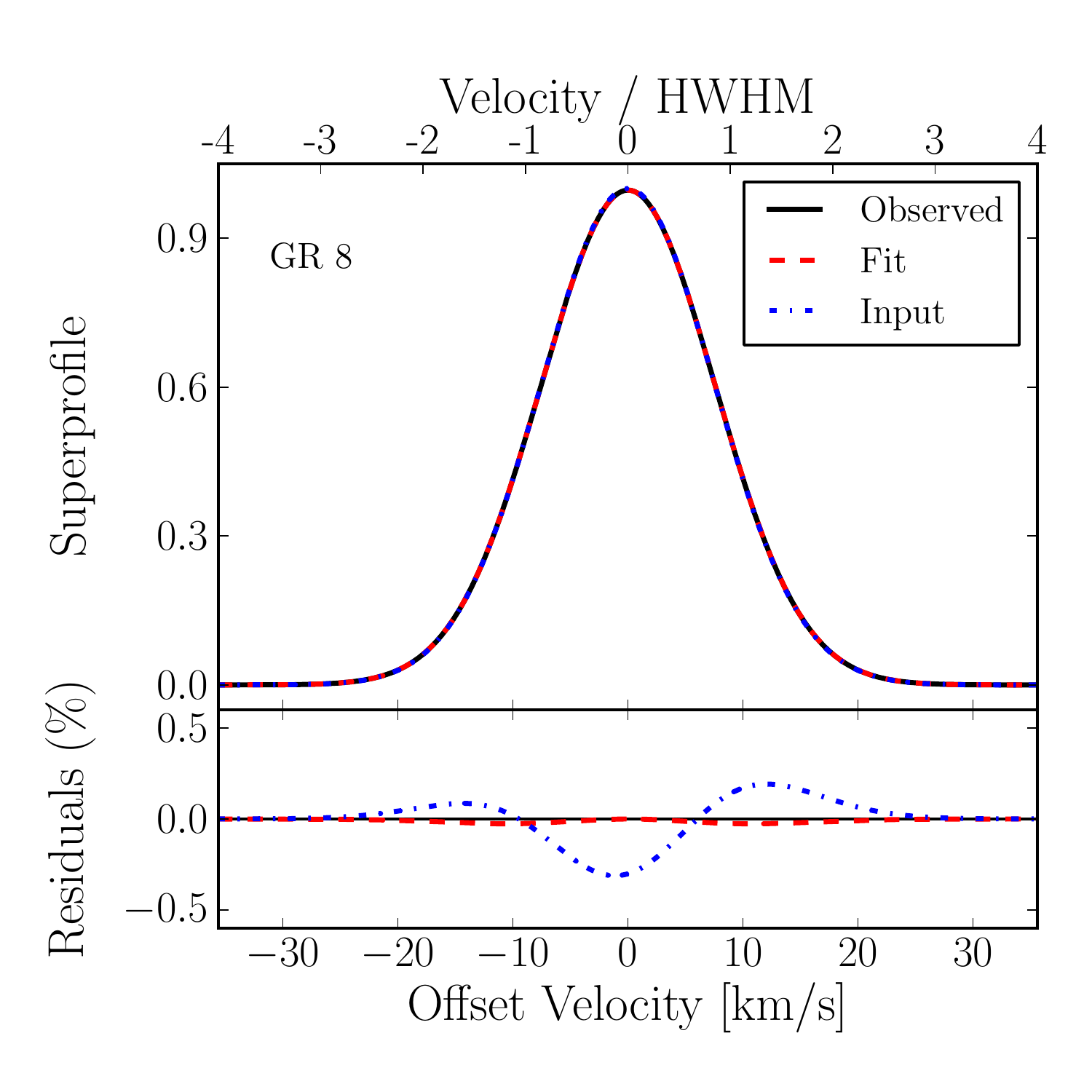}%{{figures/gr8.ro.phys200.sn.sp-offset.obs}.pdf}
  \includegraphics[width=3in]{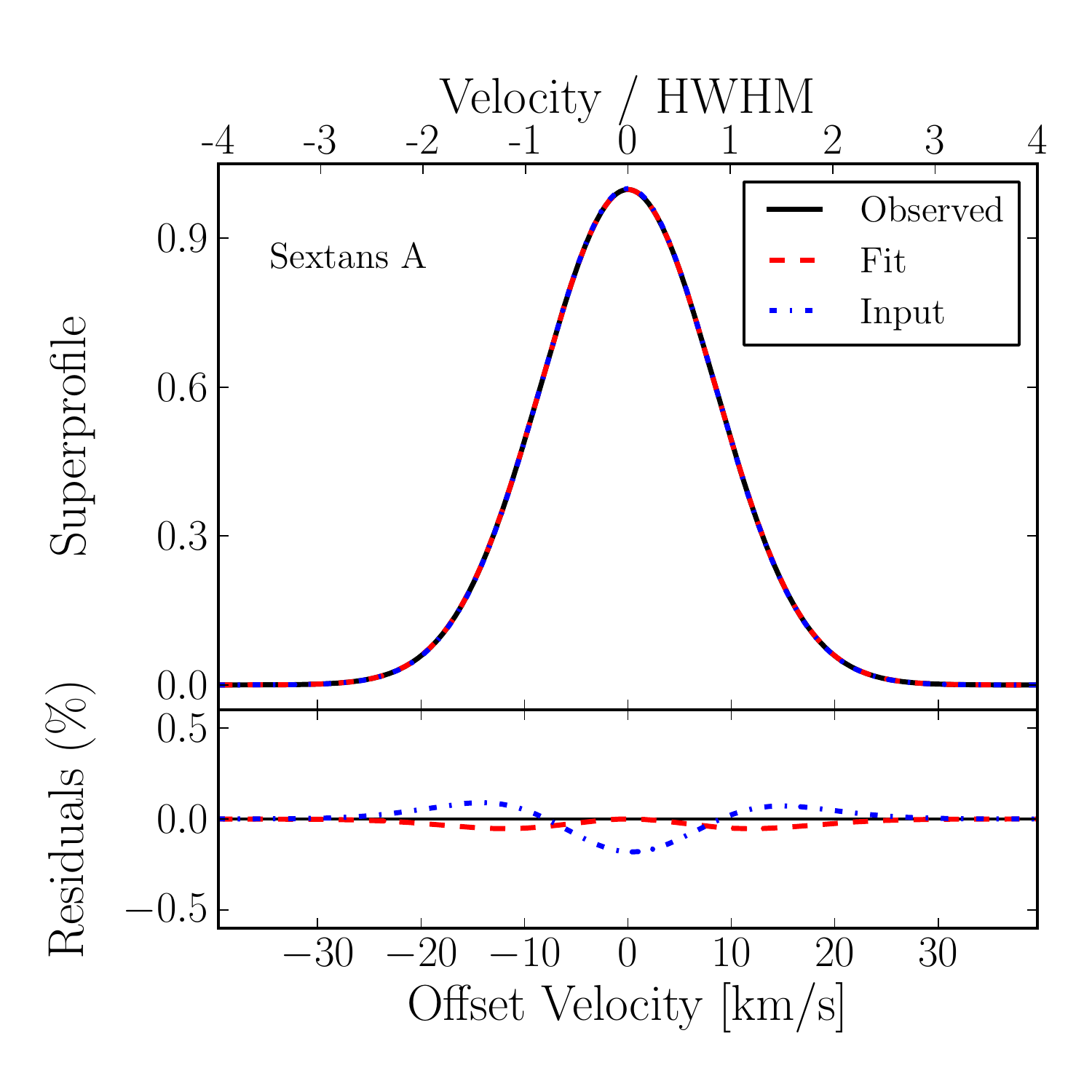}%{{figures/sexa.ro.phys200.sn.sp-offset.obs}.pdf}
  \\
  \includegraphics[width=3in]{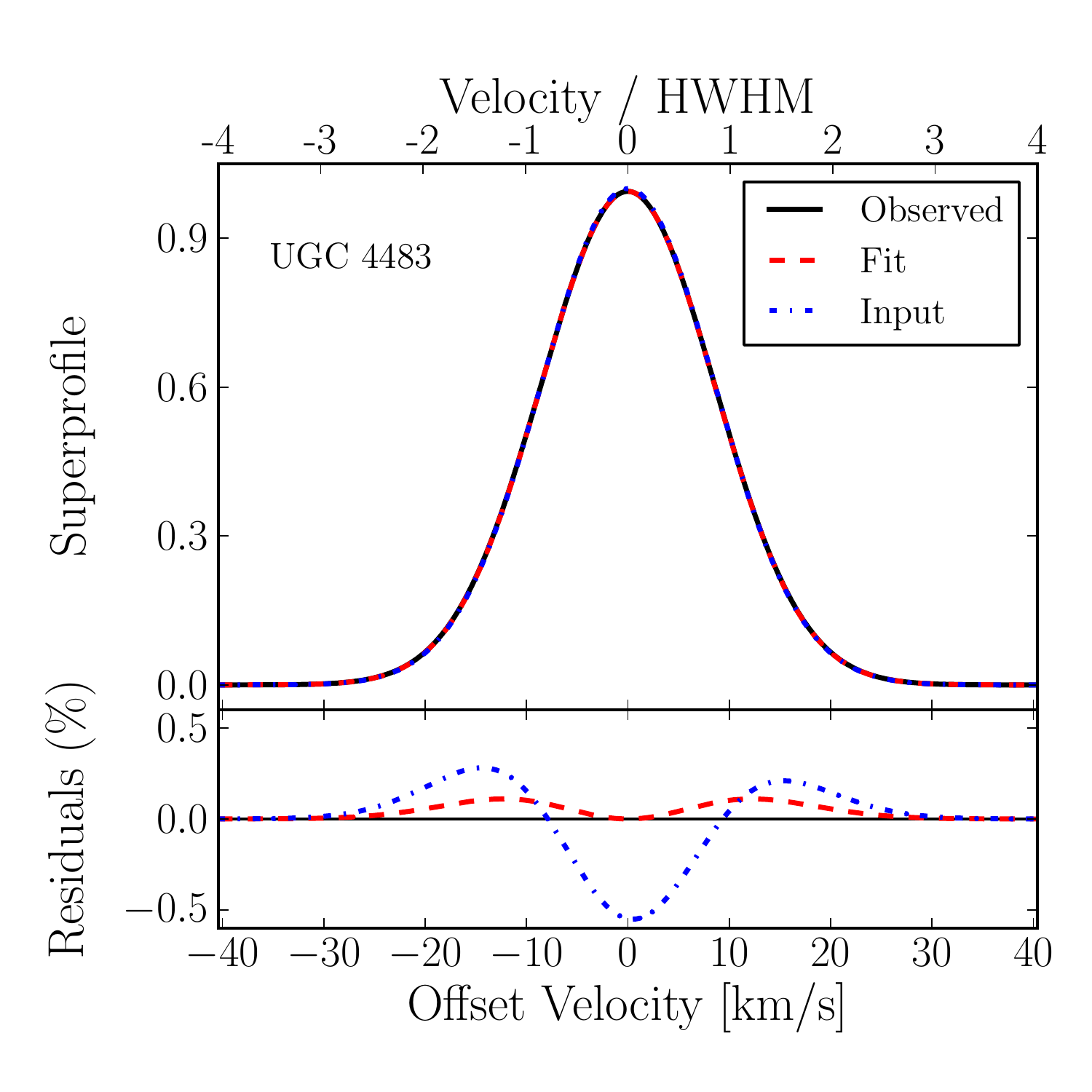}%{{figures/u4483.ro.phys200.sn.sp-offset.obs}.pdf}
  \includegraphics[width=3in]{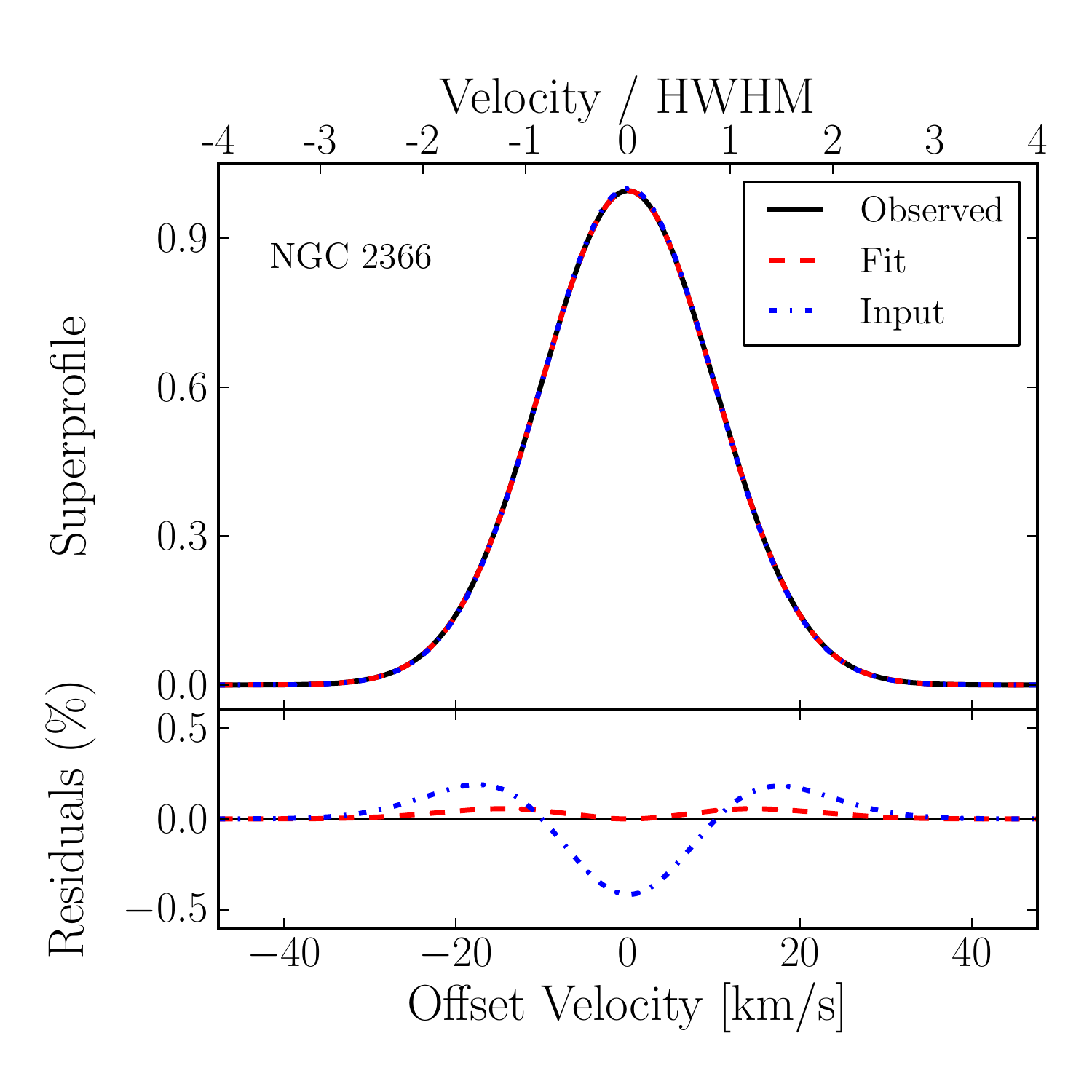}%{{figures/n2366.ro.phys200.sn.sp-offset.obs}.pdf}
  \fi
  \caption{Fake superprofiles with included uncertainties in our \vp{} measurements. Each panel shows a different galaxy. The black line represents the fake superprofile. The dashed blue line is the input Gaussian and the dashed red line is the HWHM-Gaussian fit to the fake superprofile. All lines are plotted in the top panel, but the differences are smaller than the line widths. The bottom panel shows the residuals (i.e., observed - fit and observed - input). Uncertainties in \vp{} broaden the profile slightly, but the effects are $\sim100$ times smaller than the observed amplitude of the wings. The effect due to uncertainties in \vp{} is therefore negligible. Measurement errors in \vp{} therefore do not create the shape of the observed superprofiles. \label{fig:mc-vp-offset-gaussians}}
\end{figure}

\clearpage{}
\begin{figure}
  \centering
  \ifimage
    \includegraphics[height=6in]{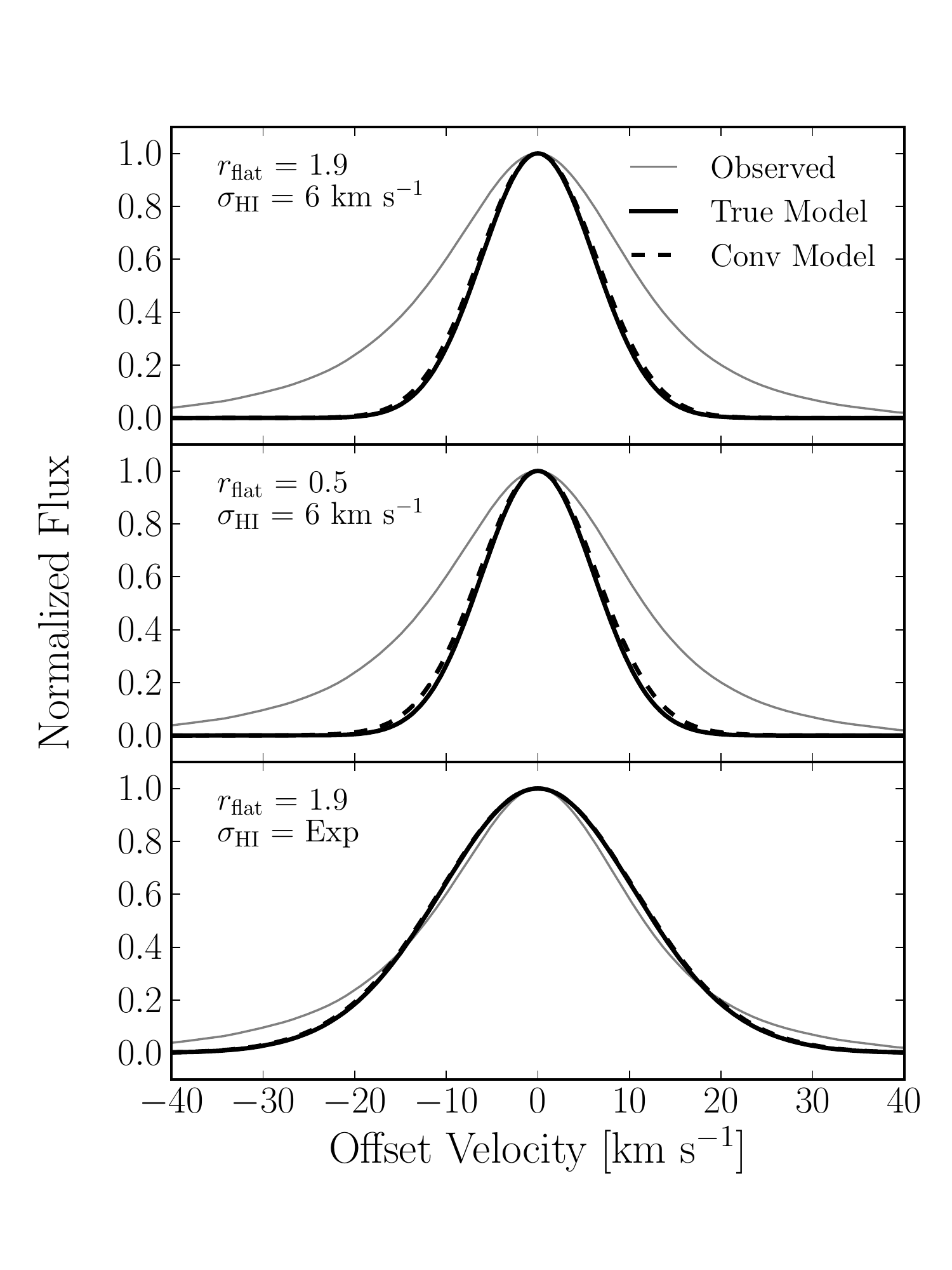}%{{figures/n2366.models.6}.pdf}
  \fi
  \caption{The resulting superprofiles for three NGC 2366 models without spatial smoothing applied (thick black line) and smoothed to 200 pc (thick dashed line). The observed superprofile for NGC 2366 is shown in grey. The upper panel is for the model using the observed rotation curve parameters and fixed velocity dispersion. The middle panel shows a rotation curve with a more extreme rise in the center, but is the larger amount of beam smearing produces no noticeable effect on the profile. The bottom panel shows a model with velocity dispersion that declines with radius based on the second moment map. In all cases, the effects of beam smearing are not strong enough to produce the wings or to substantially widen the intrinsic superprofile. 
  \label{fig:n2366-models}}
\end{figure}

\clearpage
\begin{figure}
  \centering
  \ifimage
    \includegraphics{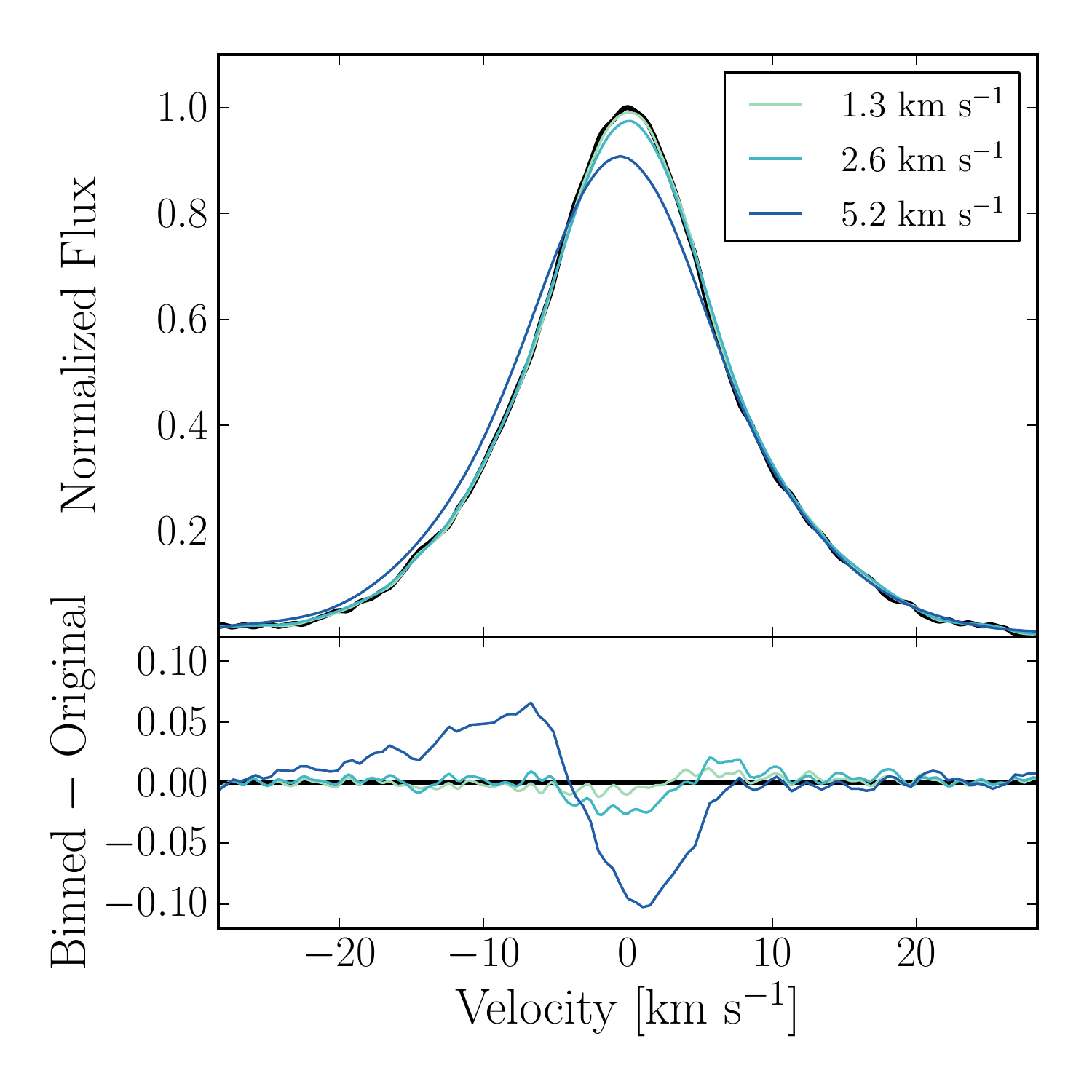}%{{figures/ddo125.superprofiles.binned}.pdf}
  \fi
  \caption{
Superprofiles generated from binned cubes for DDO~125. The thick black line represents the original superprofile. Colored lines represent superprofiles generated from cubes binned to $\Delta v = 1.3, 2.6$, and 5.2 \kms{}. While the differences from the original superprofile are small for the $\Delta v = 1.3$ and 2.6 \kms{} observations, the superprofile generated from the $\Delta v = 5.2$ \kms{} cube is often different by more than 5\% compared to the original. In the $\Delta v = 5.2$ \kms{} cubes, the superprofiles are noticeably wider and shorter, thus leading to artificially inflated \scentral{} values and decreased \fw{} values.  \label{fig:vel-res-sps} }
\end{figure}

\clearpage{}
\begin{figure}
  \centering
  \ifimage
    \includegraphics{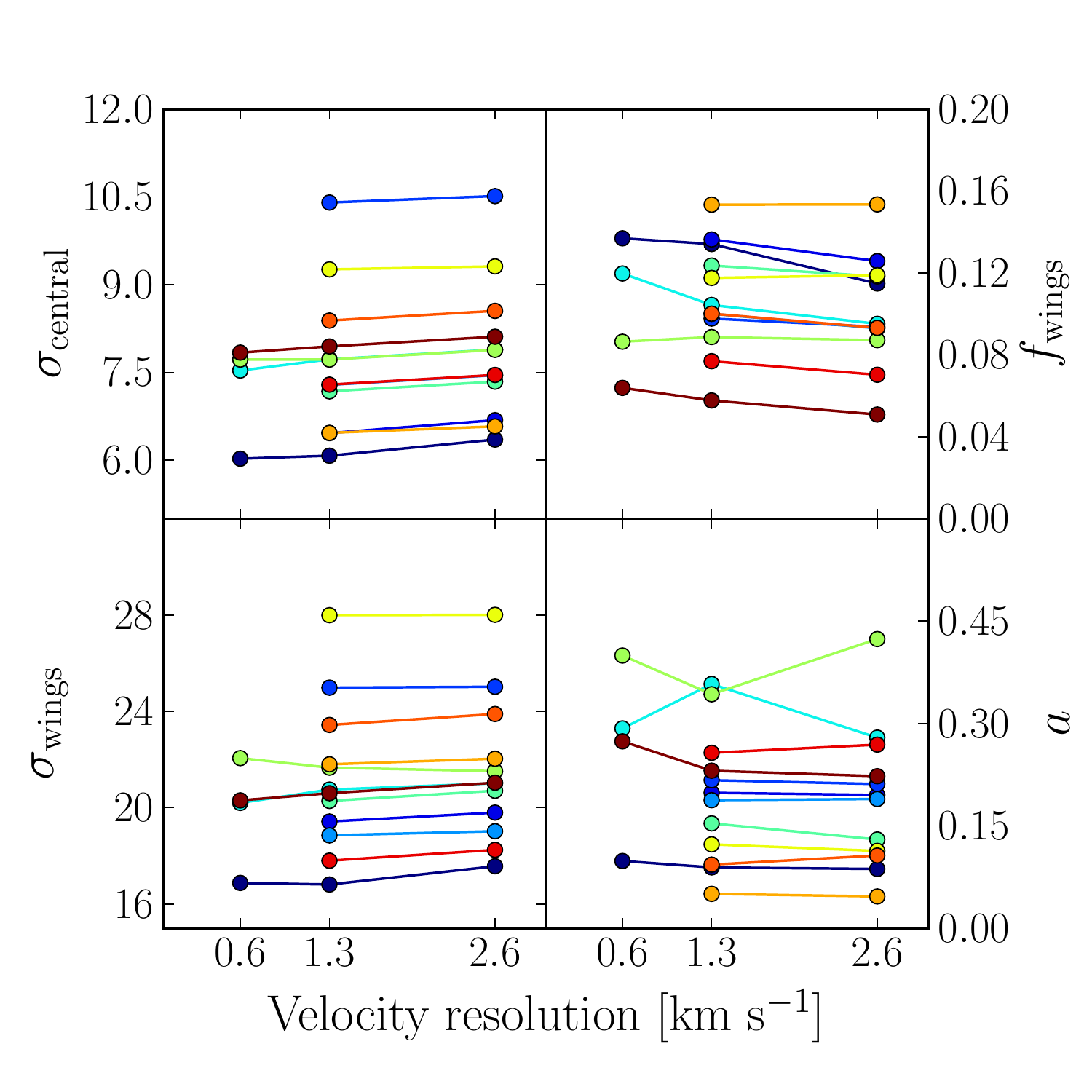}%{{figures/vel_res}.pdf}
  \fi
  \caption{The change in each parameter (\scentral{}, \swing{}, \fw{}, and \aw{}) with increasing velocity resolution. Each colored line represents a single galaxy. The parameters generally show only modest changes with velocity resolution. The most variable is \aw{}, whose variation has been accounted for in the uncertainties. \label{fig:vel-res-params}
}
\end{figure}

\clearpage
\begin{figure}
  \centering
  \ifimage
  \includegraphics[width=6in]{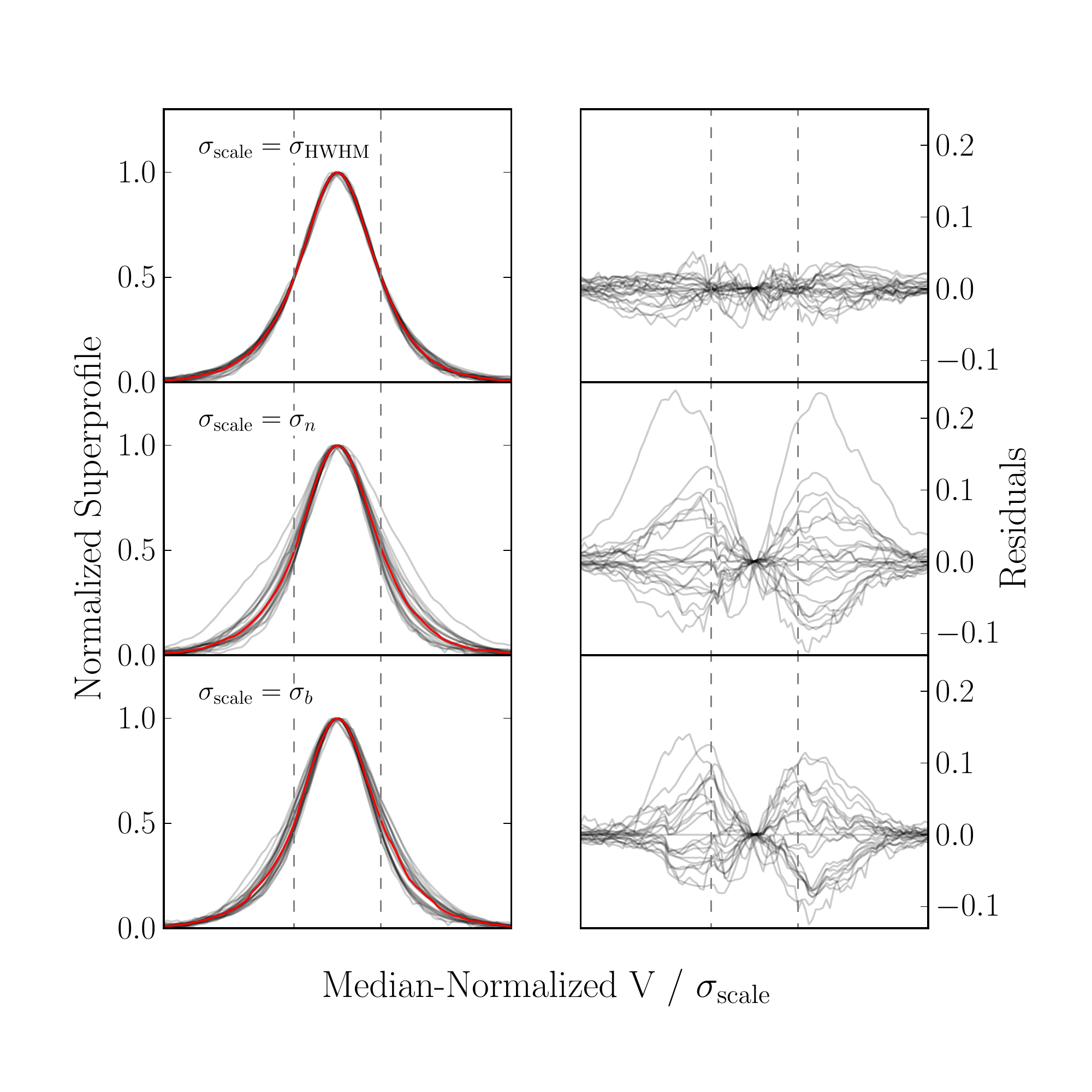}%{{figures/superprofiles.scaled.double_gauss_scaled.resid.sample}.pdf}
  \fi
  \caption{ Scaled superprofiles for all the galaxies. The top panels scale the velocity axis by \scentral{}, the middle panels by \snarrow{}, and the bottom panel by \sbroad{}. The scaled velocity axes are then normalized such that the median scaled superprofile has the same width in all panels to better show the variation in shape. The median superprofile is shown in red. On the left we show the normalized superprofiles, and on the right we show the differences from the median superprofile. The HWHM scaling provides the best overall description of the shape of the superprofiles.  \label{fig:sps-scaled-2gauss-scale} }
\end{figure}

\begin{figure}
  \centering
  \ifimage
    \includegraphics[width=3in]{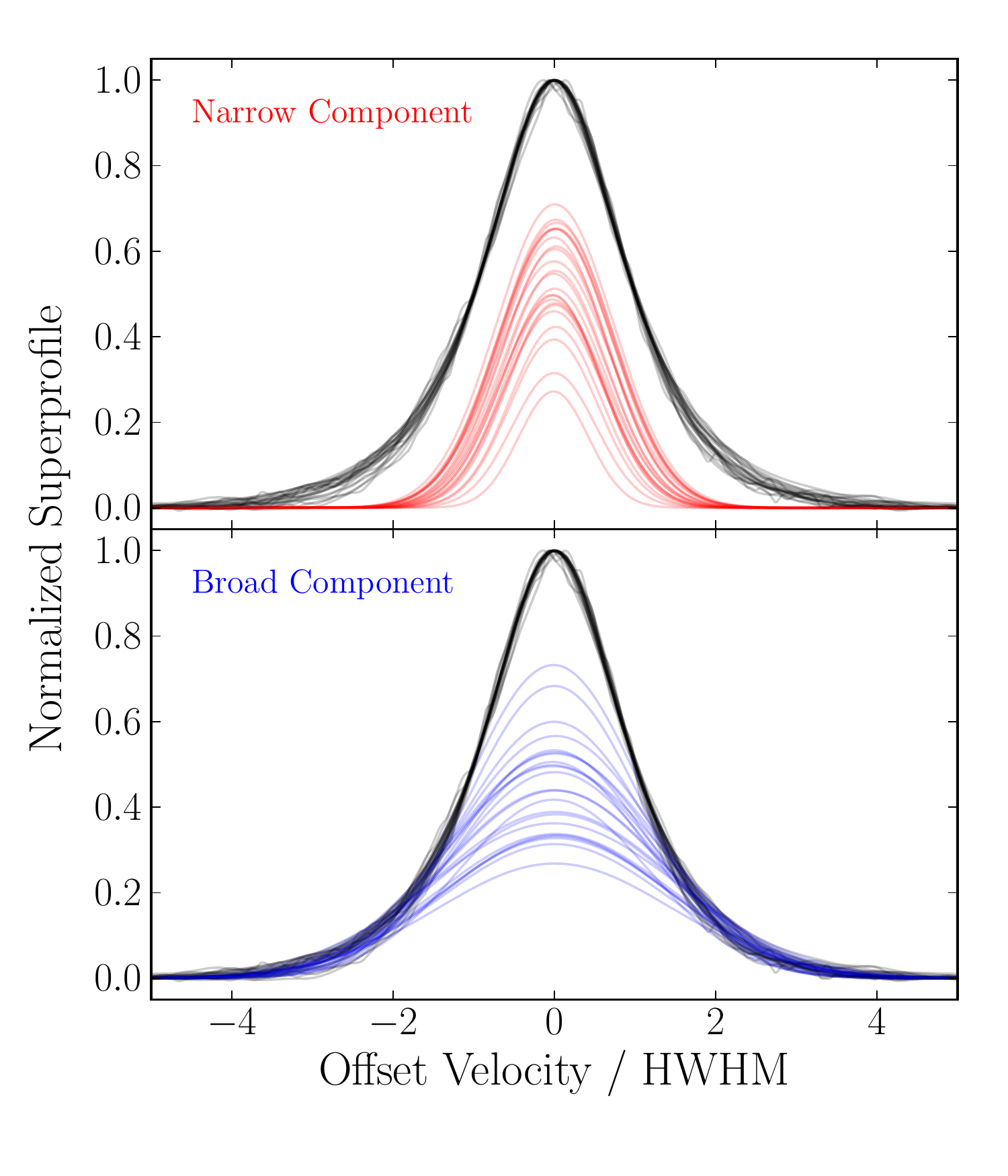}%{{figures/superprofiles.scaled.double_gauss_comp.sample}.pdf}
    \includegraphics[width=3in]{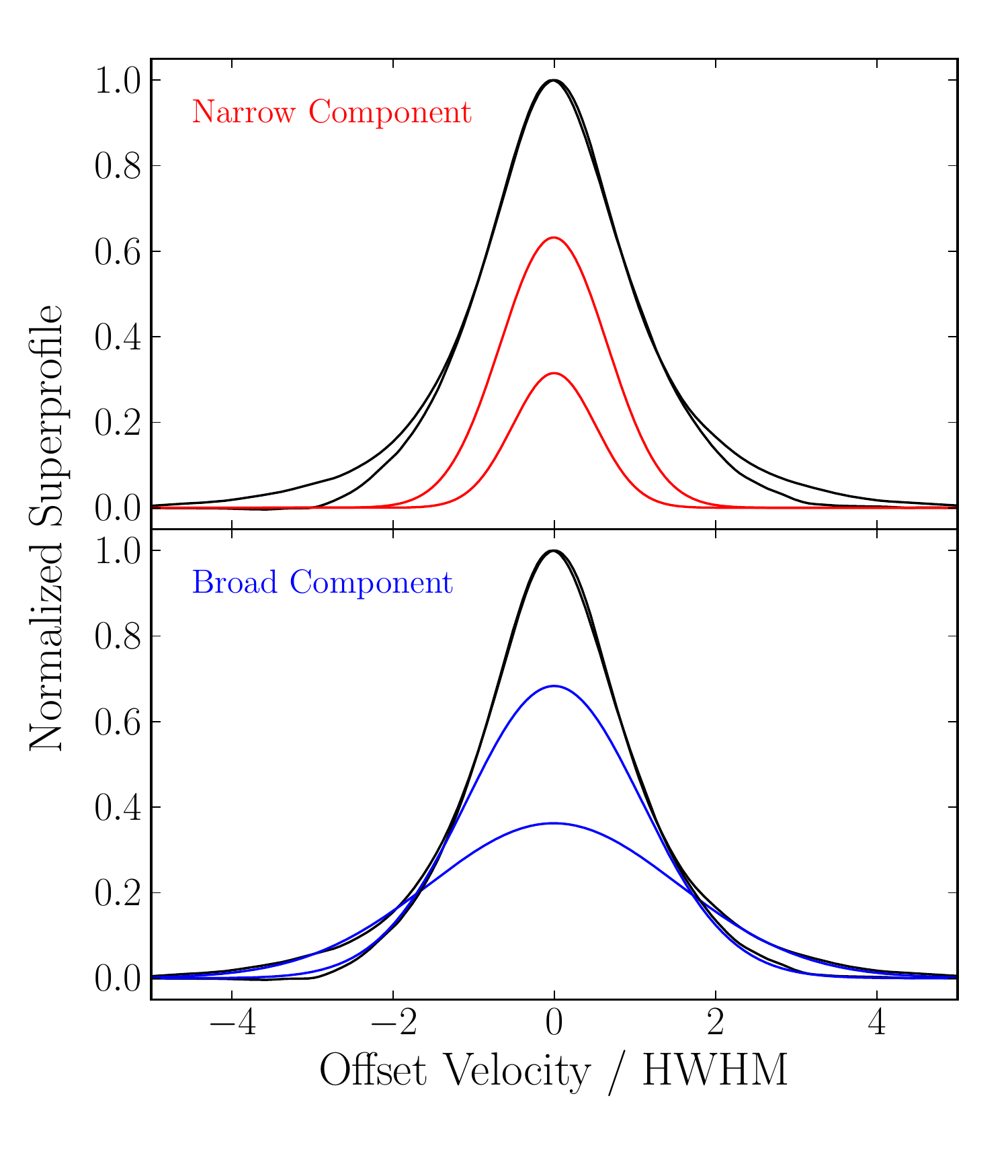}%{{figures/superprofiles.scaled.double_gauss_comp.showcase}.pdf}
  \fi
  \caption{Double Gaussian fit shapes compared to the overall superprofile shapes. The black lines are the full superprofile, while the red and blue lines represent that narrow and broad Gaussian components of a double Gaussian fit. Both the superprofiles and Gaussian components have been scaled to the HWHM of that specific superprofile. The left panel shows the full sample, while the right panel highlights NGC 7793 and Sextans B to show the drastic difference in Gaussian components for superprofiles whose wing fluxes differ by only 9\% when scaled to the same HWHM. This small change drives large differences in the best-fit double Gaussian components. \label{fig:superprofiles-scaled-2gauss-comp}}
\end{figure}

\begin{figure}
  \centering
  \ifimage
  \includegraphics[width=6in]{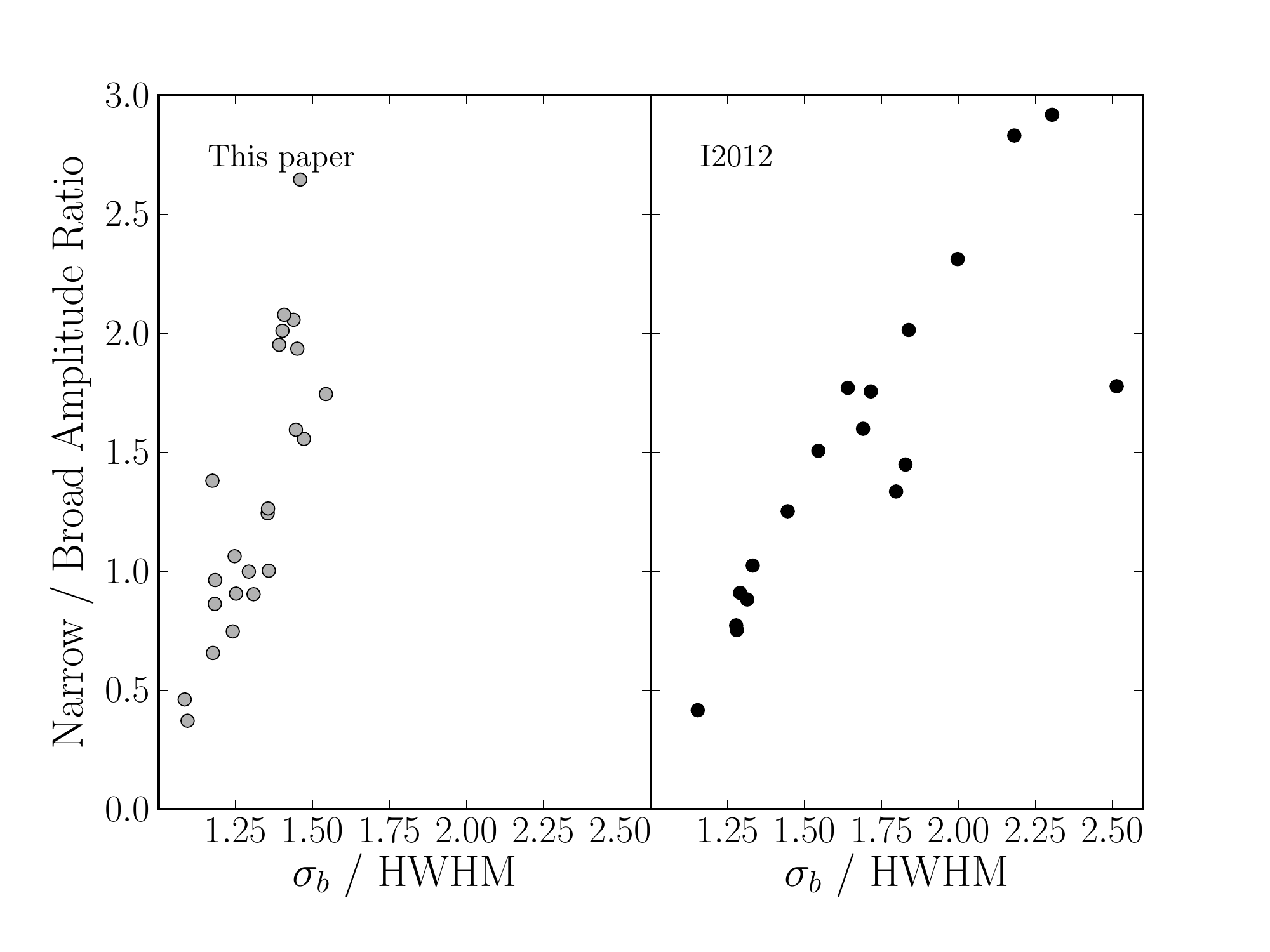}%{{figures/ianj2012.sb-hwhm.amp}.pdf}
  \fi
  \caption{The relationship between Narrow / Broad Amplitude of double Gaussian fits compared to the width of the wings relative to the characteristic superprofile width measured by the HWHM. The grey points in the left panel are galaxies from our sample, while those in right panel (black) are taken from \citetalias{Ianj2012}. In both cases, there is a clear trend that galaxies with broader wings have relatively lower broad amplitudes. \label{fig:comp-sb-amp}}
\end{figure}

\begin{figure}
  \centering
  \ifimage
  \includegraphics[width=6in]{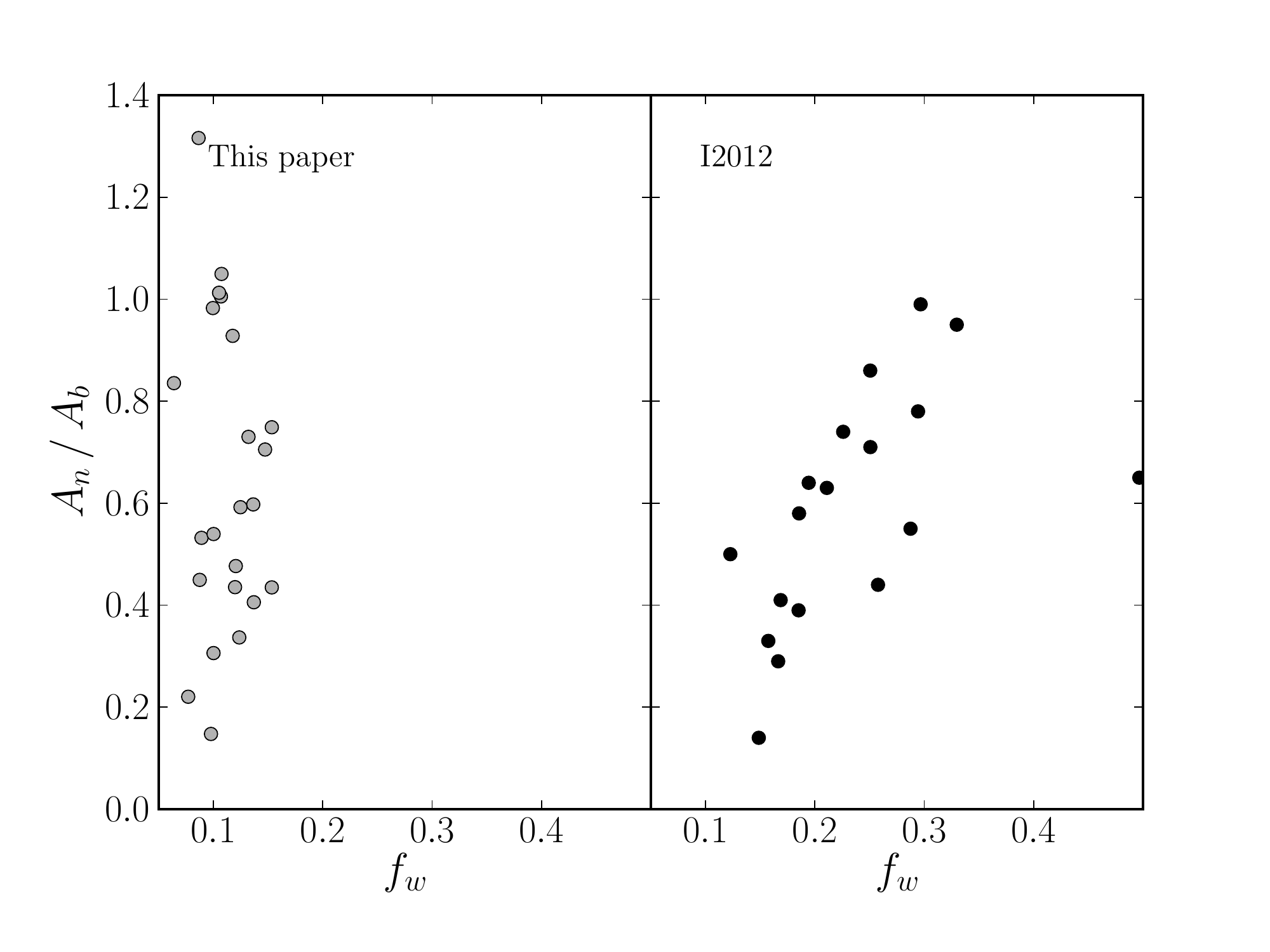}%{{figures/ianj2012.fw.anab}.pdf}
  \fi
  \caption{The relationship between $A_n / A_b$ compared to the fraction of gas in the wings, \fw{}. For galaxies in the \citetalias{Ianj2012} sample, we find that galaxies with more gas in the wings %paradoxically 
  have higher fraction of gas in the narrow component. If the \anab{} parameter were tracing the ratio between the mass of  CNM to WNM, we would on average expect galaxies with more flux in the wings of the profile to have a more \hi{} in the WNM, and therefore smaller \anab{} ratios. This expectation is the opposite of how the \anab{} parameter behaves. It is instead likely that \anab{}, like the amplitudes of the double Gaussian components, is driven by the relative broadness of the wings. \label{fig:comp-fw-anab}}
\end{figure}

\begin{figure}
  \centering
  \ifimage
  \includegraphics[width=6in]{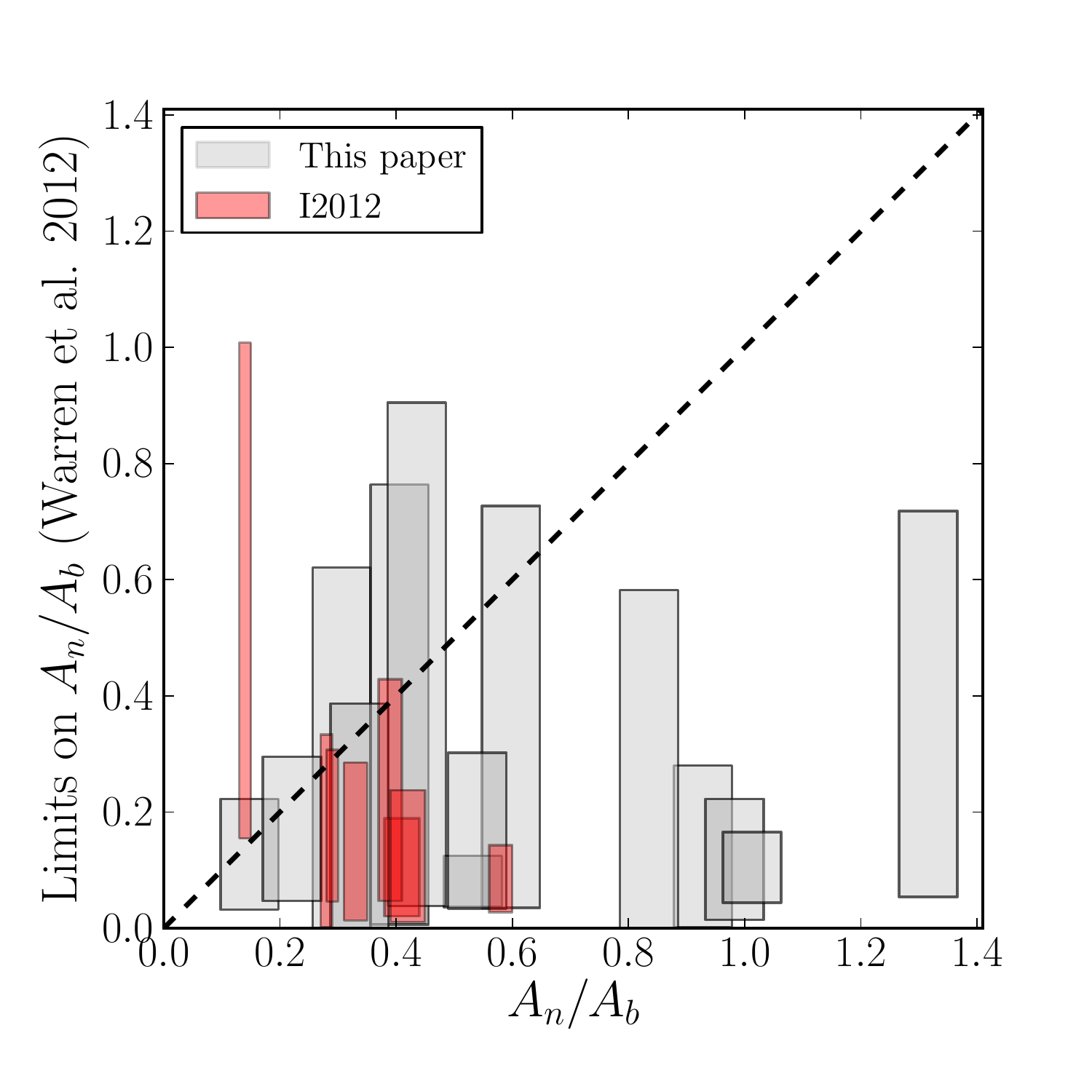}%{{figures/ianj2012.warren.anab}.pdf}
  \fi
  \caption{ A comparison between \anab{} values from double Gaussian fits and independent limits on \anab{} from \citet{Warren2012}. Each box represents an individual galaxy. The $x$-position and width are determined by double Gaussian \anab{} values and associated errors. The top and bottom of each box is the limit placed on \anab{} by \citet{Warren2012}. The line of equality is shown as a thick dashed line. The \anab{} values measured by double Gaussian fits do not match the limits placed by \citet{Warren2012}.  \label{fig:anab-warren}}
\end{figure}

%
% Tables
%
\include{table-sample}
\input{table-observations}
\input{table-derived-properties}
\input{table-measurements}
\input{table-double-gaussian}
\input{table-correlations}
\input{table-measurements-extend_400}
\input{table-scale-height}

%\bibliographystyle{apj}
%\bibliography{global_energy_sub}

\end{document}

%% file: table-sample.tex
\begin{deluxetable}{lclcccccccc}
\tabletypesize{\scriptsize}
\rotate
\tablewidth{0pt}
\tablecaption{The Sample \label{tab:sample}}
\tablehead{
  \colhead{Galaxy} &
  \colhead{Alt. Name} &
  \colhead{Survey} &
  \colhead{RA} &
  \colhead{Dec} &
  \colhead{Distance} &
  \colhead{i} &
  \colhead{$M_{\mathrm{HI,tot}}$} &
  \colhead{$r_{25}$} &
  \colhead{$w_{20}$} &
  \colhead{Type} \\

  \colhead{} &
  \colhead{} &
  \colhead{} &
  \colhead{(hh:mm:ss)} &
  \colhead{(dd:mm:ss)} &
  \colhead{(Mpc)} &
  \colhead{(\degrees{})} &
  \colhead{(log \msun{})} &
  \colhead{(kpc)} &
  \colhead{(\kms{})} &
  \colhead{} \\

  \colhead{1} &
  \colhead{2} &
  \colhead{3} &
  \colhead{4} &
  \colhead{5} &
  \colhead{6} &
  \colhead{7} &
  \colhead{8} &
  \colhead{9} &
  \colhead{10} &
  \colhead{11}
}
\startdata
  NGC 7793 &                 &     THINGS &   23:57:49.7 &    -32:35:28 & 3.90\tablenotemark{\dag} & 50 \tablenotemark{a} &  8.9 &  5.9\tablenotemark{\ddagger} &  249 &  7 \\
   IC 2574 & UGC 5666; DDO 81 &     THINGS &   10:28:27.7 &    +68:24:59 & 3.79 & 55 \tablenotemark{b} &  9.1 &  7.1\tablenotemark{\ddagger} &  160 &  9 \\
  NGC 4214 &        UGC 7278 &     THINGS &   12:15:39.2 &    +36:19:37 & 3.04 & 44 \tablenotemark{c} &  8.6 &  3.0\tablenotemark{\ddagger} &  129 & 10 \\
     Ho II &        UGC 4305 &     THINGS &   08:19:05.0 &    +70:43:12 & 3.38 & 49 \tablenotemark{b} &  8.8 &  3.2\tablenotemark{\ddagger} &   94 & 10 \\
  NGC 2366 &        UGC 3851 &     THINGS &   07:28:53.4 &    +69:12:51 & 3.21 & 63 \tablenotemark{b} &  8.8 &  2.0\tablenotemark{\ddagger} &  130 & 10 \\
   DDO 154 &        UGC 8024 &     THINGS &   12:54:05.9 &    +27:09:10 & 4.30\tablenotemark{\dag} & 66 \tablenotemark{b} &  8.6 &  1.2\tablenotemark{\ddagger} &  115 & 10 \\
      Ho I & UGC 5139; DDO 63 &     THINGS &   09:40:32.3 &    +71:10:56 & 3.90 & 13 \tablenotemark{b} &  8.2 &  1.9\tablenotemark{\ddagger} &  184 & 10 \\
  NGC 4190 &        UGC 7232 &  VLA-ANGST &   12:13:44.6 &    +36:38:00 & 3.50\tablenotemark{\dag} & 41 \tablenotemark{\ast} &  7.7 &  0.9 &  113 & 10 \\
  NGC 3741 &        UGC 6572 &  VLA-ANGST &   11:36:06.4 &    +45:17:07 & 3.24 & 64 \tablenotemark{d} &  7.9 &  0.9 &   95 & 10 \\
 Sextans A & DDO75; UGCA 205 &  VLA-ANGST &   10:11:00.8 &    -04:41:34 & 1.38 & 36 \tablenotemark{e} &  7.8 &  1.1 &  103 & 10 \\
    DDO 53 &        UGC 4459 &     THINGS &   08:34:07.2 &    +66:10:54 & 3.61 & 27 \tablenotemark{b} &  7.8 &  0.4\tablenotemark{\ddagger} &  101 & 10 \\
   DDO 190 &        UGC 9240 &  VLA-ANGST &   14:24:43.5 &    +44:31:33 & 2.79 & 30 \tablenotemark{\ast} &  7.6 &  0.7 &  126 & 10 \\
   DDO 125 &        UGC 7577 &  VLA-ANGST &   12:27:41.8 &    +43:29:38 & 2.58 & 63 \tablenotemark{f} &  7.5 &  1.6 &   48 & 10 \\
 Sextans B & DDO70; UGC 5373 &  VLA-ANGST &   10:00:00.1 &    +05:19:56 & 1.39 & 52 \tablenotemark{\ast} &  7.6 &  1.0 &   77 & 10 \\
    DDO 99 &        UGC 6817 &  VLA-ANGST &   11:50:53.0 &    +38:52:50 & 2.59 & 60 \tablenotemark{\ast} &  7.7 &  1.5 &   62 & 10 \\
   M81 DwB &        UGC 5423 &     THINGS &   10:05:30.6 &    +70:21:52 & 5.30\tablenotemark{\dag} & 44 \tablenotemark{b} &  7.4 &  0.9\tablenotemark{\ddagger} &   84 & 10 \\
  UGCA 292 &        CvnI-DwA &  VLA-ANGST &   12:38:40.0 &    +32:46:00 & 3.62 & 16 \tablenotemark{\ast} &  7.6 &  0.5 &  140 & 10 \\
  NGC 4163 &        UGC 7199 &  VLA-ANGST &   12:12:09.1 &    +36:10:09 & 2.86 & 45 \tablenotemark{\ast} &  7.0 &  0.8 &   51 & 10 \\
  UGC 4483 &                 &  VLA-ANGST &   08:37:03.0 &    +69:46:31 & 3.41 & 42 \tablenotemark{\ast} &  7.5 &  0.6 &   74 & 10 \\
   DDO 181 &        UGC 8651 &  VLA-ANGST &   13:39:53.8 &    +40:44:21 & 3.14 & 50 \tablenotemark{\ast} &  7.4 &  1.1 &   70 & 10 \\
  UGC 8833 &                 &  VLA-ANGST &   13:54:48.7 &    +35:50:15 & 3.08 & 33 \tablenotemark{\ast} &  7.1 &  0.4 &   75 & 10 \\
   DDO 187 &        UGC 9128 &  VLA-ANGST &   14:15:56.5 &    +23:03:19 & 2.21 & 55 \tablenotemark{\ast} &  7.1 &  0.5 &   59 & 10 \\
      GR 8 & DDO155; UGC 8091 &  VLA-ANGST &   12:58:40.4 &    +14:13:03 & 2.08 & 33 \tablenotemark{\ast} &  6.8 &  0.3 &   61 & 10 \\
\enddata
\tablecomments{(1) Galaxy name. (2) Alternative names. (3) \hi{} survey. (4-5) position in J200 coordinates taken from \citet{Walter2008} for THINGS galaxies or \citet{Ott2012} for VLA-ANGST galaxies. (6) Distance from \citet{Dalcanton2009} unless marked with $^{\dagger}$ indicating \citet{Karachentsev2004}. (7) Inclination, references below. (8) \m{HI,tot} in \msun{} from \citet{Walter2008} for THINGS galaxies or \citet{Ott2012} for VLA-ANGST galaxies, updated for distances given in this table. (9) $B$-band $r_{25}$ from \citet{Dalcanton2009} unless marked with $^{\ddagger}$ indicating \citet{Walter2008}. (10) inclination-corrected $w_{20}$, taken from \citet{Walter2008} for THINGS galaxies or \citet{Ott2012} for VLA-ANGST galaxies. (11) de Vaucouleurs T-type from \citet{Walter2008} for THINGS galaxies or \citet{Ott2012} for VLA-ANGST galaxies.}
\tablenotetext{\ast}{Derived from LVL 3.6\um{} images}
\tablenotetext{a}{\citet{deBlok2008}}
\tablenotetext{b}{\citet{Oh2011}}
\tablenotetext{c}{\citet{Walter2008}}
\tablenotetext{d}{\citet{Begum2005}}
\tablenotetext{e}{\citet{Skillman1988}}
\tablenotetext{f}{\citet{Swaters2009}}
\end{deluxetable}

%% file: table-observations.tex
\begin{deluxetable}{lccc}
\tablewidth{0pt}
\tablecaption{Observation Parameters \label{tab:hi-obs}}
\tablehead{
  \colhead{Galaxy} &
  \colhead{$\Delta v$} &
  \colhead{$\theta_\mathrm{200pc}$} &
  \colhead{$\sigma_\mathrm{chan}$} \\

  \colhead{} &
  \colhead{(\kms{})} &
  \colhead{($\arcsec{}$)} &
  \colhead{(mJy \per{beam})} \\

  \colhead{1} &
  \colhead{2} &
  \colhead{3} &
  \colhead{4}
}
\startdata
  NGC 7793  &   2.6  &  10.58  &  1.19    \\
   IC 2574  &   2.6  &  10.88  &  0.91    \\
  NGC 4214  &   1.3  &  13.57  &  1.08    \\
     Ho II  &   2.6  &  12.21  &  1.54    \\
  NGC 2366  &   2.6  &  12.85  &  1.00    \\
   DDO 154  &   2.6  &  9.59  &  0.60    \\
      Ho I  &   2.6  &  10.58  &  1.23    \\
  NGC 4190  &   1.3  &  11.79  &  1.50    \\
  NGC 3741  &   1.3  &  12.73  &  1.90    \\
 Sextans A  &   1.3  &  29.89  &  4.06    \\
    DDO 53  &   2.6  &  11.43  &  0.76    \\
   DDO 190  &   2.6  &  14.69  &  0.62    \\
   DDO 125  &   0.6  &  15.99  &  3.07    \\
 Sextans B  &   1.3  &  29.68  &  1.90    \\
    DDO 99  &   1.3  &  15.93  &  1.97    \\
   M81 DwB  &   2.6  &  7.78  &  0.64    \\
  UGCA 292  &   0.6  &  11.40  &  2.13    \\
  NGC 4163  &   0.6  &  14.42  &  2.16    \\
  UGC 4483  &   2.6  &  12.10  &  0.80    \\
   DDO 181  &   1.3  &  13.14  &  1.52    \\
  UGC 8833  &   2.6  &  13.39  &  0.59    \\
   DDO 187  &   1.3  &  18.67  &  2.12    \\
      GR 8  &   0.6  &  19.83  &  4.15    \\
\enddata
\tablecomments{(1) Galaxy name. (2) Channel spacing. (3) Circular beam in \arcsec{} that corresponds to 200 pc. (4) \emph{rms} noise in mJy \per{beam} in convolved, standard cube.}
\end{deluxetable}

%% file: table-derived-properties.tex
\begin{deluxetable}{lccccccc}
\tabletypesize{\scriptsize}
\rotate
\tablewidth{0pt}
\tablecaption{Derived Sample Properties \label{tab:derived-properties}}
\tablehead{
  \colhead{Galaxy} &
  \colhead{$M_\mathrm{baryon,tot}$} &
  \colhead{$M_{\mathrm{HI}}$} &
  \colhead{$M_\star$} &
  \colhead{SFR} &
  \colhead{SFR / $M_\mathrm{HI}$} &
  \colhead{$\langle \sfrsd{} \rangle$} &
  \colhead{$\langle \Sigma_\mathrm{HI} \rangle$} \\

  \colhead{} &
  \colhead{(log \msun{})} &
  \colhead{(log \msun{})} &
  \colhead{(log \msun{})} &
  \colhead{($10^{-3}$ \msun{} yr$^{-1}$)} &
  \colhead{(10$^{-10}$ yr$^{-1}$)} &
  \colhead{($10^{-3}$ \msun{} kpc$^{-2}$)} &
  \colhead{(\msun{} kpc$^{-2}$)} \\

  \colhead{1} &
  \colhead{2} &
  \colhead{3} &
  \colhead{4} &
  \colhead{5} &
  \colhead{6} &
  \colhead{7} &
  \colhead{8}
}
\startdata
  NGC 7793 &  9.7 &  8.7 &  9.4 & 193.7 & 4.01 & 3.020 & 7.53  \\
   IC 2574 &  9.4 &  9.0 &  8.8 & 62.7 & 0.66 & 0.325 & 4.91  \\
  NGC 4214 &  9.2 &  8.5 &  8.9 & 122.6 & 3.66 & 1.250 & 3.41  \\
     Ho II &  9.0 &  8.6 &  8.3 & 35.1 & 0.98 & 0.549 & 5.62  \\
  NGC 2366 &  9.0 &  8.7 &  8.3 & 55.9 & 1.17 & 0.628 & 5.39  \\
   DDO 154 &  8.7 &  8.4 &  7.8 &  6.6 & 0.29 & 0.087 & 3.01  \\
      Ho I &  8.4 &  7.8 &  7.4 &  4.9 & 0.70 & 0.654 & 9.39  \\
  NGC 4190 &  8.1 &  7.4 &  7.7 &  4.8 & 1.80 & 1.770 & 9.84  \\
  NGC 3741 &  8.1 &  7.6 &  7.2 &  2.8 & 0.72 & 0.240 & 3.32  \\
 Sextans A &  8.0 &  7.8 &  7.4 &  5.8 & 0.94 & 0.478 & 5.11  \\
    DDO 53 &  8.0 &  7.6 &  7.2 &  4.1 & 1.04 & 0.871 & 8.35  \\
   DDO 190 &  8.0 &  7.5 &  7.5 &  2.7 & 0.84 & 0.674 & 8.06  \\
   DDO 125 &  7.9 &  7.2 &  7.4 &  1.5 & 1.01 & 0.274 & 2.70  \\
 Sextans B &  7.9 &  7.6 &  7.7 &  1.7 & 0.46 & 0.091 & 1.98  \\
    DDO 99 &  7.9 &  7.4 &  7.2 &  2.0 & 0.87 & 0.237 & 2.72  \\
   M81 DwB &  7.9 &  7.1 &  7.3 &  2.1 & 1.66 & 1.676 & 10.08  \\
  UGCA 292 &  7.8 &  7.5 &  6.8 &  1.1 & 0.40 & 0.536 & 13.31  \\
  NGC 4163 &  7.7 &  6.7 &  7.3 &  1.4 & 2.84 & 1.558 & 5.49  \\
  UGC 4483 &  7.7 &  7.4 &  7.0 &  2.1 & 0.93 & 0.630 & 6.80  \\
   DDO 181 &  7.7 &  7.2 &  7.1 &  1.4 & 0.96 & 0.375 & 3.90  \\
  UGC 8833 &  7.4 &  7.0 &  6.8 &  0.6 & 0.58 & 0.356 & 6.09  \\
   DDO 187 &  7.3 &  7.0 &  6.7 &  0.4 & 0.46 & 0.230 & 5.01  \\
      GR 8 &  7.1 &  6.6 &  6.6 &  1.0 & 2.29 & 1.050 & 4.59  \\
\enddata
\tablecomments{(1) Galaxy name. (2) Total baryonic mass (\S~\ref{sec:data--other-data--mhalo}). (3) Aperture-matched \hi{} mass (\S~\ref{sec:data--other-data--mhi}). (4) Aperture-matched stellar mass (\S~\ref{sec:data--other-data--mstar}). (5) Aperture-matched SFR (\S~\ref{sec:data--other-data--sfr}). (6) Aperture-matched SFR / \mhi{} (\S~\ref{sec:data--other-data--sfr-per-mhi}). (7) Aperture-matched average star formation rate surface density (\S~\ref{sec:data--other-data--sfr}). (8) Aperture-matched average \hi{} surface density (\S~\ref{sec:data--other-data--mhi}).}
\end{deluxetable}

%% file: table-measurements.tex
\begin{deluxetable}{lcccccr}
\tablewidth{0pt}
\tablecaption{Measured Superprofile Parameters \label{tab:measurements}}
\tablehead{
  \colhead{Galaxy} &
  \colhead{$\langle \sigma_{\mathrm{m}_2} \rangle$} &
  \colhead{$\sigma_{\mathrm{central}}$} &
  \colhead{$\sigma_{\mathrm{wings}}$} &
  \colhead{$f_w$} &
  \colhead{$a$} &
  \colhead{N$_\mathrm{beams}$} \\

  \colhead{} &
  \colhead{(km s$^{-1}$)} &
  \colhead{(km s$^{-1}$)} &
  \colhead{} &
  \colhead{} &
  \colhead{} \\

  \colhead{1} &
  \colhead{2} &
  \colhead{3} &
  \colhead{4} &
  \colhead{5} &
  \colhead{6} &
  \colhead{7}
}
\startdata
  NGC 7793 & 11.3 &  $ 8.1 \pm  0.5 $ &$26.0 \pm  1.4 $ &$0.15 \pm 0.02 $ &$0.04 \pm 0.06 $ & 909 \\
   IC 2574 &  9.3 &  $ 7.2 \pm  0.5 $ &$19.8 \pm  1.0 $ &$0.12 \pm 0.02 $ &$0.11 \pm 0.06 $ &2443 \\
  NGC 4214 &  9.3 &  $ 6.5 \pm  0.5 $ &$21.8 \pm  1.2 $ &$0.15 \pm 0.02 $ &$0.05 \pm 0.06 $ &1558 \\
     Ho II &  9.0 &  $ 7.1 \pm  0.5 $ &$21.3 \pm  1.0 $ &$0.12 \pm 0.02 $ &$0.14 \pm 0.06 $ & 925 \\
  NGC 2366 & 13.2 &  $10.1 \pm  0.5 $ &$30.2 \pm  0.9 $ &$0.13 \pm 0.02 $ &$0.22 \pm 0.06 $ & 892 \\
   DDO 154 &  8.6 &  $ 7.5 \pm  0.5 $ &$19.8 \pm  0.4 $ &$0.09 \pm 0.02 $ &$0.09 \pm 0.06 $ & 682 \\
      Ho I &  9.0 &  $ 6.7 \pm  0.5 $ &$20.6 \pm  2.1 $ &$0.15 \pm 0.03 $ &$0.11 \pm 0.06 $ & 162 \\
  NGC 4190 & 11.6 &  $ 9.3 \pm  0.6 $ &$28.0 \pm  7.2 $ &$0.12 \pm 0.03 $ &$0.12 \pm 0.07 $ &  45 \\
  NGC 3741 &  8.5 &  $ 7.2 \pm  0.5 $ &$20.3 \pm  3.0 $ &$0.12 \pm 0.03 $ &$0.15 \pm 0.07 $ & 112 \\
 Sextans A &  9.5 &  $ 8.4 \pm  0.5 $ &$23.4 \pm  1.0 $ &$0.10 \pm 0.02 $ &$0.09 \pm 0.06 $ & 215 \\
    DDO 53 & 10.3 &  $ 8.2 \pm  0.5 $ &$23.8 \pm  1.6 $ &$0.11 \pm 0.03 $ &$0.18 \pm 0.07 $ &  93 \\
   DDO 190 & 10.4 &  $ 8.9 \pm  0.5 $ &$24.6 \pm  3.1 $ &$0.09 \pm 0.02 $ &$0.25 \pm 0.08 $ &  76 \\
   DDO 125 &  7.0 &  $ 6.0 \pm  0.5 $ &$16.9 \pm  2.3 $ &$0.14 \pm 0.03 $ &$0.10 \pm 0.07 $ &  53 \\
 Sextans B &  7.9 &  $ 7.3 \pm  0.5 $ &$17.8 \pm  0.7 $ &$0.08 \pm 0.02 $ &$0.26 \pm 0.07 $ & 259 \\
    DDO 99 &  8.0 &  $ 7.3 \pm  0.5 $ &$18.9 \pm  1.6 $ &$0.10 \pm 0.03 $ &$0.19 \pm 0.08 $ &  92 \\
   M81 DwB & 12.5 &  $ 9.2 \pm  0.6 $ &$30.7 \pm  5.0 $ &$0.11 \pm 0.05 $ &$0.23 \pm 0.08 $ &  20 \\
  UGCA 292 &  8.4 &  $ 7.8 \pm  0.5 $ &$20.3 \pm  4.8 $ &$0.06 \pm 0.03 $ &$0.27 \pm 0.10 $ &  45 \\
  NGC 4163 &  8.5 &  $ 7.7 \pm  0.6 $ &$22.1 \pm  3.1 $ &$0.09 \pm 0.04 $ &$0.40 \pm 0.12 $ &  14 \\
  UGC 4483 &  9.9 &  $ 8.5 \pm  0.6 $ &$24.9 \pm  2.6 $ &$0.10 \pm 0.03 $ &$0.41 \pm 0.11 $ &  55 \\
   DDO 181 &  7.8 &  $ 6.5 \pm  0.5 $ &$19.4 \pm  2.8 $ &$0.14 \pm 0.04 $ &$0.20 \pm 0.08 $ &  54 \\
  UGC 8833 &  9.6 &  $ 8.0 \pm  0.6 $ &$23.2 \pm  3.2 $ &$0.11 \pm 0.03 $ &$0.46 \pm 0.13 $ &  30 \\
   DDO 187 & 11.3 &  $10.4 \pm  0.6 $ &$25.0 \pm  3.3 $ &$0.10 \pm 0.03 $ &$0.22 \pm 0.08 $ &  23 \\
      GR 8 &  8.0 &  $ 7.5 \pm  0.5 $ &$20.2 \pm  2.4 $ &$0.12 \pm 0.04 $ &$0.29 \pm 0.10 $ &  17 \\
\enddata
\tablecomments{(1) Galaxy name. (2) Average intensity-weighted global second moment value. (3) Width of central superprofile peak. (4) Characteristic velocity of the wings. (5) Fraction of \hi{} in the wings. (6) Wing asymmetry parameter. (7) Number of independent beams contributing to superprofile.}
\end{deluxetable}

%% file: table-double-gaussian.tex
\begin{deluxetable}{lcccc}
\tablewidth{0pt}
\tablecaption{Double Gaussian Fit Parameters \label{tab:parameters--double-gaussian}}
\tablehead{
  \colhead{Name} &
  \colhead{$\sigma_n$} &
  \colhead{$\sigma_b$} &
  \colhead{$A_n / A_b$} &
  \colhead{$\sigma_n / \sigma_b$} \\

  \colhead{} &
  \colhead{(\kms{})} &
  \colhead{(\kms{})} &
  \colhead{} &
  \colhead{} \\

}
\startdata
       NGC 7793 & $ 6.3 \pm  0.1$ & $14.7 \pm  0.4$ & $0.75 \pm 0.02$ & $0.43 \pm 0.01$ \\
        IC 2574 & $ 5.3 \pm  0.1$ & $11.0 \pm  0.2$ & $0.48 \pm 0.02$ & $0.48 \pm 0.01$ \\
       NGC 4214 & $ 4.5 \pm  0.0$ & $10.3 \pm  0.2$ & $0.43 \pm 0.01$ & $0.43 \pm 0.01$ \\
          Ho II & $ 5.3 \pm  0.1$ & $11.4 \pm  0.3$ & $0.59 \pm 0.04$ & $0.47 \pm 0.01$ \\
       NGC 2366 & $ 7.9 \pm  0.1$ & $17.1 \pm  0.3$ & $0.73 \pm 0.02$ & $0.46 \pm 0.01$ \\
        DDO 154 & $ 5.4 \pm  0.2$ & $10.4 \pm  0.4$ & $0.45 \pm 0.07$ & $0.52 \pm 0.01$ \\
           Ho I & $ 5.3 \pm  0.2$ & $11.6 \pm  0.9$ & $0.71 \pm 0.10$ & $0.45 \pm 0.02$ \\
       NGC 4190 & $ 7.6 \pm  0.3$ & $15.8 \pm  2.2$ & $0.93 \pm 0.17$ & $0.48 \pm 0.04$ \\
       NGC 3741 & $ 4.7 \pm  0.4$ & $10.5 \pm  0.8$ & $0.34 \pm 0.09$ & $0.45 \pm 0.01$ \\
      Sextans A & $ 6.2 \pm  0.2$ & $12.3 \pm  0.4$ & $0.54 \pm 0.07$ & $0.51 \pm 0.01$ \\
         DDO 53 & $ 6.8 \pm  0.1$ & $13.9 \pm  1.1$ & $1.01 \pm 0.10$ & $0.49 \pm 0.03$ \\
        DDO 190 & $ 6.8 \pm  0.4$ & $12.4 \pm  0.9$ & $0.53 \pm 0.19$ & $0.55 \pm 0.01$ \\
        DDO 125 & $ 4.2 \pm  0.3$ & $ 9.3 \pm  0.9$ & $0.41 \pm 0.11$ & $0.45 \pm 0.01$ \\
      Sextans B & $ 4.5 \pm  0.1$ & $ 9.3 \pm  0.2$ & $0.22 \pm 0.03$ & $0.48 \pm 0.01$ \\
         DDO 99 & $ 4.7 \pm  0.4$ & $10.1 \pm  0.7$ & $0.31 \pm 0.08$ & $0.47 \pm 0.01$ \\
        M81 DwB & $ 7.7 \pm  0.3$ & $15.3 \pm  3.1$ & $1.05 \pm 0.06$ & $0.51 \pm 0.06$ \\
       UGCA 292 & $ 6.6 \pm  0.5$ & $10.8 \pm  1.6$ & $0.84 \pm 0.69$ & $0.61 \pm 0.03$ \\
       NGC 4163 & $ 6.6 \pm  0.2$ & $13.3 \pm  3.7$ & $1.32 \pm 0.01$ & $0.50 \pm 0.10$ \\
       UGC 4483 & $ 7.0 \pm  0.2$ & $14.0 \pm  1.4$ & $0.98 \pm 0.12$ & $0.50 \pm 0.03$ \\
        DDO 181 & $ 5.0 \pm  0.7$ & $10.3 \pm  2.0$ & $0.60 \pm 0.42$ & $0.48 \pm 0.01$ \\
       UGC 8833 & $ 6.7 \pm  0.2$ & $13.2 \pm  1.5$ & $1.01 \pm 0.11$ & $0.50 \pm 0.05$ \\
        DDO 187 & $ 5.3 \pm  0.3$ & $13.4 \pm  0.7$ & $0.15 \pm 0.03$ & $0.40 \pm 0.01$ \\
           GR 8 & $ 5.3 \pm  0.6$ & $11.1 \pm  1.6$ & $0.44 \pm 0.19$ & $0.48 \pm 0.02$ \\
\enddata
\tablecomments{All double Gaussian parameter uncertainties calculated from noise uncertainties only.(1) Galaxy  name. (2) Width of narrow Gaussian component. (3) Width of wide Gaussian component. (4) Ratio of area of narrow Gaussian component to broad Gaussian component. (5) Ratio of narrow width to broad width. }
\end{deluxetable}

%% file: table-correlations.tex
\begin{deluxetable}{lcc|cc|cc|cc}
\tablewidth{0pt}
\tablecaption{Spearman Correlation Coefficients \label{tab:correlations}}
\tablehead{
  \colhead{} &
  \multicolumn{2}{c}{ \scentral{}} &
  \multicolumn{2}{c}{ \swing{} } &
  \multicolumn{2}{c}{ \fw{} } &
  \multicolumn{2}{c}{ \aw{} } \\

  \colhead{Property} &
  \colhead{$r_S$} &
  \colhead{$p_S$} &
  \colhead{$r_S$} &
  \colhead{$p_S$} &
  \colhead{$r_S$} &
  \colhead{$p_S$} &
  \colhead{$r_S$} &
  \colhead{$p_S$} \\

}
\startdata
  $w_{20}$  &  0.079  &  0.720  &  0.261  &  0.229  &  0.262  &  0.227  &  -0.459  &  0.027  \\  
M$_\mathrm{baryon,tot}$  &  -0.189  &  0.388  &  0.083  &  0.707  &  0.491  &  0.017  &  \textbf{-0.769}  &  \textbf{$<$0.001}  \\  
 M$_\star$  &  -0.150  &  0.494  &  0.085  &  0.700  &  0.414  &  0.050  &  \textbf{-0.655}  &  \textbf{$<$0.001}  \\  
M$_\mathrm{HI}$  &  -0.152  &  0.488  &  0.047  &  0.830  &  0.377  &  0.076  &  \textbf{-0.643}  &  \textbf{$<$0.001}  \\  
       SFR  &  -0.071  &  0.747  &  0.221  &  0.310  &  0.486  &  0.019  &  \textbf{-0.721}  &  \textbf{$<$0.001}  \\  
SFR / M$_\mathrm{HI}$  &  0.066  &  0.764  &  0.390  &  0.066  &  \textbf{0.529}  &  \textbf{0.010}  &  -0.179  &  0.414  \\  
$\langle \sfrsd{} \rangle$  &  0.297  &  0.168  &  \textbf{0.626}  &  \textbf{0.001}  &  0.314  &  0.144  &  0.000  &  1.000  \\  
$\langle \Sigma_\mathrm{HI} \rangle$  &  \textbf{0.536}  &  \textbf{0.008}  &  \textbf{0.707}  &  \textbf{$<$0.001}  &  -0.053  &  0.809  &  0.200  &  0.361  \\  
$\langle \Sigma_\mathrm{baryon} \rangle$  &  0.336  &  0.117  &  \textbf{0.623}  &  \textbf{0.002}  &  0.123  &  0.578  &  0.115  &  0.603  \\  
         i  &  -0.218  &  0.318  &  -0.311  &  0.149  &  0.122  &  0.579  &  -0.278  &  0.199  \\  
\enddata
\tablecomments{Spearman correlation coefficient $r_s$ and probability $p_s$ between superprofile parameters and physical properties. Significant correlations (i.e., $p_S \leq 0.01$) are shown in bold.}
\end{deluxetable}

%% file: table-measurements-extend_400.tex
\begin{deluxetable}{lcccccr}
\tablewidth{0pt}
\tablecaption{Measured Superprofile Parameters for More Massive Galaxies \label{tab:measurements-extend400}}
\tablehead{
  \colhead{Galaxy} &
  \colhead{$\langle \sigma_{\mathrm{m}_2} \rangle$} &
  \colhead{$\sigma_{\mathrm{central}}$} &
  \colhead{$\sigma_{\mathrm{wings}}$} &
  \colhead{$f_w$} &
  \colhead{$a$} &
  \colhead{N$_\mathrm{beams}$} \\

  \colhead{} &
  \colhead{(km s$^{-1}$)} &
  \colhead{(km s$^{-1}$)} &
  \colhead{} &
  \colhead{} &
  \colhead{} \\

  \colhead{1} &
  \colhead{2} &
  \colhead{3} &
  \colhead{4} &
  \colhead{5} &
  \colhead{6} &
  \colhead{7}
}
\startdata
  NGC 5055 & 13.6 &  $ 8.8 $ &$34.3 $ &$0.18 $ &$0.05 $ &3616 \\
  NGC 2903 & 16.6 &  $ 8.8 $ &$39.5 $ &$0.21 $ &$0.06 $ &2186 \\
  NGC 5236 & 10.1 &  $ 8.8 $ &$29.0 $ &$0.12 $ &$0.18 $ & 447 \\
  NGC 3351 & 13.5 &  $ 6.6 $ &$29.0 $ &$0.17 $ &$0.05 $ & 607 \\
  NGC 4736 & 14.8 &  $ 8.6 $ &$40.6 $ &$0.20 $ &$0.04 $ & 725 \\
   NGC 628 &  7.5 &  $ 6.5 $ &$18.3 $ &$0.08 $ &$0.12 $ &2819 \\
  NGC 2403 & 10.0 &  $ 8.3 $ &$23.0 $ &$0.09 $ &$0.05 $ &2253 \\
  NGC 2976 & 14.5 &  $12.3 $ &$36.0 $ &$0.10 $ &$0.18 $ & 110 \\
\enddata
\tablecomments{Measured superprofile parameters for higher-mass THINGS spirals. Columns are the same as Table~\ref{tab:measurements}.}
\end{deluxetable}

%% file: table-scale-height.tex
\begin{deluxetable}{lc}
\tablewidth{0pt}
\tablecaption{Implied Scale Heights \label{tab:scale-height}}
\tablehead{
  \colhead{Galaxy} &
  \colhead{$\langle h_z \rangle$} \\

  \colhead{} &
  \colhead{(pc)} \\

  \colhead{1} &
  \colhead{2}
}
\startdata
  NGC 7793 & $   78 \pm   43 $ \\ 
   IC 2574 & $  336 \pm   87 $ \\ 
  NGC 4214 & $  228 \pm  103 $ \\ 
     Ho II & $  302 \pm   75 $ \\ 
  NGC 2366 & $  479 \pm   93 $ \\ 
   DDO 154 & $  708 \pm  139 $ \\ 
      Ho I & $  191 \pm   41 $ \\ 
  NGC 4190 & $  130 \pm   65 $ \\ 
  NGC 3741 & $  561 \pm  123 $ \\ 
 Sextans A & $  438 \pm   84 $ \\ 
    DDO 53 & $  259 \pm   51 $ \\ 
   DDO 190 & $  214 \pm   65 $ \\ 
   DDO 125 & $  329 \pm  135 $ \\ 
 Sextans B & $  639 \pm  224 $ \\ 
    DDO 99 & $  617 \pm  161 $ \\ 
   M81 DwB & $  140 \pm   65 $ \\ 
  UGCA 292 & $  172 \pm   29 $ \\ 
  NGC 4163 & $  120 \pm   61 $ \\ 
  UGC 4483 & $  322 \pm   67 $ \\ 
   DDO 181 & $  346 \pm  104 $ \\ 
  UGC 8833 & $  313 \pm   76 $ \\ 
   DDO 187 & $  514 \pm  108 $ \\ 
      GR 8 & $  344 \pm  100 $ \\ 
\enddata
\tablecomments{Implied scale heights for sample galaxies. (1) Galaxy name. (2) Average scale height implied by \scentral{} and \ave{\hisd{}}.}
\end{deluxetable}